\documentclass[reqno,11pt]{amsart}
\usepackage[linktocpage]{hyperref}
\usepackage[utf8]{inputenc}
\usepackage{graphicx}
\usepackage{slashed}
\usepackage{amscd}
\usepackage{amssymb}
\usepackage{ifsym}
\usepackage{esint}
\usepackage{enumitem}
\usepackage[mathscr]{eucal}
\usepackage{pstricks}
\usepackage{pst-all}

\numberwithin{equation}{section}
\allowdisplaybreaks[1]
\definecolor{labelkey}{gray}{.65}

\title[The Linearized Field Equations in Minkowski Space]
{Solving the Linearized Field Equations of the Causal Action Principle in Minkowski Space}

\author[F.\ Finster]{Felix Finster \\ \\ April 2023}
\address{Fakult\"at f\"ur Mathematik \\ Universit\"at Regensburg \\ D-93040 Regensburg}
\email{finster@ur.de}

\newtheorem{Def}{Definition}[section]
\newtheorem{Thm}[Def]{Theorem}
\newtheorem{Prp}[Def]{Proposition}
\newtheorem{Lemma}[Def]{Lemma}
\newtheorem{Remark}[Def]{Remark}
\newtheorem{Corollary}[Def]{Corollary}
\newtheorem{Example}[Def]{Example}

\newcommand{\Thanks}{\vspace*{.5em} \noindent \thanks}
\newcommand{\beq}{\begin{equation}}
\newcommand{\eeq}{\end{equation}}
\newcommand{\Proof}{\begin{proof}}
	\newcommand{\QED}{\end{proof} \noindent}
\newcommand{\QEDrem}{\ \hfill $\Diamond$}

\newcommand{\la}{\langle}
\newcommand{\ra}{\rangle}

\newcommand{\Sl}{\mbox{$\prec \!\!$ \nolinebreak}}
\newcommand{\Sr}{\mbox{\nolinebreak $\succ$}}

\newcommand{\C}{\mathbb{C}}
\newcommand{\R}{\mathbb{R}}
\newcommand{\1}{\mbox{\rm 1 \hspace{-1.05 em} 1}}
\newcommand{\Z}{\mathbb{Z}}
\newcommand{\N}{\mathbb{N}}
\newcommand{\Pdd}{\mbox{$\partial$ \hspace{-1.2 em} $/$}}

\renewcommand{\Tr}{\text{\rm{Tr}}}
\DeclareMathOperator{\tr}{tr}
\renewcommand{\O}{{\mathscr{O}}}
\renewcommand{\L}{{\mathcal{L}}}
\newcommand{\Sact}{{\mathcal{S}}}

\newcommand\B{{\mathscr{B}}}

\renewcommand{\H}{\mathscr{H}}

\newcommand{\Lin}{\text{\rm{L}}}
\newcommand{\F}{{\mathscr{F}}}

\DeclareMathOperator{\Symm}{\mbox{\rm{Symm}}}

\DeclareMathOperator{\re}{Re}

\DeclareMathOperator{\supp}{supp}

\newcommand{\scrM}{\mycal M}

\newcommand{\J}{\mathfrak{J}}
\newcommand{\s}{\mathfrak{s}}
\newcommand{\fermi}{{\mathrm{f}}}

\newcommand{\Jvary}{\mathfrak{J}^\text{\rm{\tiny{vary}}}}
\newcommand{\Jret}{\mathfrak{J}^\text{\rm{\tiny{ret}}}}
\newcommand{\Jnull}{\mathfrak{J}^\text{\rm{\tiny{null}}}}

\newcommand{\Jlin}{\mathfrak{J}^\text{\rm{\tiny{lin}}}}

\newcommand{\bu}{{\mathbf{u}}}
\newcommand{\bv}{\mathbf{v}}
\newcommand{\bw}{\mathbf{w}}
\newcommand{\bitem}{\begin{itemize}[leftmargin=2.5em]}
\newcommand{\eitem}{\end{itemize}}
\renewcommand{\div}{{\rm{div}}\,}

\newcommand{\id}{\text{id}}

\newcommand{\h}{\mathfrak{h}}
\newcommand{\scrP}{{\mathscr{P}}}
\newcommand{\scrG}{{\mathscr{G}}}
\newcommand{\scrL}{{\mathscr{L}}}
\newcommand{\scrC}{{\mathscr{C}}}
\newcommand{\scrH}{{\mycal{H}}}

\DeclareFontFamily{OT1}{rsfso}{}
\DeclareFontShape{OT1}{rsfso}{m}{n}{ <-7> rsfso5 <7-10> rsfso7 <10-> rsfso10}{}
\DeclareMathAlphabet{\mycal}{OT1}{rsfso}{m}{n}

\newcommand\Felix[1]{}

\begin{document}
\maketitle

\begin{abstract}
The linearized field equations for causal fermion systems 
in Minkowski space are analyzed systematically using methods of
functional analysis and Fourier analysis. Taking into account a direction-dependent local phase freedom,
we find a multitude of homogeneous solutions.
The time evolution of the inhomogeneous equations is studied.
It leads to the dynamical creation of retarded solutions as well as to the generation of
non-propagating perturbations.
\end{abstract}

\tableofcontents

\section{Introduction} \label{secintro}
The theory of {\em{causal fermion systems}} is a recent approach to fundamental physics
(see the basics in Section~\ref{secprelim}, the reviews~\cite{dice2014, review}, the textbooks~\cite{cfs, intro}
or the website~\cite{cfsweblink}).
In this approach, spacetime and all objects therein are described by a measure~$\rho$
on a set~$\F$ of linear operators on a Hilbert space~$(\H, \la .|. \ra_\H)$. 
The physical equations are formulated by means of the so-called {\em{causal action principle}},
a nonlinear variational principle where an action~$\Sact$ is minimized under variations of the measure~$\rho$.
A minimizing measure satisfies corresponding {\em{Euler-Lagrange (EL) equations}},
being nonlinear equations for~$\rho$.
The {\em{linearized field equations}} describe first variations of a minimizing measure
which preserve the EL equations. Similar to usual linearizations of physical equations
(like for example the equations of linearized gravity),
the linearized field equations describe the dynamics of small perturbations of a causal fermion system.
The goal of the present paper is to develop a systematic procedure for solving the
linearized field equations for causal fermion systems in Minkowski space.

The linearized field equations play a central role in the analysis of causal fermion systems,
both conceptually and computationally.
From the conceptual point of view, the analysis of the linearized field equations reveals
the causal nature of the dynamics and thereby clarifies the causal structure of spacetime itself.
From the computational point of view, being a linear equation, it becomes possible to analyze
the equations explicitly using methods of functional analysis and Fourier analysis.
Moreover, the linearized field equations are an important first step toward the analysis of
the nonlinear dynamics as described by the EL equations.

In order to put our results into context, we first note that, even before the name ``linearized field
equations'' was phrased, the computations in~\cite{pfp} and~\cite{cfs} implicitly involved
an analysis of the linearized field equations. Indeed, the methods to be developed in the present paper
are closely connected to and follow up on these computations.
The systematic study of the linearized field equations was initiated in~\cite{jet}, where these equations
were derived in the setting of causal variational principles, a mathematical generalization of
the causal action principle. In~\cite{linhyp} the analytic foundations were established
by showing that methods of hyperbolic partial differential equations can be adapted
to the linearized field equations such as to prove that the Cauchy problem is well-posed
and that solutions propagate with finite speed. This analysis was extended and refined in~\cite{dirac, localize}.
These studies clarified the analytic structure of the linearized field equations from an abstract point of view.
However, a systematic constructive method for solving these equations was still lacking.
The purpose of the present paper is to fill this gap. Combined with the perturbative
treatment in~\cite{perturb}, our methods and results also open the door for a detailed and systematic study of
nonlinear effects of the causal action principle.

Our setup is as follows. The underlying spacetime is Minkowski space~$M = \R^{1,3}$.
In this spacetime, we consider a family of four-component spinorial wave functions.
Identifying all spinor spaces, these wave functions take values in a four-dimensional
complex vector space~$V$, which we refer to as the {\em{spinor space}}. The spinor space is
endowed with an inner product of signature~$(2,2)$, which we denote
by~$\Sl .|. \Sr$ (using common notation in physics, this inner product 
can also be written as~$\Sl \psi | \phi \Sr = \overline{\psi} \phi$ with
the usual adjoint spinor~$\overline{\psi} := \psi^\dagger \gamma^0$).
Then the causal fermion system can be described by the
\[ \text{\em{wave evaluation operator}} \qquad \Psi : \H \rightarrow C^0(M, V) \:, \]
which to every vector~$u$ in an abstractly given Hilbert space~$\H$ associates the corresponding
{\em{physical wave function}}~$\psi^u := \Psi u$, being a continuous
four-component wave function in Minkowski space (for the general setup see Section~\ref{secweo}).
The EL equations can be written as (see Section~\ref{secELres})
\beq \label{ELintro}
Q \Psi = \mathfrak{r}\, \Psi \:,
\eeq
where~$\mathfrak{r}$ is a real parameter and~$Q$ is an integral operator 
whose integral kernel is denoted by~$Q(x,y)$, i.e.\
\beq \label{Qint}
(Q \psi)(x) = \int_M Q(x,y)\, \psi(y)\: d^4y \:.
\eeq
The integral kernel~$Q(x,y)$ and consequently also the corresponding integral operator~$Q$
depend nonlinearly on~$\Psi$; we write symbolically~$Q=Q(\Psi)$.
Linearizing the EL equations~\eqref{ELintro} gives the
\beq \label{lfeintro}
\text {\em{linearized field equations}} \qquad
Q\, \delta \Psi - \mathfrak{r}\, \delta \Psi + DQ|_{\Psi}(\delta \Psi) \,\Psi = \Xi \:,
\eeq
where~$DQ|_\Psi$ is the total derivative of~$Q$
(which is real-linear in~$\delta \Psi$, but not complex-linear;
for details see Section~\ref{seclinwave}).
Here~$\Xi : \H \rightarrow C^0(M, V)$ is an inhomogeneity,
which typically describes perturbations of small deviations in the EL equations~\eqref{ELintro}.
Multiplying the linearized field equations from the right by a vector~$u \in \H$, one
obtains an equation for the variation of each physical wave function~$\delta \psi^u$,
\beq \label{lfedirac}
Q\, \delta \psi^u - \mathfrak{r}\, \delta \psi^u + DQ|_{\Psi}(\delta \Psi) \,\psi^u = \Xi u \:.
\eeq
All these equations
are coupled together by the first variation of the kernel~$\delta Q = DQ|_{\Psi}(\delta \Psi)$.

From the physical perspective, the linearized field equations~\eqref{lfeintro} give a unified
description of the dynamics of all physical wave functions and of all bosonic fields
(for details see Section~\ref{secunified}).
The connection to the fermionic wave functions is made by multiplying
the EL equations~\eqref{ELintro} by a vector~$u \in \H$ to obtain
\[ (Q - \mathfrak{r})\, \psi^u = 0 \:. \]
This equation can be regarded as the analog of the Dirac equation for the wave function~$\psi^u$.
Likewise, the linearized field equation~\eqref{lfedirac} can be regarded as the equation
for the first order perturbation~$\delta \psi^u$, where~$DQ|_{\Psi}(\delta \Psi)$ takes the role of an interaction
potential. In order to get the connection to bosonic fields in the simplest possible setting, let us assume that
the unperturbed system is built up of solutions of the vacuum Dirac equation\footnote{For clarity of presentation, in the introduction we consider only one generation of elementary particles. Later on, we shall consider three generations.}
\[ 
(i \Pdd_x - m)\, \Psi(x) = 0 \]
(for the notation and the general setup see Section~\ref{secfirstorder} or the textbooks~\cite{bjorken,
peskin+schroeder, weinberg}).
Introducing an electromagnetic potential~$A$, this Dirac equation is modified to
\[ (i \Pdd_x + \slashed{A} - m)\, \Psi(x) = 0 \:, \]
giving rise to a first variation of the wave evaluation operator given by
\beq \label{delPsiintro}
\delta \Psi = -s_m\, \slashed{A}\, \Psi
\eeq
(where~$s_m$ denotes a Dirac Green's operator defined by the relation~$(i \Pdd-m) s_m = \1$; for example
one can choose the retarded Green's operator~\eqref{smdef}). We take the point of view that the
potential~$A$ merely is a device for describing the variation~$\delta \Psi$ in a convenient way.
In a more graphic language, classical potentials are a useful tool for describing collective
variations of all the physical wave functions. The linearized field equations~\eqref{lfeintro}
determine the admissible variations~$\delta \Psi$ directly, without using bosonic potentials.
Nevertheless, describing a resulting variation in terms of a bosonic potential, the
linearized field equations determine the
potential~$A$, thereby taking the role of Maxwell's equations.

Before stating our results, we make a few general comments on the mathematical structure of the linearized
field equations. As one sees immediately from~\eqref{lfeintro} and~\eqref{Qint},
the linearized field equations are {\em{integral equations}} and are therefore {\em{nonlocal}}.
In a limiting case studied in detail in~\cite{cfs}, the so-called {\em{continuum limit}},
the linearized field equations give rise to hyperbolic partial differential equations
(like the Dirac and Maxwell equations). Nevertheless,
as already observed in~\cite[Sections~3.7 and~3.10]{cfs}, even in this limiting case
the linearized field equations also involve genuinely nonlocal effects.
In the present paper, we study this phenomenon systematically
by analyzing the linearized field equations as an integral equation with a specific integral kernel.

We now summarize our constructions and results.
We assume that the causal fermion system of the vacuum is described by a
wave evaluation operator~$\Psi$ composed of all negative-energy solutions of the Dirac equation
with an ultraviolet regularization on the scale~$\varepsilon$ (for details see Section~\ref{secfirstorder}). The interacting system is described by a wave evaluation operator~$\tilde{\Psi}$
obtained by perturbing~$\Psi$ linearly, i.e.\
\[ \tilde{\Psi} = \Psi + \delta \Psi \:. \]
For the computations, it is preferable not to work directly with the wave evaluation operator, but
instead with the
\beq \label{Ptintro}
\text{\em{kernel of the fermionic projector}} \qquad \tilde{P}(x,y) := -\tilde{\Psi}(x)\, \tilde{\Psi}(y)^* \::\: V \rightarrow V
\eeq
(the connections between~$\tilde{\Psi}$ and~$\tilde{P}(x,y)$ are explained in detail in Sections~\ref{secELres},
\ref{seclinwave} and~\ref{sechom}). The fact that the vacuum configuration in Minkowski space is homogeneous
means that the vacuum kernel of the fermionic projector depends only on the difference vector~$\xi := y-x$,
making it possible to work with the Fourier representation
\beq \label{Pvacintro}
P(x,y) = \int_{\hat{M}} \frac{d^4k}{(2 \pi)^4}\: \hat{P}(k)\: e^{i k \xi}
\eeq
(here and in what follows, the hat refers to momentum space or to the Fourier transform;
moreover, $k \xi$ denotes the Minkowski inner product).
Moreover, using that also the linearized field operator on the left of~\eqref{lfeintro} is translation invariant,
in the linear perturbation~$\delta P(x,y)$ we may separate the dependence
on the variable~$(y+x)/2$ with the plane wave ansatz
\beq \label{delPintro}
\delta P(x,y) = e^{-i\frac{q}{2}\, (x+y)} \:\delta P(\xi) \qquad \text{with} \qquad
\xi := y-x \:,
\eeq
where~$q \in \hat{M}$ is a given momentum vector and~$\delta P(\xi)$
is a function of one variable~$\xi \in M$.
Employing this ansatz in the linearized field equations, we obtain separate equations for each~$q$.
With this in mind, we can fix~$q$ throughout our analysis. Clearly, we need to treat all three cases
that~$q$ is spacelike, timelike and lightlike (i.e.\ $q^2<0$, $>0$ and~$=0$,
where~$q^2 \equiv q^k q_k$ again denotes the Minkowski inner product).

In order to illustrate the ansatz~\eqref{delPintro}, we note that {\em{specific}}
perturbations of this form are obtained by considering an electromagnetic
wave of momenta~$\pm q$,
\beq \label{Agauge}
A(x) = \hat{A}(q) \: e^{-i q x} + \overline{\hat{A}(q)}\: e^{i q x}
\eeq
and perturbing according to~\eqref{delPsiintro} (for details see~\eqref{tilmuone}, \eqref{nupert}
and~\eqref{nupert2}). Such perturbations have been analyzed in detail in~\cite{firstorder, pfp, cfs}.
The goal of the present paper is to study the linearized field equations
in full generality without imposing a specific form of~$\delta P(\xi)$.
In order to understand the analytic structure of the linearized field equations,
it is important to note that, in generalization of the local gauge freedom of classical electrodynamics,
the linearized field equations are invariant under {\em{direction-dependent local phase transformations}}
\beq \label{dirlocintro}
P(x,y) \rightarrow e^{i \Lambda(x,y)}\: P(x,y)
\eeq
with a real-valued function~$\Lambda(x,y)$ which is anti-symmetric, i.e.\
\beq \label{Lsymm}
\Lambda(x,y) = -\Lambda(y,x) \qquad \text{for all~$x,y \in M$}\:.
\eeq
Therefore, in the analysis of the linearized field equations
such direction-dependent local phases must be modded out.
Another important feature of the linearized field equations is that they are of {\em{variational form}},
meaning that they are recovered as the Euler-Lagrange equations obtained by minimizing
a suitable effective action~$\Sact^\text{\rm{eff}}$. The positivity of the second variations
makes it possible to proceed hie\-rar\-chically and to solve the equations iteratively in an expansion
in powers of the regularization length.
The first set of equations in this expansion are obtained by evaluating in the so-called
{\em{continuum limit}}. In this limiting case, the perturbation~$\delta P(\xi)$ in~\eqref{delPintro}
enters the equations only if the vector~$\xi$ lies on the light cone and is not too small; i.e., more precisely, if
\beq \label{evallc}
\xi^2 = 0 \qquad \text{and} \qquad \big| \xi^0 \big| \gtrsim \ell_{\min}
\eeq
for a suitable parameter~$\ell_{\min}$ which lies between the regularization scale~$\varepsilon$ and the
length scale of macroscopic physics~$\ell_\text{macro}$,
\beq \label{ellmindef}
\varepsilon \ll \ell_{\min} \ll \ell_\text{macro}
\eeq
(for more details on the origin and what is known about the
parameter~$\ell_{\min}$ see Remark~\ref{remell} below).
Our general procedure for solving the linearized field equations in this limiting case
consists of several steps. First, we identify a space of admissible variations~$\delta \Psi$
of the wave evaluation operator which describe a {\em{retarded}} time evolution.
This space of variations is denoted by~$\Jret$. Given~$\bv \in \Jret$, the
corresponding variation of the kernel of the fermionic projector is denoted by~$\delta P =
\scrP \bv$ (see Definition~\ref{defJvary}). In order to treat the direction-dependent phase freedom~\eqref{dirlocintro},
we compensate for the direction-dependent phases by a variation~$\delta P = \scrG \bv$
which preserves the linearized field equations (see Definition~\ref{defscrG}).\
By inserting the difference of the variations~$(\scrP-\scrG)\bv$
into the linearized field equations, we obtain equations for the first variation~$\bv \in \Jret$.
These equations are integral equations, which can be solved abstractly
with functional analytic methods in Hilbert spaces.
For a more explicit analysis, it is useful to transform these equations to momentum space.
This is of advantage because the fact that~$\delta P(\xi)$ is evaluated only on the
light cone~\eqref{evallc} means in momentum space that~$\delta \hat{P}(p)$ is harmonic
(i.e., it is a solution of the scalar wave equation~$\Box_p \delta \hat{P}(p)=0$, where~$p$
is the corresponding momentum variable).
Thus we can make use of the well-posedness of the Cauchy problem,
finite propagation speed and the strong Huygens principle for waves in momentum space.
Combining these properties of harmonic functions, we find that the linearized field equations can be
solved in an expansion in momentum space, the so-called {\em{mass cone expansion}}.
From the mathematical point of view, the mass cone expansion is very similar to the
so-called light-cone expansion (as developed in~\cite{firstorder, light}).
The difference between the light cone and the mass expansions is that the roles of position
and momentum spaces are reversed. In simple terms, using analogies between the ``light cone''
and the ``mass cone'' or the ``mass shell,'' the methods developed in position space
for analyzing the behavior of distributions near the light cone can also be used in momentum
space for analyzing solutions of the linearized field equations near the mass cone.

Our analysis reveals that the solutions of the linearized field equations have a surprisingly
rich structure. It turns out that the homogeneous equations admit not one solution
(like a plane electromagnetic wave~\eqref{Agauge}), but instead a multitude of solutions,
where the number~$N$ of homogeneous fields is very large and scales like
(see~\eqref{Nscale} in Section~\ref{seciterate})
\beq \label{Nscaleintro}
N \simeq \frac{\ell_{\min}}{\varepsilon}
\eeq
(with~$\ell_{\min}$ as in~\eqref{evallc} and~\eqref{ellmindef}; note that the number of
solutions becomes infinite if the regularization length~$\varepsilon$ tends to zero).
In short, these homogeneous solutions are obtained as follows (for details see Theorem~\ref{thmhom}).
We first multiply the kernel~$\hat{P}$ in~\eqref{Pvacintro} by a cutoff function~$\eta$
which is supported in a cone-shaped subset of the lower mass shell 
of opening angle~$\vartheta$ (see the left of Figure~\ref{fighom} on page~\pageref{fighom};
the parameter~$\vartheta$ is given by~\eqref{varthetadef} with~$\omega_{\min}$ chosen arbitrarily
in the range~\eqref{omegaminrange}). Denoting the resulting kernel by
\beq \label{Peta}
P_\eta(\xi) := \int_{\hat{M}} \frac{d^4k}{(2 \pi)^4}\:\eta(k)\: \hat{P}(k)\: e^{i k \xi} \:,
\eeq
a first ansatz for~$\delta P$ is obtained by perturbing~$P_\eta$ with a classical electromagnetic
potential of the form~\eqref{Agauge}; i.e., similar to~\eqref{delPsiintro},
\beq \label{delPansatz}
\delta P := - s_m \,\slashed{A} \,P_\eta - P_\eta \,\slashed{A} \,s^*_m
\eeq
(the fact that the Green's operators act from both sides can be understood from~\eqref{Ptintro};
for more details see~\eqref{delPmink} and~\eqref{nupert}, \eqref{nupert2}).
If~$A$ is chosen as an arbitrary electromagnetic wave, this ansatz is an approximate solution to the
linearized field equations. Taking it as the starting point for an iteration scheme, we obtain an
exact solution of the linearized field equations.

The ansatz~\eqref{Peta} and~\eqref{delPansatz} can be understood intuitively as follows.
The function~$P_\eta(\xi)$ is a wave packet in the difference vector~$\xi=y-x$, propagating
with the speed of light in an arbitrary spatial direction, being localized near~$\xi=0$
on the scale~$\ell_{\min}$ (see the left of Figure~\ref{fighom}).
With~\eqref{delPansatz} this wave packet is perturbed by an electromagnetic potential.
This means that only the one-particle states forming this wave packet (i.e.\ the one-particle states inside the dark
shaded conical region on the left of Figure~\ref{fighom}) couple to the electromagnetic potential.
The one-particle states inside other conical regions, however, may couple to other electromagnetic potentials.
Due to the superposition principle, a one-particle state may be supported
many of such conical regions, in which case it couples to a superposition of the corresponding electromagnetic potentials.
In this way, the homogeneous fields can be thought of as being formed of a plethora of
electromagnetic fields coupling to different wave packets propagating in different directions,
each of them localized on the scale~$\ell_{\min}$.
The coupling of these homogeneous fields to the low-energy states (as indicated by the light
shaded region on the left of Figure~\ref{fighom}) is described by the perturbations
generated in the iteration scheme.

When solving the {\em{in}}homogeneous equations, these homogeneous solutions
are generated in a retarded way (meaning that they propagate to the future).
Typically, the most singular contribution of the resulting homogeneous solution on the light cone
describes a direction-dependent phase transformation~\eqref{dirlocintro}.
The additional perturbations obtained in the above-mentioned iteration scheme are 
less singular on the light cone. Moreover, they are {\em{non-propagating}} in the sense that they
do not have a dynamics on their own. Instead, they are
localized on the scale~$\ell_{\min}$ near the inhomogeneity and the retarded fields.

The fact that we get a multitude of solutions has far-reaching consequences for the
dynamics of the system. This will be worked out in detail in future works.
In particular, the interplay of all these fields gives rise to a quantum dynamics~\cite{fockdynamics} 
including corrections~\cite{qftcorrect} and
can explain the stochastic term in collapse theories~\cite{collapse}.

The paper is organized as follows. After the necessary preliminaries on
causal fermion systems (Section~\ref{secprelim}), we give a self-contained introduction
to the linearized field equations (Section~\ref{secintrolin}).
In contrast to the presentation in the context of causal variational principles,
we here work with the specific structures of causal fermion systems.
We also emphasize the variational structure of the equations, which will be important later on.
In Section~\ref{seclinmink} the setting is specialized to Dirac systems in Minkowski space.
We also summarize how the linearized field equations are evaluated in the continuum limit.
In Section~\ref{secabstract} a systematic procedure for solving the linearized field equations
is developed. Combining the variational structure of the linearized field equations with the positivity
of second variations, we can proceed hierarchically and solve the equations iteratively in an expansion
in powers of the regularization length.
In Section~\ref{secdirgauge} the roles of local gauge transformations and
direction-dependent local phase transformations~\eqref{dirlocintro} are worked out.
In Section~\ref{secret} it is explained how a retarded time evolution can be built in by
specifying the pole structure of~$\delta P$ in momentum space.
Based on these results, in Section~\ref{seccl} it is explained how the linearized field equations
can be solved abstractly in the continuum limit in position space.
In Section~\ref{secmom} we proceed by transforming the equations to momentum space
using the mass cone expansion.
In Section~\ref{seciterate} the formulation in momentum space is used
in order to construct solutions of the homogeneous linearized field equations.
In Section~\ref{secbeyond} it is explained how, following the abstract procedure
introduced in Section~\ref{secabstract}, one can analyze the linearized field equations
beyond the continuum limit.
Finally, in the appendices we provide additional material and develop the technical tools needed for the
mass cone expansion.

\section{Preliminaries} \label{secprelim}
This section provides the necessary abstract background on causal fermion systems.
\subsection{Causal Fermion Systems and the Reduced Causal Action Principle}
We now recall the basic setup and introduce the main objects to be used later on.
\begin{Def} \label{defcfs} (causal fermion systems) {\em{ 
Given a separable complex Hilbert space~$\H$ with scalar product~$\la .|. \ra_\H$
and a parameter~$n \in \N$ (the {\em{``spin dimension''}}), we let~$\F \subset \Lin(\H)$ be the set of all
symmetric operators on~$\H$ of finite rank, which (counting multiplicities) have
at most~$n$ positive and at most~$n$ negative eigenvalues. On~$\F$ we are given
a positive measure~$\rho$ (defined on a $\sigma$-algebra of subsets of~$\F$).
We refer to~$(\H, \F, \rho)$ as a {\em{causal fermion system}}.
}}
\end{Def} \noindent

A causal fermion system describes a spacetime together
with all structures and objects therein.
In order to single out the physically admissible
causal fermion systems, one must formulate physical equations. To this end, we impose that
the measure~$\rho$ should be a minimizer of the causal action principle.
which we now introduce. For brevity of the presentation, we only consider the
{\em{reduced causal action principle}} where the so-called boundedness constraint has been built
incorporated by a Lagrange multiplier term. This simplification is no loss of generality, because
the resulting EL equations are the same as for the non-reduced action principle
as introduced for example~\cite[Section~\S1.1.1]{cfs}.

For any~$x, y \in \F$, the product~$x y$ is an operator of rank at most~$2n$. 
However, in general it is no longer a symmetric operator because~$(xy)^* = yx$,
and this is different from~$xy$ unless~$x$ and~$y$ commute.
As a consequence, the eigenvalues of the operator~$xy$ are in general complex.
We denote these eigenvalues counting algebraic multiplicities
by~$\lambda^{xy}_1, \ldots, \lambda^{xy}_{2n} \in \C$
(more specifically,
denoting the rank of~$xy$ by~$k \leq 2n$, we choose~$\lambda^{xy}_1, \ldots, \lambda^{xy}_{k}$ as all
the non-zero eigenvalues and set~$\lambda^{xy}_{k+1}, \ldots, \lambda^{xy}_{2n}=0$).
Given a parameter~$\kappa>0$ (which will be kept fixed throughout this paper),
we introduce the $\kappa$-Lagrangian and the causal action by
\begin{align}
\text{\em{$\kappa$-Lagrangian:}} && \L(x,y) &= 
\frac{1}{4n} \sum_{i,j=1}^{2n} \Big( \big|\lambda^{xy}_i \big|
- \big|\lambda^{xy}_j \big| \Big)^2 + \kappa\: \bigg( \sum_{j=1}^{2n} \big|\lambda^{xy}_j \big| \bigg)^2 \label{Lagrange} \\
\text{\em{causal action:}} && \Sact(\rho) &= \iint_{\F \times \F} \L(x,y)\: d\rho(x)\, d\rho(y) \:. \label{Sdef}
\end{align}
The {\em{reduced causal action principle}} is to minimize~$\Sact$ by varying the measure~$\rho$
under the following constraints,
\begin{align}
\text{\em{volume constraint:}} && \rho(\F) = 1 \quad\;\; \label{volconstraint} \\
\text{\em{trace constraint:}} && \int_\F \tr(x)\: d\rho(x) = 1 \:. \label{trconstraint}
\end{align}
This variational principle is mathematically well-posed if~$\H$ is finite-dimensional.
For the existence theory and the analysis of general properties of minimizing measures
we refer to~\cite{continuum, lagrange} or~\cite[Chapter~12]{intro}.
In the existence theory one varies in the class of regular Borel measures
(with respect to the topology on~$\Lin(\H)$ induced by the operator norm),
and the minimizing measure is again in this class. With this in mind, we always assume
that~$\rho$ is a {\em{regular Borel measure}}.

\subsection{The Physical Wave Functions and the Wave Evaluation Operator} \label{secweo}
Let~$\rho$ be a {\em{minimizing}} measure. Defining {\em{spacetime}}~$M$ as the support
of this measure,
\[ 
M := \supp \rho \subset \F \:. \]
the spacetimes points are symmetric linear operators on~$\H$.
These operators contain a lot of information which, if interpreted correctly,
gives rise to spacetime structures like causal and metric structures, spinors
and interacting fields (for details see~\cite[Chapter~1]{cfs}).
Here we restrict attention to those structures needed in what follows.
We begin with a basic notion of causality.

\begin{Def} (causal structure) \label{def2}
{\em{For any~$x, y \in \F$, the product~$x y$ is an operator
of rank at most~$2n$. We denote its non-trivial eigenvalues (counting algebraic multiplicities)
by~$\lambda^{xy}_1, \ldots, \lambda^{xy}_{2n}$.
The points~$x$ and~$y$ are
called {\em{spacelike}} separated if all the~$\lambda^{xy}_j$ have the same absolute value.
They are said to be {\em{timelike}} separated if the~$\lambda^{xy}_j$ are all real and do not all 
have the same absolute value.
In all other cases (i.e.\ if the~$\lambda^{xy}_j$ are not all real and do not all 
have the same absolute value),
the points~$x$ and~$y$ are said to be {\em{lightlike}} separated. }}
\end{Def} \noindent
Restricting the causal structure of~$\F$ to~$M$, we get causal relations in spacetime.

Next, for every~$x \in \F$ we define the {\em{spin space}}~$S_xM$ by~$S_xM = x(\H)$;
it is a subspace of~$\H$ of dimension at most~$2n$.
It is endowed with the {\em{spin inner product}} $\Sl .|. \Sr_x$ defined by
\[ 
\Sl u | v \Sr_x = -\la u | x v \ra_\H \qquad \text{(for all $u,v \in S_xM$)}\:. \]
A {\em{wave function}}~$\psi$ is defined as a function
which to every~$x \in M$ associates a vector of the corresponding spin space,
\[ 
\psi \::\: M \rightarrow \H \qquad \text{with} \qquad \psi(x) \in S_xM \quad \text{for all~$x \in M$}\:. \]
A wave function~$\psi$ is said to be {\em{continuous}} at~$x$ if
for every~$\varepsilon>0$ there is~$\delta>0$ such that
\beq \label{wavecontinuous}
\big\| \sqrt{|y|} \,\psi(y) -  \sqrt{|x|}\, \psi(x) \big\|_\H < \varepsilon
\qquad \text{for all~$y \in M$ with~$\|y-x\| \leq \delta$}
\eeq
(where~$|x|$ is the absolute value of the symmetric operator~$x$ on~$\H$, and~$\sqrt{|x|}$
is the square root thereof).
Likewise, $\psi$ is said to be continuous on~$M$ if it is continuous at every~$x \in M$.
We denote the set of continuous wave functions by~$C^0(M, SM)$.

It is an important observation that every vector~$u \in \H$ of the Hilbert space gives rise to a distinguished
wave function. In order to obtain this wave function, denoted by~$\psi^u$, we simply project the vector~$u$
to the corresponding spin spaces,
\[ 
\psi^u \::\: M \rightarrow \H\:,\qquad \psi^u(x) = \pi_x u \in S_xM \:. \]
We refer to~$\psi^u$ as the {\em{physical wave function}} of~$u \in \H$.
A direct computation shows that the physical wave functions are continuous
(in the sense~\eqref{wavecontinuous}). Associating to every vector~$u \in \H$
the corresponding physical wave function gives rise to the {\em{wave evaluation operator}}
\beq \label{weo}
\Psi \::\: \H \rightarrow C^0(M, SM)\:, \qquad u \mapsto \psi^u \:.
\eeq
Every~$x \in M$ can be written as (for the derivation see~\cite[Lemma~1.1.3]{cfs})
\beq
x = - \Psi(x)^* \,\Psi(x) \label{Fid} \:.
\eeq
In words, every spacetime point operator is the {\em{local correlation operator}} of the wave evaluation operator
at this point (for details see~\cite[\S1.1.4 and Section~1.2]{cfs}).

\subsection{The Euler-Lagrange Equations}
We now state the Euler-Lagrange equations.
\begin{Prp} Let~$\rho$ be a minimizer of the reduced causal action principle.
Then the local trace is constant in spacetime, meaning that
\beq \label{loctrace}
\tr(x) = 1 \qquad \text{for all~$x \in M$} \:.
\eeq
Moreover, there are parameters~$\mathfrak{r}, \s>0$ such that
the function~$\ell$ defined by
\beq \label{elldef}
\ell \::\: \F \rightarrow \R\:,\qquad \ell(x) := \int_M \L(x,y)\: d\rho(y) - \mathfrak{r}\, \big( \tr(x) -1 \big) - \s
\eeq
is minimal and vanishes in spacetime, i.e.\
\beq \label{EL}
\ell|_M \equiv \inf_\F \ell = 0 \:.
\eeq
\end{Prp}
\Proof The proof of~\eqref{loctrace} was first given in~\cite{lagrange}; for an alternative proof
see~\cite[Proposition~1.4.1]{cfs} or~\cite[Section~6.4]{intro}).
The relation~\eqref{EL} is an extension of the EL equations derived in~\cite[Section~2]{jet}
in the setting of causal variational principles (see also or~\cite[Chapter~7]{intro}), which
for the reduced causal action principle yield minimality of~$\ell$ on the operators of fixed trace, i.e.\
(for details see~\cite[Section~6.5]{intro})
\beq \label{ELtrace}
\ell|_M \equiv \inf \big\{ \ell(x) \,\Big|\, x \in \F, \: \tr(x)= 1 \big\}  = 0 \:.
\eeq
By continuity, it suffices to derive the generalization~\eqref{EL} for an operator~$x \in \F$
with~$\tr(x) \neq 0$. We also note that, since the trace is constant, the EL equations~\eqref{EL}
still leave us the freedom to choose~$\mathfrak{r}$ arbitrarily; then~$\s$ is determined
by demanding that~$\ell$ vanishes on~$M$. Normalizing the trace by setting
\[ \hat{x} := \frac{x}{\tr(x)} \]
and using that the Lagrangian is homogeneous of degree two in each argument, we obtain
\begin{align*}
\ell(x) &= \tr(x)^2 \int_M \L(\hat{x} ,y )\: d\rho(y) - \mathfrak{r}\, \big( \tr(x) - 1 \big) - \s \\
&= \tr(x)^2 \big( \ell(\hat{x}) + \s \big) - \mathfrak{r}\, \big( \tr(x) - 1 \big) - \s \\
&\geq \s\: \tr(x)^2 - \mathfrak{r}\, \big( \tr(x) - 1 \big) - \s \:, 
\end{align*}
where in the last step we applied~\eqref{ELtrace}. Choosing~$\mathfrak{r} =2 \s$,
we obtain
\[ \ell(x) \geq \s \:\big( \tr(x) - 1 \big)^2 \geq 0\:, \]
concluding the proof.
\QED
The parameter~$\mathfrak{r}$ can be viewed as the Lagrange parameter corresponding to the
trace constraint. Likewise, $\s$ is the Lagrange parameter of
the volume constraint.

We finally comment on the rescaling freedom and its effect on the Lagrange parameters.
Clearly, setting the right side of~\eqref{volconstraint} and~\eqref{trconstraint} equal to one
is a matter of convenience. By rescaling the measure and the operators in~$\F$,
one can transform every minimizing measure to a minimizer corresponding to the constraints
\[ \rho(\F) = C \qquad \text{and} \qquad \int_\F \tr(x)\: d\rho(x) = c \]
with arbitrary parameters~$c,C>0$. Then~\eqref{loctrace} is modified to
\[ \tr(x)= \frac{c}{C} \:. \]
Moreover, the definition of~$\ell$ in~\eqref{elldef} becomes
\[ \ell(x) = \int_M \L(x,y)\: d\rho(y) - \mathfrak{r} \,\Big( \tr(x) - \frac{c}{C} \Big) - C \s 
\qquad \text{with} \qquad \mathfrak{r} = \frac{2 C}{c}\:C \s\:. \]
This rescaling freedom is of relevance for us because for the measures to be analyzed later on,
the parameters~$c$ and~$C$ will both be different from one.

\section{The Linearized Field Equations} \label{secintrolin}
In this section we derive the linearized field equations from an abstract point of view.
Our presentation differs from that in~\cite{jet} in that we make use of the structures
of causal fermion systems right from the beginning. In particular, we work exclusively with wave charts
as introduced in~\cite{gaugefix, banach}. The main goal of our constructions and considerations
is to justify that the causal fermion system can be analyzed in a description with
spinorial wave functions in Minkowski space.
The reader who wants to enter the computations 
in Minkowski space right away may skip this section in a first reading.

\subsection{The Restricted Euler-Lagrange Equations} \label{secELres}
In preparation, we want to bring the EL equations~\eqref{EL} into a form which is most suitable for
our analysis. To this end, we make the simplifying assumption that
our minimizing measure~$\rho$ is {\em{regular}}
in the sense that all spacetime point operators have the maximal possible rank
(i.e.~$\dim x(\H) = 2n$ for all~$x \in M := \supp \rho$; this assumption will indeed be satisfied
for the Dirac sea vacuum in Minkowski space to be introduced in Section~\ref{secfirstorder}).
As shown in~\cite{gaugefix, banach}, under this assumption an open neighborhood
of~$M \subset \F$ has the structure of a
smooth Banach manifold (this will be shown in more detail in Section~\ref{seclinwave} below
by the explicit construction of an atlas).
The starting point of our consideration is the formula~\eqref{Fid}, which expresses the spacetime point operator
as a local correlation operator. Using this formula, first variations of the wave evaluation operator~$\Psi(x)$ at
a given spacetime point~$x \in M$ give rise to corresponding variations of the spacetime point operator, i.e.\
\beq \label{ufermi}
\bu := \delta x = -\delta \Psi(x)^*\, \Psi(x) - \Psi(x)^*\, \delta \Psi(x) \:.
\eeq
The operator~$\bu$ can be regarded geometrically as a tangent vector to~$\F$ at~$x$.
The minimality of~$\ell$ on~$M$ as expressed by~\eqref{EL} implies that the
derivative of~$\ell$ in the direction of~$\bu$ vanishes, i.e.\
\beq \label{Dul}
D_\bu \ell(x) = 0
\eeq
for all variations of the form~\eqref{ufermi} for which the directional derivative
in~\eqref{Dul} exists. The resulting equations are also referred to as the {\em{restricted EL equations}}.

Clearly, the restricted EL equations~\eqref{Dul} contain only part of the information
of the full EL equations~\eqref{EL}. Before going on, we explain how this is to be understood
and why this restricted information
is precisely what is needed in order to describe the dynamics in Minkowski space.
\begin{Remark} {\bf{(The scalar component of the reduced EL equations)}} \label{remscal1} {\em{
The most obvious difference between~\eqref{Dul} and~\eqref{EL} is that the 
full EL equations~\eqref{EL} make a statement on the function~$\ell$ even at points~$x \in \F$
which are far away from spacetime~$M$.
At present, it is unclear how this additional information can be used or
interpreted. We take the pragmatic point of view that all the information
on the physical system must be obtained by performing observations or measurements in spacetime,
which means that the information contained in~$\ell$ away from~$M$ is inaccessible for principal reasons.
Consequently, it is sufficient to restrict attention to the function~$\ell$
in an arbitrarily small open neighborhood of~$M$ in~$\F$. A more detailed discussion of this point
can be found in~\cite[Section~7.2]{intro}.

Evaluating the EL equations~\eqref{EL} in this way, we find that the function~$\ell$ and its
first derivatives must vanish on~$M$, i.e.\
\beq \label{ELrestricted}
\ell|_M \equiv 0 \qquad \text{and} \qquad D \ell|_M \equiv 0 \:.
\eeq
These two equations are combined in the formulation of the restricted EL equations
as first introduced in~\cite[Section~4]{jet}.
Clearly, the second equation in~\eqref{ELrestricted} coincides with~\eqref{Dul}.
The first equation, however, was omitted in~\eqref{Dul}.
It turns out that the first equation can indeed be omitted in smooth spacetimes, as the following
consideration shows. Assume that the spacetime~$M$ has a {\em{smooth manifold structure}}
(for the detailed definition see~\cite[Definition~2.3]{fockbosonic}).
Moreover, assume that the wave evaluation operator is smooth, i.e.~$\Psi : \H \rightarrow C^\infty(M, SM)$.
Then to every tangent vector~$\bu \in T_xM$ we can associate a corresponding variation
\beq \label{delPsiu}
\delta \Psi(x) = D_\bu \Psi \big|_x \:.
\eeq
The corresponding variation of the spacetime point operator~\eqref{ufermi} gives us back the tangent
vector~$\bu$. Therefore, from~\eqref{Dul} we conclude that all directional derivatives of~$\ell$ vanish,
which means that~$\ell$ is constant in spacetime.
By choosing the Lagrange parameter~$\s$ in~\eqref{elldef} appropriately, this constant can always
arranged to be zero.
This argument explains why, in smooth spacetimes, we may omit the first equation in~\eqref{ELrestricted}.
\nolinebreak
}} \QEDrem
\end{Remark}

Formulating the restricted EL equations as in~\eqref{Dul} with the help of
directional derivatives of the function~$\ell$ defined on~$\F$ is most useful for
abstract considerations. However, for concrete computations, is more convenient to reformulate
the restricted EL equations in terms of variations of the kernel of the fermionic projector, as we now explain.
In preparation, we use~\eqref{elldef} in order to write~\eqref{Dul} as
\beq \label{resELL}
\int_M D_{1,\bu} \L(x,y)\: d\rho(y) = \mathfrak{r}\, D_\bu \tr(x) \:,
\eeq
where the index one means that the directional derivative acts on the first argument of the Lagrangian.
For the computation of the first variation of the Lagrangian, one can make use of the fact
that for any $p \times q$-matrix~$A$ and any~$q \times p$-matrix~$B$,
the matrix products~$AB$ and~$BA$ have the same non-zero eigenvalues, with the same
algebraic multiplicities. As a consequence, applying again~\eqref{Fid},
\beq
x y 
= \Psi(x)^* \,\big( \Psi(x)\, \Psi(y)^* \Psi(y) \big)
\simeq \big( \Psi(x)\, \Psi(y)^* \Psi(y) \big)\,\Psi(x)^* \:, \label{isospectral}
\eeq
where $\simeq$ means that the operators are isospectral (in the sense that they
have the same non-trivial eigenvalues with the same algebraic multiplicities). Thus, introducing
the {\em{kernel of the fermionic projector}} $P(x,y)$ by
\beq \label{Pxydef}
P(x,y) := -\Psi(x)\, \Psi(y)^* \::\: S_yM \rightarrow S_xM \:,
\eeq
we can write~\eqref{isospectral} as
\[ x y \simeq P(x,y)\, P(y,x) \::\: S_xM \rightarrow S_xM \:. \]
In this way, the eigenvalues of the operator product~$xy$ as needed for the computation of
the Lagrangian~\eqref{Lagrange} are recovered as
the eigenvalues of a $2n \times 2n$-matrix. Since~$P(y,x) = P(x,y)^*$,
the Lagrangian~$\L(x,y)$ in~\eqref{Lagrange} can be expressed in terms of the kernel~$P(x,y)$.
Consequently, the first variation of the Lagrangian can be expressed in terms
of the first variation of this kernel. Being real-valued and real-linear in~$\delta P(x,y)$,
it can be written as
\beq \label{delLdef}
\delta \L(x,y) = 2 \re \Tr_{S_xM} \!\big( Q(x,y)\, \delta P(x,y)^* \big)
\eeq
(where~$\Tr_{S_xM}$ denotes the trace on the spin space~$S_xM$)
with a kernel~$Q(x,y)$ which is again symmetric (with respect to the spin inner product), i.e.
\[ Q(x,y) \::\: S_yM \rightarrow S_xM \qquad \text{and} \qquad Q(x,y)^* = Q(y,x) \:. \]
More details on this method and many computations can be found in~\cite[Sections~1.4 and~2.6
as well as Chapters~3-5]{cfs}. From these computations, we know that, in Minkowski space,
the kernel~$Q(x,y)$ is well-defined as a bi-distribution (its explicit form in the vacuum will be
given in Section~\ref{secdQcont} below). With this in mind, here we may disregard
all differentiability issues. Expressing the variation of~$P(x,y)$ in terms of~$\delta \Psi$,
the first variations of the Lagrangian can be written as
\begin{align*}
D_{1,\bu} \L(x,y) = 2\, \re \tr \big( \delta \Psi(x)^* \, Q(x,y)\, \Psi(y) \big) \\
D_{2,\bu} \L(x,y) = 2\, \re \tr \big( \Psi(x)^* \, Q(x,y)\, \delta \Psi(y) \big)
\end{align*}
(where~$\tr$ denotes the trace of a finite-rank operator on~$\H$).
Using these formulas, the restricted EL equation~\eqref{resELL} becomes
\[ \re \int_M \tr \big( \delta \Psi(x)^* \, Q(x,y)\, \Psi(y) \big)\: d\rho(y) = \mathfrak{r}\,\re \tr \big( \delta \Psi(x)^* \,\Psi(x) \big)  \:. \]
Using that the variation can be arbitrary at every spacetime point, we obtain
\[ \int_M Q(x,y)\, \Psi(y) \:d\rho(y) = \mathfrak{r}\, \Psi(x) \qquad \text{for all~$x \in M$}\:, \]
where~$\mathfrak{r} \in \R$ is the Lagrange parameter of the trace constraint.
Denoting the integral operator with kernel~$Q(x,y)$ by~$Q$, the
restricted EL equations can be written in the shorter form
\beq \label{ELQ}
Q \Psi = \mathfrak{r}\, \Psi \:.
\eeq

We conclude by pointing out that first variations of a causal fermion systems can be described in two different ways.
The first method is to consider variations of the spacetime point operators~\eqref{ufermi}.
This method has the advantage of being gauge invariant. Indeed, if we consider a local
phase transformation~$\Psi(x) \rightarrow e^{i \Lambda(x)}\, \Psi(x)$, then this local phase
drops out of~\eqref{ufermi}.
The second method is to work with variations of the kernel of the fermionic projector~\eqref{Pxydef}.
This has the main advantage that~$P(x,y)$ can be represented by a $2n \times 2n$-matrix, which can
be computed in detail. This advantage outweighs the disadvantage that~$P(x,y)$ is {\em{not}} gauge invariant.
In the present paper, we will mainly work with variations of the kernel of the fermionic projector,
leading us to the restricted EL equations in the form~\eqref{ELQ}.

\subsection{The Linearized Field Equations in Wave Charts} \label{seclinwave}
The linearized field equations describe variations of the measure~$\rho$ which preserve the EL equations.
The simplest way to derive these equations is to vary the spacetime point operators again
according to~\eqref{ufermi} by varying the wave evaluation operator in~\eqref{Fid}.
Then preserving the restricted EL equations~\eqref{ELQ} means that
\beq \label{lfehom}
(DQ|_{\Psi}(\delta \Psi)) \,\Psi + Q\, \delta \Psi - \mathfrak{r}\, \delta \Psi = 0\:,
\eeq
where~$DQ|_{\Psi}(\delta \Psi)$ is the variational derivative of the kernel~$Q(x,y)$ under the first variation
of the wave evaluation operator~$\delta \Psi$. For clarity, we point out that this linearization is
real-linear but in general not complex-linear in~$\delta \Psi$ (i.e., it is linear in~$\delta \Psi$
and its conjugate; for details see Lemma~\ref{lemmaQ2} below). The equations~\eqref{lfehom} are the {\em{homogeneous linearized field
equations}}. It is useful to allow for an inhomogeneity on the right side of the equations.
Thus we write the {\em{inhomogeneous linearized field equations}} as
\beq \label{lfe}
(DQ|_{\Psi}(\delta \Psi)) \,\Psi + Q\, \delta \Psi - \mathfrak{r}\, \delta \Psi = \Xi \:,
\eeq
where the inhomogeneity is a given mapping
\[ \Xi \::\: \H \rightarrow C^0(M, SM) \:. \]
These equations were formulated and analyzed computationally in~\cite{pfp, cfs}.
In preparation of our detailed study of these equations, it is important to formulate them
more carefully in so-called wave charts as introduced in~\cite{gaugefix, banach}.
These wave charts will also be essential in the next section (Section~\ref{secpositive}),
where the variational structure of the linearized field equations will be worked out.
Moreover, they will be convenient in order
to write the first variation of~$Q$ in~\eqref{lfe} more explicitly in spinor components.
We refer to the resulting equations as the {\em{linearized field equations in wave charts}}.

For the construction of the wave charts, we let~$(\H, \F, \rho)$ be a causal fermion system. 
We again assume that the measure~$\rho$ is regular.
We choose a spacetime point~$z \in M$, which will serve as the base point of our chart.
We decompose the Hilbert space as
\[ \H = I_z \oplus J_z \]
with~$I_z := z(\H)$ and~$J_z := I_z^\perp$. We let~$\Symm(S_zM) \subseteq \Lin(S_zM)$
be the real vector space of all operators~$A$ on~$S_zM$ which are
symmetric with respect to the spin inner product, i.e.\
\[ \Sl  \phi | A \tilde{\phi} \Sr_z = \Sl A \phi | \tilde{\phi} \Sr_z \qquad
\text{for all~$\phi, \tilde{\phi} \in S_zM$}\:. \]
We consider the mapping
\[ 
R_z \::\: \Symm(S_zM) \oplus \Lin(J,S_zM) \rightarrow \F \:,\qquad R_z(\psi) = -\psi^* \psi \]
(where the star denotes the adjoint with respect to the corresponding inner products).
Then there is an open neighborhood~$W_z$ of~$(\id_{S_zM}, 0) \in \mathrm{Symm}(S_zM)\oplus \Lin(J,S_zM)$
such that the restriction of~$R_z$ to~$W_z$ defines a local chart of~$\F$ at~$z$.
More precisely, it is shown in~\cite[Theorem~6.5]{gaugefix} and~\cite[Theorem~3.2]{banach} 
that the mapping
\[ R_z|_{W_z} \::\: W_z \rightarrow \Omega_z
\overset{\text{\tiny{\rm{open}}}}{\subset} \F \]
is a homeomorphism to its image (with respect to the topology induced by
the operator norm on~$\Lin(\H)$). Moreover, taking two such parametrizations, the transition maps
are Fr{\'e}chet-smooth. These results mean that an open neighborhood of~$M \subset \F$
is a smooth Banach manifold (more precisely, this open neighborhood can be chosen even as
the set of all regular points of~$\F$).

We denote the chart corresponding to~$R_z$ by
\[ \phi_z := (R_z|_{W_z})^{-1} \::\: \Omega_z \rightarrow \Symm(S_zM) \oplus \Lin(J,S_zM) \:. \]
It is referred to as the {\em{symmetric wave chart}} around~$z$
(here ``symmetric'' refers to the fact that the first direct summand are the symmetric
linear operators on~$S_zM$). For every~$x \in W_z$,
we can regard~$\phi_z(x)$ as a mapping
\[ \phi_z(x) : \H \rightarrow S_zM \:, \]
satisfying the additional gauge condition~$\phi_z(x)|_{S_zM} \in \Symm(S_zM)$
(the connection to local gauge transformations and gauge fixing is worked out in~\cite{gaugefix}).
For every~$x \in W_z \cap M$,
\beq \label{xrep}
x = R_z\big( \phi_z(x) \big) = -\phi_z(x)^* \,\phi_z(x) \:.
\eeq
This identity is very similar to the representation of a spacetime point operator as
a local correlation operator~\eqref{Fid}. The only difference is that~$\phi_z(x)$ maps to the spin
space~$S_zM$, whereas~$\Psi(x)$ maps to the spin space~$S_xM$.
However, this difference is irrelevant in view of the following identification:
\begin{Lemma} Choosing the domain~$\Omega_z$ of the chart$~\phi_z$ sufficiently small, for every~$x \in \Omega_z$
the mapping
\beq \label{spiniso}
\phi_z|_{S_xM} \::\: S_xM \rightarrow S_zM
\eeq
is an isomorphism of the spin spaces~$S_xM$ and~$S_zM$.
\end{Lemma}
\Proof Similar as in the proof of~\cite[Proposition~1.2.6]{cfs}, for every~$\phi, \psi \in S_xM$,
\[ \Sl \phi | \psi \Sr_x = - \la \phi \,|\, x\, \psi \ra_\H \overset{\eqref{xrep}}{=} 
\la \phi \,|\, \phi_z(x)^* \,\phi_z(x)\, \psi \ra_\H =
\la \phi_z(x) \phi \,|\, \phi_z(x) \psi \ra_\H \:. \]
This shows that the mapping~\eqref{spiniso} is a unitary embedding. Moreover, by
choosing~$\Omega_z$ sufficiently small, we can arrange that this mapping is also surjective.
\QED
Working with this identification, it follows that
\[ \Psi|_{\Omega_z \cap M} = \phi_z|_{\Omega_z \cap M} \::\:
\Omega_z \cap M \rightarrow \Symm(S_zM) \oplus \Lin(J,S_zM)\:. \]
Formulating the EL equations~\eqref{ELQ} in these charts, we can vary the mapping~$\Psi$
while keeping the target space fixed. In this way, we can give~\eqref{lfehom} a well-defined meaning
with
\[ \delta \Psi(x) \in \Symm(S_zM) \oplus \Lin(J,S_zM) \]
for all~$x \in \Omega_z \cap M$.
In order to work out the variation in more detail, we let~$F_\tau : M \rightarrow \F$ be a variation
of the spacetime point operators. We choose an open neighborhood~$U \subset M$
of~$z$ such that~$F_\tau(U)$ is contained in~$W_z$ for all~$\tau$. Then the variation
is represented in the symmetric wave chart~$(\phi_z, W_z)$ by a mapping denoted by
\[ \Psi_{z,\tau} := \phi_z \circ F_\tau|_U \::\: U \subset M \rightarrow \Symm(S_zM) \oplus \Lin(J,S_zM) \:. \]
This notation suggests that~$\Psi_{z,\tau}$ is a variation of the wave evaluation
operator, as can be made precise using the identification~\eqref{spiniso}.
This point of view is very useful for the computations, because we can use the formula
\[ F(x) = - \Psi_{z,\tau}(x)^*\, \Psi_{z,\tau}(x) \]
in order to define higher derivatives in a straightforward way. For ease in notation, we write the
variations as~$\delta \Psi(x), \delta^2 \Psi(x), \ldots$, omitting the subscript~$z$. This is unambiguous
because all our formulas hold independent of the choice of base points of the local charts.
Another advantage of working in the symmetric wave charts is that we are working
in a fixed spin space~$S_zM$. Therefore, choosing a basis~$\mathfrak{f}_\alpha$
with~$\alpha=1,\ldots, 2n$, we can write out all formulas in components.
We always use the Einstein summation convention for spinor indices.
We write the spinor indices as upper indices, whereas the lower indices are dual indices.
In order to clarify the base point of the spin space, we work with a vertical line,
which separates the spinor indices corresponding to the first and second argument.
Thus we write the kernel of the fermionic projector as~$P^\alpha|_\beta(x,y)$.
The relation~$P(x,y)^*=P(y,x)$ takes the form
\[ \overline{P^\alpha|_\beta(x,y) } = P^\beta|_\alpha(y,x) \:, \]
where the overline stands for the adjoint with respect to the spin inner product. In particular,
in this formalism,
\[ \overline{\psi_\alpha(x)}\, \phi^\alpha(x) = \Sl \psi(x) | \phi(x) \Sr_x \:. \]

\begin{Lemma} \label{lemmaQ2} The first variation of~$Q(x,y)$ can be written as
\beq \label{Qvary}
\delta Q^\alpha|_\beta(x,y) = K^{\alpha \gamma}|_{\beta \delta}(x,y)\, \delta P^\delta|_\gamma(y,x)
+ K^\alpha_{\;\;\,\gamma}|_\beta^{\;\;\,\delta}(x,y)\, \delta P^\gamma|_\delta(x,y)
\eeq
with kernels~$K^{\alpha \gamma}|_{\beta \delta}(x,y)$ and~$K^\alpha_{\;\;\,\gamma}|_\beta^{\;\;\,\delta}(x,y)$,
which have the symmetry properties
\begin{align}
K^{\alpha \gamma}|_{\beta \delta}(x,y) &= K^{\gamma \alpha}|_{\delta \beta}(x,y)
= \overline{K^{\beta \delta}|_{\alpha \gamma}(y,x)} \label{Krel1} \\
K^\alpha_{\;\;\,\gamma}|_\beta^{\;\;\,\delta}(x,y) &= 
K^\delta_{\;\;\,\beta}|_\gamma^{\;\;\,\alpha}(y,x)
= \overline{K^\gamma_{\;\;\,\alpha}|_\delta^{\;\;\,\beta}(x,y) } \label{Krel2} \:.
\end{align}
\end{Lemma}
\Proof Clearly, the kernel~$\delta Q(x,y)$ is linear in~$P(x,y)$ and its adjoint. Therefore, the formula~\eqref{Qvary}
merely is a convenient form of writing the variation. It serves as the definition of the
kernels~$K^{\alpha \gamma}|_{\beta \delta}(x,y)$ and~$K^\alpha_{\;\;\,\gamma}|_\beta^{\;\;\,\delta}(x,y)$.
In order to prove~\eqref{Krel1} and~\eqref{Krel2} we make use of the fact that second variations
of the Lagrangian are real-valued and symmetric. These second variations are computed by
\begin{align}
\delta \L(x,y) &= 2 \re \big( Q^\alpha|_\beta(y,x)\, \delta P^\beta|_\alpha(x,y) \big)
= 2 \re \big( Q^\alpha|_\beta(x,y)\, \delta P^\beta|_\alpha(y,x) \big) \notag \\
&= Q^\alpha|_\beta(x,y)\, \delta P^\beta|_\alpha(y,x) + Q^\beta|_\alpha(y,x)\, \delta P^\alpha|_\beta(x,y)
\label{delL} \\
\delta_2 \big( \delta_1 \L(x,y) \big) &=
Q^\alpha|_\beta(x,y)\, \delta_{12}^2 P^\beta|_\alpha(y,x) + Q^\beta|_\alpha(y,x)\, \delta^2_{12}
P^\alpha|_\beta(x,y) \notag \\
&\quad\:+ K^{\alpha \gamma}|_{\beta \delta}(x,y)\, \delta_1 P^\beta|_\alpha(y,x) \: \delta_2
P^\delta|_\gamma(y,x) \notag \\
&\quad\:+ K^\alpha_{\;\;\,\gamma}|_\beta^{\;\;\,\delta}(x,y)\,
\delta_1 P^\beta|_\alpha(y,x) \: \delta_2 P^\gamma|_\delta(x,y) \notag \\
&\quad\:+ K^\beta_{\;\;\,\delta}|_\alpha^{\;\;\,\gamma}(y,x)\, \delta_1 P^\alpha|_\beta(x,y) \,\delta_2
P^\delta|_\gamma(y,x) \notag \\
&\quad\:+ K^{\beta \delta}|_{\alpha \gamma}(y,x)\, \delta_1 P^\alpha|_\beta(x,y) \,\delta_2 P^\gamma|_\delta(x,y) \:,
\label{del2L}
\end{align}
where
\begin{align}
\delta P^\alpha|_\beta(x,y) &= - \big( \delta \Psi(x)^\alpha \big) \Psi(y)^*_\beta - \Psi(x)^\alpha
\, \big( \delta \Psi(y)^*_\beta \big) \label{delP} \\
\delta^2_{12} P^\alpha|_\beta(x,y) &= - \big( \delta_1 \Psi(x)^\alpha \big) \big( \delta_2 \Psi(y)^*_\beta \big)
- \big( \delta_2 \Psi(x)^\alpha \big) \big( \delta_1 \Psi(y)^*_\beta \big) \notag \\
&\quad\: - \big( \delta^2_{12} \Psi(x)^\alpha \big) \Psi(y)^*_\beta - \Psi(x)^\alpha \, \big( \delta^2_{12} \Psi(y)^*_\beta
\big) \:. \label{del2P}
\end{align}
The fact that this second variation is real implies that
\begin{align*}
\overline{K^{\alpha \gamma}|_{\beta \delta}(x,y)} &= K^{\beta \delta}|_{\alpha \gamma}(y,x) \\
\overline{K^\alpha_{\;\;\,\gamma}|_\beta^{\;\;\,\delta}(x,y)} &= K^\beta_{\;\;\,\delta}|_\alpha^{\;\;\,\gamma}(y,x) \:.
\end{align*}
The symmetry of the second variations yields
\begin{align*}
K^{\alpha \gamma}|_{\beta \delta}(x,y) &= K^{\gamma \alpha}|_{\delta \beta}(x,y) \\
K^{\beta \delta}|_{\alpha \gamma}(y,x) &= K^{\delta \beta}|_{\gamma \alpha}(y,x) \\
K^\alpha_{\;\;\,\gamma}|_\beta^{\;\;\,\delta}(x,y) &= K^\delta_{\;\;\,\beta}|_\gamma^{\;\;\,\alpha}(y,x) 
= \overline{K^\gamma_{\;\;\,\alpha}|_\delta^{\;\;\,\beta}(x,y) }\:.
\end{align*}
These relations can be written more compactly in the form~\eqref{Krel1} and~\eqref{Krel2}.
\QED

Compared to the linearized field equations as introduced in~\cite{jet}
(see also~\cite[Chapter~8]{intro}), the variation~$\delta \Psi$ in~\eqref{lfe}
is more special because we left out the possibility of multiplying the measure~$\rho$
by a weight function. We finally justify this simplification.

\begin{Remark} {\bf{(Omitting the scalar component of the variation)}} \label{remscal2} {\em{
In the formulation of the linearized field equations~\eqref{lfehom}, we only considered a variation
of the wave evaluation operator. According to~\eqref{ufermi}, the variation can be described
equivalently by a vector field~$\bv \in \Gamma(M, T\F)$ on~$\F$ along~$M$.
The question arises why we did not allow for a variation of the weight of the measure~$\rho$.
In the jet formalism introduced in~\cite{jet}, such a variation of the weight is described by
the scalar component~$b \in C^\infty(M, \R)$ of a corresponding jet~$(b, \bv)$.
We now explain why variations of the weight may be disregarded.
Our argument has some similarity with Remark~\ref{remscal1}, where we explained why,
assuming that spacetime has a smooth manifold structure,
the scalar component of the reduced EL equations (i.e.\ the first equation in~\eqref{ELrestricted})
can be omitted. Now we need to make the stronger assumption that~$M:=\supp \rho$
is a {\em{Minkowski-type spacetime}} as defined in~\cite[Section~2.8]{fockfermionic}:
We assume that~$M$ is diffeomorphic to Minkowski
space~$\scrM = \R^{1,3} \simeq \R^4$. Moreover, we assume that, under the identification of this diffeomorphism,
the measure~$\rho$ is absolutely continuous with respect
to the Lebesgue measure with a smooth weight function, i.e.\
\beq \label{rhoh}
d\rho = h(x)\: d^4x \qquad \text{with} \quad h \in C^\infty(\scrM, \R^+) \:.
\eeq
We also assume that~$h$ is bounded from above and below, i.e.\ there should be a constant~$C>1$ with
\[ 
\frac{1}{C} \leq h(x) \leq C \qquad \text{for all~$x \in \scrM$} \:. \]

Under these assumptions, for every compactly supported function~$a \in C^\infty_0(M, \R)$
there is a smooth vector field~$\bu$ in~$M$ whose divergence coincides with~$a$, i.e.\
\[ a  = \div \bu := \frac{1}{h}\: \partial_j \big( h\, \bu^j \big) \]
(following the Einstein summation convention, we sum over~$j=0,\ldots,3$;
this result can be regarded as a version of Moser's theorem proved in~\cite[Proposition~3.7]{fockbosonic}).
As shown in~\cite[Section~3]{fockbosonic}, the corresponding variation of the measure defined by
\[ \delta \int_\F g(x)\: d\rho(x) := \int_M \Big( \big( 1 + a(x) + \bu^j \partial_j \big) g(x) \Big)\: d\rho(x) \]
satisfies the linearized field equations, which in generalization of~\eqref{lfe} take the form
\[ \big( (DQ|_{\Psi}(\delta \Psi)) \,\Psi \big)(x) + \big( Q\, \delta \Psi \big)(x)
+ \int_M Q(x,y)\, \Psi(x)\: a(y)\, d\rho(y)  = \mathfrak{r}\, \delta \Psi(x) + \s\, a(x)\:, \]
where~$\delta \Psi$ is the variation~\eqref{delPsiu} which realizes the vector field~$\bu$ according
to~\eqref{ufermi}. The corresponding pair~$(a, \bu)$ is referred to as an {\em{inner solution}}
of the linearized field equations.

This argument shows that an infinitesimal change of the weight of~$\rho$ as described by~$a$
can be compensated by an inner solution of the linearized field equations.
With this in mind, it is no loss of generality to restrict attention to variations which do not change
the weight of~$\rho$. For a more detailed explanation we refer to~\cite[Section~3.3]{fockbosonic}.
}} $\hspace*{0,5em}$ \QEDrem
\end{Remark}

\subsection{The Variational Structure of the Linearized Field Equations} \label{secpositive}
In order to make the variational structure of the linearized field equations more apparent,
it is useful to also work out the bilinear form corresponding to second variations of the causal action
as first considered in~\cite{positive}. Our starting point is the observation that the
EL equations~\eqref{EL} with~$\ell$ according to~\eqref{elldef} can be obtained by minimizing the
{\em{effective action}}~$\Sact^\text{\rm{eff}}$ defined by
\[ \Sact^\text{\rm{eff}}(\rho) := \int_\F d\rho(x) \int_\F d\rho(y)\: \L(x,y) 
- 2 \int_\F \Big( \mathfrak{r}\, \big( \tr (x) -1 \big) + \s \Big) \,d\rho(x) \]
under variations of the form~$\tilde{\rho}_\tau = \tilde{\rho} + \delta_{x_0}$ with~$x_0 \in \F$.
In other words, both the volume and the trace constraints can be treated with Lagrange multipliers.
As explained in the previous section (see Remarks~\ref{remscal1} and~\ref{remscal2}),
we again restrict attention to variations of the wave evaluation operator~$\Psi$.
Since these variations respect the volume constraint, we may simplify the effective action to
\beq \label{Seff}
\Sact^\text{\rm{eff}}(\rho) := \int_\F d\rho(x) \int_\F d\rho(y)\: \L(x,y) 
- 2 \mathfrak{r} \int_\F \tr (x)\,d\rho(x) \:.
\eeq
Now we can work out first and second variations, for convenience again in symmetric wave charts.
Rewriting the local trace as
\beq \label{trid}
\tr(x) = \tr \big( \pi_x x \big) = \Tr_{S_xM} \big( P(x,x) \big) \:,
\eeq
the first variation can be computed with the help of~\eqref{delL} and~\eqref{delP},
\begin{align*}
&\delta \Sact^\text{\rm{eff}}(\rho) = \int_\F d\rho(x) \int_\F d\rho(y)\: \delta \L(x,y) 
- 2 \mathfrak{r} \int_\F \Tr_{S_xM} \big( \delta P(x,x) \big)\,d\rho(x) \\
&= 2 \re \int_\F d\rho(x) \int_\F d\rho(y)\: \big( Q^\alpha|_\beta(x,y)\, \delta P^\beta|_\alpha(y,x) \big)
- 2 \mathfrak{r} \int_\F \delta P^\alpha|_\alpha(x,x)\,d\rho(x) \\
&= -4 \re \bigg( \int_\F d\rho(x) \int_\F d\rho(y)\: Q^\alpha|_\beta(x,y)\, \Psi(y)^\beta \, \delta \Psi(x)^*_\alpha
- \mathfrak{r} \int_\F \Psi(x)^\alpha \, \delta \Psi(x)^*_\alpha\,d\rho(x) \bigg) \:.
\end{align*}
Since~$\delta \Psi(x)$ is arbitrary, we obtain
\[ \int_\F Q^\alpha|_\beta(x,y)\, \Psi(y)^\beta \:d\rho(y)
= \mathfrak{r} \,\Psi(x)^\alpha \:, \]
giving us back the Euler-Lagrange equation~\eqref{ELQ}.

\begin{Prp} \label{prpsecond}
Let~$\rho$ be a minimizer of the causal action principle. Then the second variation of the
effective action~\eqref{Seff} takes the form
\begin{align*}
\delta^2 \Sact^\text{\rm{eff}}(\rho) &= - 4\, \bigg( \int_\F d\rho(x) \int_\F d\rho(y) \: Q^\alpha|_\beta(x,y)\, 
\big( \delta \Psi(y)^\beta \big) \big( \delta \Psi(x)^*_\alpha \big) \notag \\
&\qquad\qquad\qquad\quad\qquad\qquad- \mathfrak{r} \int_\F \big( \delta \Psi(x)^\alpha \big) \big( \delta \Psi(x)^*_\alpha \big) \,d\rho(x) \bigg) \\
&\quad\:+ 2 \re \int_\F d\rho(x) \int_\F d\rho(y)
\:\Big\{ K^{\alpha \gamma}|_{\beta \delta}(x,y)\, \delta P^\beta|_\alpha(y,x) \: \delta
P^\delta|_\gamma(y,x) \notag \\
&\quad\qquad\qquad\qquad\qquad\qquad\quad\:+ K^\alpha_{\;\;\,\gamma}|_\beta^{\;\;\,\delta}(x,y)\,
\delta P^\beta|_\alpha(y,x) \: \delta P^\gamma|_\delta(x,y)  \Big\} \:.
\end{align*}
\end{Prp}
\Proof A direct computation using again~\eqref{trid} gives
\[ \delta^2 \Sact^\text{\rm{eff}}(\rho) = \int_\F d\rho(x) \int_\F d\rho(y)\: \delta^2 \L(x,y) 
- 2 \mathfrak{r} \int_\F \Tr_{S_xM} \big( \delta^2 P(x,x) \big)\,d\rho(x) \:. \]
Using~\eqref{del2L} and~\eqref{del2P}, all the terms involving second variations of~$\Psi$
vanish in view of the EL equations~\eqref{ELQ}. We thus obtain
\begin{align*}
\delta^2 \Sact^\text{\rm{eff}}(\rho) &= 4 \mathfrak{r} \int_\F \big( \delta \Psi(x)^\alpha \big) \big( \delta \Psi(x)^*_\alpha \big) \,d\rho(x)\\
&\quad\:+ 2 \re \int_\F d\rho(x) \int_\F d\rho(y) \: \Big\{ Q^\alpha|_\beta(x,y)\, \delta^2 P^\beta|_\alpha(y,x) \\
&\quad\qquad\qquad\qquad\qquad\qquad\quad\:+ K^{\alpha \gamma}|_{\beta \delta}(x,y)\, \delta P^\beta|_\alpha(y,x) \: \delta
P^\delta|_\gamma(y,x) \\
&\quad\qquad\qquad\qquad\qquad\qquad\quad\:+ K^\alpha_{\;\;\,\gamma}|_\beta^{\;\;\,\delta}(x,y)\,
\delta P^\beta|_\alpha(y,x) \: \delta P^\gamma|_\delta(x,y)  \Big\} \\
&= 4 \mathfrak{r} \int_\F \big( \delta \Psi(x)^\alpha \big) \big( \delta \Psi(x)^*_\alpha \big) \,d\rho(x)\\
&\quad\:- 4 \re \int_\F d\rho(x) \int_\F d\rho(y) \: Q^\alpha|_\beta(x,y)\, 
\big( \delta \Psi(y)^\beta \big) \big( \delta \Psi(x)^*_\alpha \big) \\
&\quad\:+ 2 \re \int_\F d\rho(x) \int_\F d\rho(y)
\:\Big\{ K^{\alpha \gamma}|_{\beta \delta}(x,y)\, \delta P^\beta|_\alpha(y,x) \: \delta
P^\delta|_\gamma(y,x) \\
&\quad\qquad\qquad\qquad\qquad\qquad\quad\:+ K^\alpha_{\;\;\,\gamma}|_\beta^{\;\;\,\delta}(x,y)\,
\delta P^\beta|_\alpha(y,x) \: \delta P^\gamma|_\delta(x,y)  \Big\} \:.
\end{align*}
This concludes the proof.
\QED

Writing the second variations as in the last proposition has the advantage that the symmetry
in the two first variations is apparent; this is why we could use the short notation~$\delta^2 \Sact
= \cdots\: \delta(\cdots)\, \delta(\cdots)$. In order to see the connection to the linearized field equations~\eqref{lfe}
it is preferable to distinguish the two variations by writing~$\delta_1$ and~$\delta_2$. Using Lemma~\ref{lemmaQ2},
we obtain
\begin{align*}
\delta^2_{12} \Sact^\text{\rm{eff}}(\rho) &= - 4 \re\, \bigg( \int_\F d\rho(x) \int_\F d\rho(y) \: Q^\alpha|_\beta(x,y)\, 
\big( \delta_1 \Psi(y)^\beta \big) \big( \delta_2 \Psi(x)^*_\alpha \big) \notag \\
&\qquad\qquad\qquad\quad\qquad\qquad- \mathfrak{r} \int_\F \big( \delta_1 \Psi(x)^\alpha \big) \big( \delta_2 \Psi(x)^*_\alpha \big) \,d\rho(x) \bigg) \\
&\quad\:+ 2 \re \int_\F d\rho(x) \int_\F d\rho(y)
\:\Big\{ \delta_1 Q^\alpha|_\beta(x,y) \, \delta_2 P^\beta|_\alpha(y,x)  \Big\} \\ 
&= -4 \re \int_\F d\rho(x) \int_\F d\rho(y)\: \tr \Big( \delta_2 \Psi(x)^*\, \big(
\delta_1 Q(x,y)\, \Psi(y) + Q(x,y)\, \delta_1 \Psi(y) \big) \Big)  \\
&\quad\:+4 \mathfrak{r}  \re\int_\F \tr \Big( \delta_2 \Psi(x)^* \:\delta_1 \Psi(x) \Big) \,d\rho(x) \:.
\end{align*}
Using that~$\delta \Psi_2$ is arbitrary, we get back~\eqref{lfe}.

\subsection{Why the Linearized Field Equations Comprise Both the Bosonic and Fermionic Equations}
\label{secunified}
It is worth pointing out that the linearized field equations comprise both bosonic and fermionic
equations. More concretely, for systems in Minkowski space, the linearized field equations
gives rise to both the Maxwell equations (the ``bosonic equation'') and the
Dirac equation (the ``fermionic equation''). This connection has been worked out in~\cite{cfs}
in the so-called continuum limit analysis. More abstractly, the bosonic equations were studied
in~\cite{linhyp}, whereas the fermionic equation was studied in~\cite{dirac}.
Here we shall not enter any details, but we merely explain in general terms how the
unified description of the bosonic and fermionic equations comes about.

Our starting point are the inhomogeneous linearized field equations in the form~\eqref{lfe}.
Given an inhomogeneity~$\Xi$, these equations determine the first variation~$\delta \Psi$
of the wave evaluation operator.
The resulting variation~$\delta \Psi$ has two contributions. First, it consists of a collective variation
of all the wave functions. Such a variation is typically described by a {\em{bosonic field}}. A typical example
for systems of Dirac wave functions is to perturb by a classical potential~$A$,
as explained in~\eqref{delPsiintro} in the introduction (for more details on this example
see~\cite[Section~7]{perturb}).
Indeed, the linearized field equations allow for much more general collective perturbations,
as will be worked out in detail later in this paper.
In addition to considering the collective behavior of the wave functions,
one can also ask how individual wave functions are changed. The corresponding equation is the
{\em{fermionic wave equation}}, also referred to as the {\em{dynamical wave equation}}~\cite{dirac}.
In order to derive the fermionic wave equation from~\eqref{lfe}, one multiplies from the right
by a projection operator~$\pi_\fermi$
\beq \label{diranalog}
\big( Q - \mathfrak{r} \big) \, \delta \Psi \pi_\fermi = \big( \Xi - (\delta Q) \,\Psi \big) \pi_\fermi \:.
\eeq
Here the image of the operator~$\pi_\fermi$ can be spanned either by a single vector in~$\H$
(in case we are interested in the physical wave function of this one vector) or by a finite number
of vectors of~$\H$ (in case we are interested in many such wave functions).
The left side is a complex linear equation for the corresponding physical wave functions.
The right side involves both the inhomogeneity~$\Xi$ and the variation~$\delta Q$.
In analogy to the term~$A_j \gamma^j \psi$ in the Dirac equation,
the last summand describes the coupling of the bosonic field to the fermionic wave functions.

To summarize, in the setting of causal fermion systems there is no clear distinction
between the bosonic and fermionic equations. Instead, both equations merely are different manifestations
of the linearized field equations, which describe a mutual interaction of all wave functions.
Nevertheless, in many situations it is admissible and useful to distinguish between the collective behavior
of all physical wave functions (as described by the bosonic equations) and the behavior of individual wave
functions (as described by the fermionic equations).

\subsection{Identification with Objects in Minkowski Space} \label{secident}
In the previous sections, the linearized field equations were introduced under the only
assumptions that spacetime is regular (see the beginning of Section~\ref{secELres}),
has a smooth manifold structure (Remark~\ref{remscal1})
and that~$M$ is a Minkowski-type spacetime (Remark~\ref{remscal2}).
We now specify the setting by assuming that {\em{spacetime can be identified with Minkowski space}}
in the following sense. First, we again assume that spacetime~$M:= \supp \rho$ is diffeomorphic
to Minkowski space~$\scrM = \R^{1,3}$. Moreover, we assume that there is a {\em{measure-preserving
diffeomorphism}}~$F : \scrM \rightarrow M$, i.e.\
\[ \rho = F_* \mu \:, \]
where~$d\mu = d^4x$ is the usual volume measure on Minkowski space.
Again assuming that~$\rho$ is regular, the spin spaces~$(S_xM, \Sl .|. \Sr_x)$
all have dimension four and signature~$(2,2)$. Therefore, we can identify them all
with a given four-dimensional complex vector space~$V$ endowed with an
indefinite inner product~$\Sl .|. \Sr$ of signature~$(2,2)$, referred to as the {\em{spin space}}
(clearly, this identification is not canonical; this corresponds to local gauge freedoms
as worked out in~\cite{u22}). Using this identification, every vector~$u \in \H$
is represented by a physical wave function~$\psi^u$ in Minkowski space defined by
\[ \psi^u(x) := \pi_{F(x)} u \;\in\; S_{F(x)}M \simeq V \:. \]
In this way, the causal fermion system is formed of a Hilbert space of wave functions in Minkowski space.
The formulas for the first variation of~$Q(x,y)$ in Lemma~\ref{lemmaQ2} and for the
second variation of the causal action in Proposition~\ref{prpsecond} now hold with the spinorial
indices referring to a basis~$(\mathfrak{f}_\alpha)_{\alpha=1,\ldots, 4}$ of~$V$.

\section{The Linearized Field Equations in Minkowski Space} \label{seclinmink}
In this section, we work out the linearized field equations more concretely for a Dirac sea
configuration in Minkowski space. This configuration is known to be a minimizer of the causal action principle
in the so-called continuum limit. For this reason, it is the starting point for the explicit analysis of the
causal action principle. In order to keep the setting reasonably simple, as in~\cite[Chapter~3]{cfs}
we consider a system of three Dirac seas describing the three generations of leptons.
Systems including neutrinos and quarks (as considered in~\cite[Chapters~4 and~5]{cfs})
could be analyzed similarly.

\subsection{Perturbations of Homogeneous Systems in Minkowski Space} \label{sechom}
Following the constructions in the previous section, we identify spacetime~$M := \supp \rho$
with Minkowski space and identify the spin spaces~$S_xM$ of the causal fermion system
with the spinor spaces in Minkowski space.
Moreover, as explained in the introduction and in Section~\ref{secident}, we identify all the spinor spaces in Minkowski space
with an indefinite inner product space~$(V, \Sl .|. \Sr)$ of dimension four and signature~$(2,2)$,
which we refer to as the {\em{spin space}}.
Then the wave evaluation operator~\eqref{weo}
is composed of spinorial wave functions in Minkowski space (not necessarily solutions of the Dirac equation).
We assume that its Fourier transform is a well-defined distribution, denoted by
\beq \label{weomom}
\widehat{\tilde{\Psi}} \::\: \H \rightarrow {\mathcal{D}}'(\hat{M}, V) \:,
\eeq
where~$\hat{M} \simeq \R^4$ is momentum space and~${\mathcal{D}}'(\hat{M}, V)$ denotes the
distributions on~$\hat{M}$ taking values in~$V$
(basics on the Fourier transform can be found for example in the textbooks~\cite{reed+simon, friedlander2}).
More precisely, the distribution~$\widehat{\tilde{\Psi}}$ has the property that
\[ \int_{\hat{M}} \hat{\phi}(p)\: \widehat{\tilde{\Psi}}(p)\: \frac{d^4p}{(2 \pi)^4}
= \int_M \phi(x)\: \tilde{\Psi}(x)\: d^4x \qquad \text{for all~$\phi \in C^\infty_0(M)$}\:, \]
where~$\hat{\phi}$ is the ordinary Fourier transform
\[ \hat{\phi}(p) := \int_M \phi(x)\: e^{-i p x}\: d^4x \:. \]
The {\em{fermionic projector in momentum space}}~$\hat{\tilde{P}}$ is defined by the following product,
\beq \label{tilmufac}
\hat{\tilde{P}} := -\widehat{\tilde{\Psi}} \times \widehat{\tilde{\Psi}}^* \;\in\; {\mathcal{D}}'\big(\hat{M} \times \hat{M}, \Lin(V) \big) \:,
\eeq
where we identified a pair of vectors in~$V \times V^*$ with a linear operator on~$V$.
Choosing an orthonormal basis~$(e_\ell)_\ell$ of~$\H$,
the fermionic projector in momentum space can also be written as
\beq \label{tilmudetail}
\hat{\tilde{P}}(k,p) \,u := -\sum_\ell
\big( \widehat{\tilde{\Psi}}(k) \,e_\ell \big) \;\Sl \widehat{\tilde{\Psi}}(p) \,e_\ell \,|\, u \Sr \:.
\eeq
Obviously, the fermionic projector in momentum space is {\em{negative}} in the
sense that
\beq \label{mupositive}
\int_\Omega \frac{d^4k}{(2 \pi)^4} \int_\Omega \frac{d^4p}{(2 \pi)^4}\:
\Sl u \,|\, \hat{\tilde{P}}(k,p)\, u \Sr \leq 0 \qquad \text{for all~$\Omega \in \B(\hat{M})$
and~$u \in V$}\:.
\eeq
The kernel of the fermionic projector is obtained by Fourier transformation,
\beq \label{tildeP}
\tilde{P}(x,y) := \int_{\hat{M}} \frac{d^4k}{(2 \pi)^4} \int_{\hat{M}} \frac{d^4p}{(2 \pi)^4}
\: \hat{\tilde{P}}(k,p)\: e^{-ikx+ipy} \:.
\eeq

It is a remarkable fact that the Hilbert space structure can be recovered from the fermionic projector.
This was first observed in~\cite[Section~1.2.2]{rrev} in a somewhat different formulation in Krein spaces.
We now recall this construction in our setting.
Given the fermionic projector in momentum space~\eqref{tilmufac}, on the continuous and compactly supported wave functions we introduce the sesquilinear form
\beq \label{hatH}
\begin{split}
\la .|.\ra_{\hat{\H}} \:&:\: C^0_0(\hat{M}, V) \times C^0_0(\hat{M}, V) \rightarrow \C \:, \\
\la f | g \ra_{\hat{\H}} &:= -\int_{\hat{M}} \frac{d^4k}{(2 \pi)^4} \int_{\hat{M}} \frac{d^4p}{(2 \pi)^4}
\Sl f(k) \:|\: \hat{\tilde{P}}(k,p)\: g(p) \Sr \:.
\end{split}
\eeq
In view of~\eqref{mupositive}, this sesquilinear form is positive semi-definite. Dividing out the null space
and taking the completion gives a Hilbert space denoted by~$(\hat{\H}, \la .|. \ra_{\hat{\H}})$.
In case we started from a Hilbert space~$(\H, \la .|. \ra_\H)$ and introduced the 
fermionic projector in momentum space
via~\eqref{tilmufac}, the two Hilbert spaces could be related to each other by the mapping
\[ \iota \::\: C^0_0(\hat{M}, V) \rightarrow \H \:,\qquad \iota(f) = \int_{\hat{M}} 
\frac{d^4p}{(2 \pi)^4}\: \widehat{\tilde{\Psi}}^*(p) \:f(p) \:. \]
This mapping is an isometry because
\begin{align*}
\la \iota f \,|\, \iota g \ra_\H &= \int_{\hat{M}} \frac{d^4k}{(2 \pi)^4} \int_{\hat{M}} \frac{d^4p}{(2 \pi)^4}\:
\big\la \widehat{\tilde{\Psi}}^*(k) \:f(k) \:\big|\:
\widehat{\tilde{\Psi}}^*(p) \:f(p) \big\ra_\H \\
&=\int_{\hat{M}} \frac{d^4k}{(2 \pi)^4} \int_{\hat{M}} \frac{d^4p}{(2 \pi)^4}\: \Sl f(k) \:|\:
\widehat{\tilde{\Psi}}(k) \times \widehat{\tilde{\Psi}}^*(p) \:f(p)\Sr \\
&= -\int_{\hat{M}} \frac{d^4k}{(2 \pi)^4} \int_{\hat{M}} \frac{d^4p}{(2 \pi)^4} \:\Sl f(k) \:|\:
\hat{\tilde{P}}(k,p) \:f(p)\Sr = \la f | g \ra_{\hat{\H}} \:,
\end{align*}
making it possible to identify~$\hat{\H}$ with a closed subspace of~$\H$.
It may be possible that~$\hat{\H}$ is a proper subspace of~$\H$ (as one sees immediately in the simple
example~$\hat{\tilde{P}}=0$). In order to avoid such trivialities, we always assume that~$\iota(\hat{\H})=\H$.
For notational simplicity, we often omit the hat.

For the perturbative treatment we make the ansatz
\beq \label{Psivary}
\widehat{\tilde{\Psi}} = \hat{\Psi} + \Delta \hat{\Psi} \:,
\eeq
where~$\hat{\Psi}$ is the wave evaluation operator of the vacuum, and~$\Delta \hat{\Psi}$ is a
(small but finite) perturbation.
We assume that the unperturbed system is {\em{homogeneous}}. This means that the unperturbed
fermionic projector has the form
\beq \label{nudef}
\hat{\tilde{P}}(k,p) = \hat{\Psi}(k) \times \hat{\Psi}^*(p) 
= (2 \pi)^4\:\delta^4(k-p)\: \hat{P}(k)
\eeq
with~$\hat{P} \in {\mathcal{D}}'(\hat{M}, \Lin(V))$ a distributional kernel.
In this homogeneous setting, the scalar product~\eqref{hatH} simplifies to
\beq \label{hatHhom}
\la f | g \ra_{\hat{\H}} = -\int_{\hat{M}} \frac{d^4k}{(2 \pi)^4}\: \Sl f(k)\: |\: \hat{P}(k) \:g(k) \Sr \:,
\eeq
making it possible to construct the Hilbert space~$(\hat{\H}, \la .|. \ra_{\hat{\H}})$
from the vacuum fermionic projector~$\hat{P}$. 
Omitting the hat of the Hilbert space, the corresponding wave evaluation
operator~\eqref{weomom} takes the form
\beq \label{hPsihom}
\widehat{\Psi}(f) = \hat{P} f\:.
\eeq
In order to illustrate our formalism, let us verify that this formula is consistent with~\eqref{nudef}. To this end, we use~\eqref{hPsihom} in~\eqref{tilmudetail} to obtain
\beq \label{dmuform}
\hat{\tilde{P}}(k,p) \,g(p) = \sum_\ell \big( \hat{P}(k) \,e_\ell(k) \big) \;\Sl \hat{P}(p) \,e_\ell(p) \,|\, g(p) \Sr \:.
\eeq
On the other hand, using the completeness relation, for any~$f \in C^0_0(\hat{M}, V)$,
\begin{align*}
f(k) &=  \sum_\ell e_\ell(k) \: \la e_\ell | f \ra_\H \overset{\eqref{hatHhom}}{=}
\sum_\ell e_\ell(k) \int_{\hat{M}} \frac{d^4p}{(2 \pi)^4}\:\Sl e_\ell(p) \,|\, \hat{P}(p)\: f(p) \Sr \\
&= \sum_\ell e_\ell(k) \int_{\hat{M}} \frac{d^4p}{(2 \pi)^4}\:\Sl \hat{P}(p)\: e_\ell(p) \,|\, f(p) \Sr
\end{align*}
(in the last step we used that~$\hat{P}$ maps to the {\em{symmetric}} linear operators on~$V$).
This relation can be written in the shorter form
\[ \sum_\ell e_\ell(k) \:\Sl \hat{P}(p)\: e_\ell(p) \,| = (2 \pi)^4\:\delta^4(k-p)\:. \]
Using this formula in~\eqref{dmuform} gives
\[ \hat{\tilde{P}}(k,p) \,g(p) = (2 \pi)^4\:\hat{P}(k)\:\delta^4(k-p)\: g(p) \:, \]
in agreement with~\eqref{nudef}.

\begin{Remark} (positive definite measures in momentum space) {\em{
In order to put the above formulas into context, we note that the vacuum fermionic projector
in momentum space~$\hat{P}$ can also be used to form a matrix-valued measure~$\nu$ by setting
\[ d\nu(p) = - \hat{P}(p)\: \frac{d^4p}{(2 \pi)^4}\:. \]
Using~\eqref{mupositive}, one sees that this measure is {\em{positive definite}} in the sense that
\[ \Sl v \,|\, \nu(\Omega)\, v \Sr \geq 0 \qquad \text{for all $v \in V$}\:. \]
Working with such positive definite measures in momentum space taking values in~$\Lin(V)$
is very useful for the existence theory in the homogeneous setting as worked out in~\cite{elhom}.
This concept can also be extended to the non-homogeneous setting. To this end, one could
form the measure
\[ d \tilde{\mu}(k,p) := -\hat{\tilde{P}}(k,p)\: \frac{d^4k}{(2 \pi)^4}\: \frac{d^4p}{(2 \pi)^4}\:. \]
According to~\eqref{tildeP}, the kernel of the fermionic projector is the Fourier transform of this measure.

Although this description is very useful for the existence theory, it is not suitable
for the perturbative description. The reason is that the first order perturbation of the fermionic projector
in general is {\em{not a measure}} (this can be seen explicitly in~\eqref{nupert}, where the pole
is well-defined as a distribution, but not as a measure).
This shortcoming can be regarded as an artifact of the perturbative treatment.
A simple analog of this situation is the
one-dimensional example of the family of Dirac measures~$\delta(x-\lambda)$ with~$\lambda \in \R$.
Expanding linearly in the ``coupling constant'' $\lambda$ gives the distribution~$\delta'(x)$, which is
non longer a measure. Due to these technical issues, we here avoid measures in momentum space
and work instead with distributions on~$\hat{M}$. \QEDrem
}} 
\end{Remark}

\subsection{First Order Perturbations of Dirac Sea Configurations} \label{secfirstorder}
In order to get into the position for doing computations, we assume more specifically that
we consider first order perturbations of {\em{regularized Dirac sea configurations}}.
To this end, we choose the fermionic projector of the vacuum as
\beq \label{nuvacthree}
\hat{P}(p) = \sum_{\alpha=1}^3 \hat{P}_{m_\alpha}
\eeq
with
\beq \label{Pmdef}
\hat{P}_m(p) :=\big( \slashed{p}+m \big) \: \delta\big(p^2-m^2 \big)\: \Theta(-\omega)\:
e^{\varepsilon \omega} \:,
\eeq
where~$m_\alpha>0$ are the masses of the tree generations of Dirac particles,
and the parameter~$\varepsilon>0$ describes the ultraviolet regularization
(for the general context of the Dirac equation in Minkowski space see for example
the textbooks~\cite{bjorken, peskin+schroeder, weinberg}).
We always assume that the three masses are pairwise distinct,
\[ m_\alpha \neq m_\beta \qquad \text{if~$\alpha \neq \beta$}\:. \]
We point out that we always work with the {\em{regularized}} Dirac seas.
We also note that we only consider the case that all seas are {\em{massive}}; the reason 
for this assumption will be explained in Remark~\ref{remmassive} below.
The Hilbert space~$(\H, \la .|. \ra_\H)$ is constructed as explained after~\eqref{hatH}
for the scalar product~\eqref{hatHhom}. 
The resulting wave evaluation operator can be written in the
form~\eqref{hPsihom}.
For clarity we remark that, rewriting the scalar product~\eqref{hatHhom} in position space,
one recovers the usual scalar product on Dirac solutions, up to small corrections due to the ultraviolet regularization.
This is worked out in detail in Appendix~\ref{appscalar}.

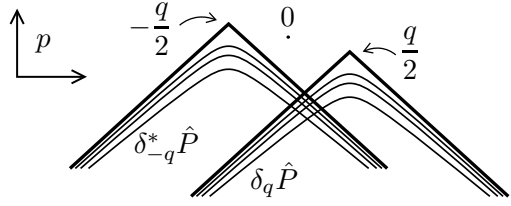
\begin{figure}
\psset{xunit=.5pt,yunit=.5pt,runit=.5pt}
\begin{pspicture}(378.17568321,143.10805433)
{
\newrgbcolor{curcolor}{0 0 0}
\pscustom[linewidth=2.54532664,linecolor=curcolor]
{
\newpath
\moveto(137.3502085,0.94065276)
\lineto(257.33430047,110.28286126)
\lineto(377.31846425,0.94065276)
}
}
{
\newrgbcolor{curcolor}{0 0 0}
\pscustom[linewidth=1.27266523,linecolor=curcolor]
{
\newpath
\moveto(373.31904378,0.94065276)
\curveto(337.32165165,33.13775638)(301.35283276,65.30931418)(280.67609575,81.41779292)
\curveto(259.99939654,97.52626788)(254.67014929,97.52626788)(234.00486425,81.42837559)
\curveto(213.33957921,65.33052111)(177.34507843,33.13596488)(141.34967811,0.94061118)
}
}
{
\newrgbcolor{curcolor}{0 0 0}
\pscustom[linewidth=1.27266523,linecolor=curcolor]
{
\newpath
\moveto(369.31954772,0.94065276)
\curveto(335.32197921,30.70783008)(301.32716976,60.47261874)(280.99610835,75.35470126)
\curveto(260.66504693,90.23678756)(253.99958551,90.23678756)(233.66939339,75.35369969)
\curveto(213.33923906,60.47061181)(179.34441449,30.70583449)(145.34914016,0.94065276)
}
}
{
\newrgbcolor{curcolor}{0 0 0}
\pscustom[linewidth=1.27266523,linecolor=curcolor]
{
\newpath
\moveto(363.32034142,0.94065276)
\curveto(331.32244535,27.06302756)(299.32738394,53.18311197)(279.99619654,66.24284237)
\curveto(260.66504693,79.30257276)(253.99958551,79.30257276)(234.66922961,66.24185969)
\curveto(215.3389115,53.18114284)(183.34381228,27.06104709)(151.34835402,0.94065276)
}
}
{
\newrgbcolor{curcolor}{0 0 0}
\pscustom[linewidth=1.51181105,linecolor=curcolor]
{
\newpath
\moveto(5.67610583,141.41779292)
\lineto(5.67610583,91.41779166)
\lineto(55.67610331,91.41779166)
}
}
{
\newrgbcolor{curcolor}{0 0 0}
\pscustom[linewidth=1.51181105,linecolor=curcolor]
{
\newpath
\moveto(0.67610835,131.41779418)
\lineto(5.67610583,141.4177967)
\lineto(10.67610709,131.41779418)
}
}
{
\newrgbcolor{curcolor}{0 0 0}
\pscustom[linewidth=1.51181105,linecolor=curcolor]
{
\newpath
\moveto(46.8531326,96.4670704)
\lineto(56.85312756,91.46707292)
\lineto(46.8531326,86.46707166)
}
}
{
\newrgbcolor{curcolor}{0 0 0}
\pscustom[linewidth=2.54532664,linecolor=curcolor]
{
\newpath
\moveto(45.69202016,22.07558567)
\lineto(165.67610457,131.41779418)
\lineto(285.66030992,22.07558567)
}
}
{
\newrgbcolor{curcolor}{0 0 0}
\pscustom[linewidth=1.27266523,linecolor=curcolor]
{
\newpath
\moveto(281.66088945,22.07558567)
\curveto(245.66349732,54.27268551)(209.69464063,86.44425465)(189.01792252,102.55272583)
\curveto(168.34120063,118.66120079)(163.01195339,118.66120079)(142.34667591,102.56335008)
\curveto(121.68139843,86.46549937)(85.68689008,54.27093937)(49.69148976,22.07558567)
}
}
{
\newrgbcolor{curcolor}{0 0 0}
\pscustom[linewidth=1.27266523,linecolor=curcolor]
{
\newpath
\moveto(277.66139339,22.07558567)
\curveto(243.66382488,51.84275544)(209.66897764,81.60755544)(189.33791244,96.48963796)
\curveto(169.00685102,111.37172048)(162.34138961,111.37172048)(142.01120882,96.4886326)
\curveto(121.68103181,81.60554473)(87.68622614,51.84077118)(53.69095181,22.07558567)
}
}
{
\newrgbcolor{curcolor}{0 0 0}
\pscustom[linewidth=1.27266523,linecolor=curcolor]
{
\newpath
\moveto(271.66218709,22.07558567)
\curveto(239.66429102,48.19795292)(207.66919181,74.31804866)(188.33801575,87.37777528)
\curveto(169.00685102,100.43750567)(162.34138961,100.43750567)(143.01105638,87.37679638)
\curveto(123.68072693,74.31608709)(91.68562394,48.19598)(59.69016567,22.07558567)
}
}
{
\newrgbcolor{curcolor}{0 0 0}
\pscustom[linestyle=none,fillstyle=solid,fillcolor=curcolor]
{
\newpath
\moveto(211.59262713,121.48835826)
\curveto(211.59262713,121.10235138)(211.23401441,120.78943097)(210.79164321,120.78943097)
\curveto(210.349272,120.78943097)(209.99065928,121.10235138)(209.99065928,121.48835826)
\curveto(209.99065928,121.87436515)(210.349272,122.18728556)(210.79164321,122.18728556)
\curveto(211.23401441,122.18728556)(211.59262713,121.87436515)(211.59262713,121.48835826)
\closepath
}
}
{
\newrgbcolor{curcolor}{0 0 0}
\pscustom[linewidth=1.06627656,linecolor=curcolor]
{
\newpath
\moveto(211.59262713,121.48835826)
\curveto(211.59262713,121.10235138)(211.23401441,120.78943097)(210.79164321,120.78943097)
\curveto(210.349272,120.78943097)(209.99065928,121.10235138)(209.99065928,121.48835826)
\curveto(209.99065928,121.87436515)(210.349272,122.18728556)(210.79164321,122.18728556)
\curveto(211.23401441,122.18728556)(211.59262713,121.87436515)(211.59262713,121.48835826)
\closepath
}
}
{
\newrgbcolor{curcolor}{0 0 0}
\pscustom[linewidth=0.99999871,linecolor=curcolor]
{
\newpath
\moveto(267.82883906,110.56572866)
\curveto(271.67979969,112.65900488)(276.17902488,113.5390074)(280.5348926,113.05090788)
\curveto(284.77941543,112.57528835)(288.87003591,110.80447008)(292.12156346,108.03506662)
}
}
{
\newrgbcolor{curcolor}{0 0 0}
\pscustom[linewidth=0.99999871,linecolor=curcolor]
{
\newpath
\moveto(272.89227213,117.07416961)
\lineto(268.04015244,110.89382567)
\lineto(275.21365795,109.51430567)
}
}
{
\newrgbcolor{curcolor}{0 0 0}
\pscustom[linewidth=0.99999871,linecolor=curcolor]
{
\newpath
\moveto(155.54552315,134.65279386)
\curveto(151.49050961,136.38142126)(146.98758425,137.04655008)(142.60597039,136.56410473)
\curveto(137.47107402,135.99871764)(132.52189228,133.84737654)(128.60568945,130.4783737)
}
}
{
\newrgbcolor{curcolor}{0 0 0}
\pscustom[linewidth=0.99999871,linecolor=curcolor]
{
\newpath
\moveto(151.04811213,140.03292488)
\lineto(155.95934362,134.65279764)
\lineto(148.43490898,132.54391575)
}
\rput[bl](205,130){$0$}
\rput[bl](90,108){$\displaystyle -\frac{q}{2}$}
\rput[bl](295,90){$\displaystyle \frac{q}{2}$}
\rput[bl](20,110){$p$}
\rput[bl](94,25){$\delta_{-q}^* \hat{P}$}
\rput[bl](182,0){$\delta_q \hat{P}$}
}
\end{pspicture}
\caption{The distribution~$\delta \hat{P} = \delta_q \hat{P}+\delta_{-q}^* \hat{P}$.}
\label{figdelP}
\end{figure}

We next consider first variations of the system. In order to keep the Hilbert space structure fixed,
it is preferable to vary the wave evaluation operator as in~\eqref{Psivary}.
Denoting first variations by a~$\delta$, we obtain
\[ \widehat{\tilde{\Psi}} = \hat{\Psi} + \delta \hat{\Psi}
\qquad \text{with} \qquad \delta \hat{\Psi} \::\: \H \rightarrow {\mathcal{D}}'(\hat{M}, V) \:. \]
The resulting fermionic projector takes the form
\[ \hat{\tilde{P}}(k,p) = (2 \pi)^4\:\delta^4(k-p)\: \hat{P}(k)
+ \big( \delta \widehat{\Psi} \big)(k) \times \widehat{\Psi}^*(p)
+ \widehat{\Psi}(k) \times \big( \delta \widehat{\Psi}^* \big)(p) \:. \]
It is convenient to parametrize by the change of momentum~$q:=k-p$.
We assume that
\beq \label{nuqdef}
\delta \widehat{\Psi}\Big( p+\frac{q}{2} \Big) \bigg) \times \widehat{\Psi}^*\Big( p-\frac{q}{2} \Big)
= \delta_q \hat{P}(p)
\eeq
with a new distribution~$\delta_q \hat{P} \in {\mathcal{D}}'(\hat{M}, V)$. Combining this formula with~\eqref{nudef}, we obtain
\beq \label{tilmuone}
\hat{\tilde{P}}\Big( p+\frac{q}{2}, p-\frac{q}{2} \Big)
= (2 \pi)^4\:\delta^4(q)\: \hat{P}(p)+ \delta_q \hat{P}(p) + \delta_{-q}^* \hat{P}(p)
\eeq
with~$\delta_{-q}^* \hat{P}(p):= (\delta_{-q} \hat{P}(p))^*$.
Using~\eqref{tildeP}, the corresponding perturbation of the kernel of the fermionic projector is computed by
\beq \label{delPmink}
\delta P(x,y) = \int_{\hat{M}} \frac{d^4q}{(2 \pi)^4} \: e^{-i\frac{q}{2}\, (x+y)}
\: \int_{\hat{M}} \frac{d^4p}{(2 \pi)^4}\: \big( \delta_q \hat{P} + \delta_{-q}^* \hat{P} \big)(p)\: e^{ip \xi }\:,
\eeq
where we set~$\xi=y-x$. The distributions~$\delta_q \hat{P}$ and~$\delta_{-q}^* \hat{P}$
are illustrated in Figure~\ref{figdelP}.
We finally point out that in the following analysis we always assume that~$q$ is non-zero,
\beq \label{qnonozero}
q \neq 0 \:.
\eeq
This is justified because our formulas for~$\delta P$ will be regular in the limit~$q \rightarrow 0$
(for details see the paragraph after Definition~\ref{defJvary}). But clearly, we must treat all three
cases that~$q$ is timelike, spacelike or lightlike.

\subsection{Evaluation in the Continuum Limit} \label{secdQcont}
In~\cite{cfs} the kernels~$Q(x,y)$ and~$\delta Q(x,y)$ as introduced in~\eqref{delLdef}, \eqref{lfe}
were computed to degree four on the light cone. The kernel~$Q(x,y)$ was shown to vanish, i.e.
\[ Q^\alpha|_\beta(x,y) = (\deg < 4) \:. \]
(however, this kernel is non-zero to lower degree on the light cone, as is worked out in some
detail in~\cite[Section~5]{noether} and~\cite{dirac}; we will come back to this point in Section~\ref{secbeyond}).
Consequently, also the parameter~${\mathfrak{r}}$ in the EL equations~\eqref{ELQ} vanishes,
\[ {\mathfrak{r}} = (\deg < 4) \:. \]
The kernel~$\delta Q(x,y)$, on the other hand, is non-zero to degree four on the light cone.
More precisely, in~\cite[Section~3.7]{cfs} it was shown for a system of three generations of leptons
that the kernels in Lemma~\ref{lemmaQ2} take the form
\begin{align}
K^\alpha_{\;\;\,\gamma}|_\beta^{\;\;\,\delta}(x,y) &= -i\,\frac{c_1}{(\varepsilon t)^3}
\: \big(\gamma^5 \slashed{\xi} \big)^\alpha_\beta \:K_0(x,y)\:
\big(\gamma^5 \slashed{\xi} \big)^\delta_\gamma + (\deg < 4) \label{cl1} \\
K^{\alpha \gamma}|_{\beta \delta}(x,y) &= -i \frac{c_2}{(\varepsilon t)^3}\:
\big( \gamma^5 \slashed{\xi} \big)^\alpha_\beta \:K_0(x,y)\:
\big( \gamma^5 \slashed{\xi} \big)^\gamma_\delta + (\deg < 4) \:, \label{cl2}
\end{align}
where~$\gamma^5$ denotes the pseudo-scalar matrix. Moreover, we
set~$t=\xi^0$, introduced two regularization
parameters~$c_1, c_2 \in \R$ and set
\beq \label{K0def}
K_0(x,y) := \frac{i}{4 \pi^2} \:\delta(\xi^2)\: \epsilon(\xi^0) =
\int_{\hat{M}} \frac{d^4p}{(2 \pi)^4} \: \delta(p^2)\: \epsilon(p^0)\: e^{-i p (x-y)}
\eeq
(note that this distribution has the remarkable property that it has, up to a constant,
the same form in position and momentum space).
Moreover, the regularization parameters satisfy the inequalities
\beq \label{c12rel}
c_1 > 0 \qquad \text{and} \qquad |c_2| \leq c_1 \:.
\eeq
These formulas hold, provided that the first variation~$\delta P(x,y)$ of the kernel of the fermionic
projector is less singular on the light cone than~$P(x,y)$; more precisely if
\beq \label{delPcond}
\Tr \big( \slashed{\xi}\: \delta P(x,y) \big) = (\deg < 1) \:.
\eeq
This condition is essential, because otherwise the formalism
of the continuum limit does not apply.

The appearance of the pseudo-scalar matrices in~\eqref{cl1} and~\eqref{cl2} corresponds to the
fact that for the system of leptons considered in~\cite[Chapter~3]{cfs} only axial gauge fields
appear. For a system including neutrinos and quarks, it was shown in~\cite[Chapter~5]{cfs}
that for electromagnetic perturbations one also gets~\eqref{cl1} and~\eqref{cl2}, but now without
the pseudo-scalar matrices. In order to treat all these different cases in a unified way,
it is useful to observe that the pseudo-scalar matrices in~\eqref{cl1} and~\eqref{cl2} can be
absorbed into the factors~$\delta P$ in~\eqref{Qvary} (by replacing an axial perturbation by a
vectorial perturbation). With this in mind, in what follows we shall omit the pseudo-scalar matrices.

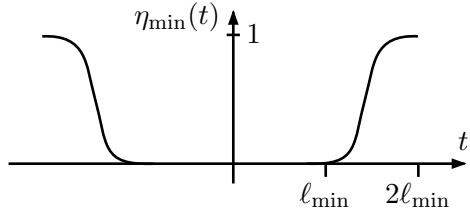
\begin{figure}
\psset{xunit=.5pt,yunit=.5pt,runit=.5pt}
\begin{pspicture}(385,170)
{
\newrgbcolor{curcolor}{0 0 0}
\pscustom[linewidth=2,linecolor=curcolor]
{
\newpath
\moveto(190,20)
\lineto(190,150)
}
}
{
\newrgbcolor{curcolor}{0 0 0}
\pscustom[linestyle=none,fillstyle=solid,fillcolor=curcolor]
{
\newpath
\moveto(190,150)
\lineto(186.67708333,140.69583333)
\lineto(190,140.69583333)
\lineto(193.32291667,140.69583333)
\closepath
}
}
{
\newrgbcolor{curcolor}{0 0 0}
\pscustom[linewidth=1,linecolor=curcolor]
{
\newpath
\moveto(190,150)
\lineto(186.67708333,140.69583333)
\lineto(190,140.69583333)
\lineto(193.32291667,140.69583333)
\closepath
}
}
{
\newrgbcolor{curcolor}{0 0 0}
\pscustom[linewidth=2,linecolor=curcolor]
{
\newpath
\moveto(20,35)
\lineto(365,35)
}
}
{
\newrgbcolor{curcolor}{0 0 0}
\pscustom[linestyle=none,fillstyle=solid,fillcolor=curcolor]
{
\newpath
\moveto(365,35)
\lineto(355.6958333,38.3229167)
\lineto(355.6958333,35)
\lineto(355.6958333,31.6770833)
\closepath
}
}
{
\newrgbcolor{curcolor}{0 0 0}
\pscustom[linewidth=1,linecolor=curcolor]
{
\newpath
\moveto(365,35)
\lineto(355.6958333,38.3229167)
\lineto(355.6958333,35)
\lineto(355.6958333,31.6770833)
\closepath
}
}
{
\newrgbcolor{curcolor}{0 0 0}
\pscustom[linewidth=2,linecolor=curcolor]
{
\newpath
\moveto(330,35)
\lineto(330,25)
}
}
{
\newrgbcolor{curcolor}{0 0 0}
\pscustom[linewidth=2,linecolor=curcolor]
{
\newpath
\moveto(185,132.5)
\lineto(195,132.5)
}
}
{
\newrgbcolor{curcolor}{0 0 0}
\pscustom[linewidth=2,linecolor=curcolor]
{
\newpath
\moveto(45.4689,131.3754)
\curveto(79.5472,133.0696)(78.663,109.3323)(84.8716,86.2339)
\curveto(94.978,48.6346)(89.5417,33.4724)(129.5,35)
}
}
{
\newrgbcolor{curcolor}{0 0 0}
\pscustom[linewidth=2,linecolor=curcolor]
{
\newpath
\moveto(329.6809,131.4135)
\curveto(295.6026,133.1077)(296.4868,109.3704)(290.2782,86.272)
\curveto(280.1718,48.6727)(285.6081,33.5105)(245.6498,35.0381)
}
}
{
\newrgbcolor{curcolor}{0 0 0}
\pscustom[linewidth=2,linecolor=curcolor]
{
\newpath
\moveto(259.913117,35)
\lineto(259.913117,25)
}
}
\rput[bl](240,2){$\ell_{\min}$}
\rput[bl](305,2){$2 \ell_{\min}$}
\rput[bl](200,126){$1$}
\rput[bl](360,45){$t$}
\rput[bl](115,135){$\eta_{\min}(t)$}
\end{pspicture}
\caption{A typical choice of the cutoff function~$\eta_{\min}(t)$.}
\label{figetamin}
\end{figure}

Moreover, in the formalism of the continuum limit, the kernel needs to be evaluated
{\em{away from the origin}}, which means that
\beq \label{lmindef}
\big| \xi^0 \big| \gg \ell_{\min}
\eeq
for a suitable parameter~$\ell_{\min}$ which lies between the regularization scale and the
length scale of macroscopic physics,
\[ \varepsilon \ll \ell_{\min} \ll \ell_\text{macro} \]
(for more details see Remark~\ref{remell} below).
Near the origin (i.e.\ if~$|\xi^0| \lesssim \ell_{\min}$), there is more freedom to satisfy the
linearized field equations. For this reason, we here simply disregard this region.
In order to implement these results in a convenient way, we regularize the kernel~$K_0$
in~\eqref{cl1} and~\eqref{cl2} by setting
\begin{align}
K^\alpha_{\;\;\,\gamma}|_\beta^{\;\;\,\delta}(x,y) &= \frac{c_1}{\varepsilon^3}
\: \frac{\slashed{\xi}^\alpha_\beta\: \eta_{\min}(t)}{t^2} \:\Big(-i t\,K_0(x,y) \Big)\:
\frac{\slashed{\xi}^\delta_\gamma\: \eta_{\min}(t)}{t^2} + (\deg < 4) \label{cn1} \\
K^{\alpha \gamma}|_{\beta \delta}(x,y) &= \frac{c_2}{\varepsilon^3}\: \frac{\slashed{\xi}^\alpha_\beta\: \eta_{\min}(t)}{t^2} \:\Big(-i t\,K_0(x,y) \Big)\:
\frac{\slashed{\xi}^\delta_\gamma\: \eta_{\min}(t)}{t^2}  + (\deg < 4) \:, \label{cn2}
\end{align}
where~$\eta_{\min} \in C^\infty(\R^+_0)$ is a non-negative smooth function with (see Figure~\ref{figetamin})
\[ \eta_{\min}|_{[-\ell_{\min}, \ell_{\min}]} \equiv 0 \qquad \text{and} \qquad
\eta_{\min}|_{\R \setminus [-2 \ell_{\min}, 2 \ell_{\min}]} \equiv 1 \:. \]

Using these results, we can write~\eqref{Qvary} more compactly as
\beq \label{delQ}
\delta Q(x,y) = \Big( c_1\: \Tr \big( \delta P(x,y)\: \slashed{\xi} \big) +
c_2\: \Tr \big( \delta P(y,x)\: \slashed{\xi} \big) \Big)\:\big(-i t\, K_0(x,y) \big)\:\frac{\eta_{\min}^2(t)}{\varepsilon^3\:t^4}
\:\slashed{\xi} \:.
\eeq
Moreover, the formula for second variations in Proposition~\ref{prpsecond}
simplifies to
\begin{align}
\delta^2 \Sact^\text{\rm{eff}}(\rho) &= 2 \int_\F d\rho(x) \int_\F d\rho(y)\:
\big(-i t\, K_0(x,y) \big)\: \frac{\eta_{\min}^2(t)}{\varepsilon^3\:t^4} \notag \\
&\qquad \times
\Big\{ c_1\: \Tr\big(\slashed{\xi}\, \delta P(y,x) \big)\: \Tr \big( \slashed{\xi}\: \delta P(x,y) \big)
+ c_2\: \re \Tr\big(\slashed{\xi}\, \delta P(y,x) \big)^2 \Big\} \:. \label{d2Scont}
\end{align}
Using~\eqref{K0def}, the kernel in~\eqref{delQ} can be written more explicitly as
\[ \big(-i t\, K_0(x,y) \big)\: \frac{\eta_{\min}^2(t)}{\varepsilon^3\: t^4}
= \frac{1}{4 \pi^2}\: \delta(\xi^2)\: \frac{\eta_{\min}^2(t)}{\varepsilon^3\, |t|^3} \:. \]
This kernel is non-negative, implying positivity also of the second variations~\eqref{d2Scont},
\[ \delta^2 \Sact^\text{\rm{eff}}(\rho) \geq 0 \:. \]
This is in agreement with the general positivity results in~\cite{positive}.

We finally explain the significance and the scaling behavior of the parameter~$\ell_{\min}$  in~\eqref{lmindef}.
\begin{Remark} {\bf{(What is known about $\ell_{\min}$?)}} {\em{ \label{remell}
The parameter~$\ell_{\min}$ is the length scale down to which the formalism of the continuum
limit applies. This length scale depends on the microscopic structure of spacetime, which is largely unknown.
One possible path for clarifying these issues is to analyze 
how the regularization affects the value of the causal action of the vacuum.
The ``optimal regularization'' obtained by minimizing this action would tell us about the microscopic
structure of physical spacetime and would determine the length scale~$\ell_{\min}$.
This minimization procedure has been carried out abstractly in~\cite{elhom}, and the existence of minimizers
has been proven. However, at present the resulting minimizers are not known in sufficient detail
for calculating or estimating~$\ell_{\min}$.
In a more computational approach,
in~\cite{reg} it was analyzed whether and how the EL equations can be
satisfied in the vacuum by adapting regularization. These extended computations
suggest that the resulting spacetimes should have a {\em{microscopic multi-layer structure}},
where spacelike and timelike separation change.
Working with cutoff functions in momentum space which decay at infinity polynomially, the
resulting so-called {\em{regularization tails}} give a lot of freedom to modify and adapt these layers.
Moreover, one can analyze which configuration of regularization tails is favorable
(in the sense that the causal action becomes small). Qualitatively speaking, it is favorable
to arrange that~$\ell_{\min}$ is large. A general result of the computations in~\cite{reg} is that the
length scale~$\ell_{\min}$ can be made at least as large as
\[ \ell_{\min} \gtrsim \frac{1}{m}\: (\varepsilon m)^\alpha \qquad \text{with} \qquad 0 < \alpha < 1 \:. \]
This implies that the number of fields~$N$ in~\eqref{Nscaleintro} really tends to infinity
as~$\varepsilon \searrow 0$.
However, at present it is not clear how small the parameter~$\alpha$ can be arranged to be.
It is also unclear whether the power law scaling can be improved to for example a logarithmic scaling.
}} $\hspace*{0,5em}$ \QEDrem
\end{Remark}

\section{Hilbert Space Setting for Solving the Linearized Field Equations} \label{secabstract}
Before delving into the detailed computations, in this section we provide the abstract framework
for solving the linearized field equations. We will make essential use of the positivity of the
second variations of the causal action. These positivity properties were already used in the existence
proof (see~\cite[Section~3.3]{linhyp}). It turns out that the resulting functional analytic setting
also sets the stage for the computational approach.

\subsection{Construction of Linearized Solutions} \label{secabslin}
The abstract setting is best described in the jet formalism as introduced in~\cite{noether};
see also~\cite[Section~2]{fockbosonic}. Here we use the same notation, but keep the setting
simpler: First, we omit the notions which specify the
conditions on differentiability and the existence of surface layer integrals, noting that in our later
computations these conditions will always be satisfied.
Second, as already explained in Remark~\ref{remscal2},
we omit the scalar components of the jets. Thus a jet is nothing but a vector field~$\bu$ which
to every point~$x$ of spacetime associates its first variation~\eqref{ufermi}.
In a differential geometric language, the vector~$\bu(x)$ lies in the tangent space~$T_x \F$,
and the vector field~$\bu \in \Gamma(M, T\F)$ is a smooth section in~$T\F$ along~$M$.
We denote the space of all such vector fields by~$\J$.
As in~\eqref{resELL}, we write variational derivatives of the Lagrangian by~$D_{1,\bu}$ and~$D_{2,\bu}$,
where the subscripts refer to the two arguments~$x$ and~$y$ of the Lagrangian, respectively.

We begin by choosing a subspace~$\Jvary \subset \J$ of jets which satisfy all the necessary regularity
and decay conditions (this and all the subsequent jet spaces will be chosen concretely in Section~\ref{secret}).
In particular, we need to ensure that the following surface layer integrals are well-defined and finite
for any time~$t \in \R$,
\begin{align}
\gamma^t \::\: \Jvary& \rightarrow \R \qquad\qquad\quad\: \text{(conserved one-form)} \notag \\
\gamma^t(\bv) &= \int_{\Omega^t} d\rho(x) \int_{M \setminus \Omega^t} d\rho(y)\:
\big( D_{1,\bv} - D_{2,\bv} \big) \L(x,y) \label{gamma} \\
\sigma^t \::\: \Jvary& \times \Jvary \rightarrow \R  \qquad \text{(symplectic form)} \notag \\
\sigma^t(\bu, \bv) &= \int_{\Omega^t} d\rho(x) \int_{M \setminus \Omega^t} d\rho(y)\:
\big( D_{1,\bu} D_{2,\bv} - D_{2,\bu} D_{1,\bv} \big) \L(x,y) \:. \label{sigma}
\end{align}
Moreover, we assume that the second variations of the action are well-defined on~$\Jvary$, giving
rise to the bilinear form
\[ \la .,. \ra := \delta^2 \Sact \::\: \Jvary \times \Jvary \rightarrow \R \:. \]
Since~$\rho$ is assumed to be a minimizing measure, the second variations are non-negative.
Therefore, the bilinear form~$\delta^2 \Sact$ is positive semi-definite. 
Dividing out the null space and forming the completion, we obtain a Hilbert space
denoted by~$(\h, \la .,. \ra)$.

We now let~$\bw$ be a linear functional on~$\Jvary$ which is bounded with respect
to the Hilbert space norm on~$\h$, i.e.\
\[ \bw : \Jvary \rightarrow \R \qquad \text{with} \qquad \big| \bw(\bu) \big| \leq C\: \|\bu\| 
\qquad \text{for all~$\bu \in \Jvary$} \:. \]
Then the Fr{\'e}chet-Riesz theorem yields a unique
vector~$\bv \in \h$ such that (for the necessary background in functional analysis
we refer for example to the textbooks~\cite{reed+simon, lax})
\beq \label{weaksol}
\delta^2 \Sact \big( \bv, \bu \big) = \bw(\bu) \qquad \text{for all~$\bu \in \Jvary$}\:.
\eeq

Before going on, we make a few remarks.
\begin{Remark} {\em{
\bitem
\item[{\rm{(i)}}] It is important to note that the solutions of the homogeneous linearized field equations
lie in the null space of the second variations. In order to make this statement mathematically precise,
for a solution~$\delta \Psi$ of the linearized field equations~\eqref{lfe}, we 
let~$\bu$ be the jet corresponding to the variation~$\bu(x) \in T_x\F$ of the
spacetime point operators~\eqref{ufermi}. The space of all these jets is denoted by~$\Jlin$.
Then
\[ \Jlin \cap \Jvary \subset \ker \delta^2 S := \big\{ \bu \in \Jvary \:\big|\:
\delta^2 S(\bu, \bv)=0 \quad \text{for all~$\bv \in \Jvary$} \big\} \:. \]
Therefore, the linearized solutions are modded out when forming the Hilbert space~$\h$.
In other words, the vectors of~$\h$ are equivalence classes of solutions obtained by adding homogeneous
solutions. As an immediate consequence of this fact, it is not possible to define the
surface layer integrals~\eqref{gamma} and~\eqref{sigma} on~$\h$: in general being non-zero on
homogeneous solutions, they are ill-defined on the above equivalence classes.
\item[{\rm{(ii)}}] This method of working modulo homogeneous solutions means in particular that
vectors in~$\h$ do not distinguish retarded solutions from advanced solutions.
Our strategy for constructing for example retarded solutions is to choose a distinguished representative
of the equivalence class. This will be explained in detail in the following section.
\QEDrem
\eitem
}}
\end{Remark}

\subsection{Distinguishing Retarded Solutions} \label{secretarded}
As already mentioned in the previous remark, the solution~$\bv \in \h$
constructed in~\eqref{weaksol} is an equivalence class of jets, formed by modding out
homogeneous solutions. In this section we show that there is a unique representative
which vanishes in the past, in a sense which will be made precise.
We again denote the space of homogeneous solutions of the linearized field equations~\eqref{lfe} by~$\Jlin$.
The linearized solutions do not necessarily need to be in~$\Jvary$, but we demand (and shall
see later) that the above surface layer integrals~\eqref{gamma} and~\eqref{sigma}
are well-defined on~$\Jlin$.
Then on the linearized solutions, the surface layer integrals are conserved in time
(see~\cite{jet, osi}), i.e.\ for all~$\bu, \bv \in \Jlin$,
\beq \label{conserve}
\gamma^t(\bu) \text{ and } \sigma^t(\bu, \bv) \qquad \text{are independent of~$t$} \:.
\eeq
We next introduce the corresponding null space by
\[ \Jnull := \big\{ \bu \in \Jlin \:\big|\: \gamma^t(\bu) = 0 = \sigma^t(\bu, \bv) \;\;
\forall \,\bv \in \Jlin \big\} \subset \Jlin \:. \]
The fact that all surface layer integrals vanish for jets in~$\Jnull$
can be understood physically that these solutions cannot be detected by measurements.
With this in mind, we may consider them as unphysical and disregard them.

Given an inhomogeneous solution~$\bv \in \h$ constructed in~\eqref{weaksol}, we want
to choose a representative with the property that for all~$\bu \in \Jlin$,
\beq \label{zerolimit}
\lim_{t \rightarrow -\infty} \gamma^t(\bv) = 0 = \lim_{t \rightarrow -\infty} \sigma^t(\bv, \bu) \:.
\eeq
This condition makes the solution unique up to homogeneous null solutions,
\[ \bv \mod \Jnull \:. \]
In other words, the obtained representative~$\bv$ is unique up to unphysical jets which we may disregard.
In the remainder of this paper, we shall always work with these distinguished representatives.
We refer to these distinguished representatives as {\em{retarded jets}}. The corresponding jet space
is denoted by~$\Jret \subset \J$.

It remains to prove the existence of a representative with the property~\eqref{zerolimit}.
Since the focus of the present paper is on the computational aspects, we shall not give this proof in detail,
but merely explain the method in words. Assume that the inhomogeneity~$\bw$ has compact support
in spacetime. Then, using the decay properties of the Lagrangian,
the solution~$\bv \in \h$ is a homogeneous solution in the distant past of the support of the inhomogeneity,
i.e.\ for all~$(t, \vec{x})$ with~$t<t_0$ and sufficiently small~$t_0$.
We choose an arbitrary representative of~$\bv$ which, for ease in notation, we denote again by~$\bv$.
Now there are two alternative methods.
The first method is to take the homogeneous solution at time~$t_0$ as initial data, and to solve the
Cauchy problem for the homogeneous equation using energy methods
(as worked out in~\cite{linhyp}). Denoting the resulting solution by~$\bu \in \Jlin$, the
difference~$\bv-\bu$ is the desired inhomogeneous solution which vanishes in the past~\eqref{zerolimit}.
The second method is to make use of the conservation law~\eqref{conserve}:
The time evolution operator acting from, say, time~$t$ to time~$t+\Delta t$ is a mapping on~$\Jlin$ which preserves the conserved one-form and the symplectic form. Using that Minkowski space is static,
we can iterate this mapping to obtain a corresponding
time evolution operator for arbitrary large times. This shows that every homogeneous solution
defined on a time strip can be extended to a global homogeneous solution.
In particular, the homogeneous solution~$\bv$ defined for~$t<t_0$ can be extended to a
global homogeneous solution~$\bu$. Then the difference~$\bv -\bu$ is again the desired inhomogeneous
solution which vanishes in the past.

\subsection{A Hierarchical Analysis in Orders of the Regularization Length} \label{sechierarchy}
In Section~\ref{secdQcont} we computed the second variations to leading order in
an expansion for small~$\varepsilon$. This raises the question how to treat the higher orders
in this expansion. We now give a systematic procedure for constructing linearized solutions to
every order in~$\varepsilon$. We begin by expanding the second variations as
\beq \label{d2Shierarchy}
\delta^2 \Sact(.,.) = \frac{1}{\varepsilon^3}\: \la .,. \ra^{(0)} +
\frac{1}{\varepsilon^2}\: \la .,. \ra^{(1)} +
\frac{1}{\varepsilon}\: \la .,. \ra^{(2)} +  \la .,. \ra^{(3)} \:.
\eeq
Clearly, $\la .,. \ra^{(0)}$ was computed in~\eqref{d2Scont}. The higher orders~$\la .,. \ra^{(p)}$
with~$p \in \{1,2,3\}$ have not yet been computed systematically, but could be calculated in a similar way.

We want to treat~$\varepsilon$ as a small parameter. This suggests that we can treat~$\la .,. \ra^{(1)}$
as a ``small perturbation'' of~$\la .,. \ra^{(0)}$, etc.\ However, there is the difficulty that~$\la .,. \ra^{(1)}$
might be non-zero on the kernel of~$\la .,. \ra^{(0)}$.
For this reason, we proceed inductively as follows.
We begin with the bilinear form~$\la .,. \ra^{(0)}$ on~$\Jret \times \Jret$.
Dividing out the null space and forming the completion exactly as explained for~$\la .,. \ra$
in the previous section gives a Hilbert space denoted by~$(\h^{(0)}, \la .,. \ra^{(0)})$.

Next, we want to construct the Hilbert space~$(\h^{(1)}, \la .,. \ra^{(1)})$ on the null space of~$\la .,. \ra^{(0)}$.
To this end, we consider sequences~$(u_n)_{n \in \N}$ in~$\Jret$ which are Cauchy sequences
with respect to~$\|.\|^{(1)}$ and converge to zero in~$\h^{(0)}$, i.e.\
\[ \|u_l - u_m \|^{(1)} \xrightarrow{l,m \rightarrow \infty} 0 \qquad
\text{and} \qquad \| u_n \|^{(0)} \xrightarrow{n \rightarrow \infty} 0 \:. \]
The Hilbert space~$\h^{(1)}$ is then defined as the space of all such sequences with the scalar product
\[ \la (u_l), (u_m) \ra^{(1)} := \lim_{l,m \rightarrow \infty} \la u_l, u_m \ra^{(1)} \:. \]
It is by construction a Hilbert space which is orthogonal to~$\h^{(0)}$.
Next, we construct~$\h^{(2)}$ from Cauchy sequences with respect to~$\|.\|^{(2)}$ which
tend to zero in both~$\h^{(0)}$ and~$\h^{(1)}$. Proceeding inductively, we get
mutually orthogonal Hilbert spaces which altogether span~$\h$, i.e.\
\[ \h := \h^{(0)} \oplus \h^{(1)} \oplus \h^{(2)} \oplus \h^{(3)}  \:. \]

Our next task is to expand the solution~$\bv$ in powers of~$\varepsilon$.
In preparation, we note that, in the applications, the inhomogeneity~$\bw$
arises from a perturbation of the Lagrangian by a first variation~$\delta P(x,y)$ of the
kernel of the fermionic projector (for example by introducing an additional Dirac wave function).
When computing the perturbation of the Lagrangian, the resulting dual jet~$\bw$ also has
a scaling in~$\varepsilon$. More precisely, testing with jets~$\bu^{(p)} \in \h^{(p)}$
(for~$p \in \{0,\ldots, 3\}$), the inhomogeneity has an expansion of the form
\[ \bw\big( \bu^{(p)} \big) = \frac{1}{\varepsilon^3}
\sum_{p=0}^3 \varepsilon^p\: \bw^{(p)} \big( \bu^{(p)} \big) . \]
Next, we decompose the solution~$\bv$ as
\[ \bv = \sum_{p=0}^3 \bv^{(p)} \qquad \text{with} \qquad \bv^{(p)} \in \h^{(p)} \]
and expand in power of~$\varepsilon$,
\[ \bv^{(p)} = \bv^{(p)}_0 + \varepsilon\: \bv^{(p)}_1 + \varepsilon^2\: \bv^{(p)}_2 + \cdots \:. \]
Using that, by construction, $\la \bu^{(q)}, . \ra^{(p)} = 0$ for~$q>p$, the inhomogeneous
equation~\eqref{weaksol} has the following expansion,
\begin{align*}
\bw^{(0)} \big( \bu^{(0)} \big) &= \la \bv^{(0)}, \bu^{(0)} \ra^{(0)} + \varepsilon\: \la \bv^{(1)}, \bu^{(0)} \ra^{(1)}
+ \varepsilon^2\: \la \bv^{(2)}, \bu^{(0)} \ra^{(2)} + \cdots \\
\bw^{(1)} \big( \bu^{(1)} \big) &= \la \bv^{(0)}, \bu^{(1)} \ra^{(1)} +
\la \bv^{(1)}, \bu^{(1)} \ra^{(1)} + \varepsilon\: \la \bv^{(0)}, \bu^{(1)} \ra^{(2)} +
\varepsilon\: \la \bv^{(1)}, \bu^{(1)} \ra^{(2)} + \cdots \:,
\end{align*}
and similarly for~$\bw^{(2)}, \ldots, \bw^{(3)}$. These equations can be solved iteratively order by order
in~$\varepsilon$. To zeroth and first order, we obtain
\begin{align*}
\bw^{(0)} \big( \bu^{(0)} \big) &= \la \bv^{(0)}_0, \bu^{(0)} \ra^{(0)} \\
\bw^{(1)} \big( \bu^{(1)} \big) &= \la \bv^{(0)}_0, \bu^{(1)} \ra^{(1)} + \la \bv^{(1)}_0, \bu^{(1)} \ra^{(1)} \:,
\end{align*}
or, in general,
\[ \la \bv^{(p)}_0, \bu^{(1)} \ra^{(p)} = \bw^{(p)} \big( \bu^{(1)} \big) 
- \sum_{q=0}^{p-1} \la \bv^{(q)}_0, \bu^{(1)} \ra^{(1)} \:. \]
Proceeding for increasing~$p$, the right hand side has already been computed in the previous steps.
Therefore, the method introduced in Section~\ref{secabslin} yields~$\bv^{(p)}_0$.
The higher orders in~$\varepsilon$ can be treated in the same way by expanding the equations
to the desired order in~$\varepsilon$ and solving iteratively for~$\bv^{(\circ)}_1$, $\bv^{(\circ)}_2$, etc.

The above perturbation scheme justifies why the linearized field equations can be analyzed
order by order in~$\varepsilon$. In the present paper, we will mainly restrict attention to the
leading order in~$\varepsilon$ as obtained by evaluating in the continuum limit.
In Section~\ref{secbeyond} will will explain how the higher orders in~$\varepsilon$ can be treated.

\section{Local Gauge Freedom and Direction-Dependent Phase Freedom} \label{secdirgauge}
In the previous sections we formulated the linearized field equations for Dirac sea configurations
and explained how these equations can be evaluated in the continuum limit.
An important point to keep in mind for the following constructions is that the evaluation formula
in the continuum limit~\eqref{delQ} applies only if the perturbation~$\delta P(x,y)$ of the kernel of
the fermionic projector is less singular on the light cone than the
unperturbed kernel~$P(x,y)$ (see~\eqref{delPcond}).
This is a major restriction, because most physically relevant perturbations do {\em{not}} have
this property. For example, an electromagnetic potential~$A$
gives rise to local gauge transformations of the form
(for the detailed derivation see~\cite{light})
\beq \label{Pgauge}
P(x,y) \rightarrow \exp \bigg( -i \int_0^1 A_j|_{\alpha y + (1-\alpha) x}\, \xi^j\: d\alpha \bigg)\: P(x,y) + \cdots
\eeq
(with~$\xi := y-x$, where the dots stand for other contributions of lower degree on the light cone).
Thus, to first order,
\beq \label{Pgauge1}
\delta P(x,y) = -i \int_0^1 A_j|_{\alpha y + (1-\alpha) x}\, \xi^j\: d\alpha \; P(x,y) + \cdots \:,
\eeq
showing that, in this example, the singularities of~$\delta P(x,y)$ and~$P(x,y)$
are of the same degree on the light cone.

Our strategy for removing this restriction is to compensate for the 
contributions by the gauge phases by hand, making use of the gauge invariance of the causal action principle.
After this has been done, the remaining singularities of~$\delta P(x,y)$ on the light cone will be of
lower degree, making it possible to apply~\eqref{delQ}.
It is one of the main goals of this section to show that this procedure of compensating for
gauge phases applies not only to gauge potentials, but more generally to certain
contributions by nonlocal potentials. This generalization is based on the observation that the causal
action principle is invariant not only under local gauge transformations, but more generally under
direction-dependent phase transformations.

In order to clarify the underlying concepts, we proceed step by step: After a short review of
first order perturbations by gauge potentials (Section~\ref{secpert}) and of
local gauge freedom (Section~\ref{seclocgauge}), general nonlocal vector potentials are introduced (Section~\ref{secnonloc}). Then we present the direction-dependent
phase freedom of the causal action principle (Section~\ref{secphase})
and explain how this phase freedom can be used to compensate for the most singular contributions
by the nonlocal potentials
on the light cone (Section~\ref{secphasemom}). In order to get a detailed description of all
contributions, it will be most convenient to work in momentum space.

\subsection{Perturbations by Gauge Potentials} \label{secpert}
We now specify the first order perturbations introduced in Section~\ref{secfirstorder}
by considering gauge potentials. For notational simplicity, in this section we only consider one generation
and simplify the distribution~\eqref{nuvacthree} to
\beq \label{nuvacone}
\hat{P}(p) = \big( \slashed{p}+m \big) \: \delta\big(p^2-m^2 \big)\: \Theta(-\omega)\:
e^{\varepsilon \omega}\:.
\eeq
The resulting Hilbert space~$(\hat{\H}, \la .|. \ra_{\hat{\H}}$ with scalar product
given by~\eqref{hatHhom}
is the space of all negative-energy solutions
of the Dirac equation of mass~$m$.

For the first order perturbation we choose
\beq \label{sAPsi}
\delta \hat{\Psi} \Big( p+\frac{q}{2} \Big) = -\hat{s}_{m} \Big( p+\frac{q}{2} \Big)\:
\,\,\hat{\!\!\slashed{A}}(q) \: \hat{\Psi} \Big( p-\frac{q}{2} \Big) \:,
\eeq
where~$\hat{A}(q)$ is the Fourier transform of the electromagnetic potential of momentum~$q$.
The fact that the electromagnetic potential is real implies that
\[ \hat{A}(q) = \overline{\hat{A}(-q)} \:. \]
Here~$\hat{s}_m$ denotes the Green's operator which satisfies the identity
\[ \big( \slashed{k} - m \big) \: \hat{s}_m(k) = \1_{\C^4} \:. \]
More precisely, having a time evolution into the future in mind, we shall always work with the
{\em{retarded Green's operator}}, which we denote for clarity with a hat indicating the past light cone,
\beq \label{smdef}
\hat{s}_m(k) = \hat{s}_m^\wedge(k) 
:= \lim_{\mu \searrow 0} \frac{\slashed{k} + m}{k^{2}-m^{2}+i \mu k^{0}} \:.
\eeq
Its adjoint (with respect to the spin inner product) is the advanced Green's operator denoted by
\[ \hat{s}_m^*(k) = \hat{s}_m^\vee(k) := \lim_{\mu \searrow 0} \frac{\slashed{k} + m}{k^{2}-m^{2}-i \mu k^{0}} \:. \]

With the above ansatz, the distribution~$\delta_q \hat{P}$ and its adjoint (as defined by~\eqref{tilmuone},
but now for one Dirac sea) take the form
\begin{align}
\delta_q \hat{P}(p) &= -\hat{s}_{m} \Big( p+\frac{q}{2} \Big)\: \,\,\hat{\!\!\slashed{A}}(q)\: \hat{P}\Big( p-\frac{q}{2} \Big) 
\label{nupert} \\
\delta_{-q}^* \hat{P}(p) &= -\hat{P}\Big( p+\frac{q}{2} \Big)\: \,\,\hat{\!\!\slashed{A}}(-q)^*
\: \hat{s}^*_{m} \Big( p-\frac{q}{2} \Big) \:. \label{nupert2} 
\end{align}
Here the star at~$\,\,\hat{\!\!\slashed{A}}(-q)^*$ denotes the adjoint with respect to the spin inner
product~$\Sl .|. \Sr$. Since the Dirac matrices are symmetric, this simply means that the complex conjugate of the potential is taken, i.e.~$\hat{A}(-q)^* = \overline{A_j(-q)} \,\gamma^j$).
These formulas coincide with the usual first order perturbation expansion of the Dirac equation
in the presence of an external electromagnetic potential~$A$ (for details see for example~\cite{firstorder}
or the more systematic treatment in~\cite{norm}).

\begin{Remark} {\bf{(Why only massive Dirac particles)}} \label{remmassive} {\em{
We now explain in which sense the perturbation formula~\eqref{nupert} is mathematically well-defined.
This will also clarify why we restrict attention to massive Dirac particles.
We first note that~\eqref{nupert} can be understood as an operator product which is
known to be well-defined provided that the electromagnetic potential~$A(x)$ is smooth and
decays sufficiently fast near infinity, both in the massive and massless cases (for details see for example~\cite[Lemma~2.1.2]{cfs}). In other words, the product~\eqref{nupert} is well-defined
after taking the Fourier transform to position space and integrating over the momentum variable~$q$.
In the present paper, we take a different point of view by fixing~$q$ and evaluating the
perturbation pointwise for this given~$q$. Put differently, we perturb by a potential~$A \sim e^{-i q x}$,
being a plane wave with no decay properties at infinity. Doing so, \eqref{nupert} involves a pointwise
product of the distributions~$\hat{P}_m$ and~$s_m$, which are both singular on the mass shells.
If~$q$ is timelike or spacelike, then the mass shells intersect transversely (see the left and middle graphic
in Figure~\ref{figdelP2}).
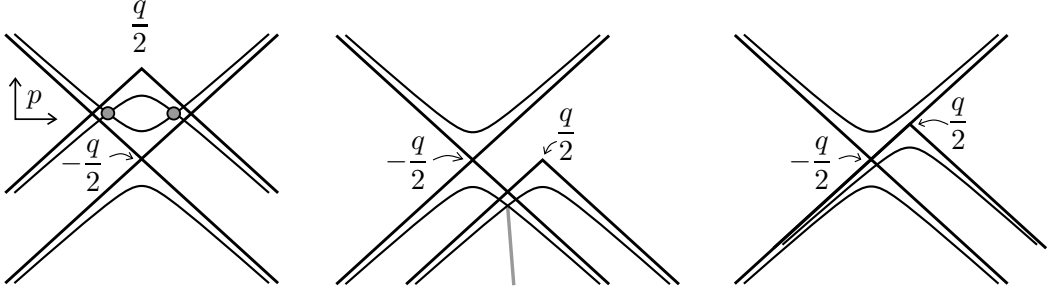
\begin{figure}
\psset{xunit=.43pt,yunit=.43pt,runit=.43pt}
\begin{pspicture}(917.34340397,221.69023168)
{
\newrgbcolor{curcolor}{0.60000002 0.60000002 0.60000002}
\pscustom[linewidth=3.02362209,linecolor=curcolor]
{
\newpath
\moveto(442.74872693,69.734912)
\lineto(448.20005291,0.06033058)
}
}
{
\newrgbcolor{curcolor}{0 0 0}
\pscustom[linewidth=2.54532664,linecolor=curcolor]
{
\newpath
\moveto(0.85723843,81.33653468)
\lineto(120.84133039,190.67874318)
\lineto(240.82549417,81.33653468)
}
}
{
\newrgbcolor{curcolor}{0 0 0}
\pscustom[linewidth=1.88976378,linecolor=curcolor]
{
\newpath
\moveto(232.82657764,81.33653468)
\curveto(198.82902047,111.103712)(164.83419213,140.86850066)(144.50313449,155.75058318)
\curveto(124.17207685,170.63266948)(117.50661543,170.63266948)(97.17643465,155.7495816)
\curveto(76.84625764,140.86649373)(42.85144441,111.10171641)(8.85617008,81.33653468)
}
}
{
\newrgbcolor{curcolor}{0 0 0}
\pscustom[linewidth=2.54532664,linecolor=curcolor]
{
\newpath
\moveto(0.85723087,1.86445027)
\lineto(120.84131528,111.20664365)
\lineto(240.82552063,1.86445027)
}
}
{
\newrgbcolor{curcolor}{0 0 0}
\pscustom[linewidth=1.88976378,linecolor=curcolor]
{
\newpath
\moveto(232.82660409,1.86445027)
\curveto(198.82905071,31.63159357)(164.83418835,61.39640869)(144.50312315,76.27848743)
\curveto(124.17206173,91.16056995)(117.50660031,91.16056995)(97.17641953,76.27748208)
\curveto(76.84623874,61.3943942)(42.85142929,31.62962822)(8.85616252,1.86445027)
}
}
{
\newrgbcolor{curcolor}{0 0 0}
\pscustom[linewidth=2.54532664,linecolor=curcolor]
{
\newpath
\moveto(0.85723087,220.74957909)
\lineto(120.84131528,111.40738192)
\lineto(240.82552063,220.74957909)
}
}
{
\newrgbcolor{curcolor}{0 0 0}
\pscustom[linewidth=1.88976378,linecolor=curcolor]
{
\newpath
\moveto(232.82660409,221.21148271)
\curveto(198.82905449,191.44432428)(164.83418457,161.67951672)(144.50312315,146.79743798)
\curveto(124.17206173,131.91535546)(117.50660031,131.91535546)(97.17642331,146.79844334)
\curveto(76.84624252,161.68153121)(42.85142929,191.44630853)(8.85616252,221.21148271)
}
}
{
\newrgbcolor{curcolor}{0 0 0}
\pscustom[linewidth=2.54532664,linecolor=curcolor]
{
\newpath
\moveto(353.52946394,0.94065814)
\lineto(473.51358992,110.28284397)
\lineto(593.4977537,0.94065814)
}
}
{
\newrgbcolor{curcolor}{0 0 0}
\pscustom[linewidth=1.88976378,linecolor=curcolor]
{
\newpath
\moveto(585.49879937,0.94065814)
\curveto(551.50126866,30.70780145)(517.50642142,60.47260523)(497.17536,75.35468775)
\curveto(476.84429858,90.23677027)(470.17883717,90.23677027)(449.84864504,75.35368239)
\curveto(429.51849071,60.47059452)(395.52368126,30.70583609)(361.52841827,0.94065814)
}
}
{
\newrgbcolor{curcolor}{0 0 0}
\pscustom[linewidth=2.54532664,linecolor=curcolor]
{
\newpath
\moveto(292.08617197,0.94065814)
\lineto(412.07026394,110.28282885)
\lineto(532.05446551,0.94065814)
}
}
{
\newrgbcolor{curcolor}{0 0 0}
\pscustom[linewidth=1.88976378,linecolor=curcolor]
{
\newpath
\moveto(524.05554898,0.94065814)
\curveto(490.05798047,30.70780145)(456.06313323,60.47259767)(435.73207181,75.35467641)
\curveto(415.40101039,90.23675515)(408.73554898,90.23675515)(388.40535685,75.35367105)
\curveto(368.07516472,60.47059074)(334.08035528,30.70583609)(300.0850885,0.94065814)
}
}
{
\newrgbcolor{curcolor}{0 0 0}
\pscustom[linewidth=2.54532664,linecolor=curcolor]
{
\newpath
\moveto(292.08617197,219.82576428)
\lineto(412.07026394,110.48356712)
\lineto(532.05446551,219.82576428)
}
}
{
\newrgbcolor{curcolor}{0 0 0}
\pscustom[linewidth=1.88976378,linecolor=curcolor]
{
\newpath
\moveto(524.05554898,220.2876679)
\curveto(490.05798047,190.52050948)(456.06313323,160.75570192)(435.73207181,145.8736194)
\curveto(415.40101039,130.99154066)(408.73554898,130.99154066)(388.40535685,145.87462475)
\curveto(368.07520252,160.75771263)(334.08035528,190.52248617)(300.0850885,220.2876679)
}
}
{
\newrgbcolor{curcolor}{0 0 0}
\pscustom[linewidth=2.54532664,linecolor=curcolor]
{
\newpath
\moveto(676.51791118,32.55266444)
\lineto(796.50203717,141.8948616)
\lineto(916.48620094,32.55266444)
}
}
{
\newrgbcolor{curcolor}{0 0 0}
\pscustom[linewidth=1.88976378,linecolor=curcolor]
{
\newpath
\moveto(908.48724661,35.67493216)
\curveto(874.48967811,65.44212082)(840.49486866,95.20690948)(820.16380724,110.08899578)
\curveto(799.83274583,124.97108208)(793.16728441,124.97108208)(772.83709228,110.08799042)
\curveto(752.50693795,95.20489877)(718.5121285,65.44011767)(684.51686551,35.67493216)
}
}
{
\newrgbcolor{curcolor}{0 0 0}
\pscustom[linewidth=2.54532664,linecolor=curcolor]
{
\newpath
\moveto(642.61815685,1.4025542)
\lineto(762.60224504,110.74474003)
\lineto(882.58644661,1.4025542)
}
}
{
\newrgbcolor{curcolor}{0 0 0}
\pscustom[linewidth=1.88976378,linecolor=curcolor]
{
\newpath
\moveto(874.58753008,1.4025542)
\curveto(840.58996157,31.16969751)(806.59511433,60.93450885)(786.26405291,75.81658759)
\curveto(765.9329915,90.69866633)(759.26753008,90.69866633)(738.93733795,75.81557846)
\curveto(718.60718362,60.93249058)(684.61237417,31.16773216)(650.61711118,1.4025542)
}
}
{
\newrgbcolor{curcolor}{0 0 0}
\pscustom[linewidth=2.54532664,linecolor=curcolor]
{
\newpath
\moveto(642.61815685,220.28767546)
\lineto(762.60224504,110.9454783)
\lineto(882.58644661,220.28767546)
}
}
{
\newrgbcolor{curcolor}{0 0 0}
\pscustom[linewidth=1.88976378,linecolor=curcolor]
{
\newpath
\moveto(874.58753008,220.74957909)
\curveto(840.58996157,190.98242066)(806.59511433,161.2176131)(786.26405291,146.33553058)
\curveto(765.9329915,131.45345184)(759.26753008,131.45345184)(738.93733795,146.33653972)
\curveto(718.60718362,161.21962759)(684.61237417,190.98441247)(650.61711118,220.74957909)
}
}
{
\newrgbcolor{curcolor}{0.60000002 0.60000002 0.60000002}
\pscustom[linestyle=none,fillstyle=solid,fillcolor=curcolor]
{
\newpath
\moveto(96.71960938,151.4402059)
\curveto(96.71960938,148.4586953)(94.2182966,146.0417034)(91.13277092,146.0417034)
\curveto(88.04724524,146.0417034)(85.54593246,148.4586953)(85.54593246,151.4402059)
\curveto(85.54593246,154.4217165)(88.04724524,156.8387084)(91.13277092,156.8387084)
\curveto(94.2182966,156.8387084)(96.71960938,154.4217165)(96.71960938,151.4402059)
\closepath
}
}
{
\newrgbcolor{curcolor}{0 0 0}
\pscustom[linewidth=1.6344756,linecolor=curcolor]
{
\newpath
\moveto(96.71960938,151.4402059)
\curveto(96.71960938,148.4586953)(94.2182966,146.0417034)(91.13277092,146.0417034)
\curveto(88.04724524,146.0417034)(85.54593246,148.4586953)(85.54593246,151.4402059)
\curveto(85.54593246,154.4217165)(88.04724524,156.8387084)(91.13277092,156.8387084)
\curveto(94.2182966,156.8387084)(96.71960938,154.4217165)(96.71960938,151.4402059)
\closepath
}
}
{
\newrgbcolor{curcolor}{0.60000002 0.60000002 0.60000002}
\pscustom[linestyle=none,fillstyle=solid,fillcolor=curcolor]
{
\newpath
\moveto(154.63980519,151.26984573)
\curveto(154.63980519,148.28833512)(152.13849241,145.87134323)(149.05296673,145.87134323)
\curveto(145.96744105,145.87134323)(143.46612827,148.28833512)(143.46612827,151.26984573)
\curveto(143.46612827,154.25135633)(145.96744105,156.66834823)(149.05296673,156.66834823)
\curveto(152.13849241,156.66834823)(154.63980519,154.25135633)(154.63980519,151.26984573)
\closepath
}
}
{
\newrgbcolor{curcolor}{0 0 0}
\pscustom[linewidth=1.6344756,linecolor=curcolor]
{
\newpath
\moveto(154.63980519,151.26984573)
\curveto(154.63980519,148.28833512)(152.13849241,145.87134323)(149.05296673,145.87134323)
\curveto(145.96744105,145.87134323)(143.46612827,148.28833512)(143.46612827,151.26984573)
\curveto(143.46612827,154.25135633)(145.96744105,156.66834823)(149.05296673,156.66834823)
\curveto(152.13849241,156.66834823)(154.63980519,154.25135633)(154.63980519,151.26984573)
\closepath
}
}
{
\newrgbcolor{curcolor}{0 0 0}
\pscustom[linewidth=0.99999871,linecolor=curcolor]
{
\newpath
\moveto(92.50194898,113.0046972)
\curveto(96.07988409,114.4813057)(99.65722961,115.95767231)(103.06470047,115.73007294)
\curveto(106.47217512,115.50246979)(109.7088378,113.57183798)(112.94437039,111.64187137)
}
}
{
\newrgbcolor{curcolor}{0 0 0}
\pscustom[linewidth=0.99999871,linecolor=curcolor]
{
\newpath
\moveto(109.87800945,117.09318223)
\lineto(112.77401953,111.64187137)
\lineto(106.98199937,110.44940019)
}
}
{
\newrgbcolor{curcolor}{0 0 0}
\pscustom[linewidth=0.99999871,linecolor=curcolor]
{
\newpath
\moveto(377.16262299,110.27904176)
\curveto(381.19477417,112.1531394)(385.22639622,114.02699137)(389.48569701,114.1686405)
\curveto(393.7449978,114.3102972)(398.23088126,112.72035578)(402.71566866,111.13081499)
}
}
{
\newrgbcolor{curcolor}{0 0 0}
\pscustom[linewidth=0.99999871,linecolor=curcolor]
{
\newpath
\moveto(398.1161348,117.09318601)
\lineto(403.05639307,111.30116586)
\lineto(394.36835528,109.25692901)
}
}
{
\newrgbcolor{curcolor}{0 0 0}
\pscustom[linewidth=0.99999871,linecolor=curcolor]
{
\newpath
\moveto(475.79733165,116.07105814)
\curveto(476.98962142,118.79624491)(478.18206236,121.52184365)(479.88602457,123.28254035)
\curveto(481.58998677,125.04323704)(483.80452535,125.83820019)(486.0185348,126.63298192)
}
}
{
\newrgbcolor{curcolor}{0 0 0}
\pscustom[linewidth=0.99999871,linecolor=curcolor]
{
\newpath
\moveto(474.49339465,122.03343672)
\lineto(475.51568126,115.73035641)
\lineto(481.30765606,119.13743042)
}
}
{
\newrgbcolor{curcolor}{0 0 0}
\pscustom[linewidth=0.99999871,linecolor=curcolor]
{
\newpath
\moveto(734.05320567,111.30116208)
\curveto(737.46082772,112.60741594)(740.86780724,113.91343168)(744.01974425,113.88465058)
\curveto(747.17171906,113.85592617)(750.06763087,112.49307767)(752.96244661,111.13081499)
}
}
{
\newrgbcolor{curcolor}{0 0 0}
\pscustom[linewidth=0.99999871,linecolor=curcolor]
{
\newpath
\moveto(750.74790803,116.24142035)
\lineto(753.30317102,110.96046035)
\lineto(748.19256945,108.74586885)
}
}
{
\newrgbcolor{curcolor}{0 0 0}
\pscustom[linewidth=0.99999871,linecolor=curcolor]
{
\newpath
\moveto(805.2609789,142.13514885)
\curveto(809.57712378,143.83889184)(813.89266394,145.54238916)(818.06673638,145.42842885)
\curveto(822.24084661,145.31447609)(826.27243087,143.38383672)(830.30291906,141.4537416)
}
}
{
\newrgbcolor{curcolor}{0 0 0}
\pscustom[linewidth=0.99999871,linecolor=curcolor]
{
\newpath
\moveto(807.02341039,146.39398853)
\lineto(804.63849071,141.96479798)
\lineto(809.74909228,139.75020649)
}
}
{
\newrgbcolor{curcolor}{0 0 0}
\pscustom[linewidth=1.51181105,linecolor=curcolor]
{
\newpath
\moveto(9.71015055,181.31645279)
\lineto(9.71013921,146.22363011)
\lineto(45.14367118,146.22361877)
}
}
{
\newrgbcolor{curcolor}{0 0 0}
\pscustom[linewidth=1.51181105,linecolor=curcolor]
{
\newpath
\moveto(4.5995452,175.18372712)
\lineto(9.72122457,182.00129184)
\lineto(14.65040126,175.18372712)
}
}
{
\newrgbcolor{curcolor}{0 0 0}
\pscustom[linewidth=1.51181105,linecolor=curcolor]
{
\newpath
\moveto(38.49988913,151.33423546)
\lineto(45.48780094,146.23127987)
\lineto(37.98883276,141.73005531)
}
\rput[bl](20,155){$p$}
\rput[bl](110,205){$\displaystyle \frac{q}{2}$}
\rput[bl](49,83){$\displaystyle -\frac{q}{2}$}
\rput[bl](485,110){$\displaystyle \frac{q}{2}$}
\rput[bl](335,85){$\displaystyle -\frac{q}{2}$}
\rput[bl](830,120){$\displaystyle \frac{q}{2}$}
\rput[bl](690,85){$\displaystyle -\frac{q}{2}$}
}
\end{pspicture}
\caption{Intersection of singular supports (gray) in the cases~$q$ timelike, spacelike and lightlike.}
\label{figdelP2}
\end{figure}%
This ensures that the product of
distributions is well-defined (for the general context see
for example~\cite{friedlander2}). However, if~$q$ is lightlike, then the mass cones intersect tangentially
along the ray~$p \in \R q$. 
As a consequence, the product of distributions in~\eqref{nupert} is
ill-defined. However, in the massive case, the mass shells do not intersect, so that~\eqref{nupert}
is again well-defined and finite (see the right graphic in Figure~\ref{figdelP2}).
This is the reason why massive Dirac fields are easier to handle.
In order to avoid technical subtleties, here we always restrict attention to the massive case.
On a technical level, the assumption of massive Dirac particles will be used
in Lemma~\ref{lemma82}.
 }} $\hspace*{0,5em}$ \QEDrem
\end{Remark}

\subsection{Local Gauge Freedom and its Perturbative Description} \label{seclocgauge}
We now work out the underlying local gauge freedom.
To this end, we consider an electromagnetic potential~$A$ which is a
total derivative, i.e.
\beq \label{Apot}
\slashed{A}(x) = \big( \Pdd \Lambda \big)(x)
\eeq
with a real-valued function~$\Lambda$ (for the basics on gauge transformations
in electrodynamics and quantum mechanics we recommend~\cite[Chapter~3]{landau2}
and~\cite[Section~2.7]{sakurai}; see~\cite{u22} for the context of the Dirac equation).
In this case, the interacting Dirac equation can be written as
\[ i \Pdd + (\Pdd \Lambda) - m = e^{i \Lambda(x)} \big(i \Pdd - m \big)\: e^{-i \Lambda(x)} \:. \]
Consequently, the interacting wave evaluation operator merely picks up a gauge phase,
\beq \label{Psiphase}
\tilde{\Psi}(x) = e^{i \Lambda(x)}\: \Psi(x) \:.
\eeq
Likewise, the kernel of the fermionic projector becomes
\[ \tilde{P}(x,y) = e^{i \Lambda(x)-i \Lambda(y)}\: P(x,y) \:. \]
This is in agreement with~\eqref{Pgauge} if one keeps in mind that, for a pure gauge potential~\eqref{Apot},
the line integral in~\eqref{Pgauge} can be simplified using integration by parts,
\[ \int_0^1 \partial_j \Lambda|_{\alpha y + (1-\alpha) x}\, \xi^j\: d\alpha
= \int_0^1 \frac{d}{d\alpha}\Lambda|_{\alpha y + (1-\alpha) x}\, \xi^j\: d\alpha
= \Lambda(y) - \Lambda(x) \:. \]

Let us verify by explicit computation that the gauge phases in~\eqref{Psiphase}
also arise in the perturbative description~\eqref{nupert}. To this end, we
first take the Fourier transform of~\eqref{Psiphase}
\[ \widehat{\tilde{\Psi}}(p) = \int_M \tilde{\Psi}(x)\: e^{i p x}\: d^4x
= \int_M e^{i \Lambda(x)}\: \Psi(x)\: e^{i p x}\: d^4x \:. \]
Expanding to first order, we obtain
\[ \delta \hat{\Psi}(p) =
\int_M \bigg( i \int_{\hat{M}} \frac{d^4q}{(2 \pi)^4}\: \hat{\Lambda}(q)\: e^{-i q x} \bigg)\: \Psi(x)\: e^{i p x}\:d^4x
= i \int_{\hat{M}} \hat{\Lambda}(q)\: \hat{\Psi}(p-q)\: \frac{d^4q}{(2 \pi)^4} \:. \]
Comparing with~\eqref{nuqdef} and~\eqref{tilmuone}, we conclude that
\beq \label{nugauge}
\delta_q \hat{P}(p) = i \hat{\Lambda}(q)\: \hat{P}\Big( p - \frac{q}{2} \Big) \:.
\eeq
This formula describes the simple fact that multiplication by a plane wave in position space corresponds to
a translation in momentum space. 
In order to verify consistency with~\eqref{nupert}, we evaluate the operator product~\eqref{nupert} for
the pure gauge potential
\[ \,\,\hat{\!\!\slashed{A}}(q) = -i \slashed{q}\: \hat{\Lambda}(q) \:. \]
In this case, the operator product can be simplified as follows,
\begin{align*}
\delta_q \hat{P}(p) &= \hat{s}_{m} \Big( p+\frac{q}{2} \Big)\: \big(
i \hat{\Lambda}(q)\: \slashed{q} \big)\: \hat{P} \Big( p-\frac{q}{2} \Big) \\
&= i \hat{\Lambda}(q)\: \hat{s}_{m} \Big( p+\frac{q}{2} \Big)\: \Big(
\big( \slashed{p} + \frac{\slashed{q}}{2} - m \big) - \big( \slashed{p} - \frac{\slashed{q}}{2} - m  \big) \Big)\: \hat{P} \Big( p-\frac{q}{2} \Big) \\
&= i \hat{\Lambda}(q)\: \Big(\hat{P} \Big( p-\frac{q}{2} \Big) -
\hat{s}_{m} \Big( p+\frac{q}{2} \Big)\: 0 \Big)
= i \hat{\Lambda}(q)\: \hat{P} \Big( p-\frac{q}{2} \Big) \:,
\end{align*}
in agreement with~\eqref{nugauge}.

\subsection{Nonlocal Vector Potentials} \label{secnonloc}
We now generalize the perturbation theory for gauge potentials introduced in Section~\ref{secpert}.
For notational simplicity, we again consider one Dirac sea~\eqref{nuvacone}.
We again use the formalism in Section~\ref{secpert}, but now for a potential~$\hat{A}_q(p)$ defined for~$p$ on the lower mass shell, i.e.\
\[ \,\,\hat{\!\!\slashed{A}}_q : \supp \hat{P} \subset \hat{M} \rightarrow \Lin(V) \:. \]
If~$\hat{A}_q$ does not depend on~$p$, we get back an electromagnetic potential
acting on the Dirac sea of mass~$m$.
Due to the additional $p$-dependence, the potential~$\hat{A}_q(p)$ can be regarded as an integral
operator with a nonlocal integral kernel. For this reason, we refer to~$\hat{A}_q$ as a {\em{nonlocal vector potential}}.
In generalization of~\eqref{sAPsi}, the first order perturbation takes the form
\[ \delta \hat{\Psi}\Big( p+\frac{q}{2} \Big) = -\hat{s}_{m} \Big( p+\frac{q}{2} \Big)\:
\,\,\hat{\!\!\slashed{A}}_q \Big( p-\frac{q}{2} \Big)\: \hat{\Psi} \Big( p-\frac{q}{2} \Big) \:. \]
Moreover, the distribution~$\delta_q \hat{P}$ and its adjoint (defined again by simplifying~\eqref{tilmuone}
for one Dirac sea) are given in generalization of~\eqref{nupert} and~\eqref{nupert2} by
\begin{align*}
\delta_q \hat{P}(p)
&= -\hat{s}_{m} \Big( p+\frac{q}{2} \Big)\: \,\,\hat{\!\!\slashed{A}}_q \Big( p-\frac{q}{2} \Big)\: \hat{P}\Big( p-\frac{q}{2} \Big) \\
\delta_{-q}^* \hat{P}(p) &= -\hat{P}\Big( p+\frac{q}{2} \Big)\: \big(\,\,\hat{\!\!\slashed{A}}(-q) \big)^*
\Big( p+\frac{q}{2} \Big)\: s^*_{m} \Big( p-\frac{q}{2} \Big) \:.
\end{align*}

Let us compute~$\delta_q \hat{P}(p)$ in more detail. Working with the retarded Green's operator~\eqref{smdef}
and using~\eqref{nuvacone}, we obtain
\begin{align}
&\hat{s}_{m} \Big( p+\frac{q}{2} \Big)\: \,\,\hat{\!\!\slashed{A}}_q\: \hat{P}_{m} \Big( p-\frac{q}{2} \Big) \notag \\
&= \lim_{\mu \searrow 0}
\Big( \slashed{p}+\frac{\slashed{q}}{2} + m \Big) \:
\frac{\text{1}}{\displaystyle \Big(p+\frac{q}{2} \Big)^2-m^2
+ i \mu \:\Big(p^0 + \frac{q^0}{2} \Big)}\: \,\,\hat{\!\!\slashed{A}}_q \:\hat{P}_{m} \Big( p-\frac{q}{2} \Big) \\
&=  \lim_{\mu \searrow 0}\Big( \slashed{p}+\frac{\slashed{q}}{2} + m \Big) \: 
\frac{\text{1}}{2 p q+ i \mu \:\Big(p^0 + \frac{q^0}{2} \Big)} \: \,\,\hat{\!\!\slashed{A}}_q
\:\hat{P}_{m}\Big( p-\frac{q}{2} \Big) \label{pqform} \\
&=  \lim_{\mu \searrow 0}\frac{\text{1}}{2 p q+ i \mu \:\Big(p^0 + \frac{q^0}{2} \Big)} \notag \\
&\qquad \times \bigg( \Big( \slashed{p}+\frac{\slashed{q}}{2} + m \Big) \: \,\,\hat{\!\!\slashed{A}}_q
+ \,\,\hat{\!\!\slashed{A}}_q\: \Big( \slashed{p}-\frac{\slashed{q}}{2} - m \Big)\bigg) \: 
\hat{P}_{m}\Big( p-\frac{q}{2} \Big) \label{zusatz} \:.
\end{align}
In~\eqref{pqform} we used that~$\hat{P}_{m}$ is supported on the mass shell, and thus
\[ \Big(p+\frac{q}{2} \Big)^2-m^2 = \Big(p-\frac{q}{2} \Big)^2-m^2 + 2 pq = 2 pq \:. \]
Moreover, in~\eqref{zusatz} we inserted a term using the Dirac equation in the form
\[ \Big( \slashed{p}-\frac{\slashed{q}}{2} - m \Big)\: 
\hat{P}_{m}\Big( p-\frac{q}{2} \Big) = 0 \:. \]
Finally, for ease in notation we omitted the argument of~$\hat{A}_q$. Using the 
Dirac anti-commutation relations, we conclude that
\begin{align}
\delta_q \hat{P}(p) &= -\lim_{\mu \searrow 0}\frac{1}{2 p q+ i \mu \:\Big(p^0 + \frac{q^0}{2} \Big)}
\notag \\
&\qquad \times
\bigg( 2 p \hat{A}_q \Big( p-\frac{q}{2} \Big)  +
\frac{1}{2}\: \Big[\slashed{q}, \,\,\hat{\!\!\slashed{A}}_q \Big(p-\frac{q}{2} \Big) \Big] \bigg)\: \hat{P}\Big( p-\frac{q}{2} \Big) \:.
\label{nuqgauge}
\end{align}
The distribution~$\delta_{-q}^* \hat{P}$ is obtained by taking the adjoint and replacing~$q$ by~$-q$,
\begin{align}
\delta_{-q}^* \hat{P}(p) &= \lim_{\mu \searrow 0}\frac{1}{2 p q+ i \mu \:\Big(p^0 - \frac{q^0}{2} \Big)}
\notag \\
&\qquad \times
\hat{P}\Big( p-\frac{q}{2} \Big) \:\bigg( 2p (\hat{A}_{-q})^*(p-\frac{q}{2}) +
\frac{1}{2}\: \Big[\slashed{q}, (\,\,\hat{\!\!\slashed{A}}(-q)^*)\Big(p+\frac{q}{2} \Big) \Big] \bigg)\:.
\label{numqgauge}
\end{align}
Finally, the distribution~$\hat{\tilde{P}}$ is obtained by substituting~\eqref{nuqgauge}
and~\eqref{numqgauge} into~\eqref{tilmuone}.

\subsection{A First Connection to Direction-Dependent Local Phase Transformations} \label{secphase}
Choosing a pure gauge potential,
\[ \,\,\hat{\!\!\slashed{A}}_q(p)=-i \slashed{q}\: \hat{\Lambda}(q) \:, \] 
the formula~\eqref{nuqgauge} gives back the contribution by the gauge transformation~\eqref{nugauge}.
The point of interest for what follows is that, even for general nonlocal vector potentials,
the contribution to~\eqref{nuqgauge}
\beq \label{pertcont}
\delta_q \hat{P}(p) \asymp -\lim_{\mu \searrow 0}\frac{2 p \hat{A}_q(p-\frac{q}{2})}{2 p q+ i \mu \:\Big(p^0 + \frac{q^0}{2} \Big)}
\hat{P}\Big( p-\frac{q}{2} \Big)
\eeq
resembles a contribution by a gauge transformation, because it is a prefactor times
the distribution~$\hat{P}$ translated in momentum space
(here the symbol ``$\asymp$'' indicates that we restrict attention to a specific contribution to~$\delta_q\hat{P}$).
The only difference compared to~\eqref{nugauge} is that the prefactor has a non-trivial $p$-dependence.
This raises the question whether the contribution~\eqref{pertcont} can be understood as 
a generalized gauge transformation.

More specifically, by a ``generalized gauge transformation''
we mean a {\em{direction-dependent
local phase transformation}} already mentioned in the introduction (see~\eqref{dirlocintro}).
Thus we consider a phase transformation of the kernel of the fermionic projector
\beq \label{gaugedirection}
\tilde{P}(x,y) = e^{i \Lambda(x,y)}\: P(x,y)
\eeq
with a real-valued and smooth function~$\Lambda \in C^\infty(M \times M, \R)$.
Since the transformed kernel must be again symmetric (meaning that~$\tilde{P}(x,y)^* = \tilde{P}(y,x)$),
we need to assume that~$\Lambda$ is anti-symmetric in the sense~\eqref{Lsymm}.
The phase factor in~\eqref{gaugedirection} drops out when forming the closed chain.
Therefore, the causal action principle is indeed invariant under the phase transformation~\eqref{gaugedirection}.
To first order, the transformation~\eqref{gaugedirection} becomes
\beq \label{gaugelin}
\delta P(x,y) = i \Lambda(x,y)\: P(x,y) \:.
\eeq

In order to get a closer connection between~\eqref{pertcont} and~\eqref{gaugelin}, it is useful
to also consider the direction-dependent local phase transformation for a fixed momentum transfer~$q$.
To this end, we take the Fourier transform in the variable~$(x+y)/2$,
\beq \label{hatLambdadef}
\Lambda(x,y) = \int_{\hat{M}} \frac{d^4q}{(2 \pi)^4}\: e^{-i\frac{q}{2}\, (x+y)}\:
\hat{\Lambda}(q,\xi)\:.
\eeq
The fact that~$\Lambda(x,y)$ is real-valued and anti-symmetric means that
\beq \label{Lamrealpos}
-\hat{\Lambda}(q,-\xi) = \hat{\Lambda}(q, \xi) = \overline{\hat{\Lambda}(-q, \xi)} \:.
\eeq
Next, it is convenient to choose the ansatz
\beq \label{hatLambdadecomp}
\hat{\Lambda}(q,\xi) = \Lambda_q(\xi)\: e^{i \frac{q}{2} \xi} - \overline{\Lambda_{-q}(-\xi)}\: e^{-i \frac{q}{2} \xi} \:.
\eeq
Then the symmetry properties~\eqref{Lamrealpos} simplify to the single relation
\beq \label{Lamreal}
\Lambda_q(-\xi) = \overline{\Lambda_{-q}(\xi)} \:.
\eeq
Taking the Fourier transform with the usual ansatz
\[ \Lambda_q(\xi) = \int_{\hat{M}} \frac{d^4k}{(2 \pi)^4}\: \hat{\Lambda}_q(k)\: e^{i k \xi} \:,\qquad
\overline{\Lambda_{-q}(-\xi)} = \int_{\hat{M}} \frac{d^4k}{(2 \pi)^4}\: \hat{\Lambda}_{-q}^*(k)\: e^{i k \xi} \:, \]
the perturbation for momentum transfer~$q$ can be written
again in the form~\eqref{delPmink} with
\begin{align}
\delta_q \hat{P}(p) &= \frac{i}{(2 \pi)^4}\: \big(\hat{\Lambda}_q * \hat{P})\Big(p - \frac{q}{2} \Big) \label{convolve} \\
\delta_{-q}^* \hat{P}(p) &= -\frac{i}{(2 \pi)^4}\: \big(\hat{\Lambda}_{-q}^* * \hat{P})\Big(p + \frac{q}{2} \Big) \:. \label{convolve2}
\end{align}
In this formulation in momentum space,
the connection to classical gauge potentials~\eqref{nugauge} is obtained simply by choosing
\[ \hat{\Lambda}_q(k) = \hat{\Lambda}(q)\; (2 \pi)^4\, \delta^4(k) \qquad \text{and} \qquad
\hat{\Lambda}_{-q}^*(k) = \hat{\Lambda}(-q)^*\; (2 \pi)^4\,\delta^4(k) \:. \]
Now the direction-dependent local phase freedom means that the causal action
is invariant under first order perturbations~\eqref{convolve} and~\eqref{convolve2}
involving the convolution with a function~$\hat{\Lambda}_q$ having the
property~\eqref{Lamreal}.

Let us compare~\eqref{pertcont} with~\eqref{convolve}. At first sight, these formulas look
very different, because~\eqref{pertcont} is a multiplication in momentum space, whereas~\eqref{convolve}
is a convolution in momentum space.
Likewise, in position space, \eqref{pertcont} corresponds to a convolution, whereas~\eqref{convolve}
becomes a multiplication (see also~\eqref{gaugelin}).
Nevertheless, the following argument gives a simple connection between multiplication in momentum
space and multiplication in position space. This argument was already used in~\cite[Chapter~4]{pfp}
and is the starting point of the continuum limit analysis in~\cite[Sections~2.2 and~2.4]{cfs}.
The basic idea can be understood from the following simple consideration.
In preparation, noting that a translation~$p \rightarrow p+q/2$
in momentum space simply corresponds to multiplication by a
smooth phase factor, instead of~\eqref{pertcont} we may consider the product
\[ -\lim_{\mu \searrow 0}\frac{(2 p+q) \hat{A}_q(p-\frac{q}{2})}{(2 p+q) q  + i \mu \:\Big(p^0 + q^0 \Big)}
\hat{P}(p) \:. \]
Next, we simplify the prefactor. First of all, we here disregard the poles (these will be treated
in Section~\ref{secmasspole} below).
Moreover, having the situation in mind that~$\hat{A}_q( \lambda p)$ has a limit as~$\lambda \rightarrow \infty$,
for simplicity we replace the prefactor by a function~$f \in C^\infty(S^2, \C)$
on the unit sphere~$S^2 \subset \R^3$ which depends only on the direction of the vector~$\vec{p}$, i.e.\
\[ f(\hat{\vec{p}}\,) \: \hat{P}(p) \qquad \text{with} \qquad \hat{\vec{p}} := \frac{\vec{p}}{|\vec{p}|} \]
(clearly, we will have to treat the errors of this approximation, as will be done systematically in Section~\ref{secmce}
below). In order to analyze the effect of the function~$f$ in position space, we need to take the
Fourier transform
\beq \label{fP}
\int_{\hat{M}} \frac{d^4p}{(2 \pi)^4} \: f(\hat{\vec{p}}\,)\: \hat{P}(p)\: e^{ip \xi} \:.
\eeq
Being interested in the behavior on the light cone, we choose~$\xi$ as a lightlike vector.
As a consequence of the oscillating phase factor~$\exp(i p \xi)$, the leading contribution
to the Fourier integral~\eqref{fP} is obtained by restricting attention to those momenta for which~$p \xi$ is small.
More precisely, for determining the leading scaling behavior for small~$\varepsilon$, it suffices to integrate
over the set
\[ \big\{ p \in \hat{M} \:\big|\:  |p \xi| \leq 1 \big\} \]
and leave out the phase factor.
On the other hand, the leading contribution for small~$\varepsilon$ to the integral is obtained when
the absolute value of the frequency~$|p^0|$ is as large as~$1/\varepsilon$.
As a consequence, the inner product~$p \xi$ is small only if the vectors~$p$ and~$\xi$ are almost
collinear in the sense that
\[ \hat{\vec{p}} = -\epsilon(\xi^0)\, \hat{\vec{\xi}} \;\Big( 1 +  \O \Big( \frac{\varepsilon}{t} \Big) \bigg)  \:. \]
Applying this equation to the argument of the function~$f$ in~\eqref{fP}, we may pull this function
out of the integral to obtain
\beq \label{Fouriersimp}
\int_{\hat{M}} \frac{d^4p}{(2 \pi)^4} \: f(\hat{\vec{p}}\,)\: \hat{P}(p)\: e^{ip \xi}
= f\Big(-\epsilon(\xi^0)\, \hat{\vec{\xi}}\,\Big) \:P(x,y) \;\Big( 1 +  \O \Big( \frac{\varepsilon}{t} \Big) \bigg) \:.
\eeq
This formula shows that, to leading order for small~$\varepsilon$, multiplication in momentum space
is the same as multiplication in position space. This simple consideration
explains why contributions of the form~\eqref{pertcont} can indeed be compensated
at least partially by direction-dependent phase transformations~\eqref{gaugelin}.
However, we cannot expect that the contributions~\eqref{pertcont} and~\eqref{gaugelin} will
cancel each other perfectly. Instead, we will pick up error terms, which need to be worked out
systematically.

Before entering these constructions in Section~\ref{seccl}, in the next section we
need to extend the previous constructions to more general perturbations and to three generations.

\section{Specifying the Retarded Jets} \label{secret}
In this section, we shall specify the jet space~$\Jret$; see Definition~\ref{defJvary} below.
This jet space can be regarded as a a concrete realization of the space formed abstractly
in Section~\ref{secretarded} by choosing retarded representatives of~$\Jvary / \Jnull$ .
Our method is a generalization of the standard method of prescribing the retardation by a suitable
choice of the poles in momentum space.

\subsection{A General Ansatz for the Retarded Perturbation}
We want to extend the first order perturbation expansion of the Section~\ref{secnonloc} to several generations.
Moreover, since the formula for the perturbation by a nonlocal potential~\eqref{nuqgauge} is somewhat complicated,
we shall simplify the setup using a more convenient notation.
The general structure of~\eqref{nuqgauge} is
\beq \label{onegen}
\lim_{\mu \searrow 0}\frac{1}{2 p q+ i \mu \:\Big(p^0 + \frac{q^0}{2} \Big)}
\: g_q\Big( p-\frac{q}{2} \Big) \: \hat{P}\Big( p-\frac{q}{2} \Big)\:,
\eeq
where~$g_q$ is a smooth, matrix-valued function. The structure of the pole implements retardation
(as can be verified in the standard way by considering the Fourier integral and carrying out the
$\omega$-integral with residues). When generalizing the resulting formula to three generations, we want to
allow for the possibility that the interaction maps solutions on one Dirac sea to another Dirac sea.
In this case, the denominator in~\eqref{onegen} involves the differences of the squares of the
two masses. This leads us to the ansatz
\beq \label{nuqgen}
\delta_q \hat{P}(p) = \sum_{\alpha, \beta=1}^3
\lim_{\mu \searrow 0}\frac{1}{2 p q + m_\alpha^2 - m_\beta^2 +i \mu \:\Big(p^0 + \frac{q^0}{2} \Big)}
\: g_q^{\beta, \alpha} \Big( p-\frac{q}{2} \Big) \: \hat{P}_{m_\alpha}\Big( p-\frac{q}{2} \Big) \:,
\eeq
where~$\hat{P}_{m_\alpha}$ as given by~\eqref{Pmdef} is the summand in~\eqref{nuvacthree} corresponding to the generation~$\alpha$. Similar to~\eqref{numqgauge}, $\delta_{-q}^* \hat{P}$ is obtained by taking the adjoint
and replacing~$q$ by~$-q$,
\begin{align}
&\delta_{-q}^* \hat{P}(p) \notag \\
&= -\sum_{\alpha, \beta=1}^3
\lim_{\mu \searrow 0}
 \: \hat{P}_{m_\beta}\Big( p+\frac{q}{2} \Big)\: \big(g_{-q}^{\alpha,\beta}\big)^* \Big( p+\frac{q}{2} \Big)\:
\frac{1}{2 p q + m_\alpha^2  - m_\beta^2 +i \mu \:\Big(p^0 - \frac{q^0}{2} \Big)} \:.  \label{numqgen}
\end{align}
Finally, the distribution~$\hat{\tilde{P}}$ is obtained by substituting these formulas into~\eqref{tilmuone}.
Note that the poles of~\eqref{Pmdef} and~\eqref{numqgen} lie precisely on the intersection points of the
mass shells in Figures~\ref{figdelP} and~\ref{figdelP2}.

\subsection{The Causal Compatibility Conditions} \label{secccc}
Causal compatibility conditions were first introduced in~\cite{sea} as conditions
for the potential in the Dirac operator which ensure that the causal perturbation expansion
is well-defined and that the resulting light-cone expansion involves only bounded line integrals
(for details see also~\cite[Section~2.2]{pfp} or~\cite[Section~4.2]{cfs}).
In~\cite[Appendix~F]{cfs}, the connection was made to the regularity of the perturbation expansion
in momentum space. The latter perspective is the starting point for the following consideration,
which can be regarded as a continuation and extension of the considerations in~\cite[Appendix~F]{cfs}.

The formulas~\eqref{nuqgen} and~\eqref{numqgen} have
poles on the intersections of the mass shells.
More precisely, in~\eqref{nuqgen} and~\eqref{numqgen} we have a pole if~$2pq = m_\beta^2 - m_\alpha^2$,
which can be written equivalently as
\[ \Big( p-\frac{q}{2} \Big)^2 - m_\alpha^2 = \Big( p+\frac{q}{2} \Big)^2 - m_\beta^2 \:. \]
With our method of treating the poles in the distributional sense, both
terms~\eqref{nuqgen} and~\eqref{numqgen} are mathematically well-defined for any~$q \neq 0$.
However, it is important to note that divergences appear in the limit~$q \rightarrow 0$.
Again, these singularities are unproblematic from the mathematical point of view, because
integrating over~$q$ gives well-defined formulas for~$P(x,y)$.
Nevertheless, the singularity as~$q \rightarrow 0$ points to an infrared issue,
which can be best described in position space by the effect that the
light-cone expansion involves unbounded line integrals.
In order to avoid such unbounded line integrals, we need to ensure that the
poles in~\eqref{nuqgen} and~\eqref{numqgen} cancel each other.
In preparation of stating the resulting condition, we introduce the abbreviation
\beq \label{habdef}
h_q^{\beta, \alpha}\Big( p-\frac{q}{2} \Big) := g_q^{\beta, \alpha}\Big( p-\frac{q}{2} \Big) \: \Big( \slashed{p} - \frac{\slashed{q}}{2} - m_\alpha \Big)
\eeq
and denote the vector component of this function with a tensor index and a slash, i.e.\
\[ \slashed{h}_q^{\beta, \alpha}
:= \frac{1}{4} \: \gamma_j \:\Tr \big( \gamma^j h_q^{\beta, \alpha} \big) \:. \]

\begin{Def} \label{defJvary}
Given~$q \in \hat{M}$, we define~$\Jret$ as all retarded perturbations of the form~\eqref{nuqgen},
where the functions~$g^{\beta, \alpha}_q$ have the following properties:
\bitem
\item[{\rm{(i)}}] The functions~$g^{\beta, \alpha}_q$ are smooth.
\item[{\rm{(ii)}}] The perturbation behaves quadratically at infinity in the sense that for every
unit vector~$\vec{k} \in S^2 \subset \R^3$ we have the following expansion for large~$\lambda$,
\beq \label{infdef}
\slashed{h}_q^{\beta, \alpha}\Big( p-\frac{q}{2} \Big) \bigg|_{p = \big(-\sqrt{\lambda^2 |\vec{k}|^2 +m_\alpha^2}, \,\lambda \vec{k} \big)} = \lambda^2\:
\Phi_q^{\beta, \alpha}\big( \vec{k} \,\big)\: \hat{\slashed{p}} + \O(\lambda) \:,
\eeq
where~$\Phi_q^{\beta, \alpha} \in C^\infty(S^2, \C)$ is a smooth function on the unit sphere and
\beq \label{hatpdef}
\hat{p} := (-1, \vec{k} \,\big) \:.
\eeq
\item[{\rm{(iii)}}] The following {\bf{causal compatibility conditions}} hold,
\begin{gather}
\slashed{h}_q^{\beta, \alpha}\Big( p-\frac{q}{2} \Big)
= \big(\slashed{h}_{-q}^{\alpha, \beta}\big)^*\Big( p+\frac{q}{2} \Big) \qquad \text{for all~$p \in \hat{M}$
with~$2 p q = m_\beta^2 - m_\alpha^2$}\:, \label{ccc0} \\
\Big(\Phi_q^{\beta, \alpha} - \big(\Phi_{-q}^{\alpha, \beta} \big)^* \Big)
\big( \hat{\vec{k}} \,\big) \:\frac{1}{\hat{p} q} \;\in\; C^\infty(S^2, \C) \:. \label{ccc}
\end{gather}
\eitem
Under the above assumptions, we introduce the {\bf{perturbation map}}~$\hat{\scrP}$ by
\[ \hat{\scrP} \::\: \Jret \rightarrow {\mathcal{D}}'(\hat{M}, \Lin(V)) \:,\qquad
\scrP(\bv) := \delta_q \hat{P} + \delta_{-q}^* \hat{P} \:. \]
with~$\delta_q \hat{P}$ and~$\delta_{-q}^* \hat{P}$ according to~\eqref{nuqgen} and~\eqref{numqgen}.
\end{Def}

We point out that these causal compatibility conditions indeed ensure that the perturbation map
is regular in the limit~$q \rightarrow 0$ when the momentum transfer tends to zero.
Namely, combining~\eqref{nuqgen} and~\eqref{numqgen}
with~\eqref{ccc0}, we find
\begin{align*}
&\lim_{q \rightarrow 0} \big( \delta_q \hat{P} + \delta_{-q}^* \hat{P} \big) \\
&= \sum_{\alpha=1}^3 \Theta(-\omega)\:e^{\varepsilon \omega} \;\lim_{q \rightarrow 0} \bigg\{ 
\frac{1}{2 p q} \\
&\quad\;\; \times \bigg(
h_q^{\alpha, \alpha} \Big( p-\frac{q}{2} \Big) \: \delta\Big( p - pq + \frac{q^2}{4} -m^2 \Big)
- \big(h_{-q}^{\alpha,\alpha}\big)^* \Big( p+\frac{q}{2} \Big)   \: \delta\Big( p + pq + \frac{q^2}{4} -m^2 \Big) \bigg) 
\bigg\} \\
&\quad\; + \sum_{\alpha \neq \beta}
\frac{1}{m_\alpha^2 - m_\beta^2} \:\bigg(g_0^{\beta, \alpha} (p) \: \hat{P}_{m_\alpha}(p)
- \hat{P}_{m_\beta}(p)\: \big(g_0^{\alpha,\beta}\big)^*(p) \bigg) \:,
\end{align*}
and the remaining limit can be be taken in the distributional sense
using again~\eqref{ccc0} and l'Hospital's rule.
With this result in mind, we may assume throughout this paper that~$q$ is non-zero.
This justifies our earlier assumption~\eqref{qnonozero}.

\section{Solving the Linearized Field Equations in the Continuum Limit} \label{seccl}
Having specified the space~$\Jret$ of all admissible retarded jets, we are now in the position
to solve the linearized field equations. After compensating the direction-dependent phases
using the method explained in Section~\ref{secphase} (Section~\ref{seccompdir}), we can rewrite the
formula~\eqref{d2Scont} for the second variations in the continuum limit
as a positive semi-definite sesquilinear form on a weighted $L^2$-space on the light cone
(see~\eqref{splight} in Section~\ref{secweighthilbert}).
In this setting, the linearized field equations can be solved abstractly by applying the Fr{\'e}chet-Riesz theorem.
This formulation is also the starting point for the more detailed computational approach
to be developed in Section~\ref{secmom}.

\subsection{Compensating for Direction-Dependent Phases} \label{seccompdir}
The perturbation of the kernel of the fermionic projector~$\delta P = \scrP(\bv)$ introduced in Definition~\ref{defJvary}
has the shortcoming that in general it does {\em{not}} satisfy the condition~\eqref{delPcond}.
A simple example already explained in detail in Section~\ref{secdirgauge}
is to consider a classical gauge potential,
in which case~$\delta P(x,y)$ has singularities of the same degree on the light cone as the vacuum kernel
(cf.~\eqref{Pgauge}).
Our general strategy for dealing with this problem is to introduce a kernel~$\scrG(\bv)$ which is obtained
from~$P(x,y)$ by multiplication with a function~$\hat{\Lambda}(q, \xi)$, i.e.\
\beq \label{scrGpresent}
\scrG : \Jret \rightarrow (C^\infty \cap {\mathcal{S}}')(M, \Lin(V)) \qquad \text{with} \qquad
\big( \scrG \bv \big)(x,y) = i \hat{\Lambda}(q, \xi) \: P(x,y) \:,
\eeq
where~$\hat{\Lambda}(q, \xi)$ is the function in~\eqref{hatLambdadef} having the properties~\eqref{Lamrealpos}
(here~$C^\infty \cap {\mathcal{S}}'$ denotes smooth solutions which increase at most polynomially
at infinity). As explained in Section~\ref{secphase}, the causal action
is invariant under the transformation~\eqref{scrGpresent}.
Our task is to choose~$\hat{\Lambda}(q, \xi)$ as a function of~$\bv$ in such a way
that the leading singularity of~$\scrP \bv$ is compensated on the light cone, i.e.\
\beq \label{leadsing}
\Tr \big( \slashed{\xi}\: \big((\scrG - \scrP)(\bv)\big)(x,y) \big) = (\deg < 1) \:.
\eeq
Once this has been accomplished, 
the formalism of the continuum limit applies to~$\delta P := (\scrG - \scrP)(\bv)$,
making it possible to solve the linearized field equations.

The correct choice of~$\scrG \bv$ can be derived from the simple Fourier transformation formula~\eqref{Fouriersimp}.
Before we can apply this formula, we need to determine the behavior of the prefactors in~\eqref{nuqgen} near infinity.
Using~\eqref{infdef}, for the vector component of~$\delta_q \hat{P}(p)$ we obtain the asymptotic formula
\[ \delta_q \hat{P}(p) \Big|_{p = \big(-\sqrt{\lambda^2 |\vec{k}|^2 +m_\alpha^2}, \lambda \vec{k} \big)}
= \sum_{\alpha, \beta=1}^3
\frac{\Phi_q^{\beta, \alpha}\big( \vec{k} \,\big)}{2 \hat{p} q}
\: \hat{P}_{m_\alpha}\Big( p-\frac{q}{2} \Big) 
\bigg( 1 + \O \Big( \frac{1}{\lambda} \Big) \bigg) \:. \]
Now we can take the Fourier transform with the help of~\eqref{Fouriersimp}.
Using~\eqref{hatpdef}, we obtain
\[ \delta_q P(\xi) = \sum_{\alpha, \beta=1}^3
\frac{\Phi_q^{\beta, \alpha}\Big( -\epsilon(\xi^0)\, \hat{\vec{\xi}} \,\Big)}{2\, \big(-1, -\epsilon(\xi^0)\, \hat{\vec{\xi}}\;\big) q}
\:e^{\frac{i}{2} q \xi}\: P_{m_\alpha}(\xi) + (\deg < 2) \:, \]
where the factor~$e^{\frac{i}{2} q \xi}$ describes a translation in momentum space (cf.\ \eqref{delPmink}
and~\eqref{hatLambdadecomp}).
To leading order for small~$\varepsilon$, the masses of the Dirac seas are irrelevant.
We thus obtain
\begin{align*}
\delta_q P(\xi) &= -\frac{1}{6} \sum_{\alpha, \beta=1}^3
\Phi_q^{\beta, \alpha}\Big( -\epsilon(\xi^0)\, \hat{\vec{\xi}} \,\Big)\:\xi^0
\:\frac{e^{\frac{i}{2} q \xi}}{q \xi}\: P(\xi) \Big( 1 +  \O \Big( \frac{\varepsilon}{t} \Big) \bigg)  + (\deg < 2) \:.
\end{align*}
Finally, using that
\[ \int_{-\infty}^\infty \epsilon(\tau)\:  e^{-i (\tau-\frac{1}{2}) \,q \xi}\: d\tau \\
= e^{\frac{i}{2} q \xi} \int_{-\infty}^\infty \epsilon(\tau)\:  e^{-i \tau \,q \xi}\: d\tau \\
= -2i \: \frac{e^{\frac{i}{2} q \xi}}{q \xi} \:, \]
we conclude that
\[ \delta_q P(\xi) = 
-\frac{i}{12} \sum_{\alpha, \beta=1}^3
\Phi_q^{\beta, \alpha}\Big( -\epsilon(\xi^0)\, \hat{\vec{\xi}} \,\Big)\:\xi^0
\:\int_{-\infty}^\infty \epsilon(\tau)\:  e^{-i \tau \,q \xi}\: d\tau\: P(\xi)  + (\deg < 2) \:. \]
This perturbation is of the form~\eqref{gaugelin}, making it possible to compensate for it by
the following direction-dependent gauge transformation.

\begin{Def} \label{defscrG} 
Given~$\bv \in \Jret$, using the notation~\eqref{hatLambdadef} and~\eqref{hatLambdadecomp},
we introduce~$\scrG \bv$ as
\beq \label{defG0}
\scrG \bv := \big( i \hat{\Lambda}_q(\xi) - i \hat{\Lambda}^*_{-q}(\xi) \big)\: P(x,y)
\eeq
with
\begin{align}
\hat{\Lambda}_q(\xi) &:=
-\frac{1}{12} \sum_{\alpha,\beta=1}^3
\Phi_q^{\beta, \alpha}\Big( -\epsilon(\xi^0)\, \hat{\vec{\xi}} \,\Big)\:
\xi^0 \int_{-\infty}^\infty \epsilon(\tau)\:
 e^{-i (\tau-\frac{1}{2}) \,q \xi}\: d\tau \label{defG1} \\
\hat{\Lambda}^*_{-q}(\xi) &:=
-\frac{1}{12} \sum_{\alpha,\beta=1}^3 \Big(\Phi_{-q}^{\alpha, \beta} \Big( -\epsilon(\xi^0)\, \hat{\vec{\xi}} \,\Big)\Big)^*
\:\xi^0 \int_{-\infty}^\infty \epsilon(\tau-1)\: e^{-i (\tau-\frac{1}{2}) \,q \xi}\: d\tau \label{defG2}
\end{align}
and~$\Phi_q^{\beta, \alpha}$ as defined by~\eqref{infdef}.
\end{Def} \noindent
By direct computation, one verifies that the reality condition~\eqref{Lamreal} holds.
This means that, writing~$\scrG \bv$ as in~\eqref{gaugelin}, the function~$\Lambda(x,y)$
is indeed real and anti-symmetric~\eqref{Lsymm}.
Also, our notation harmonizes with~\eqref{hatLambdadef}.
Therefore, the operator~$\scrG$ indeed
describes a direction-dependent phase transformation, which drops out of the
EL equations. It has the purpose of compensating the leading singularity of~$\scrP \bv$ on the
light cone, so that the total perturbation satisfies
the condition~\eqref{delPcond} needed for the evaluation in the continuum limit.

We conclude this section by explaining the above formulas in the simple example of
an electromagnetic potential~$\hat{A}(q)$. In this case, 
\begin{align*}
g_q^{\beta, \alpha} \Big( p-\frac{q}{2} \Big) &= -\delta^{\beta, \alpha}\:
\Big( \slashed{p}+\frac{\slashed{q}}{2} + m_\beta \Big) \:\,\,\hat{\!\!\slashed{A}}(q) \\
h_q^{\beta, \alpha}\Big( p-\frac{q}{2} \Big) &= -\delta^{\beta, \alpha}\:
\Big( \slashed{p}+\frac{\slashed{q}}{2} + m_\beta \Big) \:\,\,\hat{\!\!\slashed{A}}(q)\:
\Big( \slashed{p} - \frac{\slashed{q}}{2} - m_\alpha \Big) \\
\slashed{h}_q^{\beta, \alpha} \Big( p-\frac{q}{2} \Big) &= -\delta^{\beta, \alpha}\:
\Big( 2 \hat{A}_j(q)\: p^j\: \slashed{p} - p^2\, \,\,\hat{\!\!\slashed{A}}(q) - m_\alpha^2\, \,\,\hat{\!\!\slashed{A}}(q) \Big) \:.
\end{align*}
Using~\eqref{infdef}, we obtain
\[ \Phi_q^{\beta, \alpha}\big( \hat{\vec{k}} \,\big)
= -\delta^{\beta, \alpha}\:2 \hat{A}_j(q)\: \hat{p}^j \qquad \text{and thus} \qquad
\Phi_q^{\beta, \alpha}\Big( -\epsilon(\xi^0)\, \hat{\vec{\xi}} \,\Big) \:\xi^0 = \delta^{\beta, \alpha}\:2 \hat{A}_j(q)\: \xi^j \]
with~$\hat{\vec{k}}$ and~$\hat{p}$ as in~\eqref{hatpdef}. As a consequence,
the result of Definition~\ref{defscrG} simplifies to
\begin{align*}
\scrG \bv &= -\frac{i}{2}\:\hat{A}_j(q)\: \xi^j \int_{-\infty}^\infty \epsilon(\tau)\:
 e^{-i (\tau-\frac{1}{2}) \,q \xi}\: d\tau\: P(x,y) \\
&\quad\: +\frac{i}{2} \: \hat{A}_j(q)\: \xi^j \int_{-\infty}^\infty \epsilon(\tau-1)\: e^{-i (\tau-\frac{1}{2}) \,q \xi}\: d\tau\: P(x,y) \\
&= -i\:\hat{A}_j(q)\: \xi^j \int_0^1
 e^{-i (\tau-\frac{1}{2}) \,q \xi}\: d\tau\: P(x,y) \\
& = -i e^{i \frac{q}{2} (x+y)} \int_0^1A_j \big(\tau y + (1-\tau) x \big)\:\xi^j\: d\tau\: P(x,y) \:.
\end{align*}
This is precisely the first order contribution to the gauge phase~\eqref{Pgauge1}
(one should take into account the factor~$e^{-i \frac{q}{2} (x+y)}$ in~\eqref{hatLambdadef}).
It also agrees with the local phase transformation as obtained in the so-called
{\em{light cone expansion}} (see~\cite[Appendix~A]{firstorder},
\cite{pfp} or~\cite[\S2.2.4, \S3.6.2]{cfs}).
In this way, one sees that~$\scrG \bv$ indeed compensates for the leading singularity
on the light cone~\eqref{leadsing}.

\subsection{Solution in a Weighted $L^2$-Space on the Light Cone} \label{secweighthilbert}
The goal of this section is to reformulate the formula for the second variations in~\eqref{d2Scont}
in a way where functional analytic methods can be employed.
Using that the kernel and the integration measure~$d\rho = d^4x$ are translation invariant,
we can carry out the integral over the variable~$\zeta := (x+y)/2$ with the help of the distributional identity
\beq \label{qqprel}
\int_N e^{-i (q-q') \zeta}\: d^4 \zeta = (2 \pi)^4\: \delta^4(q-q') \:.
\eeq
With this in mind, we may fix the momentum~$q$ and leave out the $\zeta$-integration.
Thus it remains to consider the $\xi$-integration (where again~$\xi :=y-x$).
However, one needs to keep in mind that, in view of~\eqref{qqprel}, the plane waves~$e^{iq\zeta}$
of the two factors~$\delta P$ in~\eqref{d2Sv} must come with the opposite sign of~$q$.
Using the relation
\[ \delta P(\xi) = (\delta_q P + \delta_{-q}^*P)(\xi)^* = (\delta_{-q} P + \delta_q^* P)(-\xi) \:, \]
we obtain
\begin{align}
\delta^2 &\Sact^\text{\rm{eff}}(\rho) = 2 \int_M d^4 \xi\: \big(-i t\, K_0(x,y) \big)\:
\:\frac{\eta_{\min}^2(t)}{\varepsilon^3\: t^4} \notag \\
&\times
\Big\{ c_1\: \Tr\big(\slashed{\xi}\, \delta P(\xi)^* \big)\: \Tr \big( \slashed{\xi}\: \delta P(\xi) \big)
+ c_2\: \re \Tr\big(\slashed{\xi}\, \delta P(-\xi)^* \big)\: \Tr \big( \slashed{\xi}\: \delta P(\xi) \big) \Big\} \:. \label{d2Sv}
\end{align}
More specifically, we evaluate the second variation for
\beq \label{PGdiff}
\delta P = (\scrP - \scrG)(\bv) \qquad \text{with} \qquad \bv \in \Jret
\eeq
with~$\scrP \bv$ and~$\scrG \bv$ as in Definitions~\ref{defJvary} and~\ref{defscrG}.

\begin{Lemma} \label{lemma82} For any~$\bv \in \Jret$, the second variation~\eqref{d2Sv} is well-defined
and finite for~$\delta P$ according to~\eqref{PGdiff}.
\end{Lemma}
\Proof We first consider~$\delta P = \scrP \bv$. In view of the ultraviolet regularization
(see~\eqref{nuqgen} and~\eqref{Pmdef}), the distribution~$\hat{\scrP} \bv$ decays exponentially
in momentum space. Consequently, its Fourier transform~$\scrP \bv \in C^\infty(M, \Lin(V))$
is a smooth function in Minkowski space. Moreover, in view of the $\delta$-distributions in momentum
space in~\eqref{Pmdef}, it can be written as a sum of terms of functions~$\phi_\alpha$
which satisfy Klein-Gordon equations of the form
\[ \big( \Box + m_\alpha^2\big) \bigg( e^{\pm \frac{i q x}{2}} \phi_\alpha(x) \bigg) = 0 \:. \]
In~\eqref{d2Sv}, these functions are evaluated on the light cone~$\xi^0 = \pm |\vec{\xi}|$.
Therefore, in order prove that these integrals are well-defined, we need to show that
the functions~$\phi_\alpha$ decay sufficiently fast in lightlike directions.

Let us determine the decay rates. Expanding in spherical waves, a smooth solution of the scalar
wave equation decays in lightlike directions like one over the radius. Solutions of the Klein-Gordon equation
decay even faster (for detailed estimates see for example~\cite[Section~4.4]{treude}). Hence
\[ \Big| \phi_\alpha \big( \pm |\vec{\xi}|, \vec{\xi} \:\big) \Big| \lesssim \frac{1}{|\vec{\xi}|}\:. \]
Consequently, the absolute value of the integrand in~\eqref{d2Sv} can be estimated by
\[ \lesssim |t|\: \delta(\xi^2)\: \frac{\eta_{\min}^2(t)}{\varepsilon^3\:t^4} \:
\bigg( \frac{|t|}{|\vec{\xi}|} \bigg)^2 \:. \]
In order to estimate the integral, we carry out the time integration and choose polar coordinates
with~$r:= |\vec{\xi}|$. We thus obtain
\[ \delta^2 \Sact^\text{\rm{eff}}(\rho) \lesssim \frac{1}{\varepsilon^3}
\int_{\ell_{\min}}^\infty \bigg( \frac{1}{r} \:\frac{r}{r^4} \bigg) \:r^2\: dr < \infty \:. \]

For~$\delta P = \scrG \bv$ we need to estimate the line integrals in~\eqref{defG1} and~\eqref{defG2}.
To this end, it is most convenient to introduce the abbreviations
\[ A(\xi) := \frac{i}{6} \sum_{\alpha,\beta=1}^3 \Phi_q^{\beta, \alpha}\big( \hat{\vec{\xi}} \,\big)\:
\xi^0 \qquad \text{and} \qquad B(\xi) := \frac{i}{6} \sum_{\alpha,\beta=1}^3 \Big(\Phi_{-q}^{\alpha, \beta} \big( \hat{\vec{\xi}} \,\big)\Big)^* \:\xi^0 \:, \]
and to write~$\scrG \bv$ as
\begin{align}
\scrG \bv &= (A-B)(\xi) \int_{-\infty}^\infty \epsilon(\tau)\:
e^{-i \tau q \xi}\: d\tau\: e^{\frac{i}{2}q \xi}\:P(x,y) \label{G1} \\
&\quad\: + 2 B(\xi) \int_{0}^1 e^{-i (\tau-\frac{1}{2}) \,q \xi}\: d\tau\: P(x,y) \:. \label{G2}
\end{align}
We first treat~\eqref{G1}. Carrying out the line integral, we obtain
\[ \eqref{G1} = \frac{(A-B)(\xi)}{q \xi}\; i e^{\frac{i}{2} q \xi} \: P(x,y) \]
According to~\eqref{ccc}, the quotient is bounded for all~$\xi=(-1, \vec{\xi})$ and~$|\vec{\xi}|=1$.
Moreover, this quotient is homogeneous of degree zero. Therefore, it is smooth and uniformly
bounded for all~$\xi$ on the light cone. Consequently, the contribution~\eqref{G1} is a bounded function times~$P(x,y)$. Since~$P(x,y)$ satisfies the Klein-Gordon equation, we can argue exactly as for~$\scrP \bv$ above.

It remains to estimate~\eqref{G2}. Clearly, $B(\xi)$ is smooth and homogeneous of degree one.
Moreover, the integral in~\eqref{G2} can be estimated by one. Therefore, \eqref{G2} is bounded
by a linear polynomial times~$P(x,y)$. In the considered massive case, 
it was shown in~\cite[Section~5.1 and Appendix~B]{lagrange-hoelder} that
the restriction of~$P(x,y)$ to the light cone decays exponentially at infinity $\sim e^{-m^2 \varepsilon |t|}$. Therefore, the resulting
integrals in~\eqref{d2Sv} are well-defined and finite. This concludes the proof.
\QED

We finally polarize the second variation~\eqref{d2Sv} to obtain the bilinear form
\begin{align}
B &\::\: \Jret \times \Jret \rightarrow \R \notag \\
&B(\bu, \bv) \,\,\!\!:= 2 \re \int_M d^4 \xi\:\big(-i t\, K_0(x,y) \big)
\:\frac{\eta_{\min}^2(t)}{\varepsilon^3\: t^4} \notag \\
&\qquad\quad\: \times
\Big\{ c_1\: \Tr\big(\slashed{\xi}\, ( \scrP \bu - \scrG \bu)(\xi)^* \big)\: \Tr \big( \slashed{\xi}\: ( \scrP \bv - \scrG \bv)(\xi) \big) \notag \\
&\qquad\qquad
+ c_2\: \Tr\big(\slashed{\xi}\, ( \scrP \bu - \scrG \bu)(-\xi)^* \big) \: \Tr\big(\slashed{\xi}\, ( \scrP \bv - \scrG \bv)(\xi) \big)
\Big\} \\
&=  \frac{1}{2 \pi^2}\:\frac{1}{\varepsilon^3} \:\re \int_M d^4 \xi\:|t|\:\delta(\xi^2) \Big\{
c_1\:(\scrL \bu)(\xi)\: \overline{(\scrL \bv)(\xi)} + c_2\: (\scrL \bu)(\xi)\:\overline{ (\scrL \bv)(-\xi)}
\Big\} \:, \label{splight}
\end{align}
where we introduced the abbreviation
\[ (\scrL \bv)(\xi) := \Tr \big( \slashed{\xi}\: ( \scrP \bv - \scrG \bv)(\xi) \: \frac{\eta_{\min}(t)}{t^2} \:. \]
The integral in~\eqref{splight} can be regarded as a weighted $L^2$-scalar product on the light cone.
Using the Schwarz inequality and~\eqref{c12rel}, one sees that
the bilinear form~$B$ is indeed positive semi-definite. This makes it possible
to solve the linearized field equations abstractly by applying the Fr{\'e}chet-Riesz theorem
in the resulting Hilbert space.
This method has the disadvantage that it is not computational and does not give detailed information
on how the solution looks like.

\section{Continuum Limit Analysis in Momentum Space} \label{secmom}
The goal of this section is to develop a computational method for solving the linearized
field equations. It turns out to be preferable to work in momentum space.
The main advantage is that the evaluation on the light cone in~\eqref{splight}
means in momentum space that we are concerned with {\em{harmonic functions}}
(i.e.\ solutions of the wave equation~$\Box_p \phi(p) = 0$).
The {\em{strong Huygens principle}} (i.e.\ the fact that waves propagate with the speed of light)
will give us a deeper and more detailed understanding of the linearized field equations.
This will lead us to introducing the so-called {\em{mass cone expansion}} as a powerful
computational tool.

\subsection{Formulation in an Energy Hilbert Space}
In preparation, we introduce the local operator in position space
\beq \label{Cdef}
\begin{split}
\scrC \::\: \big(C^\infty \cap \mathcal{S}')(M, \Lin(V)) &\rightarrow \mathcal{S}'(M, \C) \:, \\
\delta P(\xi) &\mapsto \Tr \big( \slashed{\xi}\: \delta P(\xi) \big)\: \delta(\xi^2)\: \epsilon(t)\:
\frac{\eta_{\min}(t)}{t^2} \:.
\end{split}
\eeq
Since the image of~$\scrC$ is supported on the light cone, its Fourier transform is {\em{harmonic in
momentum space}} in the sense that it satisfies the distributional equation
\[ \Box_p \:\widehat{ \scrC\big( \delta P \big) }(p) = 0 \:. \]
Denoting the harmonic functions in momentum space by~$\scrH$, we thus obtain the mapping
\beq \label{Chatdef}
\hat{\scrC} \::\: \big(C^\infty \cap \mathcal{S}')(M, \Lin(V)) \rightarrow
(\scrH \cap \mathcal{S}')(\hat{M}, \C) \:.
\eeq

On harmonic functions in momentum space it is natural to work with the
{\em{energy scalar product}}
\beq \label{Esprod}
\begin{split}
&E \::\: \scrH \times \scrH \rightarrow \C \:, \\
&E(\hat{\phi}, \hat{\psi}) := \frac{1}{2} \int_{\R^3} \Big( \big( \overline{\partial_\omega \hat{\phi}(\omega, \vec{k})}
\:\partial_\omega \hat{\psi}(\omega, \vec{k}) + 
\overline{\vec{\nabla} \hat{\phi}(\omega, \vec{k})}\:
\vec{\nabla} \hat{\psi}(\omega, \vec{k}) \big)\: d^3k \:.
\end{split}
\eeq
Due to energy conservation, this energy scalar product does not depend on~$\omega$.
We next rewrite this energy scalar product in position space.
\begin{Lemma} Let~$\hat{\psi}$ be a complex-valued harmonic function in momentum space of the form
\beq \label{hatpsi}
\hat{\psi}(p) = \int_M \delta(\xi^2)\: \epsilon(\xi^0)\: \psi(\xi)\: e^{-i p \xi}\: d^4 \xi \:.
\eeq
Then
\beq \label{Ehat}
E \big(\hat{\psi}, \hat{\psi} \big) = 4 \pi^3 \int_M \delta(\xi^2)\: \big| \xi^0 \big|\:
\overline{\psi(\xi)}\: \psi(\xi)\: d^4 \xi \:.
\eeq
\end{Lemma}
\Proof Using~\eqref{hatpsi} in~\eqref{Esprod}, we obtain
\begin{align*}
&E \big(\hat{\psi}, \hat{\psi} \big) \\
&= \frac{1}{2} \int_{\R^3} d^3 k
\int_M d^4 \xi\:\delta(\xi^2)\: \epsilon(\xi^0)\: \overline{\psi(\xi)}
\int_M d^4 \tilde{\xi}\:\delta(\tilde{\xi}^2)\: \epsilon(\tilde{\xi}^0)\: \psi(\tilde{\xi})\:
\Big( \xi^0 \tilde{\xi}^0 + \vec{\xi} \vec{\tilde{\xi}} \,\Big) 
\: e^{i k (\xi-\tilde{\xi})} \\
&= 4 \pi^3 \int_{-\infty}^\infty dt \int_{-\infty}^\infty d\tilde{t} \: e^{i \omega (t-\tilde{t})} \int_{\R^3} d^3\xi \:
\delta \big( t^2 - |\vec{\xi}|^2 \big)\: \delta \big( \tilde{t}^2 - |\vec{\xi}|^2 \big)\: \epsilon(t)\: \epsilon(\tilde{t}) \\
&\qquad\qquad\qquad\qquad\qquad\qquad\qquad \times \Big( t \tilde{t} + \big|\vec{\xi}\,\big|^2\Big)
\overline{\psi \big(t, \vec{\xi} \:\big)}\: \psi\big(t, \vec{\tilde{\xi}} \:\big) \:,
\end{align*}
where in the last step we applied Plancherel's theorem.
Carrying out the integral over~$\tilde{t}$, we only get a contribution if~$t=\tilde{t}$. We thus obtain
\[ E \big(\hat{\psi}, \hat{\psi} \big)
= 4 \pi^3 \int_{-\infty}^\infty dt \int_{\R^3} d^3\xi \:
\delta \big( t^2 - |\vec{\xi}|^2 \big)\:|t|\:
\overline{\psi \big(t, \vec{\xi} \:\big)}\: \psi\big(t, \vec{\xi} \:\big) \:, \]
concluding the proof.
\QED
With the help of this lemma, we can in turn rewrite the second variation in momentum space.

\begin{Prp} \label{prpchat}
The second variation of the action~\eqref{d2Sv} can be written in momentum space as
\[ \delta^2 \Sact^\text{\rm{eff}}(\rho) = \frac{1}{8 \pi^5}\: \frac{1}{\varepsilon^3}
\Big( c_1\: E \big( \hat{\scrC}(\delta P), \hat{\scrC}(\delta P) \big)
+ c_2\: \re E \big( \hat{\scrC}( {\mathscr{R}}\, \delta P), \hat{\scrC}(\delta P) \big) \Big) \:, \]
where the operator~$\mathscr{R}$ flips the sign of the argument,
\beq \label{reflect}
{\mathscr{R}}\,\delta P(\xi) = \delta P(-\xi) \qquad \text{and} \qquad
{\mathscr{R}}\,\delta \hat{P}(p) = \delta \hat{P}(-p) \:.
\eeq
\end{Prp}
\Proof Using~\eqref{K0def} in~\eqref{Ehat}, we obtain
\beq \label{Eform}
E \big(\hat{\psi}, \hat{\psi} \big)
= 16 \pi^5 \int_{-\infty}^\infty dt \int_{\R^3} d^3\xi \: \big(-i t\, K_0(x,y) \big)\:
\overline{\psi \big(t, \vec{\xi} \:\big)}\: \psi\big(t, \vec{\xi} \:\big) \:.
\eeq
We now choose~$\hat{\psi}=\hat{\scrC}(\delta P)$. Then, comparing~\eqref{hatpsi} with~\eqref{Cdef},
we conclude that
\[ \psi(\xi) = \Tr \big( \slashed{\xi}\: \delta P(\xi) \big)\: \frac{\eta_{\min}(t)}{t^2} \:. \]
Using this formula in~\eqref{Eform} and comparing with~\eqref{d2Sv} gives the result.
\QED
Polarization shows that the bilinear form~\eqref{splight} now takes the form
\begin{align}
&B(\bu, \bv) = \frac{1}{8 \pi^5}\: \frac{1}{\varepsilon^3} \notag \\
&\times \re 
\Big( c_1\: E \big( \hat{\scrC}(\scrP \bu - \scrG \bu), \hat{\scrC}(\scrP \bv - \scrG \bv) \big)
+ c_2\: \big( \hat{\scrC}({\mathscr{R}}\,\scrP \bu - {\mathscr{R}}\,\scrG \bu), \hat{\scrC}(\scrP \bv - \scrG \bv) \big) \Big) \:.
 \label{BPG}
 \end{align}

\subsection{Evaluation on a Cauchy Surface Using the Huygens Principle} \label{sechuygens}
In view of Proposition~\ref{prpchat}, we need to compute the energy scalar product~$E(.,.)$
for harmonic functions of the form~$\hat{\scrC}(\delta P)$ and~$\hat{\scrC}({\mathscr{R}}\,\delta P)$.
Due to energy conservation, we can evaluate the energy scalar product~\eqref{Esprod}
for any~$\omega$. For the computations, it turns out to be preferable to evaluate at~$\omega=\underline{\omega}$ 
with~$\underline{\omega}$ negative and very small, even much smaller than the energy
scale~$\varepsilon^{-1}$ of the ultraviolet regularization, i.e.\
\[ \omega = \underline{\omega} \ll - \frac{1}{\varepsilon} \:. \]
Then we need to compute~$\hat{\scrC}(\delta P)$ and~$\hat{\scrC}({\mathscr{R}}\,\delta P)$ on the Cauchy surface~$(\underline{\omega}, \R^3) \subset \hat{M}$. We begin with the
contribution~$\delta P = \scrP \bv$ (Sections~\ref{sechuygens}--\ref{secmasspole}),
whereas the contribution by the direction-dependent
phases~$\delta P = \scrG \bv$ will be computed afterward (Section~\ref{secdirphase}).

Thus we let~$\delta P = \scrP \bv$. According to~\eqref{Cdef}, the computation of~$\scrC(\delta P)$ can
be performed in three steps: \label{E1toE3}
\begin{itemize}[leftmargin=3em]
\item[{\bf{(E1)}}] Multiply in position space by
\[ \delta(\xi^2)\: \epsilon(t) \:. \]
\item[{\bf{(E2)}}] Multiply in position space by
\beq \label{lminfunct}
\frac{\eta_{\min}(t)}{t^2} \:.
\eeq
\item[{\bf{(E3)}}] Multiply by~$\slashed{\xi}$ and take the trace. This amounts to contracting the
vectorial component with~$\xi$; we use the short notation
\[ \frac{1}{4}\: \Tr \big( \slashed{\xi}\, h(x,y) \big) = \xi_j \: h^j(x,y) \:. \]
\end{itemize}
Clearly, these three steps can be performed in an arbitrary order. Our numbering reflects the order
in which we will carry out these computation steps later on.
In preparation, we need to analyze what these steps mean in momentum space.
We begin with the easiest step~(E3).
Following the computation
\begin{align*}
\frac{1}{4}\: \Tr \big( \slashed{\xi} \,\delta P(x,y) \big) &= \int_{\hat{M}} \frac{d^4q}{(2 \pi)^4} \: e^{-i\frac{q}{2}\, (x+y)}
\: \int_{\hat{M}} \frac{d^4p}{(2 \pi)^4}\: e^{ip \xi }\:\xi_j \:\delta \hat{P}^j(p) \\
&= -i \int_{\hat{M}} \frac{d^4q}{(2 \pi)^4} \: e^{-i\frac{q}{2}\, (x+y)}
\: \int_{\hat{M}} \frac{d^4p}{(2 \pi)^4}\: \bigg( \frac{\partial}{\partial p^j} e^{ip \xi } \bigg) \:
\delta \hat{P}^j(p) \\
&= i \int_{\hat{M}} \frac{d^4q}{(2 \pi)^4} \: e^{-i\frac{q}{2}\, (x+y)}
\: \int_{\hat{M}} \frac{d^4p}{(2 \pi)^4} \: e^{ip \xi } \: \partial_j \delta \hat{P}^j(p) \:,
\end{align*}
Step~(E3) consist in taking the divergence, being a {\em{local}} operation momentum space.

\begin{figure}
\psset{xunit=.5pt,yunit=.5pt,runit=.5pt}
\begin{pspicture}(400.13764149,320.5609275)
{
\newrgbcolor{curcolor}{0.80000001 0.80000001 0.80000001}
\pscustom[linestyle=none,fillstyle=solid,fillcolor=curcolor]
{
\newpath
\moveto(0.7756989,9.9588916)
\lineto(0.83148472,0.69700813)
\lineto(11.10257008,0.56094514)
\lineto(59.99731654,46.82568089)
\lineto(40.7849915,46.6144053)
\closepath
}
}
{
\newrgbcolor{curcolor}{0.80000001 0.80000001 0.80000001}
\pscustom[linewidth=0,linecolor=curcolor]
{
\newpath
\moveto(0.7756989,9.9588916)
\lineto(0.83148472,0.69700813)
\lineto(11.10257008,0.56094514)
\lineto(59.99731654,46.82568089)
\lineto(40.7849915,46.6144053)
\closepath
}
}
{
\newrgbcolor{curcolor}{0.80000001 0.80000001 0.80000001}
\pscustom[linestyle=none,fillstyle=solid,fillcolor=curcolor]
{
\newpath
\moveto(-0.22490457,56.26522483)
\lineto(0.04748598,37.08244813)
\lineto(0.31987654,36.81005758)
\lineto(39.99996094,0.56094514)
\lineto(59.99731276,0.56094514)
\closepath
}
}
{
\newrgbcolor{curcolor}{0.80000001 0.80000001 0.80000001}
\pscustom[linewidth=0,linecolor=curcolor]
{
\newpath
\moveto(-0.22490457,56.26522483)
\lineto(0.04748598,37.08244813)
\lineto(0.31987654,36.81005758)
\lineto(39.99996094,0.56094514)
\lineto(59.99731276,0.56094514)
\closepath
}
}
{
\newrgbcolor{curcolor}{0.80000001 0.80000001 0.80000001}
\pscustom[linestyle=none,fillstyle=solid,fillcolor=curcolor]
{
\newpath
\moveto(82.3668737,85.7569653)
\lineto(92.59566236,85.7569653)
\lineto(102.54966425,85.7569653)
\lineto(353.95447181,319.47571554)
\lineto(334.9569789,319.47570798)
\closepath
}
}
{
\newrgbcolor{curcolor}{0.80000001 0.80000001 0.80000001}
\pscustom[linewidth=0,linecolor=curcolor]
{
\newpath
\moveto(82.3668737,85.7569653)
\lineto(92.59566236,85.7569653)
\lineto(102.54966425,85.7569653)
\lineto(353.95447181,319.47571554)
\lineto(334.9569789,319.47570798)
\closepath
}
}
{
\newrgbcolor{curcolor}{0 0 0}
\pscustom[linewidth=2.54532664,linecolor=curcolor]
{
\newpath
\moveto(124.52139213,100.93585412)
\lineto(244.50547654,210.27806262)
\lineto(364.48964031,100.93585412)
}
}
{
\newrgbcolor{curcolor}{0 0 0}
\pscustom[linewidth=2.54532664,linecolor=curcolor]
{
\newpath
\moveto(364.48964031,319.62027113)
\lineto(244.50547654,210.27806262)
\lineto(124.52138457,319.62027113)
}
}
{
\newrgbcolor{curcolor}{0 0 0}
\pscustom[linewidth=1.27266523,linecolor=curcolor]
{
\newpath
\moveto(128.52085417,319.62027113)
\curveto(164.5181178,287.42324687)(200.51262614,255.22869065)(221.17697386,239.13166766)
\curveto(241.84132535,223.03464467)(247.1737474,223.03464467)(267.83718047,239.13266924)
\curveto(288.50061354,255.23069002)(324.49513323,287.42523491)(360.49021984,319.62027113)
}
}
{
\newrgbcolor{curcolor}{0 0 0}
\pscustom[linewidth=1.27266523,linecolor=curcolor]
{
\newpath
\moveto(360.49021984,100.93585412)
\curveto(324.49282772,133.13295774)(288.52400882,165.30451554)(267.84727181,181.41299428)
\curveto(247.1705726,197.52146924)(241.84132535,197.52146924)(221.17604787,181.42357695)
\curveto(200.51077039,165.32572624)(164.51626205,133.13116624)(128.52086173,100.93581254)
}
}
{
\newrgbcolor{curcolor}{0 0 0}
\pscustom[linewidth=1.27266523,linecolor=curcolor]
{
\newpath
\moveto(132.52032378,319.62027113)
\curveto(166.51787717,289.85309002)(200.51268283,260.08831648)(220.84374425,245.20622262)
\curveto(241.17480567,230.32413254)(247.84022929,230.32413254)(268.17042142,245.20720908)
\curveto(288.50061354,260.09028183)(322.49546079,289.85509317)(356.49072378,319.62027113)
}
}
{
\newrgbcolor{curcolor}{0 0 0}
\pscustom[linewidth=1.27266523,linecolor=curcolor]
{
\newpath
\moveto(356.49072378,100.93585412)
\curveto(322.49315528,130.70303144)(288.49834583,160.4678201)(268.16728441,175.34990262)
\curveto(247.83622299,190.23198892)(241.17076157,190.23198892)(220.84058079,175.34890105)
\curveto(200.51040378,160.46581317)(166.51559811,130.70103963)(132.52032378,100.93585412)
}
}
{
\newrgbcolor{curcolor}{0 0 0}
\pscustom[linewidth=1.27266523,linecolor=curcolor]
{
\newpath
\moveto(138.51953764,319.62027113)
\curveto(170.51740346,293.49790766)(202.51251024,267.37780057)(221.84365606,254.31807396)
\curveto(241.17480567,241.25834735)(247.84022929,241.25834735)(267.1705852,254.31906042)
\curveto(286.5009411,267.37977349)(318.49604031,293.49987302)(350.49151748,319.62027113)
}
}
{
\newrgbcolor{curcolor}{0 0 0}
\pscustom[linewidth=1.27266523,linecolor=curcolor]
{
\newpath
\moveto(350.49151748,100.93585412)
\curveto(318.49362142,127.05822892)(286.49856,153.17831333)(267.1673726,166.23804372)
\curveto(247.83622299,179.29777412)(241.17076157,179.29777412)(221.84042835,166.23706483)
\curveto(202.5100989,153.17635554)(170.51499591,127.05624845)(138.51953764,100.93585412)
}
}
{
\newrgbcolor{curcolor}{0 0 0}
\pscustom[linewidth=1.27266523,linecolor=curcolor]
{
\newpath
\moveto(374.48826709,210.27806262)
\lineto(384.48696945,201.16619994)
\lineto(384.48696945,164.71881349)
\lineto(394.48563402,155.60695837)
\lineto(384.48696945,146.49509569)
\lineto(384.48696945,110.0476979)
\lineto(374.48826709,100.93585412)
}
}
{
\newrgbcolor{curcolor}{0 0 0}
\pscustom[linewidth=1.90898264,linecolor=curcolor]
{
\newpath
\moveto(344.28318992,319.75922924)
\lineto(92.59566236,86.45697538)
}
}
{
\newrgbcolor{curcolor}{0 0 0}
\pscustom[linewidth=1.90898264,linecolor=curcolor]
{
\newpath
\moveto(50.00000126,0.56092624)
\lineto(0.00663307,46.12019428)
}
}
{
\newrgbcolor{curcolor}{0 0 0}
\pscustom[linewidth=1.88976378,linecolor=curcolor,linestyle=dashed,dash=3 1]
{
\newpath
\moveto(0,23.39617412)
\lineto(390.00001512,23.39617412)
}
}
{
\newrgbcolor{curcolor}{0 0 0}
\pscustom[linewidth=1.27370083,linecolor=curcolor]
{
\newpath
\moveto(-5.6191937,56.96573002)
\lineto(-8.6191937,53.96572624)
\lineto(-8.6191937,48.96572498)
\lineto(-10.61919496,46.9657275)
\lineto(-8.6191937,44.96572624)
\lineto(-8.6191937,39.96572498)
\lineto(-5.6191937,36.96572876)
}
}
{
\newrgbcolor{curcolor}{0 0 0}
\pscustom[linewidth=1.51181105,linecolor=curcolor]
{
\newpath
\moveto(20.00000126,315.56093002)
\lineto(20.00000126,265.56092876)
\lineto(69.99999874,265.56092876)
}
}
{
\newrgbcolor{curcolor}{0 0 0}
\pscustom[linewidth=1.51181105,linecolor=curcolor]
{
\newpath
\moveto(15,305.5609275)
\lineto(19.99999748,315.56093002)
\lineto(24.99999874,305.5609275)
}
}
{
\newrgbcolor{curcolor}{0 0 0}
\pscustom[linewidth=1.51181105,linecolor=curcolor]
{
\newpath
\moveto(60,270.56093002)
\lineto(69.99999496,265.56092876)
\lineto(60,260.5609275)
}
}
{
\newrgbcolor{curcolor}{0 0 0}
\pscustom[linewidth=2.54532664,linecolor=curcolor]
{
\newpath
\moveto(85.01591433,111.21872026)
\lineto(204.99999874,220.56092876)
\lineto(324.98420409,111.21872026)
}
}
{
\newrgbcolor{curcolor}{0 0 0}
\pscustom[linewidth=2.54532664,linecolor=curcolor]
{
\newpath
\moveto(324.98420409,329.90313576)
\lineto(204.99999874,220.56092876)
\lineto(85.01590677,329.90313576)
}
}
{
\newrgbcolor{curcolor}{0 0 0}
\pscustom[linewidth=1.27266523,linecolor=curcolor]
{
\newpath
\moveto(89.01537638,329.90313576)
\curveto(125.01264,297.70611302)(161.00714457,265.5115568)(181.67149606,249.4145338)
\curveto(202.33584756,233.31751081)(207.66826961,233.31751081)(228.33171402,249.4155316)
\curveto(248.99517732,265.51355239)(284.98969701,297.70810861)(320.98478362,329.90313576)
}
}
{
\newrgbcolor{curcolor}{0 0 0}
\pscustom[linewidth=1.27266523,linecolor=curcolor]
{
\newpath
\moveto(320.98478362,111.21872026)
\curveto(284.9873915,143.4158201)(249.0185348,175.58738924)(228.34181669,191.69586042)
\curveto(207.6650948,207.80433538)(202.33584756,207.80433538)(181.67057008,191.70648467)
\curveto(161.0052926,175.60863396)(125.01078425,143.41407396)(89.01538394,111.21872026)
}
}
{
\newrgbcolor{curcolor}{0 0 0}
\pscustom[linewidth=1.27266523,linecolor=curcolor]
{
\newpath
\moveto(93.01484598,329.90313576)
\curveto(127.01239937,300.13595617)(161.00720504,270.37118262)(181.33826646,255.48908876)
\curveto(201.66932787,240.60699868)(208.3347515,240.60699868)(228.66493984,255.49007144)
\curveto(248.99513953,270.3731442)(282.99002457,300.13796309)(316.98528756,329.90313576)
}
}
{
\newrgbcolor{curcolor}{0 0 0}
\pscustom[linewidth=1.27266523,linecolor=curcolor]
{
\newpath
\moveto(316.98528756,111.21872026)
\curveto(282.98771906,140.98589002)(248.99287181,170.75069002)(228.66180661,185.63277254)
\curveto(208.3307452,200.51485506)(201.66528378,200.51485506)(181.33510299,185.63176719)
\curveto(161.00492598,170.74867931)(127.01012031,140.98390577)(93.01484598,111.21872026)
}
}
{
\newrgbcolor{curcolor}{0 0 0}
\pscustom[linewidth=1.27266523,linecolor=curcolor]
{
\newpath
\moveto(99.01405984,329.90313576)
\curveto(131.01192567,303.7807738)(163.00703244,277.66066672)(182.33817827,264.6009401)
\curveto(201.66932787,251.54121349)(208.3347515,251.54121349)(227.66511496,264.60192278)
\curveto(246.99546709,277.66263585)(278.99060409,303.78274294)(310.98608126,329.90313576)
}
}
{
\newrgbcolor{curcolor}{0 0 0}
\pscustom[linewidth=1.27266523,linecolor=curcolor]
{
\newpath
\moveto(310.98608126,111.21872026)
\curveto(278.9881852,137.3410875)(246.99308598,163.46118325)(227.66190992,176.52090987)
\curveto(208.3307452,189.58064026)(201.66528378,189.58064026)(182.33495055,176.51993097)
\curveto(163.0046211,163.45922168)(131.00951811,137.33911459)(99.01405984,111.21872026)
}
}
{
\newrgbcolor{curcolor}{0 0 0}
\pscustom[linestyle=none,fillstyle=solid,fillcolor=curcolor]
{
\newpath
\moveto(225.73326205,215.40753228)
\curveto(225.73326205,215.02152539)(225.37464933,214.70860498)(224.93227812,214.70860498)
\curveto(224.48990692,214.70860498)(224.1312942,215.02152539)(224.1312942,215.40753228)
\curveto(224.1312942,215.79353916)(224.48990692,216.10645957)(224.93227812,216.10645957)
\curveto(225.37464933,216.10645957)(225.73326205,215.79353916)(225.73326205,215.40753228)
\closepath
}
}
{
\newrgbcolor{curcolor}{0 0 0}
\pscustom[linewidth=1.06627656,linecolor=curcolor]
{
\newpath
\moveto(225.73326205,215.40753228)
\curveto(225.73326205,215.02152539)(225.37464933,214.70860498)(224.93227812,214.70860498)
\curveto(224.48990692,214.70860498)(224.1312942,215.02152539)(224.1312942,215.40753228)
\curveto(224.1312942,215.79353916)(224.48990692,216.10645957)(224.93227812,216.10645957)
\curveto(225.37464933,216.10645957)(225.73326205,215.79353916)(225.73326205,215.40753228)
\closepath
}
}
{
\newrgbcolor{curcolor}{0 0 0}
\pscustom[linewidth=0.99999871,linecolor=curcolor]
{
\newpath
\moveto(255.00001512,210.56093002)
\curveto(258.85101354,212.65421002)(263.35020094,213.53421254)(267.70606866,213.04610924)
\curveto(271.9505915,212.57048593)(276.04121197,210.79966766)(279.29273953,208.03026798)
}
}
{
\newrgbcolor{curcolor}{0 0 0}
\pscustom[linewidth=0.99999871,linecolor=curcolor]
{
\newpath
\moveto(260.06344819,217.06937097)
\lineto(255.2113285,210.88902703)
\lineto(262.38483402,209.50950703)
}
}
{
\newrgbcolor{curcolor}{0 0 0}
\pscustom[linewidth=0.99999871,linecolor=curcolor]
{
\newpath
\moveto(195.75398929,220.68362735)
\curveto(191.69897575,222.41225475)(187.19604661,223.07738735)(182.81443276,222.59493821)
\curveto(177.67954016,222.02954735)(172.73036598,219.87820624)(168.81416693,216.50920719)
}
}
{
\newrgbcolor{curcolor}{0 0 0}
\pscustom[linewidth=0.99999871,linecolor=curcolor]
{
\newpath
\moveto(191.25657449,226.06375837)
\lineto(196.16784,220.68363113)
\lineto(188.64340535,218.57474924)
}
}
{
\newrgbcolor{curcolor}{0 0 0}
\pscustom[linewidth=1.90898264,linecolor=curcolor]
{
\newpath
\moveto(49.91388094,45.90912656)
\lineto(0.33096189,0.69359522)
}
}
{
\newrgbcolor{curcolor}{0 0 0}
\pscustom[linewidth=1.90898264,linecolor=curcolor,linestyle=dashed,dash=2.02033997 2.02033997]
{
\newpath
\moveto(86.99779276,81.00057349)
\lineto(53.52668598,50.3404412)
}
\rput[bl](400,13){$\underline{\omega}$}
\rput[bl](-59,23){$\displaystyle \frac{1}{\ell_{\min}}$}
\rput[bl](406,134){$\displaystyle \frac{1}{\varepsilon}$}
\rput[bl](205,120){$\delta \hat{P}$}
\rput[bl](195, 290){${\mathscr{R}}\,\delta \hat{P}$}
\rput[bl](280, 187){$\displaystyle \frac{q}{2}$}
\rput[bl](135, 197){$\displaystyle -\frac{q}{2}$}
\rput[bl](29, 305){$k^0$}
\rput[bl](58, 276){$\vec{k}$}
\rput[bl](23, 45){$p$}
}
\end{pspicture}
\caption{Evaluation in momentum space}
\label{figconvolve}
\end{figure}
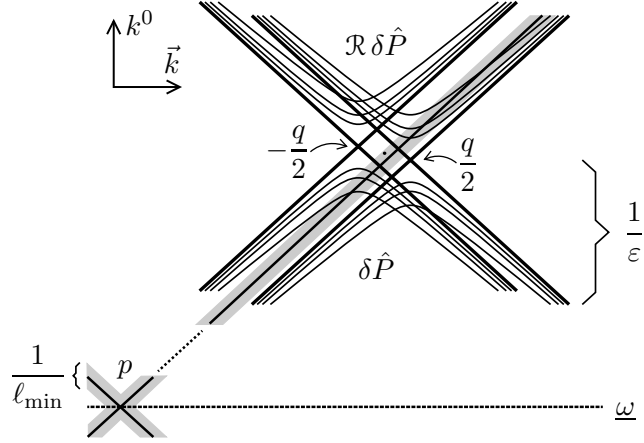

Steps~(E1) and~(E2), on the other hand, give rise to convolutions in momentum space, being
{\em{nonlocal}} operations. More precisely, using the explicit Fourier transform in~\eqref{K0def},
in Step~(E1) we obtain the convolution with~$K_0$, i.e.\
\beq \label{K0convolve}
-\int_{\hat{M}} \delta \hat{P}(k)\: K_0(p-k)\: d^4k \:,
\eeq
where we set
\beq \label{K0pdef}
K_0(p) := \frac{i}{4 \pi^2} \:\delta(p^2)\: \epsilon(p^0) \:.
\eeq
This convolution integral is illustrated in Figure~\ref{figconvolve}
(note that operating by~${\mathscr{R}}$ gives a reflection at the origin,
leading to the mass shells opening for large~$k^0$).
Step~(E2) gives rise to the convolution with the Fourier transform of the function in~\eqref{lminfunct}.
This convolution gives rise to a ``smearing'' in $\omega$-direction on the scale~$1/\ell_{\min}$.
Combining Steps~(E1) and~(E2) yields a convolution with a distribution supported in a strip around the mass
cone, as is indicated in Figure~\ref{figconvolve} by the gray stripes.

\begin{figure}
\psset{xunit=.5pt,yunit=.5pt,runit=.5pt}
\begin{pspicture}(641.76038847,225.6431904)
{
\newrgbcolor{curcolor}{0.69803923 0.69803923 0.69803923}
\pscustom[linestyle=none,fillstyle=solid,fillcolor=curcolor]
{
\newpath
\moveto(325.67952,0.25819789)
\lineto(325.67952,17.33290908)
\lineto(482.51373732,158.55489207)
\lineto(500.06333102,158.43179286)
\lineto(610.4787137,57.57148293)
\lineto(610.4787137,41.29133049)
\lineto(491.29293732,150.1973664)
\closepath
}
}
{
\newrgbcolor{curcolor}{0 0 0}
\pscustom[linewidth=0,linecolor=curcolor]
{
\newpath
\moveto(325.67952,0.25819789)
\lineto(325.67952,17.33290908)
\lineto(482.51373732,158.55489207)
\lineto(500.06333102,158.43179286)
\lineto(610.4787137,57.57148293)
\lineto(610.4787137,41.29133049)
\lineto(491.29293732,150.1973664)
\closepath
}
}
{
\newrgbcolor{curcolor}{0.80000001 0.80000001 0.80000001}
\pscustom[linestyle=none,fillstyle=solid,fillcolor=curcolor]
{
\newpath
\moveto(325.67952,0.25820167)
\lineto(345.86737512,0.25820167)
\lineto(499.86774047,142.57899915)
\lineto(482.56691528,142.57899915)
\lineto(610.4787137,25.83693301)
\lineto(610.4787137,41.38226215)
\lineto(491.03970898,150.1973664)
\closepath
}
}
{
\newrgbcolor{curcolor}{0.80000001 0.80000001 0.80000001}
\pscustom[linewidth=0,linecolor=curcolor]
{
\newpath
\moveto(325.67952,0.25820167)
\lineto(345.86737512,0.25820167)
\lineto(499.86774047,142.57899915)
\lineto(482.56691528,142.57899915)
\lineto(610.4787137,25.83693301)
\lineto(610.4787137,41.38226215)
\lineto(491.03970898,150.1973664)
\closepath
}
}
{
\newrgbcolor{curcolor}{0 0 0}
\pscustom[linewidth=1.51181105,linecolor=curcolor]
{
\newpath
\moveto(25.51968,0.81691034)
\lineto(145.50376441,110.15911884)
\lineto(265.48793575,0.81691034)
}
}
{
\newrgbcolor{curcolor}{0 0 0}
\pscustom[linewidth=1.51181105,linecolor=curcolor]
{
\newpath
\moveto(265.48793575,219.50132734)
\lineto(145.50376441,110.15911884)
\lineto(25.51967244,219.50132734)
}
}
{
\newrgbcolor{curcolor}{0 0 0}
\pscustom[linewidth=0.94488189,linecolor=curcolor]
{
\newpath
\moveto(29.51914205,219.50132734)
\curveto(65.51640567,187.30430309)(101.51091402,155.10974687)(122.17526173,139.01272388)
\curveto(142.83961323,122.91570089)(148.17203528,122.91570089)(168.83547591,139.01372545)
\curveto(189.49890898,155.11175002)(225.49342866,187.30629112)(261.48851528,219.50132734)
}
}
{
\newrgbcolor{curcolor}{0 0 0}
\pscustom[linewidth=0.94488189,linecolor=curcolor]
{
\newpath
\moveto(261.48851528,0.81691034)
\curveto(225.49112315,33.01401396)(189.52230425,65.18557175)(168.84556724,81.29404671)
\curveto(148.16886047,97.40252545)(142.83961323,97.40252545)(122.17433575,81.30467475)
\curveto(101.50905827,65.20682404)(65.51454992,33.01226404)(29.51914961,0.81691034)
}
}
{
\newrgbcolor{curcolor}{0 0 0}
\pscustom[linewidth=0.94488189,linecolor=curcolor]
{
\newpath
\moveto(33.51861165,219.50132734)
\curveto(67.51616504,189.73414624)(101.51097071,159.9693727)(121.84203213,145.08727884)
\curveto(142.17309354,130.20518876)(148.83851717,130.20518876)(169.16871685,145.0882653)
\curveto(189.49890898,159.97133805)(223.49375622,189.73614939)(257.48901921,219.50132734)
}
}
{
\newrgbcolor{curcolor}{0 0 0}
\pscustom[linewidth=0.94488189,linecolor=curcolor]
{
\newpath
\moveto(257.48901921,0.81691034)
\curveto(223.49145071,30.58408766)(189.49664126,60.34887632)(169.16557984,75.23095884)
\curveto(148.83451087,90.11304514)(142.16904945,90.11304514)(121.83886866,75.22995727)
\curveto(101.50869165,60.34686939)(67.51388598,30.58209585)(33.51861165,0.81691034)
}
}
{
\newrgbcolor{curcolor}{0 0 0}
\pscustom[linewidth=1.27370083,linecolor=curcolor]
{
\newpath
\moveto(55.47212598,168.78218782)
\lineto(45.4832126,158.09202845)
\lineto(45.47187402,113.16129207)
\lineto(35.47320945,104.04943695)
\lineto(45.47187402,94.93757427)
\lineto(45.4832126,47.24480829)
\lineto(55.27483465,37.97854813)
}
}
{
\newrgbcolor{curcolor}{0 0 0}
\pscustom[linewidth=1.51181105,linecolor=curcolor]
{
\newpath
\moveto(89.62202079,219.85677528)
\lineto(89.62202457,189.60166246)
\lineto(119.18569323,189.60166246)
}
}
{
\newrgbcolor{curcolor}{0 0 0}
\pscustom[linewidth=1.51181105,linecolor=curcolor]
{
\newpath
\moveto(84.63017197,209.99775065)
\lineto(89.63016945,219.9977543)
\lineto(94.63017071,209.99775065)
}
}
{
\newrgbcolor{curcolor}{0 0 0}
\pscustom[linewidth=1.51181105,linecolor=curcolor]
{
\newpath
\moveto(109.12541858,194.60877301)
\lineto(119.12541354,189.60877175)
\lineto(109.12541858,184.60877049)
}
}
{
\newrgbcolor{curcolor}{0 0 0}
\pscustom[linewidth=1.51181105,linecolor=curcolor]
{
\newpath
\moveto(5.05244598,6.40006624)
\lineto(125.03653039,115.74227475)
\lineto(245.02073575,6.40006624)
}
}
{
\newrgbcolor{curcolor}{0 0 0}
\pscustom[linewidth=1.51181105,linecolor=curcolor]
{
\newpath
\moveto(245.02073575,225.08448174)
\lineto(125.03653039,115.74227475)
\lineto(5.05243843,225.08448174)
}
}
{
\newrgbcolor{curcolor}{0 0 0}
\pscustom[linewidth=0.94488189,linecolor=curcolor]
{
\newpath
\moveto(9.05190803,225.08448174)
\curveto(45.04917165,192.887459)(81.04367622,160.69290278)(101.70802772,144.59587978)
\curveto(122.37237921,128.49885679)(127.70480126,128.49885679)(148.36824567,144.59687758)
\curveto(169.03170898,160.69489837)(205.02622866,192.88945459)(241.02131528,225.08448174)
}
}
{
\newrgbcolor{curcolor}{0 0 0}
\pscustom[linewidth=0.94488189,linecolor=curcolor]
{
\newpath
\moveto(241.02131528,6.40006624)
\curveto(205.02392315,38.59716608)(169.05506646,70.76873522)(148.37834835,86.8772064)
\curveto(127.70162646,102.98568136)(122.37237921,102.98568136)(101.70710173,86.88783065)
\curveto(81.04182425,70.78997994)(45.04731591,38.59541994)(9.05191559,6.40006624)
}
}
{
\newrgbcolor{curcolor}{0 0 0}
\pscustom[linewidth=0.94488189,linecolor=curcolor]
{
\newpath
\moveto(13.05137764,225.08448174)
\curveto(47.04893102,195.31730215)(81.04373669,165.5525286)(101.37479811,150.67043475)
\curveto(121.70585953,135.78834467)(128.37128315,135.78834467)(148.7014715,150.67141742)
\curveto(169.03167118,165.55449018)(203.02655622,195.31930908)(237.02181921,225.08448174)
}
}
{
\newrgbcolor{curcolor}{0 0 0}
\pscustom[linewidth=0.94488189,linecolor=curcolor]
{
\newpath
\moveto(237.02181921,6.40006624)
\curveto(203.02425071,36.16723601)(169.02940346,65.93203601)(148.69833827,80.81411852)
\curveto(128.36727685,95.69620104)(121.70181543,95.69620104)(101.37163465,80.81311317)
\curveto(81.04145764,65.9300253)(47.04665197,36.16525175)(13.05137764,6.40006624)
}
}
{
\newrgbcolor{curcolor}{0 0 0}
\pscustom[linewidth=0.94488189,linecolor=curcolor]
{
\newpath
\moveto(243.34084913,5.92577333)
\curveto(206.52224126,40.40527317)(169.72759559,74.86232262)(148.78002898,92.10956923)
\curveto(127.83246992,109.35681207)(122.7759874,109.35681207)(101.83730268,92.11774813)
\curveto(80.89861417,74.87868419)(44.08343433,40.40237805)(7.26793323,5.92577333)
}
}
{
\newrgbcolor{curcolor}{0 0 0}
\pscustom[linestyle=none,fillstyle=solid,fillcolor=curcolor]
{
\newpath
\moveto(136.20447936,112.7362152)
\curveto(136.20447936,112.35020832)(135.84586664,112.03728791)(135.40349543,112.03728791)
\curveto(134.96112422,112.03728791)(134.60251151,112.35020832)(134.60251151,112.7362152)
\curveto(134.60251151,113.12222209)(134.96112422,113.4351425)(135.40349543,113.4351425)
\curveto(135.84586664,113.4351425)(136.20447936,113.12222209)(136.20447936,112.7362152)
\closepath
}
}
{
\newrgbcolor{curcolor}{0 0 0}
\pscustom[linewidth=1.06627656,linecolor=curcolor]
{
\newpath
\moveto(136.20447936,112.7362152)
\curveto(136.20447936,112.35020832)(135.84586664,112.03728791)(135.40349543,112.03728791)
\curveto(134.96112422,112.03728791)(134.60251151,112.35020832)(134.60251151,112.7362152)
\curveto(134.60251151,113.12222209)(134.96112422,113.4351425)(135.40349543,113.4351425)
\curveto(135.84586664,113.4351425)(136.20447936,113.12222209)(136.20447936,112.7362152)
\closepath
}
}
{
\newrgbcolor{curcolor}{0 0 0}
\pscustom[linewidth=0.94488189,linecolor=curcolor]
{
\newpath
\moveto(263.5403074,0.71733868)
\curveto(226.72169953,35.19683475)(189.92705386,69.65388419)(168.97947591,86.90113081)
\curveto(148.03189039,104.14837742)(142.97541165,104.14837742)(122.03673449,86.90944577)
\curveto(101.09805354,69.67038183)(64.28286992,35.19407569)(27.46736882,0.71747475)
}
}
{
\newrgbcolor{curcolor}{0 0 0}
\pscustom[linewidth=0.94488189,linecolor=curcolor]
{
\newpath
\moveto(243.07303181,225.16012823)
\curveto(206.25442394,190.68062687)(169.45977827,156.22357364)(148.51221921,138.97633081)
\curveto(127.56466394,121.72908797)(122.50818142,121.72908797)(101.56949669,138.96815191)
\curveto(80.63081197,156.20721585)(43.81562835,190.68352199)(7.00012724,225.16012823)
}
}
{
\newrgbcolor{curcolor}{0 0 0}
\pscustom[linewidth=0.94488189,linecolor=curcolor]
{
\newpath
\moveto(263.54026961,218.28890026)
\curveto(226.72166173,183.80939664)(189.92701606,149.35233963)(168.97943811,132.10509679)
\curveto(148.03189039,114.85785774)(142.97541165,114.85785774)(122.03673449,132.09678183)
\curveto(101.09805732,149.33598939)(64.2828737,183.81229553)(27.4673726,218.28890026)
}
}
{
\newrgbcolor{curcolor}{0 0 0}
\pscustom[linewidth=1.32322773,linecolor=curcolor,linestyle=dashed,dash=2.80082989 1.40041006]
{
\newpath
\moveto(67.54628117,168.00334974)
\lineto(210.16287754,168.00334974)
\lineto(210.16287754,37.18480101)
\lineto(67.54628117,37.18480101)
\closepath
}
}
{
\newrgbcolor{curcolor}{0 0 0}
\pscustom[linewidth=0.94488189,linecolor=curcolor]
{
\newpath
\moveto(263.5403074,0.71733868)
\curveto(226.72169953,35.19683475)(189.92705386,69.65388419)(168.97947591,86.90113081)
\curveto(148.03189039,104.14837742)(142.97541165,104.14837742)(122.03673449,86.90944577)
\curveto(101.09805354,69.67038183)(64.28286992,35.19407569)(27.46736882,0.71747475)
}
}
{
\newrgbcolor{curcolor}{0 0 0}
\pscustom[linewidth=1.51181105,linecolor=curcolor]
{
\newpath
\moveto(429.92444598,94.87495128)
\lineto(491.03967118,150.34605301)
\lineto(553.11429165,94.00555034)
}
}
{
\newrgbcolor{curcolor}{0 0 0}
\pscustom[linewidth=0.75590552,linecolor=curcolor,linestyle=dashed,dash=0.60000002 0.2]
{
\newpath
\moveto(482.55814218,158.43014888)
\lineto(500.02767806,158.43014888)
\lineto(500.02767806,142.45054184)
\lineto(482.55814218,142.45054184)
\closepath
}
}
{
\newrgbcolor{curcolor}{0 0 0}
\pscustom[linewidth=1.51181105,linecolor=curcolor]
{
\newpath
\moveto(429.69801449,205.2837349)
\lineto(490.81323969,149.81263317)
\lineto(552.88786016,206.15313585)
}
}
{
\newrgbcolor{curcolor}{0 0 0}
\pscustom[linewidth=0.99999871,linecolor=curcolor]
{
\newpath
\moveto(413.35586646,71.18151254)
\lineto(394.77586016,55.1642527)
}
}
{
\newrgbcolor{curcolor}{0 0 0}
\pscustom[linewidth=0.99999871,linecolor=curcolor]
{
\newpath
\moveto(397.19275465,61.75813364)
\lineto(394.94045858,55.53129396)
\lineto(401.82974362,56.59117553)
}
}
{
\newrgbcolor{curcolor}{0 0 0}
\pscustom[linewidth=0.99999871,linecolor=curcolor]
{
\newpath
\moveto(394.80783496,70.65156986)
\lineto(376.22782866,54.63431002)
}
}
{
\newrgbcolor{curcolor}{0 0 0}
\pscustom[linewidth=0.99999871,linecolor=curcolor]
{
\newpath
\moveto(378.37974047,60.83073207)
\lineto(376.12744441,54.60389238)
\lineto(383.01672945,55.66377396)
}
}
{
\newrgbcolor{curcolor}{0 0 0}
\pscustom[linewidth=1.88976378,linecolor=curcolor]
{
\newpath
\moveto(236.9043515,112.04524671)
\curveto(264.80505071,133.42765553)(292.70499402,154.80946341)(328.19683654,162.07349994)
\curveto(363.68871685,169.33753648)(406.76856567,162.48348041)(449.84822551,155.62945459)
}
}
{
\newrgbcolor{curcolor}{0 0 0}
\pscustom[linewidth=1.88976378,linecolor=curcolor]
{
\newpath
\moveto(439.12449638,163.43590246)
\lineto(450.32025071,155.97206498)
\lineto(437.69912315,151.37202845)
}
}
{
\newrgbcolor{curcolor}{0 0 0}
\pscustom[linewidth=0.99999871,linecolor=curcolor]
{
\newpath
\moveto(579.7699011,32.22207443)
\curveto(584.11529953,31.59229931)(588.45956409,30.96269049)(592.35111685,32.33281459)
\curveto(596.24266961,33.70293112)(599.68234205,37.07248199)(603.12212787,40.44212356)
}
}
{
\newrgbcolor{curcolor}{0 0 0}
\pscustom[linewidth=0.99999871,linecolor=curcolor]
{
\newpath
\moveto(596.5386822,106.0592801)
\curveto(594.96239244,98.91299097)(593.38629165,91.76759758)(590.49933732,87.25724923)
\curveto(587.61238299,82.74690089)(583.41419717,80.87102687)(579.21544441,78.99491475)
}
}
{
\newrgbcolor{curcolor}{0 0 0}
\pscustom[linewidth=0.99999871,linecolor=curcolor]
{
\newpath
\moveto(597.92818772,39.89119695)
\lineto(603.46364598,40.83276041)
\lineto(602.49627591,35.30194719)
}
}
{
\newrgbcolor{curcolor}{0 0 0}
\pscustom[linewidth=0.99999871,linecolor=curcolor]
{
\newpath
\moveto(579.80765858,82.8502823)
\lineto(578.68211528,78.72355978)
\lineto(584.05932472,77.97325931)
}
\rput[bl](100,200){$k$}
\rput[bl](-35,80){$\displaystyle \sim \frac{1}{\ell_{\min}}$}
\rput[bl](420,70){wave}
\rput[bl](430,50){propagation}
\rput[bl](395,17){contribution by $\delta \hat{P}$}
\rput[bl](550,132){contribu-}
\rput[bl](570,107){tion by ${\mathscr{R}}\,\delta \hat{P}$}
}
\end{pspicture}
\caption{Wave propagation by the strong Huygens principle.}
\label{figeval}
\end{figure}
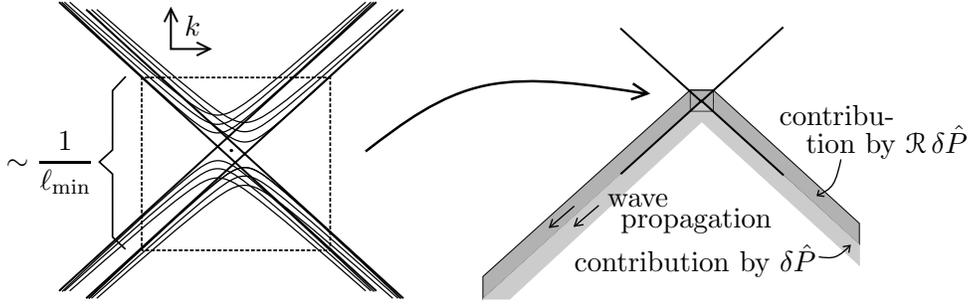

We next discuss the general structure of the evaluation in momentum space.
The parameter~$\ell_{\min}$ gives us a minimal length scale on which
to evaluate the linearized field equations. This means that 
it suffices to evaluate these equations for momenta~$p$ inside a cube of side length~$\sim 1/\ell_{\min}$.
In order to simplify the discussion, we begin with the
typical situation that the momentum~$q$ is much smaller than the corresponding energy scale, i.e.\
\beq \label{qsmall}
|q^0|, |\vec{q}| \ll \frac{1}{\ell_{\min}} \:.
\eeq
Then this cube is shown on the left of Figure~\ref{figeval}.
We begin with the result of Step~(E1) on page~\pageref{E1toE3}. 
As explained above, the convolution with~$K_0$ is nonlocal.
It gives a harmonic function in momentum space. It is important to note that the
distribution~$K_0$ is supported on the mass cone (i.e.\ $K_0(p)=0$ unless $p^2=0$).
This is a manifestation of the {\em{strong Huygens principle}} for solutions of the wave equation.
More precisely, the strong Huygens principle states that the wave obtained in Step~(E1)
propagates backwards with characteristic speed,
as is shown on the right of Figure~\ref{figeval}.
Therefore, when evaluating on the Cauchy surface~$\omega=\underline{\omega}$,
we may restrict attention to the annular region
\[ -\frac{1}{\ell_{\min}} \;\lesssim\; |\vec{k}| - |\underline{\omega}| \;\lesssim\; \frac{1}{\ell_{\min}} \:. \]
Moreover, the contributions by~$\delta P$ and~${\mathscr{R}}\,\delta P$ are disjoint on the Cauchy surface,
being supported inside and outside the sphere~$|\vec{k}| = |\underline{\omega}|$ respectively,
as is again indicated in Figure~\ref{figeval} by the two different shades of gray.
If~$q$ is not small in the sense that~\eqref{qsmall} is violated, then the above picture is modified
in that the two gray stripes overlap on the scale~$\sim q$.

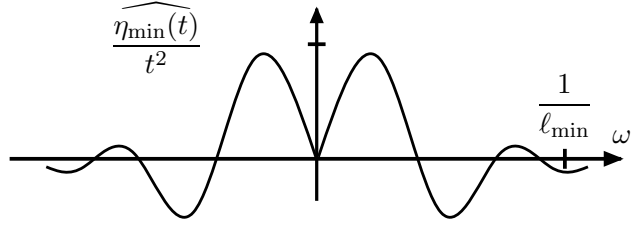
\begin{figure}
\psset{xunit=.5pt,yunit=.5pt,runit=.5pt}
\begin{pspicture}(385,170)
{
\newrgbcolor{curcolor}{0 0 0}
\pscustom[linewidth=2.45047236,linecolor=curcolor]
{
\newpath
\moveto(206.198308,11.40140375)
\lineto(206.198308,156.93436247)
}
}
{
\newrgbcolor{curcolor}{0 0 0}
\pscustom[linestyle=none,fillstyle=solid,fillcolor=curcolor]
{
\newpath
\moveto(206.198308,156.93436247)
\lineto(201.74235019,146.51849398)
\lineto(206.198308,146.51849398)
\lineto(210.65426581,146.51849398)
\closepath
}
}
{
\newrgbcolor{curcolor}{0 0 0}
\pscustom[linewidth=1.22523618,linecolor=curcolor]
{
\newpath
\moveto(206.198308,156.93436247)
\lineto(201.74235019,146.51849398)
\lineto(206.198308,146.51849398)
\lineto(210.65426581,146.51849398)
\closepath
}
}
{
\newrgbcolor{curcolor}{0 0 0}
\pscustom[linewidth=2.84302714,linecolor=curcolor]
{
\newpath
\moveto(-26.117135,43.22748)
\lineto(436.5202405,43.22748)
}
}
{
\newrgbcolor{curcolor}{0 0 0}
\pscustom[linestyle=none,fillstyle=solid,fillcolor=curcolor]
{
\newpath
\moveto(436.5202405,43.22748)
\lineto(424.04355858,48.23473665)
\lineto(424.04355858,43.22748)
\lineto(424.04355858,38.22022335)
\closepath
}
}
{
\newrgbcolor{curcolor}{0 0 0}
\pscustom[linewidth=1.42151357,linecolor=curcolor]
{
\newpath
\moveto(436.5202405,43.22748)
\lineto(424.04355858,48.23473665)
\lineto(424.04355858,43.22748)
\lineto(424.04355858,38.22022335)
\closepath
}
}
{
\newrgbcolor{curcolor}{0 0 0}
\pscustom[linewidth=2.84302714,linecolor=curcolor]
{
\newpath
\moveto(393.935214,50.76191)
\lineto(393.935214,35.69305)
}
}
{
\newrgbcolor{curcolor}{0 0 0}
\pscustom[linewidth=2.84302714,linecolor=curcolor]
{
\newpath
\moveto(199.4934185,129.988045)
\lineto(212.9031975,129.988045)
}
}
{
\newrgbcolor{curcolor}{0 0 0}
\pscustom[linewidth=2.1322701,linecolor=curcolor]
{
\newpath
\moveto(207.5058,44.6854)
\curveto(222.971,90.645)(238.4363,136.6051)(253.9,118.8962)
\curveto(269.3637,101.1872)(284.8246,19.8113)(300.1623,2.6495)
\curveto(315.4999,-14.5122)(330.7133,32.54)(344.8332,47.219)
\curveto(358.9531,61.8979)(371.9758,44.2034)(382.889,37.1294)
\curveto(393.8022,30.0554)(402.6018,33.6029)(411.4031,37.1514)
}
}
{
\newrgbcolor{curcolor}{0 0 0}
\pscustom[linewidth=2.1322701,linecolor=curcolor]
{
\newpath
\moveto(205.4849,44.6854)
\curveto(190.0197,90.645)(174.5544,136.6051)(159.0907,118.8962)
\curveto(143.627,101.1872)(128.1661,19.8114)(112.8284,2.6496)
\curveto(97.4908,-14.5122)(82.2774,32.54)(68.1575,47.2189)
\curveto(54.0376,61.8979)(41.0149,44.2034)(30.1017,37.1294)
\curveto(19.1885,30.0554)(10.3889,33.6029)(1.5876,37.1514)
}
\rput[bl](372,60){$\displaystyle \frac{1}{\ell_{\min}}$}
\rput[bl](430,55){$\omega$}
\rput[bl](50,108){$\displaystyle \widehat{\frac{\eta_{\min}(t)}{t^2}}$}
}
\end{pspicture}
\caption{A typical plot of the Fourier transform of~$\eta_{\min}(t)/t^2$.}
\label{figetahat}
\end{figure}

We next discuss the effect of the convolution in Step~(E2) on page~\pageref{E1toE3}.
Recall that~$\eta_{\min}$ is a smooth cutoff function for times smaller than~$\ell_{\min}$
(see Figure~\ref{figetamin}). Let us consider what this means for the Fourier transform of the
function~$\eta_{\min}(t)/t^2$. Clearly, this function~$\eta_{\min}(t)/t^2$ is smooth, meaning that
its Fourier transform has rapid decay.
More quantitatively, the derivatives of the function~$\eta_{\min}(t)/t^2$ 
scale in powers of~$1/\ell_{\min}$, which means that its Fourier transform 
decays on the scale~$1/\ell_{\min}$, as shown in Figure~\ref{figetahat}.
Next, the fact that the function~$\eta_{\min}(t)/t^2$ decays only quadratically at infinity
implies that the first derivative of its Fourier transform has a discontinuity,
as is indicated by the ``cusp'' in the plot of Figure~\ref{figetahat}.
Finally, the fact that the function~$\eta_{\min}(t)/t^2$ vanishes near~$t=0$ implies that
its Fourier transform has vanishing moments, i.e.\
\beq \label{zeromom}
\int_{-\infty}^\infty \omega^p\: \Big(\widehat{\frac{\eta_{\min}(t)}{t^2}} \Big) (\omega)\: d\omega = 0 
\qquad \text{for all~$p \in \N$}\:.
\eeq
This means that this function has an oscillatory behavior, as is also indicated in Figure~\ref{figetahat}.

We next discuss the effect of the multiplication by~$\eta_{\min}(t)/t^2$
in Step~(E2). Recall that the function~$\eta_{\min}$
vanishes if~$|t|<\ell_{\min}$ (see Figure~\ref{figetamin}).
Therefore, multiplying by the function~$\eta_{\min}(t)/t^2$
means that we work modulo contributions to~$\delta P(\xi)$ which are localized near the origin
in the sense that they
\beq \label{modpos}
\text{decay in position space on the scale} \quad\lesssim\ell_{\min}\:.
\eeq
In momentum space, in Step~(E2) we take the convolution with the Fourier transform
of the function~$\eta_{\min}(t)/t^2$, i.e.\
\beq \label{geconvolve}
\int_{-\infty}^\infty g(\omega)\: \Big(\widehat{\frac{\eta_{\min}(t)}{t^2}} \Big)\big(\underline{\omega}-\omega \big)\:.
d\omega
\eeq
If~$g$ is chosen as a smooth function, expanding~$g$ as a Taylor polynomial around~$\omega=0$
we can use~\eqref{zeromom} to conclude that the Taylor polynomial drops out, leaving us with the
remainder term of the Taylor expansion. This remainder term can be made arbitrarily small provided that
the function~$g$ has the property that its
\beq \label{modmom}
\text{derivatives scale in momentum space} \quad\lesssim \frac{1}{\ell_{\min}}\:.
\eeq
The relations~\eqref{modpos} and~\eqref{modmom} are equivalent statements
of the same property, expressed either in position or in momentum space.
The connection between decay in position space and smoothness in momentum space
is seen most easily using integration-by-parts
\beq \label{integration-by-parts}
\begin{split}
t^n \int_{-\infty}^\infty f(\omega)\: e^{-i \omega t} \:d\omega
&= \int_{-\infty}^\infty f(\omega)\: \Big( i \frac{d}{d\omega} \Big)^n e^{-i \omega t} \:d\omega \\
&= \int_{-\infty}^\infty (- i)^n\: f^{(n)}(\omega)\:  e^{-i \omega t} \:d\omega \:,
\end{split}
\eeq

We conclude that contributions to~$\delta \hat{P}$ drop out of the linearized field equations
provided that ~\eqref{modmom} hold. However, if the contribution is not smooth,
then the convolution~\eqref{geconvolve} will not not small.
This explains why on the right of Figure~\ref{figeval} we may restrict attention to all
non-smooth or singular contributions. The cusp-like singularity in Figure~\ref{figetahat}
means that the convolution improves the order of differentiability only by two.

\subsection{The Mass Cone Expansion} \label{secmce}
At the end of the previous section, we concluded that on the right of Figure~\ref{figeval},
we may restrict attention to all non-smooth or singular contributions to the convolution
integral~\eqref{K0convolve}.
This raises the question how these contributions can be computed.
In order to answer this question, we now develop the so-called {\em{mass cone expansion}},
which gives a systematic procedure for computing
all non-smooth and singular contributions to the convolution integral~\eqref{K0convolve}.

We begin with a general formula of the mass cone expansion. For~$n \in \N$ and~$m^2 \geq 0$, we introduce
the distributions
\begin{align}
T_{m^2}^{[0]}(p) &= \delta(p^2 - m^2)\: \Theta(-p^0) \label{Tm0def} \\
T_{m^2}^{[n]}(p) &= \frac{(-1)^n}{(n-1)!}
\:\frac{1}{4^n}\: \big( p^2 - m^2 \big)^{n-1}\: \Theta(p^2-m^2)\: \Theta(-p^0) \qquad \text{if~$n>0$} \label{Tmpdef} \:.
\end{align}
Moreover, we set
\beq \label{Tbardef}
\overline{T_{m^2}^{\bullet}}(p) := T_{m^2}^{\bullet}(-p)
\eeq
(this notation is motivated by the fact that taking the complex conjugate in position space
flips the sign of~$p$).
Note that~$T_{m^2}^{[0]}(p)$ is a distribution supported on the lower mass shell.
For~$n>0$, the distributions~$T_{m^2}^{[n]}(p)$ are regular and supported in the inside the
lower mass shell (i.e.\ for $p^2 \geq m^2$ and~$p^0<0$).
Incrementing~$n$ gives an additional factor~$p^2-m^2$ which vanishes on the mass shell.
The following theorem gives an expansion of a convolution integral as a sum of
terms with increasing~$n$. The larger~$n$ gets, the faster the summands decay near the mass shell.
In this sense, the mass cone expansion gives information on the behavior of a distribution
near the mass shell. 

\begin{Thm} {\bf{(mass cone expansion)}} \label{thmTVK} Let~$V \in C^\infty(\hat{M})$ be a smooth (possibly matrix-valued) function which for a suitable constant~$c>0$ satisfies the bounds
\beq \label{Vbound}
\big| \Box^n V(k) \big| \leq \bigg( c\:\Big( \varepsilon^2 +  \frac{\ell_{\min}}{|k^0|+m} \Big) \bigg)^n
\:e^{\varepsilon |k^0|} \qquad \text{for all~$k \in \hat{M}$ and~$n \in \N_0$}\:.
\eeq
Then, choosing~$p^0=\underline{\omega} \ll -1/\varepsilon$, the following expansions hold,
\begin{align}
&\int_{\hat{M}} T_{m^2}^{[0]}(k)\: V(k)\: K_0(p-k)\: d^4k \notag \\
&= \frac{i}{2 \pi} \sum_{n=0}^\infty \frac{1}{n!} \int_0^\infty (\alpha-\alpha^2)^n \:\Box^n V|_{\alpha p}\: d\alpha\:
T_{m^2/\alpha}^{[n+1]}(p) + \O\Big( \frac{1}{\varepsilon \underline{\omega}} \Big) \label{TVK1} \\
&\int_{\hat{M}} \overline{T_{m^2}^{[0]}}(k)\: V(k)\: K_0(p-k)\: d^4k \notag \\
&= -\frac{i}{2 \pi} \sum_{n=0}^\infty \frac{1}{n!} \int_{-\infty}^0 (\alpha-\alpha^2)^n \:\Box^n V|_{\alpha p}\: d\alpha\:
T_{m^2/|\alpha|}^{[n+1]}(p) \label{TVK3} \\
&\quad\, + \big( \text{decay in position space on the scale $\ell_{\min}$} \big)\:. \label{TVK4}
\end{align}
\end{Thm} \noindent
In order to clarify the structure of this expansion, we point out that, due to the factors~$T^{[0]}_{m^2}$
and~$\overline{T^{[0]}_{m^2}}$, the left side of the mass cone expansion involves the potential~$V$
evaluated only on the mass cone. Acting with the wave operator, however, makes it necessary to
specify~$V$ in an open neighborhood of the mass cone.
The way to understand this seeming inconsistency is that the mass cone expansion holds
for any any smooth extension of~$V$ from the mass shell to~$\hat{M}$.
For two different extensions, the individual summands in~\eqref{TVK1} and~\eqref{TVK3} will
in general be different. But the whole series are still the same.

This mass cone expansion is closely related to the so-called {\em{light-cone expansion}}
used in order to analyze the behavior of distributions near the light cone
(see~\cite{firstorder, light} or~\cite[Section~2.2]{cfs}). The main difference is that, compared to the
light-cone expansion, in the mass cone expansion the roles of position and momentum variables
are interchanged. This difference also made it necessary to adapt and generalize the light-cone expansion
in two ways: First, instead of considering the mass cone (i.e.\ the set~$p^2=0$,
being the analog of the light cone), we allow for a non-zero mass and consider the mass shell
(i.e.\ the set~$p^2=m^2$; see~\eqref{Tm0def} and~\eqref{Tmpdef}).
Second, in contrast to the light-cone expansion in~\cite{firstorder, light}, our mass cone expansion involves
unbounded line integrals (see for example~\eqref{TVK1}).
The corresponding generalizations of the light-cone expansion are developed in Appendix~\ref{applightcone}.
Formulating these results in momentum space gives our mass cone expansions, as is worked out
in Appendix~\ref{appproof}. More precisely, the proof of Theorem~\ref{thmTVK} is given in Appendix~\ref{appproof} on page~\pageref{proofTVK}.
We remark that in Appendix~\ref{appexconv} the mass cone expansion~\ref{TVK1} is illustrated
by computing the convolution integral on the left side of~\eqref{TVK1} explicitly in the simple example~$V(k)=e^{\varepsilon k^0}$.

In order to clarify how the above theorem fits to the previous considerations, we note that the expansion~\eqref{TVK1}
applies to the convolution of~$K_0$ with~$\delta \hat{P}$, whereas the expansion~\eqref{TVK3}
and~\eqref{TVK4} applies to the convolution of~$K_0$ with~${\mathscr{R}}\, \delta \hat{P}$
(see Figure~\ref{figconvolve}). The summands of the mass cone expansion~\eqref{TVK1}
are supported inside the lower mass cone, in agreement with the wave propagation
on the right of Figure~\ref{figeval}. At first sight, it might be surprising that the
summands of the mass cone expansion~\eqref{TVK3} are also supported inside the lower mass cone,
although the corresponding wave on the right of Figure~\ref{figeval} propagates outside the mass cone.
However, this is no contradiction if we keep in mind the error term~\eqref{TVK4}:
As explained at the end of the previous section, contributions which are smooth in momentum
space in the sense~\eqref{modmom} drop out when taking the
convolution in Step~(E2). According to~\eqref{modpos}, these such smooth contributions in
momentum space can be absorbed precisely in
the error term~\eqref{TVK4}. As is made precise in the proof of Theorem~\ref{thmTVK},
by subtracting such smooth terms in momentum space, one can indeed
compensate for the wave outside the lower mass cone
in Figure~\ref{figeval}, but gets instead a wave which is supported inside the lower mass cone.
This explains the expansion~\eqref{TVK3}.

Our goal is to apply the mass cone expansions of Theorem~\ref{thmTVK} to the perturbations
of the form~\eqref{nuqgen} and~\eqref{numqgen} as specified in Definition~\ref{defJvary}.
These perturbations can indeed be written in the form~$T_{m^2}^{[0]}(k)\: V(k)$, but with a potential~$V$
which has poles. For this reason, before we can apply Theorem~\ref{thmTVK}, we need to
remove these poles. In the next section, we give a method for doing so.

\subsection{Removing the Poles on the Mass Shells} \label{secmasspole}
Making use of the causal compatibility conditions, we can remove the poles
of the functions~\eqref{nuqgen} and~\eqref{numqgen},
as we now explain. For ease in notation, given~$\alpha, \beta \in \{1,2,3\}$,
we consider the corresponding summands of~\eqref{nuqgen} and~\eqref{numqgen} separately.
Rewriting the poles with the help of the relation
\[ \lim_{\mu \searrow 0} \frac{1}{x+ i \mu} = \frac{\text{PP}}{x} + i \pi \delta(x) \:, \]
the resulting $\delta$-contributions cancel each other in view of the causal compatibility condition~\eqref{ccc0}.
Therefore, in the following argument we may replace the poles by principal parts.
For ease in notation, the symbol ``PP'' for the principal part will be omitted.

We again fix~$\alpha, \beta \in \{1,2,3\}$. We extend the function~$\slashed{h}^{\alpha, \beta}_q$
introduced in~\eqref{habdef} to all~$p \in \hat{M}$. Moreover, we combine this function with the
regularizing factor and introduce the short notation
\[ \slashed{h}^{\varepsilon, \alpha, \beta}_q(p) :=
\slashed{h}_q^{\beta, \alpha}(p) \, e^{\varepsilon p^0} \:. \]

\begin{Lemma} \label{lemmaline}
For any perturbation~$\bv \in \Jret$ and every~$p \in \hat{M}$ with~$|p^0| > |q^0|/2$,
\begin{align}
&\delta^{\varepsilon, \alpha, \beta}_q \hat{P}(p) + \big(\delta^{\varepsilon, \alpha, \beta}_{-q} \hat{P} \big)^*(p) \notag \\
&= \frac{1}{4} \int_{-1}^1 \bigg( (1+\tau)\: \big(\slashed{h}_{-q}^{\varepsilon, \alpha, \beta} \big)^*\Big( p + \frac{q}{2} \Big) +
(1-\tau)\: \slashed{h}_q^{\varepsilon, \beta, \alpha} \Big( p - \frac{q}{2} \Big) \bigg) \notag \\
&\qquad\qquad \times
\: \delta'\Big( p^2 + \frac{q^2}{4} - \frac{m_\alpha^2+m_\beta^2}{2} + \frac{\tau}{2}\: \big(
2 pq + m_\alpha^2 - m_\beta^2 \big) \Big)\:  \Theta(-\omega)\:d\tau \label{line1} \\
&\quad\:- \frac{1}{2}\: \frac{\big(\slashed{h}_{-q}^{\varepsilon, \alpha, \beta} \big)^*\Big( p + \frac{q}{2} \Big) - \slashed{h}_q^{\varepsilon, \beta, \alpha} \Big( p - \frac{q}{2} \Big)}{2 pq + m_\alpha^2 - m_\beta^2} \notag \\
&\qquad\qquad \times
\int_{-1}^1 \delta\bigg( p^2 + \frac{q^2}{4} - \frac{m_\alpha^2+m_\beta^2}{2} + \frac{\tau}{2}\: \big(
2 pq + m_\alpha^2 - m_\beta^2 \big) \bigg)\: \Theta(-\omega)\: d\tau \:. \label{line2}
\end{align}
\end{Lemma}
\Proof Omitting the factor~$\Theta(-\omega)$, we rewrite the integral in~\eqref{line1} as
\begin{align*}
&\int_{-1}^1 \bigg( (1+\tau)\: \big(\slashed{h}_{-q}^{\varepsilon, \alpha, \beta} \big)^*\Big( p + \frac{q}{2} \Big) +
(1-\tau)\: \slashed{h}_q^{\varepsilon, \beta, \alpha} \Big( p - \frac{q}{2} \Big) \bigg) \\
&\qquad \times
\: \delta'\Big( p^2 + \frac{q^2}{4} - \frac{m_\alpha^2+m_\beta^2}{2} + \frac{\tau}{2}\: \big(
2 pq + m_\alpha^2 - m_\beta^2 \big) \Big)\: d\tau \\
&=\frac{2}{2 pq + m_\alpha^2 - m_\beta^2}
\int_{-1}^1 \bigg( (1+\tau)\: \big(\slashed{h}_{-q}^{\varepsilon, \alpha, \beta} \big)^*\Big( p + \frac{q}{2} \Big) +
(1-\tau)\: \slashed{h}_q^{\varepsilon, \beta, \alpha} \Big( p - \frac{q}{2} \Big) \bigg) \\
&\qquad \times \frac{d}{d\tau}
\delta\Big( p^2 + \frac{q^2}{4} - \frac{m_\alpha^2+m_\beta^2}{2} + \frac{\tau}{2}\: \big(
2 pq + m_\alpha^2 - m_\beta^2 \big) \Big)\: d\tau \:,
\end{align*}
we can integrate by parts,
\begin{align*}
&=\frac{4}{2 pq + m_\alpha^2 - m_\beta^2}\:
\big(\slashed{h}_{-q}^{\varepsilon, \alpha, \beta} \big)^*\Big( p + \frac{q}{2} \Big) \:
\delta\Big( p^2 + \frac{q^2}{4} - \frac{m_\alpha^2+m_\beta^2}{2} + \frac{1}{2}\: \big(
2 pq + m_\alpha^2 - m_\beta^2 \big) \Big) \\
&\quad\:-\frac{4}{2 pq + m_\alpha^2 - m_\beta^2}\:
 \slashed{h}_q^{\varepsilon, \beta, \alpha} \Big( p - \frac{q}{2} \Big) \bigg) \:
 \delta\Big( p^2 + \frac{q^2}{4} - \frac{m_\alpha^2+m_\beta^2}{2} - \frac{1}{2}\: \big(
2 pq + m_\alpha^2 - m_\beta^2 \big) \Big) \\
&\quad\: -\frac{2}{2 pq + m_\alpha^2 - m_\beta^2}
\int_{-1}^1 \bigg( \big(\slashed{h}_{-q}^{\varepsilon, \alpha, \beta} \big)^*\Big( p + \frac{q}{2} \Big) - \slashed{h}_q^{\varepsilon, \beta, \alpha} \Big( p - \frac{q}{2} \Big) \bigg) \\
&\qquad\qquad\qquad\qquad\qquad\qquad \times
\delta\Big( p^2 + \frac{q^2}{4} - \frac{m_\alpha^2+m_\beta^2}{2} + \frac{\tau}{2}\: \big(
2 pq + m_\alpha^2 - m_\beta^2 \big) \Big)\: d\tau \\
&=\frac{4}{2 pq + m_\alpha^2 - m_\beta^2}\:
\big(\slashed{h}_{-q}^{\varepsilon, \alpha, \beta} \big)^*\Big( p + \frac{q}{2} \Big) \:\delta \bigg( \Big(p +\frac{q}{2} \Big)^2 - m_\beta^2 \bigg) \\
&\quad\:-\frac{4}{2 pq + m_\alpha^2 - m_\beta^2}\:
 \slashed{h}_q^{\varepsilon, \beta, \alpha} \Big( p - \frac{q}{2} \Big) \bigg) \:\delta\bigg( \Big(p - \frac{q}{2} \Big)^2 - 
 m_\alpha^2 \bigg) \\
&\quad\: -\frac{2}{2 pq + m_\alpha^2 - m_\beta^2}
\int_{-1}^1 \bigg( \big(\slashed{h}_{-q}^{\varepsilon, \alpha, \beta} \big)^*\Big( p + \frac{q}{2} \Big) - \slashed{h}_q^{\varepsilon, \beta, \alpha} \Big( p - \frac{q}{2} \Big) \bigg) \\
&\qquad\qquad\qquad\qquad\qquad\qquad \times
\delta\bigg( p^2 + \frac{q^2}{4} - \frac{m_\alpha^2+m_\beta^2}{2} + \frac{\tau}{2}\: \big(
2 pq + m_\alpha^2 - m_\beta^2 \big) \bigg)\: d\tau \:.
\end{align*}
This gives the result.
\QED
\noindent
We remark that the appearance of line integral in~\eqref{line1} can be understood in analogy to the
light-cone expansion in momentum space as worked out in~\cite{firstorder}.

Applying this lemma, the poles have disappeared. On the other hand, a $\delta'$-distri\-bu\-tion
supported on the lower mass shell arises. Its convolution with~$K_0$ can be computed
with the following modification of Theorem~\ref{thmTVK}. For~$m^2 \geq 0$ we define
\beq \label{Tmmdef}
T_{m^2}^{[-1]}(p) := -4\: \delta'(p^2 - m^2)\: \Theta(-p^0) \:.
\eeq
\begin{Thm} \label{thmTVK2} Under the assumptions of Theorem~\ref{thmTVK}
and assuming in addition that~$V$ vanishes identically in a neighborhood of~$p=0$,
the following mass cone expansions hold for~$p^0=\underline{\omega} \ll -\varepsilon^{-1}$,
\begin{align}
&\int_{\hat{M}} T_{m^2}^{[-1]}(k)\: V(k)\: K_0(p-k)\: d^4k \notag \\
&= \frac{i}{2 \pi} \sum_{n=0}^\infty \frac{1}{n!} \int_0^\infty \frac{1}{\alpha}\,
(\alpha-\alpha^2)^n \:\Box^n V|_{\alpha p}\: d\alpha\:
T_{m^2/\alpha}^{[n]}(p) + \O\Big( \frac{1}{\varepsilon \underline{\omega}} \Big) \label{TVK1n} \\
&\int_{\hat{M}} \overline{T_{m^2}^{[-1]}}(k)\: V(k)\: K_0(p-k)\: d^4k \notag \\
&= -\frac{i}{2 \pi} \sum_{n=0}^\infty \frac{1}{n!} \int_{-\infty}^0 \frac{1}{\alpha}\,(\alpha-\alpha^2)^n \:\Box^n V|_{\alpha p}\: d\alpha\:
T_{m^2/|\alpha|}^{[n]}(p) \label{TVK3n} \\
&\quad\, + \big( \text{decay in position space on the scale $\ell_{\min}$} \big)\:. \label{TVK4n}
\end{align}
\end{Thm} \noindent
The proof of this theorem is given in Appendix~\ref{appproof} on page~\pageref{proofTVK2}.

When applying the method of Lemma~\ref{lemmaline} and Theorem~\ref{thmTVK2} we must
be careful for two reasons:
\bitem
\item[{\rm{(a)}}] The potential~$V(k)$ obtained as the line integral in~\eqref{line1} does not
necessarily need to vanish in a neighborhood of~$k=0$.
\item[{\rm{(b)}}] Rewriting the argument of the $\delta$- and $\delta'$-distributions in Lemma~\ref{lemmaline}
according to
\begin{align*}
&p^2 + \frac{q^2}{4} - \frac{m_\alpha^2+m_\beta^2}{2} + \frac{\tau}{2}\: \big(
2 pq + m_\alpha^2 - m_\beta^2 \big) \\
&= \Big( p + \frac{\tau}{2}\: q \Big)^2 + 
\frac{q^2}{4} - \frac{m_\alpha^2+m_\beta^2}{2} + \frac{\tau}{2}\: \big( m_\alpha^2 - m_\beta^2 \big)
-\frac{\tau^2}{4}\: q^2 \\
&= \Big( p + \frac{\tau}{2}\: q \Big)^2 - m_\text{\rm{eff}}^2
\end{align*}
with the ``effective mass''~$m_\text{\rm{eff}}$ given by
\beq \label{meff2}
m_\text{\rm{eff}}^2 := - (1-\tau^2)\: \frac{q^2}{4} 
+ \frac{1-\tau}{2}\: m_\alpha^2 + \frac{1+\tau}{2}\: m_\beta^2 \:,
\eeq
this effective mass could be imaginary, in which case the mass cone expansion cannot be applied.
\eitem
These issues can be treated as follows.
In order to deal with~(a), assume that~$V(k)$ does not vanish at~$k=0$. Then, in the case~$n=0$,
the integrals in~\eqref{TVK1n} and~\eqref{TVK3n} diverge at~$\alpha=0$. This divergence can be removed
by subtracting a counter term supported at~$k=0$. This counter term changes the convolution integral by a term
which is proportional to~$K_0(p)$ and thus already has the desired form of the mass cone expansion.
In order to treat~(b), we distinguish two cases.
 If~$q$ is timelike (i.e.\ if~$q^2 > 0$), then the poles in~\eqref{nuqgen} and~\eqref{numqgen}
lie on a spacelike hypersurface, which intersects the mass shell only on a compact set
(see the left of Figure~\ref{figdelP2}). Therefore, these poles can be treated by modifying~$V(k)$
near~$k=0$ as explained under~(a). The remaining potential is smooth, making it possible to
apply Theorem~\ref{thmTVK} immediately (without applying Lemma~\ref{lemmaline}).
In the remaining case that~$q$ is spacelike or lightlike (i.e.\ $q^2 \leq 0$), one sees from~\eqref{meff2}
that the effective mass squared is non-negative, making it possible to apply Theorem~\ref{thmTVK2}.
We note that, in the case~$q^2<0$, the result of Lemma~\ref{lemmaline} holds even for all~$p \in \hat{M}$
if the Heaviside functions in~\eqref{line1} and~\eqref{line2} are replaced
by~$\Theta(-\omega - \tau q^0/2)$.

We remark that, as an alternative to the just-described procedure, one can also expand
the above formulas in a Taylor series in the momentum~$q$ and in the mass parameters (i.e.\ 
in the parameters~$m$, $m_\alpha$ and~$m_\text{\rm{eff}}$).
One advantage of this alternative procedure is that the mass cone expansions of the resulting
Taylor components can also be used in the case~$m_{\text{\rm{eff}}}^2<0$, making it unnecessary
to distinguish the cases~$q^2>0$ and~$q^2 \leq 0$.
Indeed, a Taylor expansion in the mass parameter corresponds precisely
to the light-cone expansion in momentum space (for details see~\cite{firstorder}). Using that
the higher orders of the light-cone expansion vanish on the light cone, they drop out
when multiplying by~$\delta(\xi^2)\: \epsilon(t)$ in Step~(E1). Therefore, one can even work with a
truncated expansion. For clarity, we do not follow this alternative procedure in this paper\footnote{
We remark for clarity that, in our context, the error terms of the light cone expansion
are of the order
\[ \times \bigg( 1 + \O \Big( \big( |q^0| + |\vec{q}| \big)\, \ell_{\min} \Big) \bigg) \:. \]
This error term can be computed exactly as in the proof of Theorem~\ref{thmTVK} in Appendix~\ref{appproof}
on page~\pageref{proofTVK}. For classical potentials,
the error in the above expansion would be even much better, namely of the order
\[ \times \bigg( 1 + \O \Big( \big( |q^0| + |\vec{q}| \big)\, \varepsilon \Big) + \O \big( m \varepsilon \big) \bigg) \:. \]
This can be understood from the fact that the higher orders of the light-cone expansion
vanish when evaluated on the light cone. The reason for the worse error term in~\eqref{lightcone}
comes from the $p$-derivatives of the nonlocal potentials, as specified by~\eqref{Vbound}.}.

We conclude this section by identifying those contributions to the mass cone expansion
which correspond to the leading singularity of~$\scrP \bv$ on the light cone.
As we shall see, these contributions are closely related to the direction-dependent phase transformations in
Definition~\ref{defscrG}. We begin with the case~$q^2>0$ in which, as just explained,
we can work directly with the mass cone expansion of Theorem~\ref{thmTVK}.
In this case, we can treat~$\delta_q \hat{P}$ and~$\delta_{-q}^* \hat{P}$ separately.
We only consider~$\delta_q \hat{P}$, because the other term can be treated similarly.
We write the vectorial component of a summand in~\eqref{nuqgen} symbolically as
\beq \label{delqtime}
\delta_q \hat{P}(p) \asymp \,\slashed{\!b}(p_\text{\rm{eff}})\: T_{m^2}^{[0]}(p_\text{\rm{eff}}) \: e^{\varepsilon p_\text{\rm{eff}}^0}
\eeq
with~$p_\text{\rm{eff}} := p - q/2$ (here and in what follows, the symbol ``$\asymp$'' again indicates that we restrict attention to a specific contribution). Taking the convolution~\eqref{K0convolve},
we can apply Theorem~\ref{thmTVK} as well as~\eqref{infdef}
to compute the leading contribution of the mass cone expansion to be
\begin{align}
&\int_{\hat{M}} T_{m^2}^{[0]}(k)\: e^{\varepsilon k^0}\: \,\slashed{\!b}(k)\: K_0(p-k)\: d^4k
\asymp \frac{i}{2 \pi} \int_0^\infty \slashed{\!b}(\alpha p)\: e^{\varepsilon \alpha p^0} d\alpha\:+ \O\Big( \frac{1}{\varepsilon \underline{\omega}} \Big) \notag \\
&= \frac{i}{2 \pi} \int_0^\infty \frac{\alpha^2 \omega \:\Phi(\hat{\vec{p}}\,)\: \slashed{p}}{2 \alpha p q}\: e^{\varepsilon \alpha p^0}\: d\alpha\:
T_{m^2/\alpha}^{[1]}(p) + \O\Big( \frac{1}{\varepsilon \underline{\omega}} \Big) \notag \\
&= \frac{i}{2 \pi} \:\frac{\omega}{2 p q}\:\Phi(\hat{\vec{p}}\,) \int_0^\infty
\alpha \slashed{p} \: e^{\varepsilon \alpha p^0}\: d\alpha\:
T_{m^2/\alpha}^{[1]}(p) + \O\Big( \frac{1}{\varepsilon \underline{\omega}} \Big) \notag \\
&= \frac{i}{2 \pi} \:\frac{1}{2 p q}\:\Phi(\hat{\vec{p}}\,) 
\: \frac{\slashed{p}}{\varepsilon^2 \omega}\:
T_0^{[1]}(p) + \O\Big( \frac{1}{\varepsilon \underline{\omega}} \Big) \label{ph3}
\end{align}
and similarly
\begin{align}
&\int_{\hat{M}} \overline{T_{m^2}^{[0]}}(k)\: e^{-\varepsilon k^0}\: \,\slashed{\!b}(k)\: K_0(p-k)\: d^4k
\asymp -\frac{i}{2 \pi} \:\frac{1}{2 p q}\:\Phi(\hat{\vec{p}}\,) \:
\frac{\slashed{p}}{\varepsilon^2 \omega}\:T_0^{[1]}(p) \label{ph4} \\
&\qquad + \big( \text{decay in position space on the scale $\ell_{\min}$} \big)\:, \notag
\end{align}
to be evaluated at~$p=p_\text{\rm{eff}}$.

In the case~$q^2<0$, we first remove the poles by applying Lemma~\ref{lemmaline}.
Therefore, our task is to compute the convolution of~$K_0$ with the integrands of both~\eqref{line1}
and~\eqref{line2}. The integrand in~\eqref{line2} is of the same form as the right side in~\eqref{delqtime},
but now with~$p_\text{\rm{eff}} = p + \tau q/2$; it can be treated exactly as explained above.
It remains to treat the integrand in~\eqref{line1}. For a compact notation, we write it symbolically as
\begin{align*}
\delta^{\varepsilon, \alpha, \beta}_q \hat{P}(p_\text{\rm{eff}}) + \big(\delta^{\varepsilon, \alpha, \beta}_{-q} \hat{P} \big)^*(p_\text{\rm{eff}})
&\asymp \frac{1}{4} \int_{-1}^1
\slashed{h}_\tau(p_\text{\rm{eff}})\: T_{m_\text{\rm{eff}}^2}^{[-1]}(p_\text{\rm{eff}}) \: e^{\varepsilon p_\text{\rm{eff}}^0} \:d\tau \\
\text{with}\qquad p_\text{\rm{eff}} &= p_\text{\rm{eff}}(\tau) := p + \tau q/2
\end{align*}
(and~$m_\text{\rm{eff}}^2$ is again the effective mass~\eqref{meff2}).
Applying Theorem~\ref{thmTVK2} and again~\eqref{infdef}, we obtain
\begin{align}
&\int_{\hat{M}} T_{m^2}^{[-1]}(k)\: e^{\varepsilon k^0}\: \slashed{h}_\tau(k)\: K_0(p-k)\: d^4k \notag \\
&\asymp \frac{i}{2 \pi} \int_0^\infty \frac{1}{\alpha}\,
\slashed{h}_\tau(\alpha p)\: e^{\varepsilon \alpha p^0}\: d\alpha\:
T_{m^2/\alpha}^{[0]}(p) + \O\Big( \frac{1}{\varepsilon \underline{\omega}} \Big) \notag \\
&= \frac{i}{2 \pi} \:\omega \Phi(\hat{\vec{p}}\,) \int_0^\infty \alpha \slashed{p}\:
e^{\varepsilon \alpha p^0}\: d\alpha\:
T_{m^2/\alpha}^{[0]}(p) + \O\Big( \frac{1}{\varepsilon \underline{\omega}} \Big) \notag \\
&= \frac{i}{2 \pi} \:\Phi(\hat{\vec{p}}\,) \:
\frac{\slashed{p}}{\varepsilon^2 \omega}\:T_0^{[0]}(p) + \O\Big( \frac{1}{\varepsilon \underline{\omega}} \Big) \label{ph1}
\end{align}
and similarly
\begin{align}
&\int_{\hat{M}} \overline{T_{m^2}^{[-1]}}(k)\: e^{-\varepsilon k^0}\: \slashed{h}_\tau(k)\: K_0(p-k)\: d^4k 
\asymp -\frac{i}{2 \pi} \:\Phi(-\hat{\vec{p}}\,) \:
\frac{\slashed{p}}{\varepsilon^2 \omega}\:T_0^{[0]}(p) \label{ph2} \\
&\qquad + \big( \text{decay in position space on the scale $\ell_{\min}$} \big)\:, \notag
\end{align}
to be evaluated at~$p=p_\text{\rm{eff}}(\tau)$.

The connection between the contributions~\eqref{ph3}--\eqref{ph2} and the direction-dependent
phases as computed in Section~\ref{seccompdir} is made as follows.
We first point out that the contributions~\eqref{ph3}--\eqref{ph2} all involve factors~$1/\varepsilon^2$
and thus diverge in the limit~$\varepsilon \searrow 0$.
This divergence can be understood immediately from the fact that, multiplying the
contributions by the direction-dependent gauge phases (as stated in Definition~\ref{defscrG})
by~$\delta(\xi^2)\: \epsilon(t)$ in Step~(E1), one gets singularities on the light cone
of the form~$\delta(\xi^2)\: \delta'(\xi^2)$. These singular terms are well-defined due to
the ultraviolet regularization, but diverge quadratically as~$\varepsilon \searrow 0$.
Combining this fact with the formula for the Fourier transform~\eqref{Fouriersimp},
the contributions~\eqref{ph3} and~\eqref{ph4} correspond in position space
precisely to the term~\eqref{defG1} (the factor~$1/(\hat{p}q)$ arises when carrying out
the line integral in~\eqref{defG1} and using that, according to~\eqref{Fouriersimp},
$\hat{\vec{\xi}}$ is to be replaced by~$\hat{\vec{p}}$).
The contributions~\eqref{ph1} and~\eqref{ph2}, on the other hand,
describe both terms~\eqref{defG1} and~\eqref{defG2} combined.
The fact that~\eqref{ph1} and~\eqref{ph2} do not diverge as~$q \rightarrow 0$
corresponds to the fact that, as a consequence of the causal compatibility conditions,
the combination of~\eqref{defG1} and~\eqref{defG2} no longer involves unbounded line integrals.

\subsection{Compensating for the Direction-Dependent Phases} \label{secdirphase}
It remains to compute the phase compensation function~$\scrG \bv$ introduced in Definition~\ref{defscrG}
in momentum space. In view of~\eqref{BPG} and~\eqref{Cdef}, we need to compute the product in position space
\[ \scrC \scrG \bv = \Tr \big( \slashed{\xi}\: (\scrG \bv)(\xi) \big)\: \delta(\xi^2)\: \epsilon(t)\:
\frac{\eta_{\min}}{t^2} 
= i \hat{\Lambda}(q, \xi)\: \Tr \big( \slashed{\xi}\: P(x,y) \big)\; \delta(\xi^2)\: \epsilon(t)\;
\frac{\eta_{\min}(t)}{t^2} \:, \]
where~$\Lambda(x,y)$ describes the direction-dependent phase~\eqref{gaugelin},
and~$\hat{\Lambda}(q, \xi)$ is given as in~\eqref{hatLambdadef} and~\eqref{Lamrealpos}.
Clearly, we may take the products in arbitrary order. It is preferable to proceed as follows:
\label{G01}
\begin{itemize}[leftmargin=3em]
\item[{\bf{(G0)}}] Take the product
\[ \hat{\Lambda}(q,\xi)\: \delta(\xi^2)\: \epsilon(t)\:. \]
\item[{\bf{(G1)}}] Multiply the result of~(G0) with the distribution~$P(x,y)$.
\end{itemize}
Then we proceed with Steps~(E2) and~(E3) on page~\pageref{E1toE3}.

We begin with Step~(G0). Using~\eqref{K0def}, we thus consider the product
\[ -4 \pi^2 i\:\hat{\Lambda}(q,\xi)\: K_0(x,y)\:. \]
Since multiplication in position space corresponds to convolution in momentum space, 
using the ansatz~\eqref{hatLambdadecomp} and disregarding the plane waves
(which merely describe a translation in momentum space), we need to compute
the convolution
\beq \label{LamK0}
-\big( \hat{\Lambda}_q * K_0 \big)(p)
\eeq
(again with~$K_0(p)$ according to~\eqref{K0pdef}; see also~\eqref{K0def}).
The basic question is how to choose the function~$\hat{\Lambda}_q$. The most obvious choice
would be to take the Fourier transform of the function~$\hat{\Lambda}_q(\xi)$ introduced in Definition~\ref{defscrG}.
But this is not the only possible choice, because we have the freedom to modify
this function according to~\eqref{modpos} or~\eqref{modmom}.
Making use of this freedom, we can arrange that the convolution~\eqref{LamK0} can be computed
explicitly with a variant of the mass cone expansion. We first give the construction and discuss the
connection to Definition~\ref{defscrG} afterward. In preparation, it is useful to set
\beq \label{Tmmmdef}
T_{m^2}^{[-2]}(p) := 16\: \delta''(p^2 - m^2)\: \Theta(-p^0) \:.
\eeq
and
\beq \label{Kndef}
K_{m^2}^{[n]}(p) := \frac{i}{4 \pi^2} \:\Big( \overline{T_{m^2}}^{[n]}(p) - T_{m^2}^{[n]}(p) \Big)
\qquad \text{for~$n \geq -2$}
\eeq
(with~$T_{m^2}^{[n]}$ and~$\overline{T_{m^2}^{[n]}}$ as defined in~\eqref{Tm0def}, \eqref{Tmpdef},
\eqref{Tmmdef}, \eqref{Tmmmdef} and~\eqref{Tbardef}; note that the notation~\eqref{Kndef}
harmonizes with~\eqref{K0pdef}).
We make the general ansatz
\beq \label{Lamansatz}
\hat{\Lambda}_\bullet(p) := K^{[-s]}_0(p) \:g(p) \qquad \text{with~$s=1$ or~$s=2$}\:.
\eeq
Before specifying~$g$, we state a general mass cone expansion.
\begin{Thm} \label{thmKgK} Let~$\omega_0$ be a parameter in the range
\[ \frac{1}{\ell_{\min}} \lesssim \omega_0 \lesssim \frac{1}{\varepsilon} \:. \]
Moreover, let~$g \in C^\infty(\hat{M}, \C)$ be a smooth function
which is supported on the scale~$\omega_0$ in the sense that
\[ g(p) = 0 \qquad \text{unless} \qquad \frac{\omega_0}{2} < \big|p^0| < \omega_0 \:. \]
Moreover, we assume that~$g$ satisfies for a suitable constant~$c>0$ the bounds
\beq \label{gbound}
\sup_{k \in \hat{M}} \big| D^n g(k) \big| \leq \bigg( c\: \frac{\ell_{\min}}{\omega_0} \bigg)^{\frac{n}{2}}
\qquad \text{for all~$n \in \N_0$}\:.
\eeq
Then for all~$s \in \{1,2\}$ and all~$p \in \hat{M}$ with~$p^0 \ll -\omega_0$ the following mass cone expansion holds,
\begin{align}
&\int_{\hat{M}} K_0^{[-s]}(k)\: g(k)\: K_0(p-k)\: d^4k \notag \\
&= \frac{1}{8 \pi} \sum_{n=0}^\infty \frac{1}{n!} \int_{-\infty}^\infty \frac{1}{\alpha^s}\,
(\alpha-\alpha^2)^n \:\Box^n g|_{\alpha p}\: d\alpha\:
T_0^{[n-s+1]}(p) \label{KgK1} \\
&\quad\, + \O\Big( \frac{\omega_0}{\omega} \Big) + \big( \text{decay in position space on the scale $\ell_{\min}$} \big)\:. \label{KgK4}
\end{align}
\end{Thm} \noindent
The proof of this theorem is given in Appendix~\ref{appproof} on page~\pageref{proofKgK}.

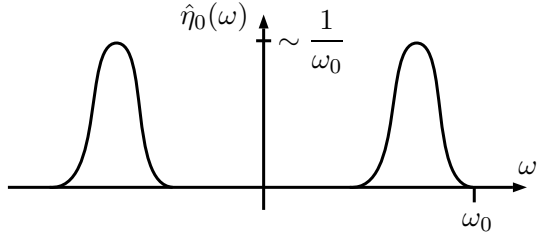
\begin{figure}
\psscalebox{1.0 1.0} 
{
\begin{pspicture}(0,28.614866)(6.9497313,31.264597)
\psline[linecolor=black, linewidth=0.04, arrowsize=0.05291667cm 2.0,arrowlength=1.4,arrowinset=0.0]{->}(3.41,28.624866)(3.41,31.224865)
\psline[linecolor=black, linewidth=0.04, arrowsize=0.05291667cm 2.0,arrowlength=1.4,arrowinset=0.0]{->}(0.01,28.924866)(6.91,28.924866)
\psline[linecolor=black, linewidth=0.04](6.21,28.924866)(6.21,28.724865)
\psline[linecolor=black, linewidth=0.04](3.31,30.874866)(3.51,30.874866)
\psbezier[linecolor=black, linewidth=0.04](4.555,28.924866)(5.2992945,28.873209)(4.995,30.829866)(5.44,30.84486572265625)(5.885,30.859865)(5.5961313,28.968449)(6.19,28.924866)
\psbezier[linecolor=black, linewidth=0.04](0.565,28.924866)(1.3092947,28.873209)(1.005,30.829866)(1.45,30.84486572265625)(1.895,30.859865)(1.6061313,28.968449)(2.2,28.924866)
\rput[bl](6.8,29.1){$\omega$}
\rput[bl](2.3,31){$\hat{\eta}_0(\omega)$}
\rput[bl](6.05,28.35){$\omega_0$}
\rput[bl](3.6,30.45){$\displaystyle \sim \frac{1}{\omega_0}$}
\end{pspicture}
}
\caption{A typical choice of the cutoff function~$\hat{\eta}_0(\omega)$.}
\label{figeta}
\end{figure}

We choose the function~$g$ as follows. We let~$\hat{\eta}_0 \in C^\infty(\R)$ be a ``smeared $\delta$-distribution'' on the scale~$\omega_0$, meaning that this function should vanish for large arguments and in a
neighborhood of the origin. More precisely, we assume that
\beq \label{etasupp}
\supp \hat{\eta}_0 \subset \Big[ -\omega_0, -\frac{\omega_0}{2} \Big] \cup 
\Big[ \frac{\omega_0}{2}, \omega_0 \Big] \:.
\eeq
Moreover, this function should be symmetric, and its integral should be equal to one,
\[ 
\qquad \hat{\eta}_0(-\omega)=\hat{\eta}_0(\omega) \qquad \text{and} \qquad \int_{-\omega_0}^{\omega_0} \hat{\eta}_0(\omega)\:d\omega = 1 \:. \]
Finally, we assume that its derivatives scale like
\[ \sup_\omega \big| \hat{\eta}_0^{(n)}(\omega) \big| \leq \bigg( \frac{c}{\omega_0} \bigg)^n
\qquad \text{for all~$n \in \N$} \]
(where~$c$ is again the constant in~\eqref{gbound}). 
A typical example of the function~$\hat{\eta}_0$ is shown in Figure~\ref{figeta}.

Our next task is to choose the parameter~$s$ and the function~$g$ in~\eqref{Lamansatz} in such a way
that we can compensate for the contributions~\eqref{ph3}--\eqref{ph4}.
To this end, two different choices are of interest.
The first choice is
\beq \label{galtdef}
s=-1 \qquad \text{and} \qquad g(p) = \frac{2}{\pi}\: \frac{\omega^2}{2pq}\: \hat{\eta}_0(\omega)\, \Phi\big( -\epsilon(\omega)\: \hat{\vec{p}}\,\big) \:,
\eeq
where~$\Phi$ is the direction-dependent phase function in~\eqref{ph1} and~\eqref{ph2}.
Following the ansatz~\eqref{hatLambdadecomp}, we
denote the resulting function by~$\hat{\Lambda}_q$,
\beq \label{Lamqchoice}
\hat{\Lambda}_q(p) := K^{[-1]}_0(p) \:\frac{2}{\pi}\: \frac{\omega^2}{2pq}\: \hat{\eta}_0(\omega)\, \Phi\big( -\epsilon(\omega)\: \hat{\vec{p}}\,\big) \:.
\eeq
Then the smoothness condition~\eqref{gbound} leads us to impose that
\beq \label{Lamcond}
\sup_{\hat{\vec{p}} \in S^2} \big| D^n \Phi( \hat{\vec{p}}\,) \big|
\lesssim \big( \omega_0 \,\ell_{\min} \big)^\frac{n}{2}
\eeq
(here the derivatives are computed in local coordinates on the sphere).
In this case, the summand for~$n=0$ in~\eqref{KgK1} is
\begin{align}
(\hat{\Lambda}_q * K_0)(p) &\asymp
\frac{1}{8 \pi} \int_{-\infty}^\infty \frac{1}{\alpha}\, g|_{\alpha p}\: d\alpha\: T_0^{[0]}(p)
=\frac{1}{4 \pi^2}\: \Phi(\hat{\vec{p}}\,)\,\frac{\omega^2}{2 pq}
\int_{-\infty}^\infty \hat{\eta}_0(\alpha \omega)\: d\alpha\: T_0^{[0]}(p) \notag \\
&=\frac{1}{4 \pi^2}\: \Phi(\hat{\vec{p}}\,)\: \frac{\omega^2}{2 pq}
\:\frac{1}{|\omega|}\: T_0^{[0]}(p) = \frac{1}{4 \pi^2}\: \frac{1}{2 \hat{p} q} \:\Phi(\hat{\vec{p}}\,)\: T_0^{[0]}(p) \:.
\label{galtres}
\end{align}

The second choice of interest is
\beq \label{gneudef}
s=-2 \qquad \text{and} \qquad
g(p) = \frac{2}{\pi}\:\omega^2\, \hat{\eta}_0(\omega)\, \Phi\big( -\epsilon(\omega)\: \hat{\vec{p}}\,\big)
\eeq
(where~$\Phi$ is again the direction-dependent phase function in~\eqref{ph1} and~\eqref{ph2}).
We denote the resulting function by~$\hat{\Lambda}_\tau$,
\beq \label{Lamtauchoice}
\hat{\Lambda}_\tau(p) = K^{[-2]}_0(p) \:
\frac{2}{\pi}\:\omega^2\, \hat{\eta}_0(\omega)\, \Phi\big( -\epsilon(\omega)\: \hat{\vec{p}}\,\big)\:.
\eeq
The smoothness condition~\eqref{gbound} again means that~\eqref{Lamcond} must hold.
In this case, the summand for~$n=0$ in~\eqref{KgK1} becomes
\begin{align}
(\hat{\Lambda}_\tau * K_0)(p) &\asymp
\frac{1}{8 \pi}\:  \int_{-\infty}^\infty \frac{1}{\alpha^2}\, g|_{\alpha p}\: d\alpha\: T_0^{[-1]}(p) =\frac{1}{4 \pi^2}\:\Phi(\hat{\vec{p}}\,)\,\omega^2
\int_{-\infty}^\infty \:\hat{\eta}_0(\alpha \omega)\: d\alpha\: T_0^{[-1]}(p) \notag \\
&=\frac{1}{4 \pi^2}\: \Phi(\hat{\vec{p}}\,)\,\omega^2
\:\frac{1}{|\omega|}\: T_0^{[-1]}(p) = -\frac{1}{4 \pi^2}\: \omega\,\Phi(\hat{\vec{p}}\,)\: T_0^{[-1]}(p) \:. \label{gneures}
\end{align}
These mass cone expansions conclude the computation in Step~(G0).

In Step~(G1) we need to take the convolution of the resulting mass cone expansion with~$\hat{P}$.
This convolution can be computed by iterating the mass cone expansion with the help of the following
result.
\begin{Thm} \label{thmTVT} Let~$V \in C^\infty(\hat{M})$ be a smooth (possibly matrix-valued) function which
for a suitable constant~$c>0$ satisfies the bounds~\eqref{Vbound}.
Then, choosing~$p^0=\underline{\omega} \ll -\varepsilon^{-1}$, for any~$r \geq -1$
the following mass cone expansions hold,
\begin{align}
&\int_{\hat{M}} T_{m^2}^{[0]}(k)\: V(k)\: T_0^{[r]}(p-k)\: d^4k \notag \\
&= -2 \pi \sum_{n=0}^\infty \frac{1}{n!} \int_0^\infty (1-\alpha)^r\,
(\alpha-\alpha^2)^n \:\Box^n V|_{\alpha p}\: d\alpha\:
T_{m^2/\alpha}^{[n+r+1]}(p) + \O\Big( \frac{1}{\varepsilon \underline{\omega}} \Big) \label{TVK1nn} \\
&\int_{\hat{M}} \overline{T_{m^2}^{[0]}}(k)\: V(k)\: T_0^{[r]}(p-k)\: d^4k \notag \\
&= 2 \pi \sum_{n=0}^\infty \frac{1}{n!} \int_{-\infty}^0 
(1-\alpha)^r\,(\alpha-\alpha^2)^n \:\Box^n V|_{\alpha p}\: d\alpha\:
T_{m^2/|\alpha|}^{[n+r+1]}(p) \label{TVK3nn} \\
&\quad\, + \big( \text{decay in position space on the scale $\ell_{\min}$} \big)\:. \label{TVK4nn}
\end{align}
\end{Thm} \noindent
The proof of this theorem is given in Appendix~\ref{appproof} on page~\pageref{proofTVT}.

We now go through the different cases and show that, by a suitable ansatz we can 
indeed compensate for the leading contribution to the mass expansion as computed
in~\eqref{ph3}--\eqref{ph2}. For notational brevity, we omit the error terms, which are always
of the form
\beq \label{error} \O\Big( \frac{1}{\varepsilon \underline{\omega}} \Big)
+ \big( \text{decay in position space on the scale $\ell_{\min}$} \big)\:.
\eeq
In order to compensate for~\eqref{ph3} and~\eqref{ph4},
we take the choice~\eqref{Lamqchoice}.
Applying Theorem~\ref{thmKgK} together with~\eqref{galtres}, and then applying Theorem~\ref{thmTVT}, 
we obtain
\begin{align}
i &\big( (\hat{\Lambda}_q * \hat{P}_m) * K_0 \big)(p) =
i \big( \hat{P}_m * (\hat{\Lambda}_q * K_0) \big)(p) \label{G10} \\
&=\frac{i}{4 \pi^2} \int_{\hat{M}} \hat{P}_m(k)\: \frac{p^0-k^0}{2 (p-k)q}\: \Phi \big( \widehat{\vec{p} - \vec{k}} \,\big)\:
T_0^{[0]}(p-k)\: d^4k \notag \\
&=\frac{i}{4 \pi^2} \int_{\hat{M}} T^{[0]}_{m^2}(k)\: (\slashed{k}+m)\:e^{\varepsilon k^0}\: 
\frac{p^0-k^0}{2 (p-k)q}\: \Phi \big( \widehat{\vec{p} - \vec{k}} \,\big)\: T_0^{[0]}(p-k)\: d^4k \notag \\
&=-\frac{i}{2 \pi}\: \frac{1}{2 \hat{p} q} \: \Phi \big( \hat{\vec{p}} \,\big) \int_0^\infty
(\alpha \slashed{p}+m)\:e^{\varepsilon \alpha \omega}\: d\alpha\:
T_{m^2/\alpha}^{[1]}(p) \notag \\
&=-\frac{i}{2 \pi}\: \frac{1}{2 \hat{p} q} \: \Phi \big( \hat{\vec{p}} \,\big)\:\frac{\slashed{p}}{\varepsilon^2 \omega^2}\: T_{0}^{[1]}(p) = \frac{i}{2 \pi}\: \frac{1}{2 p q} \: \Phi \big( \hat{\vec{p}} \,\big)\:\frac{\slashed{p}}{\varepsilon^2 \omega}\: T_{0}^{[1]}(p)\:. \label{Gneu1}
\end{align}
This gives precisely~\eqref{ph3}
(note that the translation in momentum space $p \rightarrow p_\text{\rm{eff}} = p - q/2$ is already taken into
account by the plane wave in our ansatz~\eqref{hatLambdadecomp}).
Similarly, again using the notation~\eqref{reflect},
\begin{align}
i &\Big( \big({\mathscr{R}} (\hat{\Lambda}_q * \hat{P}_m) \big) * K_0 \Big)(p) =
i \Big( \big( ({\mathscr{R}} \hat{\Lambda}_q) * ({\mathscr{R}} \hat{P}_m) \big) * K_0 \Big)(p) \notag \\
&= i \Big( ({\mathscr{R}} \hat{P}_m) * \big( ({\mathscr{R}} \hat{\Lambda}_q) * K_0 \Big)(p) 
= -\frac{i}{2 \pi} \:\frac{1}{2 p q}\:\Phi(\hat{\vec{p}}\,) \:
\frac{\slashed{p}}{\varepsilon^2 \omega}\:T_0^{[1]}(p) \:,
\end{align}
giving agreement with~\eqref{ph4}

In order to compensate for~\eqref{ph1} and~\eqref{ph2}, for the function~$\hat{\Lambda}(q, \xi)$
in~\eqref{hatLambdadef} we make the ansatz
\[ \hat{\Lambda}(q, \xi) = \frac{1}{4} \int_{-1}^1 \Lambda_\tau(\xi)\: e^{-i \tau \frac{q}{2}\: \xi} \:. \]
We choose the Fourier transform of the function~$\Lambda_\tau(\xi)$ according to~\eqref{Lamtauchoice}.
Applying Theorem~\ref{thmKgK} together with~\eqref{gneures}, and then applying Theorem~\ref{thmTVT}, 
we obtain
\begin{align}
i &\big( (\hat{\Lambda}_q * \hat{P}_m) * K_0 \big)(p) =
i \big( \hat{P}_m * (\hat{\Lambda}_q * K_0) \big)(p) \notag \\
&=-\frac{i}{4 \pi^2} \int_{\hat{M}} \hat{P}_m(k)\: (p^0-k^0)\: \Phi \big( \widehat{\vec{p} - \vec{k}} \,\big)\:
T_0^{[-1]}(p-k)\: d^4k \notag \\
&=\frac{i}{4 \pi^2} \int_{\hat{M}} T^{[0]}_{m^2}(k)\: (\slashed{k}+m)\:e^{\varepsilon k^0}\: 
(p^0-k^0)\: \Phi \big( \widehat{\vec{p} - \vec{k}} \,\big)\: T_0^{[-1]}(p-k)\: d^4k \notag \\
&=\frac{i}{2 \pi} \: \omega \Phi \big( \hat{\vec{p}} \,\big) \int_0^\infty
(\alpha \slashed{p}+m)\:e^{\varepsilon \alpha \omega}\: d\alpha\:
T_{m^2/\alpha}^{[0]}(p) \notag \\
&=\frac{i}{2 \pi} \:  \omega \Phi \big( \hat{\vec{p}} \,\big)\:\frac{\slashed{p}}{\varepsilon^2 \omega^2}\: T_{0}^{[0]}(p)
= \frac{i}{2 \pi} \:  \Phi \big( \hat{\vec{p}} \,\big)\:\frac{\slashed{p}}{\varepsilon^2 \omega}\: T_{0}^{[0]}(p) \:,\label{Gneu2}
\end{align}
giving~\eqref{ph1}. Similarly,
\beq
i \Big( \big({\mathscr{R}} (\hat{\Lambda}_q * \hat{P}_m) \big) * K_0 \Big)(p)
= -\frac{i}{2 \pi} \:\Phi(-\hat{\vec{p}}\,) \:
\frac{\slashed{p}}{\varepsilon^2 \omega}\:T_0^{[0]}(p) \:,
\eeq
being in agreement with~\eqref{ph2}.

\subsection{Completing the Construction and Summary} \label{secsummary}
In this section, we complete the construction, obtaining to a 
systematic computational procedure for solving the linearized field equations.
For clarity, we begin with a brief summary of the previous constructions.
Given a retarded jet~$\bv \in \Jret$, the perturbation map~$\scrP$
gives us a variation of the kernel of the fermionic projector (see Definition~\ref{defJvary}).
In order to evaluate the second variation of~$\Sact^\text{\rm{eff}}$ in momentum space,
we must compute~$\hat{\scrC}(\scrP \bv)$ (see Proposition~\ref{prpchat} and~\eqref{Cdef},
\eqref{Chatdef}). To this end, we need to perform the computation steps~(E1), (E2) and~(E3)
on page~\pageref{E1toE3} in momentum space.
Step~(E2) is a convolution which removes all smooth contributions in momentum space
(in the sense~\eqref{modmom}).

The remaining non-smooth contributions can be computed with the
help of the mass cone expansion: In the case~$q^2>0$, we can apply Theorem~\ref{thmTVK}.
The leading contribution to the resulting mass cone expansions is given by~\eqref{ph3} and~\eqref{ph4}.
In the case~$q^2 \leq 0$, on the other hand, we first remove the poles in momentum space
with the help of Lemma~\ref{lemmaline} and then perform the mass cone expansion
by applying Theorem~\ref{thmTVK2}.
The leading contribution to the resulting mass cone expansion is given by~\eqref{ph1} and~\eqref{ph2}.

In the next step, we can compensate for these leading contributions to the mass cone expansions
by suitably chosen direction-dependent local phase transformations~$\scrG \bv$.
In the case~$q^2>0$, this was accomplished by the ansatz~\eqref{Lamansatz} and~\eqref{galtdef},
giving~\eqref{Gneu1}. Likewise, in the case~$q^2 \leq 0$, we choose the ansatz~\eqref{Lamansatz}
and~\eqref{gneudef}, giving~\eqref{Gneu2}.

The result of these construction is a mass cone expansion of the following structure.
In the case~$q^2>0$, we obtain a contribution of the form
\beq \label{lc1}
\begin{split}
&\Big( K_0 * \big( \scrP(\bv) - \scrG(\bv) \big) \Big) (p) \\
&= \sum_{\pm} \frac{1}{\varepsilon^2} 
\frac{1}{p_\text{\rm{eff}} \,q} \sum_{k=1}^\infty
h_k^\pm(p_\text{\rm{eff}})\: T_\bullet^{[k+1]}(p_\text{\rm{eff}}) \quad \text{with} \quad p_\text{\rm{eff}} := p \pm 
\frac{q}{2} \:.
\end{split}
\eeq
Here the functions~$h_{k,\pm}$ are smooth on the hypersurface~$\omega = \underline{\omega}$.
Moreover, the causal compatibility conditions ensure that the two summands combine in such a way
that the total expression is regular in the limit~$q \rightarrow 0$.
Likewise, in the case~$q^2 \leq 0$, we obtain
\begin{align*} 
&\Big( K_0 * \big( \scrP(\bv) - \scrG(\bv) \Big) \big)(p) \\
&=\frac{1}{\varepsilon^2} \sum_{k=1}^\infty \int_{-1}^1
h_{k,\tau}(p_\text{\rm{eff}})\: T_\bullet^{[k]}(p_\text{\rm{eff}})\: d\tau  \quad \text{with} \quad p_\text{\rm{eff}} := p + 
\frac{\tau}{2}\:q \:,
\end{align*}
where the functions~$h_{k,\tau}$ are again smooth on the hypersurface~$\omega = \underline{\omega}$.
Here it is useful to transform the integrals such as to obtain an expansion which is again of the form~\eqref{lc1}.
To this end, we use the relation
\[ T_\bullet^{[k]} \Big( p + \frac{\tau}{2}\:q \Big)
= -\frac{4}{pq}\: \frac{d}{d\tau} T_\bullet^{[k+1]}\Big( p + \frac{\tau}{2}\:q \Big) \:,\qquad k \geq 1 \]
(which follows immediately from~\eqref{Tmpdef}) and integrate by parts.
The boundary terms are of the desired form as in~\eqref{lc1}. Proceeding iteratively, we
obtain an expansion of the form~\eqref{lc1}. For clarity, we point out that the factors~$1/pq$
generated in this procedure may lead to poles of the resulting functions~$h_k^\pm$. But this is unproblematic
for the following construction steps (because these terms will not be differentiated).

Now we can perform the computation steps~(E2) and~(E3) (see page~\pageref{E1toE3}).
This gives rise to a mollification of each summand of the mass cone expansion~\eqref{lc1} on
the scale~$1/\ell_{\min}$. We thus obtain the expansion
\beq \label{lightcone}
\hat{\scrC} \big( \scrP(\bv) - \scrG(\bv) \big)(p) = \frac{1}{\varepsilon^2} \sum_{\pm}
\sum_{k=1}^\infty h^\pm_{k}\Big(p \pm \frac{q}{2} \Big)\:
\bigg( \widehat{\frac{\eta_{\min}(t)}{t^2}} * T_\bullet^{[k]} \bigg)
\Big(p \pm \frac{q}{2} \Big)
\eeq
with new functions~$h^\pm_{k}$ (which may have poles at the points~$p$ with~$pq=0$).
The next step is to compute the second variation of the causal action by
using the above formulas for~$\hat{\scrC} (\scrP(\bv) - \scrG(\bv))$ in the formula
of Proposition~\ref{prpchat}. It suffices to test with variations~$\scrP(\bu)$ where~$\bu \in \Jret$
is a jet of the form as in Definition~\ref{defJvary}, but with potentials~$g_q^{\beta, \alpha}$ which decay
at infinity in the sense that their limit~$\Phi^{\beta, \alpha}_q$ in~\eqref{hatpdef} vanishes.
Then the corresponding phase compensation function~$\scrG \bu$ vanishes.
Considering the mass cone expansion of Theorem~\ref{thmTVK}, one sees that there are enough
degrees of freedom to conclude that the mass cone expansion~\eqref{lightcone}
of the perturbation must vanish to every order on the mass cone, i.e.\
\[ h^\pm_{k}\Big(-|\underline{\omega}|, \sqrt{\underline{\omega}^2-m^2}\,\hat{\vec{p}} \,\Big) = 0 \qquad
\text{for all~$k \in \N$ and~$\hat{\vec{p}} \in S^2$} \:.\]
This equation can be computed and evaluated iteratively order by order on the mass cone.
In this way, we can solve the linearized field equations in the continuum limit
explicitly order by order on the mass cone.
\Felix{Betone, dass die erzeugten Potentiale im allgemeinen nicht symmetrisch sind;
dies hat weitreichende Konsequenzen in~\cite{collapse}.}%

\section{Construction of Homogeneous Solutions} \label{seciterate}
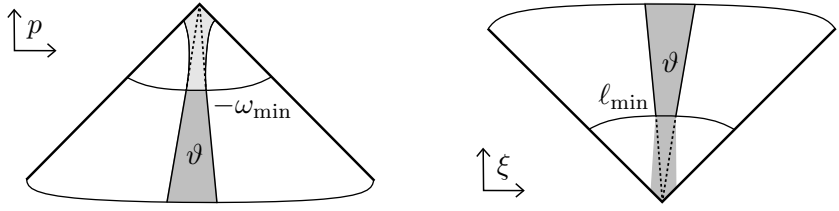
\begin{figure}
\psset{xunit=.5pt,yunit=.5pt,runit=.5pt}
\begin{pspicture}(651.19917261,152.44141586)
{
\newrgbcolor{curcolor}{0.74901962 0.74901962 0.74901962}
\pscustom[linestyle=none,fillstyle=solid,fillcolor=curcolor]
{
\newpath
\moveto(495.62945386,66.56308137)
\lineto(491.84452157,9.70674122)
\lineto(499.84786016,2.03233492)
\lineto(510.44607118,11.77172389)
\lineto(509.99173417,67.09300893)
\closepath
}
}
{
\newrgbcolor{curcolor}{0.74901962 0.74901962 0.74901962}
\pscustom[linewidth=1.00157475,linecolor=curcolor]
{
\newpath
\moveto(495.62945386,66.56308137)
\lineto(491.84452157,9.70674122)
\lineto(499.84786016,2.03233492)
\lineto(510.44607118,11.77172389)
\lineto(509.99173417,67.09300893)
\closepath
}
}
{
\newrgbcolor{curcolor}{0.90196079 0.90196079 0.90196079}
\pscustom[linestyle=none,fillstyle=solid,fillcolor=curcolor]
{
\newpath
\moveto(155.06352378,85.42316578)
\lineto(150.47458394,85.35468074)
\lineto(139.99536378,85.56013586)
\lineto(141.57066331,96.38181523)
\lineto(142.18709291,104.94327665)
\lineto(142.46107087,114.32663791)
\lineto(141.98162268,123.70999539)
\lineto(140.88576,129.8742461)
\lineto(139.24196031,136.17547444)
\lineto(138.07760126,138.98363696)
\lineto(150.13213228,150.90118657)
\lineto(162.11817071,139.18911476)
\lineto(160.26890079,136.99737429)
\lineto(158.96756031,134.2577157)
\lineto(157.66621984,130.01122752)
\lineto(156.57034961,124.53189523)
\lineto(155.33750173,113.50473807)
\lineto(154.72107213,102.95701775)
\lineto(154.72107213,94.66952673)
\closepath
}
}
{
\newrgbcolor{curcolor}{0 0 0}
\pscustom[linewidth=0,linecolor=curcolor]
{
\newpath
\moveto(155.06352378,85.42316578)
\lineto(150.47458394,85.35468074)
\lineto(139.99536378,85.56013586)
\lineto(141.57066331,96.38181523)
\lineto(142.18709291,104.94327665)
\lineto(142.46107087,114.32663791)
\lineto(141.98162268,123.70999539)
\lineto(140.88576,129.8742461)
\lineto(139.24196031,136.17547444)
\lineto(138.07760126,138.98363696)
\lineto(150.13213228,150.90118657)
\lineto(162.11817071,139.18911476)
\lineto(160.26890079,136.99737429)
\lineto(158.96756031,134.2577157)
\lineto(157.66621984,130.01122752)
\lineto(156.57034961,124.53189523)
\lineto(155.33750173,113.50473807)
\lineto(154.72107213,102.95701775)
\lineto(154.72107213,94.66952673)
\closepath
}
}
{
\newrgbcolor{curcolor}{0.74901962 0.74901962 0.74901962}
\pscustom[linestyle=none,fillstyle=solid,fillcolor=curcolor]
{
\newpath
\moveto(139.98041197,85.73092515)
\lineto(147.15999496,85.73092515)
\lineto(154.80277417,85.73092515)
\lineto(163.37195339,0.73395381)
\lineto(125.38964409,1.19713492)
\closepath
}
}
{
\newrgbcolor{curcolor}{0.74901962 0.74901962 0.74901962}
\pscustom[linewidth=0,linecolor=curcolor]
{
\newpath
\moveto(139.98041197,85.73092515)
\lineto(147.15999496,85.73092515)
\lineto(154.80277417,85.73092515)
\lineto(163.37195339,0.73395381)
\lineto(125.38964409,1.19713492)
\closepath
}
}
{
\newrgbcolor{curcolor}{0.74901962 0.74901962 0.74901962}
\pscustom[linestyle=none,fillstyle=solid,fillcolor=curcolor]
{
\newpath
\moveto(510.27380031,66.73111917)
\lineto(503.09420976,66.73111917)
\lineto(495.45143811,66.73111917)
\lineto(486.88222866,151.72813586)
\lineto(524.86455307,151.26495476)
\closepath
}
}
{
\newrgbcolor{curcolor}{0.74901962 0.74901962 0.74901962}
\pscustom[linewidth=0,linecolor=curcolor]
{
\newpath
\moveto(510.27380031,66.73111917)
\lineto(503.09420976,66.73111917)
\lineto(495.45143811,66.73111917)
\lineto(486.88222866,151.72813586)
\lineto(524.86455307,151.26495476)
\closepath
}
}
{
\newrgbcolor{curcolor}{0 0 0}
\pscustom[linewidth=1.88976378,linecolor=curcolor]
{
\newpath
\moveto(19.20785575,18.23524515)
\lineto(150.22211528,151.06549397)
\lineto(282.43786961,17.41901838)
}
}
{
\newrgbcolor{curcolor}{0 0 0}
\pscustom[linewidth=1.88976378,linecolor=curcolor]
{
\newpath
\moveto(368.18476724,132.53475759)
\lineto(500.15838614,1.3340672)
\lineto(631.29476409,132.59196452)
}
}
{
\newrgbcolor{curcolor}{0 0 0}
\pscustom[linewidth=0.99999871,linecolor=curcolor]
{
\newpath
\moveto(95.16674646,95.68640862)
\curveto(101.88206362,92.23444137)(108.59604661,88.783162)(123.52059213,87.05454972)
\curveto(138.44513764,85.32593744)(161.57880567,85.31951224)(176.47161449,87.1083324)
\curveto(191.36441953,88.89715255)(198.01542425,92.48090059)(204.66617953,96.06451255)
}
}
{
\newrgbcolor{curcolor}{0.74901962 0.74901962 0.74901962}
\pscustom[linestyle=none,fillstyle=solid,fillcolor=curcolor]
{
\newpath
\moveto(140.02987087,85.43166594)
\lineto(125.50708913,1.00573964)
}
}
{
\newrgbcolor{curcolor}{0 0 0}
\pscustom[linewidth=1.13385831,linecolor=curcolor]
{
\newpath
\moveto(140.02987087,85.43166594)
\lineto(125.50708913,1.00573964)
}
}
{
\newrgbcolor{curcolor}{0.74901962 0.74901962 0.74901962}
\pscustom[linestyle=none,fillstyle=solid,fillcolor=curcolor]
{
\newpath
\moveto(525.00643654,150.76422783)
\lineto(510.48363969,66.33831287)
}
}
{
\newrgbcolor{curcolor}{0 0 0}
\pscustom[linewidth=1.13385831,linecolor=curcolor]
{
\newpath
\moveto(525.00643654,150.76422783)
\lineto(510.48363969,66.33831287)
}
}
{
\newrgbcolor{curcolor}{0 0 0}
\pscustom[linewidth=1.13385831,linecolor=curcolor]
{
\newpath
\moveto(155.23703055,85.26285334)
\lineto(163.36340787,0.40097744)
}
}
{
\newrgbcolor{curcolor}{0 0 0}
\pscustom[linewidth=1.32283462,linecolor=curcolor,linestyle=dashed,dash=2.0999999 2.0999999]
{
\newpath
\moveto(150.23986772,150.73956641)
\lineto(140.0438589,85.60244389)
}
}
{
\newrgbcolor{curcolor}{0 0 0}
\pscustom[linewidth=1.32283462,linecolor=curcolor,linestyle=dashed,dash=2.0999999 2.0999999]
{
\newpath
\moveto(150.2532548,150.64684326)
\lineto(155.0305663,85.44726027)
}
}
{
\newrgbcolor{curcolor}{0 0 0}
\pscustom[linewidth=0.94488189,linecolor=curcolor]
{
\newpath
\moveto(140.21201386,85.96252704)
\curveto(140.98388409,91.1340735)(141.75588283,96.30646279)(142.14194268,103.13864673)
\curveto(142.52800252,109.97082689)(142.52800252,118.46263539)(141.79483087,124.71572011)
\curveto(141.06165165,130.96880862)(139.59485858,134.98319224)(138.12761953,138.99878531)
}
}
{
\newrgbcolor{curcolor}{0 0 0}
\pscustom[linewidth=0.94488189,linecolor=curcolor]
{
\newpath
\moveto(155.22360567,85.52545114)
\curveto(154.85181354,90.22485145)(154.47993449,94.92533271)(154.75712504,102.79555255)
\curveto(155.03431559,110.66577618)(155.96067402,121.70489933)(157.23448819,128.22880988)
\curveto(158.50830236,134.75272421)(160.12946268,136.75987633)(161.75075528,138.7671872)
}
}
{
\newrgbcolor{curcolor}{0 0 0}
\pscustom[linewidth=0.99999871,linecolor=curcolor]
{
\newpath
\moveto(445.18360819,56.60009555)
\curveto(451.89888378,60.05205145)(458.61287433,63.50332704)(473.53743496,65.23189397)
\curveto(488.46199559,66.9604609)(511.59565228,66.96673492)(526.48846488,65.17784673)
\curveto(541.38127748,63.38895854)(548.03226331,59.80521051)(554.68302236,56.22161366)
}
}
{
\newrgbcolor{curcolor}{0 0 0}
\pscustom[linewidth=1.32283462,linecolor=curcolor,linestyle=dashed,dash=2.0999999 2.0999999]
{
\newpath
\moveto(510.46262551,66.32852389)
\lineto(500.26663181,1.19142783)
}
}
{
\newrgbcolor{curcolor}{0 0 0}
\pscustom[linewidth=1.32283462,linecolor=curcolor,linestyle=dashed,dash=2.0999999 2.0999999]
{
\newpath
\moveto(495.51380031,66.86854279)
\lineto(500.29112315,1.66897114)
}
}
{
\newrgbcolor{curcolor}{0 0 0}
\pscustom[linewidth=1.13385831,linecolor=curcolor]
{
\newpath
\moveto(487.22503181,151.32139822)
\lineto(495.35143181,66.45952232)
}
}
{
\newrgbcolor{curcolor}{0 0 0}
\pscustom[linewidth=1.13385831,linecolor=curcolor]
{
\newpath
\moveto(9.85412334,146.79440484)
\lineto(10.14296617,115.90344169)
\lineto(40.3118211,115.83121492)
}
}
{
\newrgbcolor{curcolor}{0 0 0}
\pscustom[linewidth=1.13385831,linecolor=curcolor]
{
\newpath
\moveto(364.74196913,40.87775224)
\lineto(365.03080063,9.98680421)
\lineto(395.19965858,9.91461523)
}
}
{
\newrgbcolor{curcolor}{0 0 0}
\pscustom[linewidth=1.13385831,linecolor=curcolor]
{
\newpath
\moveto(5.15543546,140.96234437)
\lineto(9.92134639,146.79908011)
\lineto(15.1205212,140.96234437)
}
}
{
\newrgbcolor{curcolor}{0 0 0}
\pscustom[linewidth=1.13385831,linecolor=curcolor]
{
\newpath
\moveto(359.9112,35.72977114)
\lineto(364.67711244,41.56649555)
\lineto(369.8762948,35.72977114)
}
}
{
\newrgbcolor{curcolor}{0 0 0}
\pscustom[linewidth=1.13385831,linecolor=curcolor]
{
\newpath
\moveto(35.12290394,121.10438531)
\lineto(41.10733228,115.83300641)
\lineto(35.19511559,110.92266689)
}
}
{
\newrgbcolor{curcolor}{0 0 0}
\pscustom[linewidth=1.13385831,linecolor=curcolor]
{
\newpath
\moveto(389.37083339,15.18796326)
\lineto(395.35526173,9.91658059)
\lineto(389.44306016,5.00621838)
}
}
{
\newrgbcolor{curcolor}{0 0 0}
\pscustom[linewidth=0.99999871,linecolor=curcolor]
{
\newpath
\moveto(19.26262904,19.4324861)
\curveto(19.68062287,13.18179004)(20.09853846,6.93226563)(63.29879811,3.62982783)
\curveto(106.49905512,0.32742783)(192.47547969,-0.02864146)(236.05130079,2.86745933)
\curveto(279.62712567,5.76356011)(280.79820094,11.91232232)(281.96945386,18.06202941)
}
}
{
\newrgbcolor{curcolor}{0 0 0}
\pscustom[linewidth=0.99999871,linecolor=curcolor]
{
\newpath
\moveto(368.57991685,132.50421145)
\curveto(373.54557354,139.1042757)(378.51005858,145.70281681)(419.78654362,148.88923822)
\curveto(461.06302866,152.07565964)(538.64482394,151.8487746)(580.00578142,148.42705114)
\curveto(621.3667011,145.00532389)(626.50134047,138.39096169)(631.63650898,131.77594563)
}
\rput[bl](20,125){$p$}
\rput[bl](375,18){$\xi$}
\rput[bl](160,65){$-\omega_{\min}$}
\rput[bl](452,72){$\ell_{\min}$}
\rput[bl](140,30){$\vartheta$}
\rput[bl](500,100){$\vartheta$}
}
\end{pspicture}
\caption{A homogeneous solution in momentum and position space.}
\label{fighom}
\end{figure}

As a specific application of the previous computational procedure, we now 
construct homogeneous solutions of the linearized field equations. We choose a momentum transfer which
is not too large in the sense that
\[ |q^0| + |\vec{q}| \ll \frac{1}{\ell_{\min}} \:. \]
Moreover, we choose a parameter~$\omega_{\min}$ in the range
\beq \label{omegaminrange}
\frac{\pi^2}{\ell_{\min}} \lesssim \omega_{\min} \leq \frac{1}{\varepsilon} \:.
\eeq
We choose an angle~$\vartheta \in (0, \pi]$ as
\beq \label{varthetadef}
\vartheta = \frac{1}{\sqrt{\ell_{\min} \,\omega_{\min}}} \:.
\eeq
Therefore, by choosing~$\omega_{\min}$ larger, this angle can be made smaller.
Choosing~$\omega_{\min} = 1/\varepsilon$, we obtain the minimal opening angle
\[ \vartheta_{\min} = \sqrt{\frac{\varepsilon}{\ell_{\min}}}\:. \]
Next, we choose a cutoff function~$\eta \in C^\infty_0(\hat{M}, \R)$ which is equal to one on a
cone of opening angle~$\vartheta$ on the mass shells intersected with the region~$\omega < -\omega_{\min}$
(see the dark shaded region on the left side of Figure~\ref{fighom}).
This cutoff function should go to zero outside this region. The transition region should be smooth
in the sense that its derivatives should scale like
\beq \label{etareg}
\big| D^p \eta(p) \big| \lesssim \frac{1}{\big( \vartheta\,(\omega_{\min} + |\omega|) \big)^p}
= \frac{\big(\ell_{\min} \,\omega_{\min}\big)^\frac{p}{2}}{(\omega_{\min} + |\omega|)^p}\:  \:.
\eeq
We allow that this transition region extends to~$q=0$, as is indicated by the light gray
region in the plot in Figure~\ref{fighom}.
Finally, we choose~$\hat{A}_q$ as a classical electromagnetic potential of momentum~$q$ which
satisfies the homogeneous Maxwell equations in the Lorenz gauge, i.e.
\beq \label{lorenz}
q^j \hat{A}_j(q) = 0 = q^2\, \hat{A}_j(q) \:.
\eeq
Moreover, in order for the potential to be real-valued in position space, we set
\[ \hat{A}(-q) = \hat{A}(q)^* \:. \]
Next, we choose the potentials~$g^{\beta, \alpha}_q$ in~\eqref{nuqgen} as
\beq \label{homansatz}
g^{\beta, \alpha}_q\Big(p-\frac{q}{2} \Big) = -\delta^{\beta \alpha}\:
\Bigg( \slashed{p} + \frac{\slashed{q}}{2} +  m_\beta \bigg)\: \,\,\hat{\!\!\slashed{A}}(q)\: \eta(p)\:.
\eeq

\begin{Thm} \label{thmhom} The potential~\eqref{homansatz} describes a retarded jet~$\bv \in \Jret$
(see Definition~\ref{defJvary}) with
\beq \label{Phihom}
\Phi^{\beta \alpha}_q(\vec{\hat{k}}) = -\delta^{\beta \alpha}\: \hat{p}^j \hat{A}_j(q)
\lim_{\lambda \rightarrow \infty} \eta \big( \lambda \hat{p} \big) \:.
\eeq
Moreover, there is a jet~$\bv_{\text{\rm{err}}} \in \Jret$ for which the limit in~\eqref{infdef} vanishes,
chosen such that~$\bv + \bv_{\text{\rm{err}}}$ satisfies the linearized field equations.
\end{Thm}
\Proof Using the ansatz~\eqref{homansatz} in~\eqref{infdef}, a direct computation gives~\eqref{Phihom}.
Moreover, one readily verifies that the causal compatibility conditions~\eqref{ccc0} and~\eqref{ccc} are satisfied.
Therefore, the ansatz~\eqref{infdef} indeed gives rise to a retarded jet~$\bv \in \Jret$.

The leading summand for~$n=0$ of the mass cone expansion can be analyzed
exactly as worked out in~\cite{cfs}. In particular, as a consequence of the homogeneous
Maxwell equations~\eqref{lorenz}, the logarithmic poles of~$P(x,y)$ drop out of the
linearized field equations (for details see~\cite[\S3.7.1 and~\S4.4.3]{cfs}).
But we need to consider the higher orders in the mass cone expansion.
First, using~\eqref{etareg}, one sees that
\beq \label{Boxes}
\bigg| \Box^n g^{\beta, \alpha}_q\Big(p-\frac{q}{2} \Big) \bigg|
\lesssim \frac{\big(\ell_{\min} \,\omega_{\min}\big)^n}{\big(\omega_{\min} + |\omega|\big)^{2n}}
\leq \frac{\ell_{\min}^n}{|\omega|^n}\:,
\eeq
so that~\eqref{Vbound} is satisfied. Now the summands of the mass cone 
can be compensated order by order by potentials of the form
\beq \label{nodyn}
\delta_q \hat{P}(p) = \sum_{\alpha, \beta=1}^3 V^{\beta, \alpha}(p)\:
 \hat{P}_{m_\alpha}\Big( p-\frac{q}{2} \Big) \:,
\eeq
where~$V$ is matrix-valued and vanishes at infinity. This can be done abstractly by
applying the Fr{\'e}chet-Riesz theorem to the sesquilinear form~$\delta^2 \Sact^\text{\rm{eff}}(\rho)$
in~\eqref{d2Sv} (leaving out the mapping~$\scrG$ because~$V$ vanishes at infinity, so that we do not
pick up direction-dependent phases). 
An alternative and more computational method is to proceed iteratively in orders of the
mass-cone expansion as follows. In the order~$s$
of the iteration, our task is to find a potential~$V$
such that in~\eqref{TVK1} (and similarly in~\eqref{TVK3}) the term of order~$n=s+1$ realizes a
a prescribed asymptotic behavior for large~$|\underline{\omega}|$, whereas the terms of order~$n \leq s$
decay faster in this asymptotics than the terms constructed in the previous iteration steps.
Noting that the asymptotics for large~$|\underline{\omega}|$ is given by
\begin{align*}
\int_0^\infty (\alpha-\alpha^2)^n \:\Box^n V|_{\alpha p}\: d\alpha
&= \int_0^\infty \alpha^n \:\Box^n V|_{\alpha p}\: d\alpha + \O \Big( \frac{1}{\varepsilon \underline{\omega}} \Big) \\
&= \frac{1}{|\underline{\omega}|^{n+1}} \int_0^\infty \alpha^n \:\Box^n V|_{\alpha \hat{p}}\: d\alpha + \O \Big( \frac{1}{\varepsilon \underline{\omega}} \Big) \:,
\end{align*}
given a smooth function~$g$ on the unit sphere, we need to choose~$V$ such that
\beq \label{Vkiter}
\int_0^\infty \alpha^n \:\Box^n V|_{\alpha \hat{p}}\: d\alpha = \left\{
\begin{array}{cl} g(\hat{\vec{p}}) & \text{if~$n = s+1$} \\[0.1em] 0 & \text{if~$n \leq s$}\:.
\end{array} \right.
\eeq
Inspecting this equation for any angular momentum mode, one sees that these relations can indeed
be satisfied by choosing the behavior of the potential~$V(k)$ near~$k=0$ appropriately
(for more details see Remark~\ref{remexpmom} below).

Adding all these potentials gives~$\bv_\text{err}$. A scaling argument shows that
the resulting potential satisfies the bounds~\eqref{Vbound}, proving convergence of the mass cone
expansion.
\QED
In the next remark we explain in more detail how to choose the function~$V(k)$ in~\eqref{Vkiter}.
\begin{Remark} \label{remexpmom}
{\em{ In order to solve~\eqref{Vkiter}, we can clearly consider each angular momentum
mode~$l \in \N_0$ separately. Moreover, in view of~\eqref{nodyn}, the function~$V$
can be chosen arbitrarily outside the mass cone (see also the explanation after the
statement of Theorem~\ref{thmTVK}).
\Felix{Rechne nochmals genau nach!}%
Therefore, it is no loss of generality to take the ansatz
\beq \label{Vansatz0}
V(p) = f(\omega)\: |\vec{p}|^l\: Y_{lm}(\hat{\vec{p}})
\eeq
with a smooth function~$f \in C^\infty_0(\R)$ and parameters~$r \in \N$ and~$|m| \leq l$.
Using that the spherical harmonics are the harmonic homogeneous polynomials restricted to the sphere,
we know that
\[ \Delta_{\R^3} \big( |\vec{p}|^l\: Y_{lm}(\hat{\vec{p}}) = 0 \big) \:. \]
Hence
\[ \Box^n V(p) = f^{(2n)}(\omega)\: |\vec{p}|^l\: Y_{lm}(\hat{\vec{p}}) \]
and thus
\[ \int_0^\infty \alpha^n \:\Box^n V|_{\alpha p}\: d\alpha = \frac{1}{\omega^{2n}}
\int_0^\infty \alpha^{n} \bigg( \frac{d^{2n}}{d\alpha^{2n}} f(\alpha \omega) \bigg)\:
\alpha^l \,|\vec{p}|^l\: Y_{lm}(\hat{\vec{p}})\: d\alpha \:. \]
Now we integrate by parts iteratively in~$\alpha$.
In the case~$n \leq l$, we do this~$2n$ times to obtain
\begin{align}
&\int_0^\infty \alpha^n \:\Box^n V|_{\alpha \hat{p}}\: d\alpha = |\omega|^{n+1} \int_0^\infty \alpha^n \:\Box^n V|_{\alpha p}\: d\alpha \notag \\
&= \frac{|\vec{p}|^l\: Y_{lm}(\hat{\vec{p}}) }{\omega^{n-1}}\:
(-1)^{n+1}\: \frac{(n+l)!}{(l-n)!} \int_0^\infty \alpha^{l-n}\: f(\alpha \omega)\: d\alpha
\qquad \text{if~$n\leq l$}\:. \label{pint1}
\end{align}

In the case~$n>l$, we integrate by parts~$n+l$ times. This gives
\[ \int_0^\infty \alpha^n \:\Box^n V|_{\alpha p}\: d\alpha = \frac{|\vec{p}|^l\: Y_{lm}(\hat{\vec{p}}) }{\omega^{2n}}\:
(-1)^{n+l}\: (n+l)! \int_0^\infty \frac{d^{n-l}}{d\alpha^{n-l}} f(\alpha \omega)\: d\alpha 
\:. \]
Integrating by parts once again, only the boundary term remains,
\[ \int_0^\infty \alpha^n \:\Box^n V|_{\alpha p}\: d\alpha = \frac{|\vec{p}|^l\: Y_{lm}(\hat{\vec{p}}) }{\omega^{2n}}\:
(-1)^{n+l+1}\: (n+l)!\:  \frac{d^{n-l-1}}{d\alpha^{n-l-1}} f(\alpha \omega) \Big|_{\alpha=0} \:. \]
Hence
\beq \label{pint2}
\int_0^\infty \alpha^n \:\Box^n V|_{\alpha \hat{p}}\: d\alpha 
= \frac{|\vec{p}|^l\: Y_{lm}(\hat{\vec{p}}) }{\omega^{l}}\: (-1)^{l}\: (n+l)!\: f^{n-l-1}(0)
\qquad \text{if~$n>l$}\:.
\eeq
With the help of~\eqref{pint1} and~\eqref{pint2} one can solve the equations~\eqref{Vkiter}
explicitly. In the case~$n>0$, this procedure determines the functions~$f$ only at the origin.

In order to avoid the above case distinction, it is a bit easier to consider instead of~\eqref{Vansatz0} the
ansatz
\[ V(p) = f^{(l+1)}(\omega)\: |\vec{p}|^l\: Y_{lm}(\hat{\vec{p}}) \:, \]
again with~$f \in C^\infty_0(\R)$, $r \in \N$ and~$|m| \leq l$.
Then
\[ \int_0^\infty \alpha^n \:\Box^n V|_{\alpha p}\: d\alpha = \frac{1}{\omega^{2n+l+1}}
\int_0^\infty \alpha^{n} \bigg( \frac{d^{2n+l+1}}{d\alpha^{2n+l+1}} f(\alpha \omega) \bigg)\:
\alpha^l \,|\vec{p}|^l\: Y_{lm}(\hat{\vec{p}})\: d\alpha \:. \]
Iteratively integrating by parts~$n+l+1$ times, we obtain
\begin{align}
\int_0^\infty \alpha^n \:\Box^n V|_{\alpha \hat{p}}\: d\alpha &= |\omega|^{n+1} \int_0^\infty \alpha^n \:\Box^n V|_{\alpha p}\: d\alpha \notag \\
&= \frac{|\vec{p}|^l\: Y_{lm}(\hat{\vec{p}}) }{\omega^{n+l}}\:
(-1)^{l}\: (n+l)! \: \frac{d^n}{d\alpha^n} f(\alpha \omega) \Big|_{\alpha=0} \notag \\
&= \frac{|\vec{p}|^l\: Y_{lm}(\hat{\vec{p}}) }{\omega^{l}}\:
(-1)^{l}\: (n+l)! \: f^{(n)}(0) \:.
\end{align}
Apart from being simpler, this procedure has the advantage that
it determines the function~$f$ only at the origin.
}} \QEDrem
\end{Remark}

We now discuss the result of Theorem~\ref{thmhom}. We first note that the last inequality in~\eqref{Boxes}
is not optimal, which means that on the left side of Figure~\ref{fighom} one could work
with a cone whose opening angle gets smaller for large~$|\omega|$ according
to~$\vartheta(\omega) \simeq 1/\sqrt{|\omega|\,\ell_{\min}}$. However, this would not give anything
new because, in view of linearity, this ``thinned cone'' can also be realized with the
potentials in Theorem~\ref{thmhom} by taking superpositions of potentials supported in
thinner and thinner cones
for larger and larger values of~$\omega_{\min}$. This is the reason why we preferred to state the theorem
for cones with a fixed opening angle~$\vartheta$. Following the consideration leading to~\eqref{Fouriersimp},
the corresponding perturbation~$\delta P(x,y)$ is supported again in a cone with opening angle~$\vartheta$,
as is shown on the right side of Figure~\ref{fighom}
(note that the behavior for~$|t| \lesssim \ell_{\min}$ is irrelevant because it drops out of the linearized field equations).
With this in mind, the homogeneous solutions constructed in Theorem~\ref{thmhom} can be
thought of as an electromagnetic potential which does not couple to all wave functions in the same way,
but which couples only to the high-energy states in a cone of opening angle~$\vartheta$.
The coupling to the low-energy states is determined iteratively order by order in the mass cone expansion.
The resulting effective coupling is indicated by the light gray region on the left of Figure~\ref{fighom}.

We point out that the perturbations~\eqref{nodyn} used in order to treat the higher orders
of the mass cone expansion have no poles. Therefore, they are neither retarded nor advanced.
Instead, smoothness in momentum space means that these perturbations are of short range in
position space. In other words, they are {\em{non-propagating}}, but instead they describe small local
changes of~$P(x,y)$ needed in order to satisfy the linearized field equations.
The dynamical degrees of freedom, can be associated to direction-dependent
gauge potentials of the form~\eqref{homansatz}. In analogy to the electromagnetic potential
corresponding to local gauge freedom, the direction-dependent gauge potentials correspond to the
direction-dependent phase freedom~\eqref{dirlocintro}.

We note that the ansatz~\eqref{homansatz} can be described as in Section~\ref{secnonloc}
in terms of the
\[ \text{nonlocal vector potential} \qquad
\,\,\hat{\!\!\slashed{A}}_q\Big(p- \frac{q}{2} \Big) = - \,\,\hat{\!\!\slashed{A}}(q)\:\eta(p) \:. \]
The regularity assumption in momentum space~\eqref{etareg} implies that
the Fourier transform of~$\eta$ decays in position space
\[ \text{on the length scale} \qquad \sqrt{\frac{\ell_{\min}}{\omega_{\min}}} \lesssim \frac{1}{\ell_{\min}} \:. \]
With this in mind, the nonlocality of the potential coincides with the microscopic length scale~$\ell_{\min}$
on which the formalism of the continuum limit applies.

The critical reader may wonder why in~\eqref{homansatz} the potential~$\,\,\hat{\!\!\slashed{A}}(q)$
was chosen independent of the generation index. More generally, one could choose different
a potential for each generation. This would lead to relative phase transformations of the three Dirac
seas. As a consequence, the layer structure of the regularized Dirac sea configuration as
analyzed in detail in~\cite{reg} would be changed. This suggests that such relative phase transformations
would violate the EL equations. This is the reason why we restricted attention to
joint phase transformations of all three Dirac seas.

We finally determine the number of the direction-dependent gauge fields.
The scaling behavior can be determined by choosing~$\vartheta=\vartheta_{\min}$
and counting how many disjoint balls with this opening angle can be put on the unit sphere.
This gives the scaling behavior
\beq \label{Nscale}
N \simeq \frac{4 \pi}{\vartheta_{\min}^2} \simeq \frac{\ell_{\min}}{\varepsilon} \:.
\eeq

\section{A Hierarchical Analysis Beyond the Continuum Limit} \label{secbeyond}
With the constructions in Sections~\ref{seccl} and~\ref{secmom} we obtained a systematic
procedure for solving the linearized field equations in the continuum limit.
In the hierarchical description of Section~\ref{sechierarchy}, this corresponds to
solving the equations on the Hilbert space~$(\h^{(0)}, \la .,. \ra^{(0)})$.
As explained in general terms in Section~\ref{sechierarchy}, one
can proceed inductively by solving the equations on~$\h^{(1)}$, $\h^{(2)}$ and~$\h^{(3)}$.
Doing so gives more detailed information on the solutions.
This additional information is needed in particular for
the analysis of the {\em{dynamical wave equation}}~\eqref{diranalog}.
Namely, as observed in~\cite{noether, dirac}, the relevant kernel~$Q(x,y)$ in this equation
is regular even in the limit~$\varepsilon \searrow 0$ when the regularization is removed.
In the description in Section~\ref{sechierarchy}, this means that the dynamical wave equation
describes the solution on the Hilbert space~$(\h^{(3)}, \la .,. \ra^{(3)})$.

A systematic study of the hierarchical equations goes beyond the scope of the present paper.
But we explain how one can proceed in principle, pointing out how the analysis in the continuum limit must be
modified.
Expanding the second variation according to~\eqref{d2Shierarchy} in powers of~$\varepsilon$,
the singular contributions~$\sim \varepsilon^{-2}$ and~$\sim \varepsilon^{-1}$ are again supported
on the light cone. Therefore, they can be written in analogy to~\eqref{cl1} and~\eqref{cl2}
again in terms of the distribution~$K_0(x,y)$, multiplied by corresponding scaling factors~$\varepsilon$
and~$t$. Again restricting attention to the contributions away from the origin~\eqref{lmindef},
similar to~\eqref{cn1} and~\eqref{cn2} we may again insert factors~$\eta_{\min}$.
Then we can proceed in position space similar as in Section~\ref{seccl}.
Moreover, using again that distributions supported on the light cone correspond to harmonic functions
in momentum space, also all the methods in Section~\ref{secmom} apply.
Therefore, the linearized field equations on~$\h^{(1)}$~$\h^{(2)}$ can again be analyzed
explicitly similar as explained in Section~\ref{secsummary}.

However, the analysis of the linearized field equations on~$\h^{(3)}$ is quite different.
In this case, the second variations are regular as~$\varepsilon \searrow 0$ (see again~\eqref{d2Shierarchy}).
Consequently, the corresponding kernels of the operators~$Q$ and~$\delta Q:= (DQ|_{\Psi}(\delta \Psi))$ in~\eqref{lfehom} need {\em{not}} be supported on the light cone.
On the contrary, as shown in~\cite{noether} and~\cite{dirac}, the kernel~$Q(x,y)$ has relevant contributions
away from the light cone. But we may again restrict attention to the the contributions away from the origin~\eqref{lmindef}. Despite these differences, the methods in position space in Section~\ref{seccl} again apply.
The only difference is that the kernel in the second variations~\eqref{d2Sv} will no longer be supported
on the light cone.
In contrast, the procedure in momentum space in Section~\ref{secmom} can{\em{not}} be
used for the analysis on~$\h^{(3)}$. The basic reason is that we no longer deal with
harmonic functions in momentum space. Consequently, it is no longer possible to evaluate
the equations on the Cauchy surface~$\omega = \underline{\omega}$.
Instead, one must analyze the equations for all~$p \in \hat{M}$.
The only simplification comes from the fact that, following~\eqref{lmindef} and using the
equivalence of~\eqref{modpos} and~\eqref{modmom}, we may disregard smooth contributions
in momentum space (in the sense~\eqref{modmom}).
With this in mind, the mass cone expansion can still be used for the computation of
the phase compensation function~$\scrG \bv$, albeit with improved error terms.
We proceed by stating and explaining a corresponding mass cone expansion
(Section~\ref{secphasemom}). Then we conclude with a few remarks (Section~\ref{secconclude}).

\subsection{Mass Cone Expansion of Direction-Dependent Local Phase Transformations} \label{secphasemom}
We now explain how the mass cone expansion can again be used for the construction
of the phase compensation function~$\scrG \bv$. Compared to the procedure in Section~\ref{secdirphase},
there are the following major differences:
\bitem
\item[{\rm{(i)}}] Since we may no longer evaluate on the light cone, we are no longer allowed to
take the convolution with~$K_0$. Instead, we must first take the convolution of~$\hat{\Lambda}_q$
with~$P_m$. Thus, compared to~\eqref{G10}, we must not first compute~$\hat{\Lambda}_q * K_0$
and then take the convolution with~$\hat{P}_m$, but our task is to compute
the single convolution~$\hat{\Lambda}_q * \hat{P}_m$.
\item[{\rm{(ii)}}] Since~$\delta P$ is no longer necessarily contracted with~$\xi$
(as for example in~\eqref{d2Scont}), it is no longer sufficient to consider the vectorial component
of~$\delta P$ and to contract with~$\xi$. Instead, all the matrix entries of~$\delta P(x,y)$ must be taken
into account.
\item[{\rm{(iii)}}] The error terms must be improved.
\eitem
We now state a such-adapted version of the mass cone expansion.
We consider again the ansatz~\eqref{Lamansatz}. For brevity, we restrict attention to the
case~\eqref{gneudef}, i.e.\ the function~$\hat{\Lambda}_\tau$ given by~\eqref{Lamtauchoice}.
In order to obtain a compact statement, it is convenient
to expand the function~$\Phi$ in spherical harmonics. By linearity, it suffices to
consider one angular mode, i.e.\
\[ \Phi(\hat{\vec{k}}\,) = Y_{lm}(\hat{\vec{k}}\,) \]
with~$l \in \N$ and~$m \in \{-l, \ldots, l\}$.

\begin{Thm} \label{thmconvolve}
Choosing~$g$ as in~\eqref{gneudef}, the following mass cone expansion holds
for all~$p \in \hat{M}$,
\begin{align}
&\int_{\hat{M}} \hat{P}_m(k)\: g(p-k)\, K^{[-2]}_0(p-k)\: d^4k \label{mcm1} \\
&= -\frac{i}{\pi^2}\: \omega\, \Phi(\hat{\vec{p}}\,)  \:
\big( \slashed{p}+m \big) \: e^{\varepsilon \omega}\: T^{[-1]}_{[m^2
+ \O(\omega_0/\omega)]}(p)\:\bigg( 1 + \O \Big( \frac{\omega_0^2}{\omega^2}\Big) 
+ \O \big( \varepsilon^2 \omega_0^2 \big) \bigg) \label{mc0} \\
&\quad\: + \O \Big( \frac{1+l^2}{\omega\, \omega_0} \Big) \:\omega^2 \: e^{\varepsilon \omega}\:T^{[0]}_{[m^2 + \O(\omega_0/\omega)]}(p) \label{err1} \\
&\quad\: + \O \Big( \frac{(1+l^2)^2}{\omega^2\, \omega_0^2} \Big) \:\omega^2 \: e^{\varepsilon \omega}\:
\bigg( 1+ \O\Big( \frac{(1+l^2)^2}{\omega^2\, \omega_0^2} \:p^2 \Big) \bigg) \:. \label{err2}
\end{align}
\end{Thm} \noindent
The proof of this theorem is given in Appendix~\ref{appproof} on page~\pageref{proofconvolve}.

Rewriting the convolution integral in~\eqref{mcm1} as
\beq \label{mcform}
\int_{\hat{M}} \hat{P}(k)_m\:g(p-k)\: K^{[-2]}_0(p-k)\: d^4k
= \int_{\hat{M}} T^{[0]}_{m^2}(k)\: V(k)\: K^{[-2]}_0(p-k)\: d^4k
\eeq
with the potential
\beq \label{mcV}
V(k) := (\slashed{k}+m)\: e^{\varepsilon k^0}\: g(p-k) \:,
\eeq
the above theorem can be regarded as a variant of Theorem~\ref{thmTVK} obtained by
replacing the factor~$K_0$ in~\eqref{TVK1} by~$K^{[-2]}$. Moreover, the leading summand~\eqref{mc0}
is obtained in analogy to~\eqref{TVK1} by evaluating the line integral
\[ \frac{i}{2 \pi} \int_0^\infty \frac{1}{(1-\alpha)^2}\: V|_{\alpha p}\: d\alpha\:
T_{m^2/\alpha}^{[n-1]}(p) \:. \]
The main structural difference of Theorem~\ref{thmconvolve} compared to Theorem~\ref{thmTVK}
is that it applies to all~$p \in \hat{M}$, making it necessary to work with refined error terms.

\subsection{Concluding Remarks and Outlook} \label{secconclude}
We conclude with a few remarks.
Our analysis led to a systematic procedure for solving the linearized field equations in Minkowski
space in the continuum limit (Section~\ref{secsummary}). These methods and results extend and
improve the previous analysis in~\cite{pfp, cfs}.
We also gave a systematic procedure for going beyond the continuum limit by 
proceeding hierarchically order by order on the light cone (Section~\ref{sechierarchy}).
However, for brevity this construction is not carried out systematically in the present paper.

We now make a few remarks on how our results fit together with the computations in~\cite[Chapter~3]{cfs}.
In~\cite[Section~3.7]{cfs} the microlocal chiral transformation was used in order to treat the logarithmic poles 
of~$P(x,y)$ on the light cone (see~\cite{cfs}). The iteration scheme proposed in Section~\ref{secsummary}
should comprise this transformation in a more general and systematic way.
Likewise, we expect that this iteration scheme should reproduce the form of the nonlocal potentials
proposed in~\cite[Section~3.10]{cfs}. From a general point of view, all these methods and
results seem to fit together. But the detailed connections still need to be worked out.

The main open problem is to analyze the linearized field equations on~$\h^{(3)}$.
Apart from giving a more detailed understanding of the linearized dynamics of causal fermion systems,
this is an important point in view of {\em{current conservation}}:
In the abstract setting of causal fermion systems, current conservation is a consequence of the
conservation law for the commutator inner product (see~\cite[Section~5]{noether} or~\cite[Section~3]{dirac}).
Therefore, it is a consequence of the linearized field equations.
However, since the commutator inner product remains finite in the limit~$\varepsilon \searrow 0$,
its conservation law is to be recovered from the linearized field equations on~$\h^{(3)}$.
With this in mind, extending the methods of the present paper to~$\h^{(3)}$ is an important
project for the future. This will also build the bridge to the derivation of
current conservation follows in the context of the dynamical wave equation in~\cite{dirac}.

As already mentioned in the introduction, the methods and results developed here
open the door to a detailed and explicit analysis of the
dynamics of causal fermion systems in Minkowski space.
As concrete applications, we plan to work out the resulting dynamics of quantum fields~\cite{fockdynamics}
and collapse phenomena~\cite{collapse}.

\appendix
\section{The Regularized Scalar Product in Position Space} \label{appscalar}
In Section~\ref{seclinmink} regularized vacuum Dirac sea configurations were described
by a distribution~$\hat{P}$ in momentum space (see~\eqref{nuvacthree}). The underlying Hilbert space
scalar product was deduced from this distribution (see~\eqref{hatHhom}). This procedure has the advantage
that perturbations of the system can be described conveniently by a perturbed distribution
(see~\eqref{tilmuone}), without the need to worry about whether or how the scalar product is to be perturbed.
Clearly, in order to show that this procedure is sensible, it is important to verify that, up to regularization effects,
this scalar product coincides with the usual scalar product on solutions of the Dirac equation.
This appendix is devoted to a detailed analysis of this point. Before entering the computations, we remark that,
without referring to the Dirac equation, the scalar product~\eqref{hatHhom} can be identified with the
commutator inner product, which is time independent as a consequence of the EL equations of the
causal action principle (for details see~\cite{dirac}).

Combining~\eqref{hatHhom} with~\eqref{nuvacthree}, we obtain
\beq
\la f | g \ra_{\hat{\H}} = -\sum_{\alpha=1}^3 \int_{\hat{M}} \frac{d^4p}{(2 \pi)^4}\:
\Sl f(p) \,|\, \big( \slashed{p}+m_\alpha \big)\, g(p) \Sr \;
\delta\big(p^2-m_\alpha^2 \big)\: \Theta(-\omega)\: e^{\varepsilon \omega} \:. \label{fourprod}
\eeq
On the other hand, the physical wave functions corresponding to~$g$ are given by~\eqref{hPsihom}, i.e.\
in position space
\[ \psi^g(x) = -\int_{\hat{M}} \frac{d^4p}{(2 \pi)^4}\: e^{-i p x}\: (\hat{P} g)(p) = \sum_{\alpha=1}^3 \psi^{g,\alpha}(x) \:, \]
where~$\psi^{g,\alpha}(x)$ is the wave function of the generation~$\alpha$,
\beq \label{psigalpha}
\psi^{g,\alpha}(x) := -\int_{\hat{M}} \big( \slashed{p}+m_\alpha \big)\: g(p)\:
\delta\big(p^2-m_\alpha^2 \big)\: \Theta(-\omega)\: e^{\varepsilon \omega}\:e^{-i p x} \: \frac{d^4p}{(2 \pi)^4} \:.
\eeq
In the next lemma we compute the $L^2$-scalar product of these wave functions.
\begin{Lemma} The $L^2$-scalar product of the physical wave functions is given by
\begin{align}
&( \psi^{f,\alpha} | \psi^{g,\alpha} ) := \int_{\R^3} \Sl \psi^{f,\alpha}(t,\vec{x}) \,|\, \gamma^0\,
\psi^{g,\alpha}(t, \vec{x}) \Sr\:
d^3x \label{L2product} \\
&= -\frac{1}{2 \pi} \int_{\R^3} \frac{d^4p}{(2 \pi)^4}\:
\Sl f(p) \:|\: \big( \slashed{p}+m_\alpha \big) \,g(p) \Sr \: 
\delta\big(p^2-m_\alpha^2 \big)\: \Theta(-\omega)\: e^{2\varepsilon \omega} \:. \label{L2res}
\end{align}
\end{Lemma}
\Proof We first carry out the $\omega$-integral in~\eqref{psigalpha},
\[ \psi^{g,\alpha}(x)
= \int_{\R^3} \frac{d^3p}{(2 \pi)^4}\: \frac{1}{2 \omega}\, \big( \slashed{p}+m_\alpha \big)\: g(p)\: e^{\varepsilon \omega} \:e^{-i p x} \bigg|_{\omega = - \sqrt{|\vec{p}|^2+m_\alpha^2}} \:. \]
Using this formula in~\eqref{L2product} and applying Plancherel's theorem, we obtain
\[ ( \psi^{f,\alpha} | \psi^{g,\alpha}) = \int_{\R^3} \frac{d^3p}{(2 \pi)^5}\: \frac{1}{4 \omega^2} \:
\Sl \big( \slashed{p}+m_\alpha \big) f(p) \:|\: \gamma^0\: \big( \slashed{p}+m_\alpha \big) g(p) \Sr
\: e^{2 \varepsilon \omega} \bigg|_{\omega = - \sqrt{|\vec{p}|^2+m_\alpha^2}} \:. \]
Evaluating on the mass shell, we know that~$\slashed{p} \,(\slashed{p}+m_\alpha)
= m_\alpha\,(\slashed{p}+m_\alpha)$. Therefore,
\begin{align*}
\big(\slashed{p}+m_\alpha \big)\: \gamma^0\: \big( \slashed{p}+m_\alpha \big) 
&= \frac{1}{2 m_\alpha}\: \big(\slashed{p}+m_\alpha \big)\: \big\{ \slashed{p}, \gamma^0 \big\}\: \big( \slashed{p}+m_\alpha \big) \\
&= \frac{\omega}{m_\alpha}\: \big(\slashed{p}+m_\alpha \big)\:\big( \slashed{p}+m_\alpha \big) 
= 2 \omega\: \big(\slashed{p}+m_\alpha \big) \:.
\end{align*}
Hence
\begin{align*}
( \psi^{f,\alpha} | \psi^{g,\alpha}) &= \int_{\R^3} \frac{d^3p}{(2 \pi)^5}\: \frac{1}{2 \omega} \:
\Sl f(p) \:|\: \big( \slashed{p}+m_\alpha \big)\: g(p) \Sr
\: e^{2 \varepsilon \omega} \bigg|_{\omega = - \sqrt{|\vec{p}|^2+m_\alpha^2}} \:.
\end{align*}
This agrees with~\eqref{L2res} if in the latter equation we carry out the $\omega$-integral.
\QED

We finally compare the scalar product in momentum space~\eqref{fourprod}
with the $L^2$-scalar product on a Cauchy surface, again expressed in momentum space~\eqref{L2res}.
After summing in~\eqref{L2res} over the generation index, we get agreement with~\eqref{fourprod},
up to a factor~$e^{\varepsilon \omega}/(2 \pi)$.
If no regularization is present, the scalar products agree up to an irrelevant numerical prefactor.
With regularization, however, the factor~$e^{\varepsilon \omega}$ changes the form of the scalar
products for large energies. One way of dealing with this issue is to absorb factors of~$e^{\varepsilon \omega}$
into the definition of the physical wave functions. Details of this procedure can be found
in~\cite[Section~4]{lqg}.  With the present knowledge, it seems conceptually cleaner
to take~\eqref{fourprod} as the definition of the underlying space scalar product.
This differs slightly from the $L^2$-scalar product in position space~\eqref{L2product}.
But this difference seems unproblematic in view of the fact that the scalar product can be expressed
independent of Dirac theory as a surface layer inner product,
the commutator inner product (for details see~\cite{dirac}).

\section{An Explicit Convolution Integral} \label{appexconv}
In this section we illustrate the statement of Theorem~\ref{thmTVK} by computing
the convolution integral on the left side of~\eqref{TVK1} in the special case~$V(k)=e^{\varepsilon k^0}$.
We let~$\hat{K}_0$ be the fundamental solution of the massless Klein-Gordon equation
in momentum space, i.e.\
\[ \hat{K}_0(k) = \delta(k^2)\: \epsilon(k^0)\:. \]
We define the mass cone by
\[ \scrC := \{ p \in \hat{M} \:|\: p^2 > 0 \} \:; \]
it consist of the upper and lower mass cone defined by
\[ \scrC^\vee := \{ p \in \scrC \:|\: p^0>0 \} \qquad \text{and} \qquad
\scrC^\wedge := \{ p \in \scrC \:|\: p^0<0 \} \:. \]
\begin{Lemma} \label{lemmaexconv}
For any~$b \geq 0$,
\begin{align*}
{\mathscr{I}}(p) &:=\int_{\hat{M}} \delta(k^2-b)\: \Theta(-k^0)\: e^{\varepsilon k^0}\: \hat{K}_0(p-k)\: d^4k \\
&\,= \left\{ \begin{array}{ll} \displaystyle
\frac{\pi}{2 \varepsilon \,|\vec{p}\,|}\:\exp \Big\{ \frac{\varepsilon}{2}\: \Big( (p^0-|\vec{p}\,|)
+ \frac{b}{p^0-|\vec{p}\,|} \Big) \Big\} & \text{if~$p \not \in \scrC$} \\[1em]
\displaystyle \frac{\pi}{2 \varepsilon \,|\vec{p}\,|}\:\bigg(
\exp \Big\{ \frac{\varepsilon}{2}\: \Big( (p^0-|\vec{p}\,|) + \frac{b}{p^0-|\vec{p}\,|} \Big) \Big\} & \\
\quad\quad\;\;\,  \displaystyle - \exp \Big\{ \frac{\varepsilon}{2}\: \Big( (p^0+|\vec{p}\,|) + \frac{b}{p^0+|\vec{p}\,|} \Big) \Big\} \bigg) & \text{if~$p \in \scrC^\wedge$} \\[1em]
\quad\; 0 & \text{if~$p \in \scrC^\vee$} \:.
\end{array} \right.
\end{align*}
\end{Lemma}
\Proof We proceed similar as in~\cite[Lemma~5.4]{vacstab}.
We write the factor~$\varepsilon k^0$ in a coordinate independent form as~$ku$
with a future-directed timelike vector~$u$. We begin with the case~$p^2 \not \in \scrC$.
We choose a reference frame with
\begin{align*}
p &= (0,x,0,0) \qquad \text{with} \qquad x>0 \:, \\
u &= (\alpha, \beta, 0, 0) \qquad\! \text{with} \qquad \alpha^2-\beta^2 = \varepsilon^2\:.
\end{align*}
We set
\[ a := p^2 = -x^2 < 0 \:. \]
Moreover, we choose cylindrical coordinates
\[ k = (\omega, \rho, r \cos \varphi, r \sin \varphi) \]
with~$\omega, \rho \in \R$, $r \geq 0$ and~$\varphi \in [0, 2 \pi)$.
Then
\begin{align*}
{\mathscr{I}}(p) &= 2 \pi \int_{-\infty}^0 d\omega \int_{-\infty}^\infty d\rho \int_0^\infty r\: dr\;
\delta \big( \omega^2 - \rho^2 - r^2 - b \big)\: \delta\big( \omega^2 - (\rho-x)^2-r^2 \big)\: e^{\alpha \omega - \beta \rho} \\
&= \pi \int_{-\infty}^0 d\omega \int_{-\infty}^\infty d\rho\;
\Theta \big( \omega^2 - \rho^2 - b \big)\: \delta\big( 2 \rho x -x^2+b \big)\: e^{\alpha \omega - \beta \rho} \\
&= \frac{\pi}{\alpha} \int_{-\infty}^\infty \delta\big( 2 \rho x -x^2+b \big)\: e^{-\alpha \sqrt{\rho^2+b} - \beta \rho}\: d\rho \\
&= \frac{\pi}{2 \alpha x}\:e^{-\alpha \sqrt{K^2+b} - \beta K} \qquad \text{with} \qquad
K:= \frac{x^2-b}{2x}\:.
\end{align*}
This can be written in the shorter form
\[ {\mathscr{I}}(p) = \frac{\pi}{2 \alpha x}\:e^{-\alpha x A - \beta x B} \:, \]
where~$A$ and~$B$ are computed by
\begin{align*}
K^2 +b &= \frac{1}{4 x^2} \:\big( x^4 -2bx^2 + b^2 + 4 b x^2 \big) 
= \frac{1}{4 x^2} \:\big( x^2 +b  \big)^2 = \frac{1}{4 x^2} \:\big( -a +b  \big)^2 \\
A &= \frac{\sqrt{K^2+b}}{x} = \frac{-a+b}{2x^2} = \frac{a-b}{2a} \\
B &= \frac{K}{x} = \frac{1}{2x^2} \:\big( x^2-b \big) = -\frac{1}{2a}\:\big( -a-b \big)
= \frac{a+b}{2a} \:.
\end{align*}
In order to transform back to the original reference frame where~$u=(\varepsilon,0,0,0)$,
we write~$\alpha x$ and~$\beta x$ invariantly and the re-express them in the original reference frame,
\begin{align*}
\beta x &= -p u = -\varepsilon p^0 \\
\alpha x &= \sqrt{\varepsilon^2 + \beta^2}\:x
= \sqrt{-\varepsilon^2 p^2 + (p u)^2} \\
&= \sqrt{-\varepsilon^2 \,\big( (p^0)^2 - |\vec{p}\,|^2 \big) + \varepsilon^2 (p^0)^2}
= \sqrt{\varepsilon^2 \, |\vec{p}\,|^2} = \varepsilon\, |\vec{p}\,| \:.
\end{align*}
We thus obtain
\begin{align}
{\mathscr{I}}(p) &=
\frac{\pi}{2 \varepsilon\,|\vec{p}\,|}\:\exp \Big( -\varepsilon\,|\vec{p}\,|\:\frac{a-b}{2a} +\varepsilon p^0\:
\frac{a+b}{2a} \Big) \notag \\
&= \frac{\pi}{2 \varepsilon\,|\vec{p}\,|}\:\exp \bigg( \frac{\varepsilon}{2}\: \big( p^0 - |\vec{p}\,| \big)
+ \frac{\varepsilon}{2}\: \big( p^0 + |\vec{p}\,| \big) \:\frac{b}{p^2} \bigg) \notag \\
&= \frac{\pi}{2 \varepsilon\,|\vec{p}\,|}\:\exp \bigg( \frac{\varepsilon}{2}\: \big( p^0 - |\vec{p}\,| \big)
+ \frac{\varepsilon}{2}\:\frac{b}{p^0 - |\vec{p}\,|} \bigg) \qquad \text{if~$p \not \in \scrC$}\:. \label{pspace}
\end{align}
This proves the lemma in the case~$p \not \in \scrC$.

In order to treat the cases~$p \in \scrC^\vee$ and~$p \in \scrC^\wedge$,
we make use of the fact that~$\hat{K}_0$, and therefore also~${\mathscr{I}}$,
satisfies the wave equation momentum space, i.e.\
\[ \Box_p {\mathscr{I}}(p) = 0 \:. \]
Due to the uniqueness of solutions of the Cauchy problem with initial data on
the hyperplane~$p^0=0$, it suffices to extend the above formula~\eqref{pspace} to a harmonic
wave on~$\hat{M}$. The general spherically symmetric solution to the wave equation
can be written as
\beq \label{gharmonic}
\hat{\phi}(p) := \frac{1}{|\vec{p}\,|}\: \Big( g \big(p^0+|\vec{p}\,| \big) - g \big( p^0-|\vec{p}\,| \big) \Big)
\eeq
with an arbitrary function~$g$ on~$\R$. One way of verifying this formula is to write
a general spherically solution as a Fourier integral,
\[ \hat{\phi}(p) = \int_{M} d^4x\: \delta(x^2)\: \frac{d^4k}{(2 \pi)^4} \: \phi \big( t, |\vec{x}\,| \big) \:
e^{i p x} \:. \]
Choosing polar coordinates~$(r, \vartheta, \varphi)$ in space and carrying out the angular integrals gives
\begin{align*}
\hat{\phi}(p) &= 2 \pi \int_{-\infty}^\infty dt \int_0^\infty r^2\:dr \int_{-1}^1 d\cos \vartheta\;
\phi(t,r)\: e^{i p^0 t - i |\vec{p}\,| r \cos \vartheta} \\
&= \frac{2 \pi}{|\vec{p}\,|}\:
\int_{-\infty}^\infty d\omega \int_0^\infty r\:dr\:
\phi(t,r)\: e^{i p^0 t} \,\Big( e^{i |\vec{p}\,| r} -  e^{-i |\vec{p}\,| r} \Big) \:,
\end{align*}
which is of the desired form~\eqref{gharmonic} with
\[ g(z) = 2 \pi
\int_{-\infty}^\infty d\omega \int_0^\infty r\:dr\: \phi(t,r)\: e^{i z} \:. \]

In order to write the function~\eqref{pspace} in the form~\eqref{gharmonic} we need to choose
\[ g(z) = -\frac{\pi}{2 \varepsilon}\: \Theta(-z) \: \exp \Big\{ \frac{\varepsilon}{2} \Big( z + \frac{b}{z} \Big) \Big\}\:. \]
Evaluating the harmonic function~\eqref{gharmonic} for this choice of~$g$ gives the result.
\QED

We finally expand the formula in Lemma~\ref{lemmaexconv} and explain how it is related
to the mass cone expansion in Theorem~\ref{thmTVK} (a somewhat different connection
to the mass-cone expansion will be made in Example~\ref{exlightcone} below). First, in the massless case~$b=0$,
the formulas simplify to
\beq \label{Iexpand}
{\mathscr{I}}(p) = \left\{ \begin{array}{cl} \displaystyle
-\frac{\pi}{2 \varepsilon \,|\vec{p}\,|}\:\exp \Big\{ \frac{\varepsilon}{2}\: (p^0-|\vec{p}\,|) \Big\} & \text{if~$p \not \in \scrC$} \\[1em]
\displaystyle -\frac{\pi}{2 \varepsilon \,|\vec{p}\,|}\:\bigg(
\exp \Big\{ \frac{\varepsilon}{2}\: (p^0-|\vec{p}\,|) \Big\}
 - \exp \Big\{ \frac{\varepsilon}{2}\: (p^0+|\vec{p}\,|) \Big\} \bigg) & \text{if~$p \in \scrC^\wedge$} \\[1em]
0 & \text{if~$p \in \scrC^\vee$} \:.
\end{array} \right.
\eeq
Expanding in powers of~$\varepsilon$ gives
\begin{align*}
{\mathscr{I}}(p) &= \left\{ \begin{array}{cl} \displaystyle
-\frac{\pi}{2 \varepsilon \,|\vec{p}\,|} & \text{if~$p \not \in \scrC$} \\[1em]
0 & \text{if~$p \in \scrC$}
\end{array} \right. \\
&\quad\: +\left\{ \begin{array}{cl} \displaystyle
-\frac{\pi}{4 \,|\vec{p}\,|}\:\Big( (p^0-|\vec{p}\,|)
+ \frac{b}{p^0-|\vec{p}\,|} \Big) & \text{if~$p \not \in \scrC$} \\[1em]
\displaystyle \frac{\pi}{2}\:
\Big( 1 - \frac{b}{p^2} \Big) & \text{if~$p \in \scrC^\wedge$} \\[1em]
0 & \text{if~$p \in \scrC^\vee$}
\end{array} \right. \\
&\quad\: + \O(\varepsilon) \:.
\end{align*}
In order to see the connection with the mass cone expansion~\eqref{TVK1} in Theorem~\ref{thmTVK},
we first note that the potential~$V(k)=e^{\varepsilon k^0}$ clearly satisfies the bounds~\eqref{Vbound}.
Next, choosing~$p^0=\underline{\omega} \lesssim -1/\varepsilon$,
the exponential factor~$\exp(\frac{\varepsilon}{2}\: (p^0-|\vec{p}\,|)$ in~\eqref{Iexpand}
can be absorbed into the error term in~\eqref{thmTVK}. We thus obtain
\[ {\mathscr{I}}(p) = \left\{ \begin{array}{cl} \displaystyle \frac{\pi}{2 \varepsilon \,|\vec{p}\,|}\:
\exp \Big\{ \frac{\varepsilon}{2}\: (p^0+|\vec{p}\,|) \Big\}  & \text{if~$p \in \scrC^\wedge$} \\[1em]
0 & \text{if~$p \in \scrC \cup \scrC^\vee$} \end{array} \right. + \O\Big( \frac{1}{\varepsilon \underline{\omega}}
\Big) \:. \]
Now expanding in powers of~$\varepsilon$ gives all the summands of the mass
cone expansion~\eqref{TVK1}, as can be verified for any~$n$ by a straightforward computation
carrying out the integral in~\eqref{TVK1}.

\section{Light-Cone Expansions Involving Unbounded Line Integrals} \label{applightcone}
Light-cone expansions involving unbounded line integrals were already
derived and studied in~\cite[Appendix~F]{pfp}. The method was based on a contour integral formula
in momentum space (see the proof of~\cite[Lemma~F.3]{pfp}).
Unfortunately, this method does not seem to extend to the more general
operator products needed here. Moreover, the light-cone expansion
in~\cite[Lemma~F.3]{pfp} holds only up to smooth contributions. Such smooth error terms
are not good enough for our purposes. For these reasons, we re-derive the light-cone expansions
needed here systematically from the beginning. We work in position space,
generalizing a method developed in~\cite[Section~3]{firstorder}
(see also~\cite[Lemma~2.5.2]{pfp} or~\cite[Lemma~2.2.2]{cfs}).

A {\em{Green's kernel}} of the Klein-Gordon equation~$S_a(x,y)$ is defined by the
distributional identity
\beq \label{KGgreen}
(-\Box_x - m^2) \:S_{m^2}(x,y) = \delta^4(x-y) \:.
\eeq
Since all our computations will involve only the mass squared, it is convenient to set~$a:=m^2 \geq 0$.
The corresponding integral operator is referred to as the {\em{Green's operator}}.
For the following computations, we need different Green's operators, which will be denoted
by different superscripts. 
For ease in notation, we shall omit these superscripts in all formulas which are valid
for all Green's operators. The {\em{causal Green's operators}} have the kernels
\[ S^{\vee\!/\!\wedge}_{a}(x,y) := \lim_{\nu \searrow 0} \int \frac{d^4p}{(2 \pi)^4}
\:\frac{1}{p^2-a \mp i \nu p^0} \:e^{-ip(x-y)} \]
(here~$S^\vee_a$ is the {\em{advanced}} and~$S^\wedge_a$ the {\em{retarded}} Green's operator).
The {\em{symmetric Green's operator}} $S^\times$ is defined as the mean
of the advanced and retarded Green's operators,
\[ 
S^\times_a := \frac{1}{2} \:\big( S^\vee_a + S^\wedge_a \big) \:. \]
The corresponding kernel can be written as
\beq
S_a^\times(x,y) = -\frac{1}{4 \pi}\:\delta(\xi^2) + \Theta(\xi^2) \:H_a(x,y) \:, \label{eq:ba}
\eeq
where $H_a$ is a smooth solution of the Klein-Gordon equation with the series
expansion
\beq
H_a(x,y) = \frac{a}{16 \pi} \:\sum_{j=0}^\infty \frac{(-1)^j}{j!\:(j+1)!}
\:\frac{a^j \:\xi^{2j}}{4^j}\:. \label{eq:bb}
\eeq
Similarly, the retarded Green's operators is given explicitly by
\begin{align}
S^\wedge_a(x,y) &=-\frac{1}{2 \pi}\: \delta(\xi^2)\: \Theta(-\xi^0) + 2\, \: \Theta(\xi^2)\: \Theta(-\xi^0) \:H_a(x,y) 
\label{eq:Swedge} \\
&= -\frac{1}{2 \pi}\: \delta(\xi^2)\: \Theta(-\xi^0) + \frac{a}{8 \pi} \: \Theta(\xi^2)\: \Theta(-\xi^0) + \O\big( a^2 \big) \:,
\end{align}
whereas the advanced Green's operator is obtained from this formula by the replacement~$\Theta(-\xi^0)
\rightarrow \Theta(\xi^0)$.
Finally, we introduce the {\em{spatial Green's operator}}~\label{SJoin_a} by
\beq
S^{\bowtie}_a = S^\times_a - H_a \:. \label{eq:by}
\eeq
According to~\eqref{eq:ba}, it vanishes in time-like directions. It has the expansion
\[ S^{\bowtie}_a = -\frac{1}{2 \pi}\: \delta(\xi^2) - \frac{a}{16 \pi}\: \Theta(-\xi^2) + \O\big( a^2 \big) \:. \]
For the mass expansion we use the notation
\[ 
S^{(l)} := \left( \frac{d}{da} \right)^l S_a \big|_{a=0} \qquad \text{for~$l \in \N_0$} \:. \]

We begin by stating the light-cone expansion as first derived in~\cite{firstorder}; see also~\cite[Lemma~2.2.2]{cfs}.
\begin{Lemma} \label{lemma222} Let~$V \in C^\infty_0(M, \C)$. Assume that~$V$
vanishes in a neighborhood of both~$x$ and~$y$. Then for any~$l,r \in \Z$,
the following light-cone expansions hold,
\begin{align}
&\big( S^{\wedge, (l)} \:V\: S^{\wedge, (r)} \big)(x,y) \notag \\
&= \sum_{n=0}^\infty \frac{1}{n!} \int_0^1 \alpha^{l} \:(1-\alpha)^r\:
(\alpha - \alpha^2)^n \: (\Box^n V) \big|_{\alpha y + (1-\alpha) x} \:d\alpha\;
S^{\wedge, (n+l+r+1)}(x,y) \:. \label{C9}
\end{align}
\end{Lemma}
For the proof and a detailed explanation of this formula we refer to~\cite[\S2.2.2]{cfs}.
Here we only point out that, writing the operator product as an integral,
\[ \big( S^{\wedge, (l)} \:V\: S^{\wedge, (r)} \big)(x,y) = \int_M 
S^{\wedge, (l)}(x,z) \:V(z)\: S^{\wedge, (r)}(z,y)\: d^4z \:, \]
the integration range is compact (more precisely, it suffices to integrate over the
causal diamond generated by~$x$ and~$y$). Likewise, on the right side of~\eqref{C9}
only bounded line integrals appear; these integrals can be understood as integrals
along the line segment joining~$x$ and~$y$.

In what follows, we want to generalize Lemma~\ref{lemma222} to operator products
where the $z$-integration extends over an unbounded region of Minkowski space.
Likewise, on the right side unbounded line integrals will appear.
In preparation, we derive computation rules for the $S^{(l)}$,
which generalize those used for the proof of~\cite[Lemma~2.2.2]{cfs}.
Differentiating~\eqref{KGgreen} with respect to $a$ and setting~$a=0$ gives
\beq
-\Box_x S^{(l)}(x,y) = \delta_{l,0} \:\delta^4(x-y)
+ l \:S^{(l-1)}(x,y) \qquad \text{for~$l \geq 0$}\:. \label{l:5}
\eeq
(For $l=0$, this formula does not seem to make sense because $S^{(-1)}$ 
is undefined. However, the expression is meaningful if one keeps 
in mind that in this case the factor $l$ is zero, and thus the whole 
second summand vanishes. We shall use this convention throughout the following 
calculations.) In momentum space, the causal Green's operators have the form
\[ 
S_a(p) = \frac{1}{p^2-a} \]
with a suitable prescription for treating the pole (which we do not need to specify here).
Differentiating with respect to both~$p$ and $a$ and comparing the resulting formulas, we
obtain the relation
\[ \frac{\partial}{\partial p^k} S_{a}(p) = -2p_k \:\frac{d}{da} S_a(p) \:. \]
Expanding in the mass parameter~$a$ gives
\beq
\frac{\partial}{\partial p^k} S^{(l)}(p) = -2 p_k \:S^{(l+1)}(p) \qquad \text{for~$l \geq 0$}\: .
        \label{l:21a}
\eeq
This formula also determines the derivatives of $S^{(l)}$ in position space; namely
\begin{align}
\frac{\partial}{\partial x^k} & S^{(l)}(x,y) =
\int \frac{d^4p}{(2 \pi)^4} \:S^{(l)}(p) \:(-i p_k) \:e^{-ip(x-y)} \notag \\
&\!\!\!\!\!\stackrel{\eqref{l:21a}}{=} \frac{i}{2} \int \frac{d^4p}{(2 \pi)^4} \: 
\frac{\partial}{\partial p^k} S^{(l-1)}(p) \; e^{-ip(x-y)}
= -\frac{i}{2} \int \frac{d^4p}{(2 \pi)^4} \: 
         S^{(l-1)}(p) \;\frac{\partial}{\partial p^k} e^{-ip(x-y)} \notag \\
&=\frac{1}{2} \: \xi_k \: S^{(l-1)}(x,y) \qquad \text{for~$l \geq 1$}\:.
        \label{l:7}
\end{align}
Recall that we derived this formula for the causal Green's operators.
But this formula also holds for the Green's operator~$S^{\bowtie}_a$, as one sees directly
from~\eqref{eq:bb}, \eqref{eq:by} and the computation
\begin{align*}
&\frac{\partial}{\partial x^k} \frac{d}{da} 
\sum_{j=0}^\infty \frac{(-1)^j}{j!\:(j+1)!}
\:\frac{a^{j+1} \:\xi^{2j}}{4^j} \\
&= \sum_{j=0}^\infty \frac{(-1)^j}{j!\:(j+1)!}\: 
\:(j+1)\: \big(-2j \,\xi_k \big)\:  \frac{a^{j} \:\xi^{2j-2}}{4^j}
= \frac{1}{2}\: \xi_k \sum_{j=0}^\infty \frac{(-1)^j}{j!\:(j+1)!}
\:\frac{a^{j+1} \:\xi^{2j}}{4^j} \:.
\end{align*}

We iterate~\eqref{l:7} to calculate the Laplacian,
\begin{align*}
        -\Box_x S^{(l)}(x,y) &= -\frac{1}{2} \:\frac{\partial}{\partial 
        x^k} \left( (y-x)^k \:S^{(l-1)}(x,y) \right) \\
        &= 2 \:S^{(l-1)}(x,y) + \frac{1}{4} \:(y-x)^2 \:S^{(l-2)}(x,y)
        \qquad \text{for~$l \geq 2$}\:.
\end{align*}
After comparing with \eqref{l:5}, we conclude that
\beq \label{l:22a}
(y-x)^2 \:S^{(l)}(x,y) = -4l\: S^{(l+1)}(x,y) \qquad \text{for~$l \geq 0$}\:.
\eeq
Finally, $S^{(l)}(x,y)$ is a function of $(y-x)$ only, which 
implies that
\beq \label{l:20a}
\frac{\partial}{\partial x^k} S^{(l)}(x,y) =
-\frac{\partial}{\partial y^k} S^{(l)}(x,y) \qquad \text{for~$l \geq 0$}\:.
\eeq

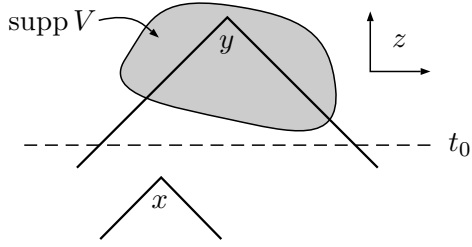
\begin{figure}
\psscalebox{1.0 1.0} 
{
\begin{pspicture}(0,27.519648)(5.4497313,30.700026)
\definecolor{colour2}{rgb}{0.8,0.8,0.8}
\psbezier[linecolor=black, linewidth=0.02, fillstyle=solid,fillcolor=colour2](2.73,30.650188)(1.7400506,30.79161)(1.66,30.410189)(1.35,29.930189208984373)(1.04,29.45019)(1.9600505,29.30161)(2.75,29.13019)(3.5399494,28.958767)(4.2172136,28.785763)(4.15,29.53019)(4.0827866,30.274616)(3.7199495,30.508768)(2.73,30.650188)
\psline[linecolor=black, linewidth=0.03](1.02,27.54019)(1.83,28.36019)(2.63,27.54019)
\psline[linecolor=black, linewidth=0.03](0.71,28.490189)(2.71,30.490189)(4.71,28.490189)
\psline[linecolor=black, linewidth=0.02, arrowsize=0.05291667cm 2.0,arrowlength=1.4,arrowinset=0.0]{<->}(4.61,30.570189)(4.61,29.77019)(5.41,29.77019)
\psline[linecolor=black, linewidth=0.02, linestyle=dashed, dash=0.17638889cm 0.10583334cm](0.01,28.79019)(5.41,28.79019)
\psbezier[linecolor=black, linewidth=0.02, arrowsize=0.05291667cm 2.0,arrowlength=1.4,arrowinset=0.0]{->}(1.0,30.45019)(1.45,30.560188)(1.59,30.400188)(1.8,30.220189208984376)
\rput[bl](1.7,27.95){$x$}
\rput[bl](2.6,30){$y$}
\rput[bl](4.9,30.1){$z$}
\rput[bl](5.65,28.65){$t_0$}
\rput[bl](-0.2,30.25){$\supp V$}
\end{pspicture}
}
\caption{The corresponding Cauchy problem.}
\label{figcauchy}
\end{figure}

We now derive our first light-cone expansion which involves unbounded line integrals.
\begin{Lemma} \label{lemmaSlS0}
Let~$V \in C^\infty_0(M, \C)$. Then for any~$l \in \N_0$ and
\[ x,y \in M \qquad \text{with} \qquad x^0 < y^0 \:, \]
the operator
product $S^{\wedge, (l)} \:V\: S^\vee_0$ has the light-cone expansion
\begin{align}
\big(&S^{\wedge, (l)} \:V\: S^\vee_0 \big)(x,y)\notag \\
&=-2 \sum_{n=0}^\infty \frac{1}{n!} \int_{-\infty}^0 \alpha^{l} \:
(\alpha - \alpha^2)^n \: (\Box^n V) \big|_{\alpha y + (1-\alpha) x} \:d\alpha \; S^{\bowtie, (n+l+1)}(x,y) \:. \label{l:4}
\end{align}
\end{Lemma}
\Proof Following the method of proof of~\cite[Lemma~2.2.2]{cfs}, we want to show that,
choosing a fixed~$y$ and viewing both sides of~\eqref{l:4} as functions of~$x$,
both of these functions are solutions of the inhomogeneous wave equation
with the same inhomogeneity and the same initial data. Then the result follows
from the uniqueness of solutions of the Cauchy problem.
In order to determine the Cauchy surface, 
we denote the operator product on the left side of~\eqref{l:4} by
\[ F(x) := \big(S^{\wedge, (l)} \:V\: S^\vee_0 \big)(x,y) = \int_M S^{\wedge, (l)}(x,z) \:V(z)\: S^\vee_0(z,y)\:
d^4z \:. \]
This operator product vanishes if~$x$ lies in the past of a surface~$\{t=t_0\}$ which lies
to the past of the support of~$V$ (see Figure~\ref{figcauchy}).
Likewise, the right hand side of~\eqref{l:4} vanishes for~$x$ in the past of
the surface~$\{t=t_0\}$, because in this case the argument~$\alpha y + (1-\alpha)x$
of the line integrals also lies in the past of this surface.
Therefore, both sides of~\eqref{l:4} vanish in a neighborhood of the Cauchy surface~$\{t=t_0\}$.

It remains to apply the wave operator to both sides of~\eqref{l:4}.
Since this computation is quite similar to that in the proof of~\cite[Lemma~2.2.2]{cfs},
we only outline the main steps and results.
On the left side of \eqref{l:4}, we compute the Laplacian with the help of \eqref{l:5},
\beq \label{l:6}
-\Box_x (S^{\wedge, (l)} \:V\: S^\vee_0)(x,y) = \delta_{l,0} 
        \:V(x) \: S^\vee_0(x,y) +
        l \: (S^{\wedge, (l-1)} \:V\: S^\vee_0)(x,y) \:.
\eeq
The Laplacian of the integral on the right side of~\eqref{l:4}, on the other hand, can be 
computed with the help of~\eqref{l:7} and~\eqref{l:5},
\begin{align*}
-&\Box_x \int_{-\infty}^0 \alpha^{l} \:(\alpha-\alpha^2)^n
\: (\Box^n V)_{|\alpha y + (1-\alpha) x} \:d\alpha \;
S^{\bowtie, (n+l+1)}(x,y) \\
& = -\int_{-\infty}^0 \alpha^{l} \:(1-\alpha)^{2} \:(\alpha-\alpha^2)^n \:
(\Box^{n+1} V)_{|\alpha y + (1-\alpha) x} \:d\alpha \;
S^{\bowtie, (n+l+1)}(x,y) \notag \\
&\quad-\int_{-\infty}^0 \alpha^{l} \:(1-\alpha) \:(\alpha-\alpha^2)^n \:
(\partial_k \Box^{n} V)_{|\alpha y + (1-\alpha) x}
\:d\alpha \; (y-x)^k \:
S^{\bowtie, (n+l)}(x,y) \notag \\
&\quad +(n+l+1) \int_{-\infty}^0 \alpha^{l} \:(\alpha-\alpha^2)^n \:
(\Box^{n} V)_{|\alpha y + (1-\alpha) x}
\:d\alpha \;S^{\bowtie, (n+l)}(x,y) \: . \notag
\end{align*}
In the second summand, we rewrite the partial derivative as a
derivative with respect to $\alpha$,
\[  (y-x)^k (\partial_k \Box^{n} V)_{|\alpha y + (1-\alpha) x}
= \frac{d}{d\alpha}  (\Box^{n} V)_{|\alpha y + (1-\alpha) x} \:, \]
and integrate by parts. A straightforward computation yields
\begin{align*}
&\int_{-\infty}^0 \alpha^{l} \:(1-\alpha) \:(\alpha-\alpha^2)^n \:
(\partial_k \Box^{n} V)_{|\alpha y + (1-\alpha) x}
\:d\alpha \; (y-x)^k \\
&= \delta_{n,0} \:\delta_{l,0}\:V(x) -n \int_{-\infty}^0 \alpha^{l} \:(1-\alpha)^2 \:(\alpha-\alpha^2)^{n-1} \:
(\Box^{n} V)_{|\alpha y + (1-\alpha) x} \:d\alpha \\
&\qquad+(n+l+1) \int_{-\infty}^0 \alpha^{l} \:(\alpha-\alpha^2)^n \:
(\Box^{n} V)_{|\alpha y + (1-\alpha) x} \:d\alpha \\
&\qquad-l \int_{-\infty}^0 \alpha^{l-1} \:(\alpha-\alpha^2)^{n} \:
(\Box^{n} V)_{|\alpha y + (1-\alpha) x} \:d\alpha \: .
\end{align*}
We substitute back into the original equation to obtain
\begin{align*}
-&\Box_x \int_{-\infty}^0 \alpha^{l} \:(\alpha-\alpha^2)^n
\: (\Box^n V)_{|\alpha y + (1-\alpha) x} \:d\alpha \;
S^{\bowtie, (n+l+1)}(x,y) \\
&= -\delta_{n,0} \:\delta_{l,0} \: V(x) \: S^{\bowtie}_0(x,y) \\
&\quad+l \int_{-\infty}^0 \alpha^{l-1} \:(\alpha-\alpha^2)^{n} \:
(\Box^{n} V)_{|\alpha y + (1-\alpha) x} 
\:d\alpha \; S^{\bowtie, (n+l)}(x,y) \\
&\quad-\int_{-\infty}^0 \alpha^{l} \:(1-\alpha)^2 \:(\alpha-\alpha^2)^{n} \:
(\Box^{n+1} V)_{|\alpha y + (1-\alpha) x} 
\:d\alpha \; S^{\bowtie, (n+l+1)}(x,y) \\
&\quad+n\int_{-\infty}^0 \alpha^{l} \:(1-\alpha)^2 \:(\alpha-\alpha^2)^{n-1} \:
(\Box^{n} V)_{|\alpha y + (1-\alpha) x} 
\:d\alpha \; S^{\bowtie, (n+l)}(x,y) \: .
\end{align*}
After dividing by $n!$ and summing over $n$, the last two summands 
are telescopic and cancel each other. We thus obtain
\begin{align}
-&\Box_x \bigg(-2 \sum_{n=0}^\infty \frac{1}{n!} \int_{-\infty}^0 
\alpha^{l} \:(\alpha-\alpha^2)^n \: (\Box^n 
V)_{|\alpha y + (1-\alpha) x} \:d\alpha \; S^{\bowtie, (n+l+1)}(x,y) \bigg) \notag \\
&= \delta_{l, 0} \:V(x)\:2\,S^{\bowtie}_0(x,y) \notag \\
&\quad\: +l \,\bigg( \!\!-2 \sum_{n=0}^\infty \frac{1}{n!} \int_{-\infty}^0 
\alpha^{l-1} \:(\alpha-\alpha^2)^n \: (\Box^n 
V)_{|\alpha y + (1-\alpha) x} \:d\alpha \; S^{\bowtie, (n+l)}(x,y) \bigg) \:.\label{l:9a}
\end{align}
We now compare the formulas~\eqref{l:6} and~\eqref{l:9a} for the Laplacian of 
both sides of \eqref{l:4}. Using that
\beq \label{Svb}
S^\vee_0(x,y)=2 \,S^{\bowtie}_0(x,y) \qquad \text{for~$x^0 \leq y^0$} \:,
\eeq
these formulas coincide if~$l=0$, proving~\eqref{l:4} in this case.
Proceeding inductively in~$l$, one sees that~\eqref{l:6} and~\eqref{l:9a}
coincide for all~$l$. This concludes the proof.
\QED
Before going on, we point out that this lemma does {\em{not}} extend to the more general
operator products
\[ 
S^{\wedge, (l)} \:V\: S^{\vee, (r)} \qquad \text{for~$r \in \N_0$}\:. \]
The reason is that the identity~\eqref{Svb} does not hold in the case~$r>0$,
\[ S^{\vee,(r)}(x,y) \neq 2 \,S^{\bowtie, (r)}(x,y) \:, \]
simply because the left side is supported in timelike directions, whereas the right side is
supported for spacelike directions. One gets equality only in the case~$r=0$, when both sides are supported
on the light cone. One way to deal with this problem is to 
work with smooth error terms, as worked out in~\cite[Lemma~F.3]{pfp}.
Alternatively, as a method for improving on these error terms, one can work with
light-cone expansions which involve the Green's operator~$S^{\bowtie}_a$ and its mass derivatives:
\begin{Lemma} \label{lemmaSlbowtie}
Let~$V \in C^\infty_0(M, \C)$. Then for any~$l,r \in \N_0$ and
\[ x,y \in M \qquad \text{with} \qquad x^0 < y^0 \:, \]
the operator
product $S^{\wedge, (l)} \:V\: S^{\bowtie, (r)}$ has the light-cone expansion
\begin{align}
\big(S^{\wedge, (l)}& \:V\: S^{\bowtie, (r)} \big)(x,y) = - \sum_{n=0}^\infty \frac{1}{n!}\:
S^{\bowtie, (n+l+r+1)}(x,y) \notag \\
& \times \int_{-\infty}^0 \alpha^{l} \:(1-\alpha)^r\:
(\alpha - \alpha^2)^n \: (\Box^n V) \big|_{\alpha y + (1-\alpha) x} \:d\alpha \:. \label{lbowtie}
\end{align}
\end{Lemma}
\Proof Exactly as explained in the proof of Lemma~\ref{lemmaSlS0}, both sides of~\eqref{lbowtie}
vanish in a neighborhood of a Cauchy surface~$t=t_0$ lying in the past of the support of~$V$.
Therefore, it remains to show that both sides satisfy the same inhomogeneous wave equation.
On the left side, we compute the Laplacian again with the help of \eqref{l:5},
\[ -\Box_x \big(S^{\wedge, (l)} \:V\: S^{\bowtie, (r)} \big)(x,y) = \delta_{l,0} 
\:V(x) \: S^{\bowtie, (r)}(x,y) + l \: (S^{\wedge, (l-1)} \:V\: S^{\bowtie, (r)}_0)(x,y) \:. \]
On the right side, on the other hand, one can follow the computation in the proof of Lemma~\ref{lemmaSlS0}
to obtain in analogy to~\eqref{l:9a} the identities
\begin{align*}
&-\Box_x \bigg(\sum_{n=0}^\infty \frac{1}{n!} \int_{-\infty}^0 
\alpha^{l} \:(\alpha-\alpha^2)^n\:(1-\alpha)^r \: (\Box^n 
V)_{|\alpha y + (1-\alpha) x} \:d\alpha \; S^{\bowtie, (n+l+r+1)}(x,y) \bigg) \notag \\
&= -\delta_{l, 0} \:V(x)\:S^{\bowtie, (r)}_0(x,y) \notag \\
&\quad\: +l \,\bigg( \sum_{n=0}^\infty \frac{1}{n!} \int_{-\infty}^0 
\alpha^{l-1}\:(1-\alpha)^r \:(\alpha-\alpha^2)^n \: (\Box^n 
V)_{|\alpha y + (1-\alpha) x} \:d\alpha \; S^{\bowtie, (n+l+r)}(x,y) \bigg) \:.
\end{align*}
Comparing these formulas and proceeding inductively for increasing~$l$ gives the result.
\QED

\begin{figure}
\psscalebox{1.0 1.0} 
{
\begin{pspicture}(0,27.639698)(11.75,31.160303)
\definecolor{colour0}{rgb}{0.5019608,0.5019608,0.5019608}
\psline[linecolor=black, linewidth=0.03](0.425,28.030302)(2.425,30.030302)(4.425,28.030302)
\psline[linecolor=black, linewidth=0.03](1.275,28.660303)(3.275,30.660303)(5.275,28.660303)
\psline[linecolor=black, linewidth=0.03](6.375,28.160303)(8.375,30.160303)(10.375,28.160303)
\psline[linecolor=black, linewidth=0.03](7.21,27.660303)(9.21,29.660303)(11.21,27.660303)
\psline[linecolor=black, linewidth=0.03, linestyle=dotted, dotsep=0.10583334cm](3.935,31.135303)(0.025,28.330303)
\psline[linecolor=black, linewidth=0.03, linestyle=dotted, dotsep=0.10583334cm](7.09,30.905302)(11.725,28.205303)
\psline[linecolor=colour0, linewidth=0.07](2.425,30.040302)(0.04,28.350304)
\psline[linecolor=colour0, linewidth=0.07](9.205,29.660303)(11.68,28.220303)
\pscircle[linecolor=black, linewidth=0.02, fillstyle=solid,fillcolor=black, dimen=outer](0.6675,28.797802){0.07}
\pscircle[linecolor=black, linewidth=0.02, fillstyle=solid,fillcolor=black, dimen=outer](9.8775,29.277803){0.07}
\psbezier[linecolor=black, linewidth=0.02, arrowsize=0.05291667cm 2.0,arrowlength=1.4,arrowinset=0.0]{->}(0.705,28.020302)(0.42,28.090303)(0.48,28.360302)(0.6,28.615302734375)
\rput[bl](2.3,30.2){$x$}
\rput[bl](8.34,30.3){$x$}
\rput[bl](3.1,30.75){$y$}
\rput[bl](9.2,29.8){$y$}
\rput[bl](0.8,27.8){$z=\alpha y + (1-\alpha) x$}
\rput[bl](9.9,29.45){$z$}
\end{pspicture}
}
\caption{Unbounded line integrals (gray) extending to the past.}
\label{figlineintegral}
\end{figure}
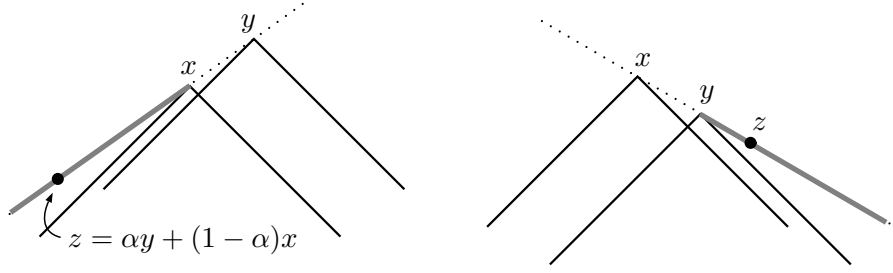

If in~\eqref{l:4} we choose~$l=0$, we can take the adjoint to obtain a light-cone expansion also
in the case~$x^0>y_0$. We thus obtain the following result.
\begin{Corollary} \label{corlight} Let~$V \in C^\infty_0(M, \C)$. Then the operator
product $S^\wedge_0 \:V\: S^\vee_0$ has the light-cone expansion
\begin{align}
\big( S^\wedge_0 \:V\: S^\vee_0 \big)(x,y)
&= -2\sum_{n=0}^\infty \frac{1}{n!} \:S^{\bowtie, (n+1)}(x,y) \notag \\
&\quad\: \times \left\{
\begin{array}{ll} 
\displaystyle \int_{-\infty}^0 (\alpha - \alpha^2)^n \: (\Box^n V) \big|_{\alpha y + (1-\alpha) x} \:d\alpha
& \text{if~$x^0\leq y^0$} \\[1em]
\displaystyle \int_1^\infty (\alpha - \alpha^2)^n \: (\Box^n V) \big|_{\alpha y + (1-\alpha) x} \:d\alpha
& \text{if~$x^0 > y^0$} \:.
\end{array} 
\right. \label{Scases}
\end{align}
\end{Corollary} \noindent
We remark for clarity that the distinction between the two cases in~\eqref{Scases} ensures
that the argument~$\alpha y + (1-\alpha) x$ of the
unbounded line integrals extends to the distant {\em{past}}; see Figure~\ref{figlineintegral}.

The next step is to extend the formula of Corollary~\ref{corlight} to the operator products
\[ \big(S^{\wedge, (l)} \:V\: S^{\vee, (r)} \big)(x,y) \qquad \text{for~$l,r \leq 0$} \]
for {\em{negative}} values of the superscripts~$l$ and~$r$.
To this end, it is most convenient to introduce a real parameter~$\nu$ which deforms
the light cone to a hyperbola.
\begin{Def} \label{defnu}
Given parameters~$a>0$ and~$\nu \in \R$, we define the distributions
\begin{align}
H_{a, [\nu]}(x,y) &:= \frac{a}{16 \pi} \:\sum_{j=0}^\infty \frac{(-1)^j}{j!\:(j+1)!}
\:\frac{a^j}{4^j}\: \big( \xi^2 - \nu \big)^j \label{Hanu} \\
S_{a, [\nu]}^\times(x,y) &:= -\frac{1}{4 \pi}\:\delta(\xi^2 - \nu) + \Theta(\xi^2-\nu) \:H_{a, [\nu]}(x,y) \label{Stanu} \\
S_{a, [\nu]}^\vee(x,y) &:= 2\,S_{a, [\nu]}^\times(x,y)\: \Theta(\xi^0) \label{Svanu} \\
S_{a, [\nu]}^\wedge(x,y) &:= 2\,S_{a, [\nu]}^\times(x,y)\: \Theta(-\xi^0) \label{Swanu} \\
S_{a, [\nu]}^{\bowtie}(x,y) &:= S_{a, [\nu]}^\times(x,y) - H_{a, [\nu]}(x,y) \label{Sbanu} \\
&= -\frac{1}{4 \pi}\:\delta(\xi^2 - \nu) - \Theta(-\xi^2+\nu) \:H_{a, [\nu]}(x,y) \:. \label{Sbanu2}
\end{align}
If the parameter~$a$ is set to zero, we omit the subscript~$a$.
\end{Def}
Setting~$\nu=0$, we get back the earlier formulas~\eqref{eq:bb}, \eqref{eq:ba}, \eqref{eq:Swedge}
and~\eqref{eq:by}. We also note that, although these definitions work any~$\nu \in \R$,
the sign of~$\nu$ changes the structure of the formulas. Indeed, the distribution~$S_{a, [\nu]}^\times$
is supported inside the mass cone iff~$\nu \geq 0$. Consequently, the distributions~$S_{a, [\nu]}^\vee$
and~$S_{a, [\nu]}^\wedge$ are Lorentz invariant iff $\nu \geq 0$.
As we shall see, our resulting formulas will involve~$S_{a, [\nu]}^\times$, $S_{a, [\nu]}^\vee$
and~$S_{a, [\nu]}^\wedge$ only for~$\nu \geq 0$, whereas~$S_{a, [\nu]}^{\bowtie}$ will
enter only for~$\nu \leq 0$.

The main purpose of the parameter~$\nu$ is that it allows us to introduce {\em{negative}}
superscripts by differentiating with respect to it. In order to explain how this comes about,
we applying~\eqref{l:7} for~$l=0$ gives
\beq \label{relS0}
\frac{\partial}{\partial x^k} S^{(0)}(x,y) =\frac{1}{2} \: \xi_k \: S^{(-1)}(x,y)
\eeq
On the other hand, differentiating~\eqref{Svanu}, we obtain for the retarded Green's operator
\[ \frac{\partial}{\partial x^k} S^\vee_{[\nu]}(x,y) = -\frac{1}{2 \pi}\:\delta' \big(\xi^2 -\nu \big) \:\Theta\big( \xi^0 \big)
\: (-2 \xi_k) = 2 \xi^k \: \frac{d}{d\nu} S^\vee_{[\nu]}(x,y) \:. \]
Comparing with~\eqref{relS0}, we find that
\[ S^{\vee, (-1)}(x,y) = 4\: \frac{d}{d\nu} S^\vee_{[\nu]}(x,y) \big|_{\nu=0} \:. \]
Iterating this method leads us to define~$S^{\vee, (-p)}$ by
\beq \label{Snudiff}
S^{\vee, (-p)}(x,y) := \Big( 4\: \frac{d}{d\nu} \Big)^p S^\vee_{[\nu]}(x,y) \Big|_{\nu=0} \:.
\eeq
We use this method for all the other Green's operators as well.
\begin{Def} For any~$p \geq 0$, we define
\begin{align}
H_{a, [\nu]}^{(p)} &:= \partial_a^p H_{a, [\nu]} \label{ppos} \\
H_{a, [\nu]}^{(-p)} &:= \big( 4 \partial_\nu \big)^p H_{a, [\nu]} \:, \label{pneg}
\end{align}
and similarly for all the other distributions introduced in Definition~\ref{defnu}.
\end{Def}
In order for this definition to be sensible, it is crucial that the two operations in~\eqref{ppos}
and~\eqref{pneg} are inverses of each other, as is made precise in the following lemma.
\begin{Lemma} All the distributions introduced in Definition~\ref{defnu} satisfy the functional equation
\[ 4 \partial_a \partial_\nu G_{a, [\nu]} = G_{a, [\nu]} \:. \]
\end{Lemma}
\Proof For the distribution~$H_{a, [\nu]}$ the functional equation follows directly from the
series representation~\eqref{Hanu},
\begin{align}
	\partial_a \partial_\nu H_{a, [\nu]}(x,y) &= \frac{a}{16 \pi} \:\sum_{j=1}^\infty \frac{(-1)^j}{j!\:(j+1)!}\:
(j+1)(-j)\:\frac{a^{j-1}}{4^j}\: \big( \xi^2 - \nu \big)^{j-1} \notag \\
&= \frac{a}{16 \pi} \:\sum_{j=0}^\infty \frac{(-1)^j}{j!\:(j+1)!}\:\frac{a^{j-1}}{4\cdot4^j}\: \big( \xi^2 - \nu \big)^{j}
= \frac{1}{4}\: H_{a, [\nu]}(x,y) \label{danH}
\end{align}
(in the last line we performed an index shift~$j \rightarrow j+1$).
For the distribution~$S^\times_{a, [\nu]}$, we proceed as follows,
\begin{align*}
&\partial_a \partial_\nu S_{a, [\nu]}^\times(x,y) =  \partial_a \partial_\nu \big( \Theta(\xi^2-\nu) \:H_{a, [\nu]}(x,y) \big)\\
&= \Theta(\xi^2-\nu) \:\partial_a \partial_\nu H_{a, [\nu]}(x,y) +
\big( \partial_\nu \Theta(\xi^2-\nu) \big) \: \partial_a H_{a, [\nu]}(x,y) \:.
\end{align*}
In the first summand, we can use~\eqref{danH}. We thus obtain
\beq \label{anS}
	\partial_a \partial_\nu S_{a, [\nu]}^\times(x,y) = \frac{1}{4}\: 
\Theta(\xi^2-\nu) \:\partial_a \partial_\nu H_{a, [\nu]}(x,y) - \delta(\xi^2-\nu)\: \partial_a H_{a, [\nu]}(x,y) \:.
\eeq
The last summand can be computed further using again the series~\eqref{Hanu}. Indeed,
evaluating for~$\xi^2=\nu$, only the summand for~$j=0$ contributes. We thus obtain
\[ \delta(\xi^2-\nu)\: \partial_a H_{a, [\nu]}(x,y) = \delta(\xi^2-\nu)\: \frac{1}{16 \pi} \:. \]
Using this relation in~\eqref{anS} and comparing with~\eqref{Stanu}, we obtain
precisely~$S^\times_{a, [\ni]}/4$.

For the distributions~$S_{a, [\nu]}^\vee$ and~$S_{a, [\nu]}^\wedge$ one argues similarly,
noting that the Heaviside functions~$\Theta(\pm \xi^0)$ are not differentiated.
Finally, for~$S_{a, [\nu]}^{\bowtie}(x,y)$ we use~\eqref{Sbanu}
and apply the functional relations for~$S_{a, [\nu]}^\times$ and~$H_{a, [\nu]}$.
\QED
As a consequence of this lemma, all the computation rules for the Green's operators
used previously (see~\eqref{l:21a}, \eqref{l:22a} and \eqref{l:20a}) hold for any~$l \in \Z$.

After these preparations, we can derive a light-cone expansion involving negative
indices~$l$ and~$r$.
\begin{Lemma} \label{lemmaneg1} Let~$V \in C^\infty_0(M, \C)$. Assume that~$V$
vanishes in a neighborhood of both~$x$ and~$y$. Then for any~$l,r \leq 0$,
the operator product $S^{\wedge, (l)} \:V\: S^{\vee, (r)}$ has the light-cone expansion
\begin{align*}
&\big( S^{\wedge, (l)} \:V\: S^{\vee, (r)} \big)(x,y)
= -2\sum_{n=0}^\infty \frac{1}{n!} \:S^{\bowtie, (n+l+r+1)}(x,y) \\
&\quad\: \times \left\{
\begin{array}{ll} 
\displaystyle \int_{-\infty}^0 \alpha^{l} \:(1-\alpha)^r\:
(\alpha - \alpha^2)^n \: (\Box^n V) \big|_{\alpha y + (1-\alpha) x} \:d\alpha
& \text{if~$x^0\leq y^0$} \\[1em]
\displaystyle \int_1^\infty \alpha^{l} \:(1-\alpha)^r\:
(\alpha - \alpha^2)^n \: (\Box^n V) \big|_{\alpha y + (1-\alpha) x} \:d\alpha
& \text{if~$x^0 > y^0$} \:.
\end{array} 
\right. 
\end{align*}
\end{Lemma}
\Proof By symmetry, it suffices to consider the case~$x^0<y^0$.
We begin with the case~$l<0$ and~$r=0$.
In this case, we proceed inductively for decreasing~$l$, beginning with~$l=0$.
Then the induction hypothesis~$l=0$ holds in view of Lemma~\ref{lemmaSlS0}.
Thus assume that the formula holds for given~$l$, i.e.\
\begin{align}
&\big( S^{\wedge, (l)} \:V\: S^{\vee, (r)} \big)(x,y) \notag \\
&= -2\sum_{n=0}^\infty \frac{1}{n!}
\int_{-\infty}^0 \alpha^{l} \:(1-\alpha)^r\:
(\alpha - \alpha^2)^n \: (\Box^n V) \big|_{\alpha y + (1-\alpha) x} \:d\alpha\:S^{\bowtie, (n+l+1)}(x,y)  \:.
\label{toprove}
\end{align}
Taking the partial derivative in the variable~$x$ and using~\eqref{l:7}, we obtain
\[ \frac{\partial}{\partial x^k} \big( S^{\wedge, (l)} \:V\: S^{\vee, (r)} \big)(x,y)
= \frac{1}{2} \big( S^{\wedge, (l-1)} \:W_k\: S^{\vee, (r)} \big)(x,y) \]
with
\[ W_k(z) := (z-x)_k\: V(z) \:. \]
On the other hand, differentiating the right side of~\eqref{toprove} with the chain rule
and again applying~\eqref{l:7}, we obtain
\begin{align*}
& \frac{\partial}{\partial x^k} \bigg( \sum_{n=0}^\infty \frac{1}{n!}
\int_{-\infty}^0 \alpha^{l} \:(1-\alpha)^r\:
(\alpha - \alpha^2)^n \: (\Box^n V) \big|_{\alpha y + (1-\alpha) x} \:d\alpha\:S^{\bowtie, (n+l+1)}(x,y) \bigg ) \\
&= \frac{1}{2} \sum_{n=0}^\infty \frac{1}{n!}
\int_{-\infty}^0 \alpha^{l-1} \:(1-\alpha)^r\:
(\alpha - \alpha^2)^n \: (z-x)_k (\Box^n V) \big|_{\alpha y + (1-\alpha) x} \:d\alpha\:S^{\bowtie, (n+l)}(x,y) \\
&\quad\: +\sum_{n=0}^\infty \frac{1}{n!}
\int_{-\infty}^0 \alpha^{l} \:(1-\alpha)^{r+1}\:
(\alpha - \alpha^2)^n \: (\partial_k \Box^n V) \big|_{\alpha y + (1-\alpha) x} \:d\alpha\:S^{\bowtie, (n+l+1)}(x,y) \\
&= \frac{1}{2} \sum_{n=0}^\infty \frac{1}{n!}
\int_{-\infty}^0 \alpha^{l-1} \:(1-\alpha)^r\:
(\alpha - \alpha^2)^n \:(\Box^n W_k) \big|_{\alpha y + (1-\alpha) x} \:d\alpha\:S^{\bowtie, (n+l)}(x,y) \\
&\quad\: -\sum_{n=0}^\infty \frac{1}{n!}
\int_{-\infty}^0 \alpha^{l-1} \:(1-\alpha)^r\:
(\alpha - \alpha^2)^n \:n\,(\partial_k \Box^{n-1} V) \big|_{\alpha y + (1-\alpha) x} \:d\alpha\:S^{\bowtie, (n+l)}(x,y) \\
&\quad\: +\sum_{n=0}^\infty \frac{1}{n!}
\int_{-\infty}^0 \alpha^{l} \:(1-\alpha)^{r+1}\:
(\alpha - \alpha^2)^n \: (\partial_k \Box^n V) \big|_{\alpha y + (1-\alpha) x} \:d\alpha\:S^{\bowtie, (n+l+1)}(x,y) \:.
\end{align*}
After performing an index shift, the last two series cancel each other.
We thus obtain~\eqref{toprove} with~$l$ replaced by~$l-1$ and~$V$ replaced by~$W_k$.

The induction for decreasing values of~$r$ works similarly.
\QED

The method of proof of the previous lemma can also be used in order to
extend the light-cone expansion of Lemma~\ref{lemma222} to arguments~$l, r \in \Z$:
\begin{Lemma} \label{lemmaneg2} Let~$V \in C^\infty_0(M, \C)$. Assume that~$V$
vanishes in a neighborhood of both~$x$ and~$y$. Then for any~$l,r \in \Z$,
the following light-cone expansions hold,
\begin{align*}
&\big( S^{\wedge, (l)} \:V\: S^{\wedge, (r)} \big)(x,y) \\
&= \sum_{n=0}^\infty \frac{1}{n!} \int_0^1 \alpha^{l} \:(1-\alpha)^r\:
(\alpha - \alpha^2)^n \: (\Box^n V) \big|_{\alpha y + (1-\alpha) x} \:d\alpha\;
S^{\wedge, (n+l+r+1)}(x,y) \:.
\end{align*}
\end{Lemma}
\Proof For~$l,r \in \N_0$, we obtain the statement of Lemma~\ref{lemma222}.
The cases~$l<0$ and~$r<0$ can be derived inductively, exactly as explained in
the proof of Lemma~\ref{lemmaneg1} above.
\QED

We finally turn attention to light-cone expansions of operator products which involve
the {\em{causal fundamental solution}}, being defined (up to a prefactor) as the difference
of the advanced and retarded Green's operators. More precisely, we set
\beq \label{Knudef}
K_{[\nu]} = \frac{1}{2 \pi i} \:\big( S^\vee_{[\nu]} - S^\wedge_{[\nu]} \big)
\eeq
and
\beq \label{Knudiff}
K^{(-p)}(x,y) := \Big( 4\: \frac{d}{d\nu} \Big)^p K_{[\nu]}(x,y) \Big|_{\nu=0} \:.
\eeq

\begin{Prp} \label{prpKfinal} Let~$V \in C^\infty_0(M, \C)$. Assume that~$V$
vanishes in a neighborhood of both~$x$ and~$y$. Then for any~$l,r \leq 0$,
the operator product $S^{\wedge, (l)} \:V\: K^{(r)}$ 
has the light-cone expansion
\begin{align*}
&2 \pi i \:\big( S^{\wedge, (l)}\:V\: K^{(r)} \big)(x,y) \notag \\
&= -2\sum_{n=0}^\infty \frac{1}{n!} \:S^{\bowtie, (n+l+r+1)}(x,y) \notag \\
&\quad\qquad \times \left\{
\begin{array}{ll} 
\displaystyle \int_{-\infty}^0 \alpha^{l} \:(1-\alpha)^{r}\:
(\alpha - \alpha^2)^n \: (\Box^n V) \big|_{\alpha y + (1-\alpha) x} \:d\alpha
& \text{if~$x^0\leq y^0$} \\[1em]
\displaystyle \int_1^\infty \alpha^{l} \:(1-\alpha)^{r}\:
(\alpha - \alpha^2)^n \: (\Box^n V) \big|_{\alpha y + (1-\alpha) x} \:d\alpha
& \text{if~$x^0 > y^0$}
\end{array} \right. \\
&\quad\, -\sum_{n=0}^\infty \frac{1}{n!} \int_0^1 \alpha^{l} \:(1-\alpha)^{r}\:
(\alpha - \alpha^2)^n \: (\Box^n V) \big|_{\alpha y + (1-\alpha) x} \:d\alpha\;
S^{\wedge, (n+l+r+1)}(x,y) \:.
\end{align*}
\end{Prp}
\Proof Using the relation
\[ K^{(r)} = \frac{1}{2 \pi i} \big( S^{\vee, (r)} - S^{\wedge, (r)} \big) \:, \]
we obtain
\[ \big( S^{\wedge, (l)} \:V\: K^{(r)} \big)(x,y)
= \frac{1}{2 \pi i} 
\big( S^{\wedge, (l)} \:V\: S^{\vee, (r)} - S^{\wedge, (l)}\:V\:S^{\wedge, (r)} \big)(x,y) \:. \]
To each summand, we can apply either Lemma~\ref{lemmaneg1} or Lemma~\ref{lemmaneg2}.
This gives the result.
\QED

\begin{Corollary} \label{corKK} Under the assumptions of Proposition~\ref{prpKfinal}, for any~$l,r \leq 0$,
the operator product $K^{(l)} \:V\: K^{(r)}$  has the light-cone expansion
\begin{align*}
-2 &\pi^2 \:\big( K^{(l)}\:V\: K^{(r)} \big)(x,y) \notag \\
&= \sum_{n=0}^\infty \frac{1}{n!} 
\int_{-\infty}^\infty \alpha^{l} \:(1-\alpha)^{r}\:
(\alpha - \alpha^2)^n \: (\Box^n V) \big|_{\alpha y + (1-\alpha) x} \:d\alpha
\:S^{\bowtie, (n+l+r+1)}(x,y) \\
&\quad\: + \sum_{n=0}^\infty \frac{1}{n!} 
\int_0^1 \alpha^{l} \:(1-\alpha)^{r}\:
(\alpha - \alpha^2)^n \: (\Box^n V) \big|_{\alpha y + (1-\alpha) x} \:d\alpha
\:H^{(n+l+r+1)}(x,y) \:.
\end{align*}
\end{Corollary}
\Proof Applying a time reflection, the result of Proposition~\ref{prpKfinal} gives a corresponding
light cone expansion for the operator product~$S^{\vee, (l)}\:V\: K^{(r)}$.
Subtracting the resulting formulas and applying~\eqref{Knudef} and~\eqref{Knudiff}
gives the result.
\QED

We now state our main result. The novel feature compared to the previous expansion is that
the sum over the index~$l$ is re-summed to obtain the distributions introduced in Definition~\ref{defnu}.
It turns out that, after this re-summation, the light cone expansion holds even for potentials~$V$
which do {\em{not}} vanish at the spacetime point~$x$.

\begin{Prp} \label{prpKfinal2}
Let~$V \in C^\infty_0(M, \C)$, $\nu \in \R^+_0$
and~$r \leq 0$. In the case~$r<0$, we assume that~$V$
vanishes in a neighborhood of the point~$y \in M$. Then
the kernel~$(S^{\wedge}_{[\nu]} \:V\: K^{(r)})(x,y)$  has the light-cone expansion
\begin{align*}
&2 \pi i \:\big( S^\wedge_{[\nu]}\:V\: K^{(r)} \big)(x,y) \\
&= -2\sum_{n=0}^\infty \frac{1}{n!} \notag \\
&\quad\qquad \times \left\{
\begin{array}{ll} 
\displaystyle \int_{-\infty}^0 (1-\alpha)^{r}\:
(\alpha - \alpha^2)^n \: (\Box^n V) \big|_{\alpha y + (1-\alpha) x}\:S^{\bowtie, (n+r+1)}_{[\nu/\alpha]}(x,y)  \:d\alpha
& \text{if~$x^0\leq y^0$} \\[1em]
\displaystyle \int_1^\infty (1-\alpha)^{r}\:
(\alpha - \alpha^2)^n \: (\Box^n V) \big|_{\alpha y + (1-\alpha) x}\:S^{\bowtie, (n+r+1)}_{[\nu/\alpha]}(x,y)  \:d\alpha
& \text{if~$x^0 > y^0$}
\end{array} \right. \\
&\quad\, -\sum_{n=0}^\infty \frac{1}{n!} \int_0^1 (1-\alpha)^{r}\:
(\alpha - \alpha^2)^n \: (\Box^n V) \big|_{\alpha y + (1-\alpha) x} \:
S^{\wedge, (n+r+1)}_{[\nu/\alpha]}(x,y)\:d\alpha \:.
\end{align*}
\end{Prp}
\Proof We first consider the case that~$V$ vanishes in a neighborhood of both~$x$ and~$y$.
Then, using~\eqref{Snudiff} we obtain
\[ 
\big( S^\wedge_{[\nu]} \:V\: K^{(r)} \big)(x,y)
= \sum_{\ell=0}^\infty \frac{1}{\ell!} \:\bigg( \frac{\nu}{4} \bigg)^\ell\:
\big( S^{\wedge, (-\ell)} \:V\:K^{(r)} \big)(x,y) \:. \]
To each summand we apply the expansion of Proposition~\ref{prpKfinal}.
Finally, we carry out the $\ell$-sum explicitly using~\eqref{Snudiff} to obtain the desired
light-cone expansion.

It remains to extend the expansion to the cases that~$V$ does not vanish at~$x$ and possibly at~$y$.
If~$r=0$, all the line integrals are bounded at~$\alpha=1$. Therefore, the condition~$V(y)=0$ can
be removed by approximation.

The argument to remove the condition~$V(x)=0$ is a bit more involved.
We may assume that~$\nu>0$, because in the case~$\nu=0$ all the integrals are regular at~$\alpha=0$.
We consider the individual terms after each other.
For the summands~$S^{\wedge, (n)}_{[\nu/\alpha]}$ we know that~$\alpha > 0$,
so that the lower argument~$\nu/\alpha$ is strictly positive. Therefore, the Heaviside function~$\Theta(\xi^2-\nu)$
in~\eqref{Swanu} and~\eqref{Stanu} cuts out a region near the light cone.
As a consequence, $S^{\wedge, (n)}_{[\nu/\alpha]}$ vanishes for small~$\alpha$.
For the factors~$S^{\bowtie, (n)}_{[\nu/\alpha]}$, on the other hand~$\alpha$ is either bounded away from zero
or is negative. In the latter case, $\nu/\alpha$ is strictly negative. Therefore, the
Heaviside function~$\Theta(\xi^2-\nu)$ in~\eqref{Sbanu2} again
cuts out a region near the light cone, implying that the~$S^{\bowtie, (n)}_{[\nu/\alpha]}$
again vanishes for small~$\alpha<0$. For this reason, the integrands stay regular
if we remove the condition~$V(x)=0$ by approximation.
\QED

In order to illustrate this result, we finally apply it to a convolution integral similar to that in
in Lemma~\ref{lemmaexconv}.
\begin{Example} \label{exlightcone} {\em{ We choose the potential~$V$ in Proposition~\ref{prpKfinal2} as an exponential,
\[ V(z) = e^{\varepsilon z^0} \:. \]
Moreover, we choose~$x=0$ and~$r=1$. We thus obtain
\begin{align}
&2 \pi i \:\big( S^\wedge_{[\nu]}\:V\: K_0 \big)(0,y) \notag \\
&= -2\sum_{n=0}^\infty \frac{1}{n!} \,\times \left\{
\begin{array}{ll} 
\displaystyle \int_{-\infty}^0
(\alpha - \alpha^2)^n \: \varepsilon^{2n}\, e^{\varepsilon \alpha t}\:S^{\bowtie, (n+1)}_{[\nu/\alpha]}(0,y)  \:d\alpha
& \text{if~$t \geq 0$} \\[1em]
\displaystyle \int_1^\infty
(\alpha - \alpha^2)^n \: \varepsilon^{2n}\, e^{\varepsilon \alpha t}\:S^{\bowtie, (n+1)}_{[\nu/\alpha]}(0,y)  \:d\alpha
& \text{if~$t<0$}
\end{array} \right. \label{exc1} \\
&\quad\, -\sum_{n=0}^\infty \frac{1}{n!} \int_0^1
(\alpha - \alpha^2)^n \: \varepsilon^{2n}\, e^{\varepsilon \alpha t} \:
S^{\wedge, (n+1)}_{[\nu/\alpha]}(0,y)\:d\alpha \:, \label{exc2}
\end{align}
where we again set~$t=\xi^0$.
Let us compute the leading contribution for small~$\varepsilon$.
Clearly, the convolution integral vanishes if~$\xi$ lies inside the future light cone.
If~$\xi$ is in the past light cone,  bounded line integral appears, making it possible to choose~$\varepsilon=0$.
We thus obtain the contribution
\begin{align*}
-\int_0^1 S^{\wedge, (1)}_{[\nu/\alpha]}(0,y) \:d\alpha
&= -2 \int_0^1 \Theta\Big(\xi^2- \frac{\nu}{\alpha} \Big) \:H^{(1)}_{[\nu/\alpha]}(0,y) \:d\alpha \\
&= -\frac{1}{8 \pi} \int_0^1 \Theta\Big(\xi^2- \frac{\nu}{\alpha} \Big) \:d\alpha
= -\frac{1}{8 \pi} \Big( 1 - \frac{\nu}{\xi^2} \Big)
\end{align*}
(where we used~\eqref{Swanu}, \eqref{Stanu} and~\eqref{Hanu}).
In order to compare with the formula in Lemma~\ref{lemmaexconv}, we note that, the corresponding
contribution for~$p \in \C^\vee$ has the leading form for small~$\varepsilon$
\[ \frac{\pi}{4\, |\vec{p}|} \Big( -2\, |\vec{p}| + \frac{b}{p^2}\: 2\,|\vec{p}| \Big)
= -\frac{\pi}{2} \:\Big( 1 - \frac{b}{p^2} \Big) \:. \]
We thus get agreement by the obvious replacements~$\xi \rightarrow p$, $\nu \rightarrow b$,
keeping in mind that the prefactors for the Green's operators and~$K$
(see~\eqref{Swanu}, \eqref{Stanu} and~\eqref{Knudef}) give an overall relative factor of~$4 \pi^2$.

If~$\xi$ is spacelike, the computation is more subtle, because the factor~$e^{\varepsilon \alpha t}$
is needed for convergence. Moreover, computing the unbounded line integrals by iterative integration by parts
yields factors of~$\varepsilon$, making the power counting in~$\varepsilon$ non-trivial.
For this reason, for simplicity we only consider the leading contribution on the light cone,
i.e.\ the summand~$n=0$. Then, if~$\xi$ is spacelike and~$t<0$, the leading contribution for
small~$\varepsilon$ is given by
\begin{align*}
-2 &\int_1^\infty e^{\varepsilon \alpha t}\:S^{\bowtie, (1)}_{[\nu/\alpha]}(0,y) \:d\alpha
= 2 \int_1^\infty e^{\varepsilon \alpha t}\:
\Theta \Big(-\xi^2+\frac{\nu}{\alpha} \Big) \:H^{(1)}_{[\nu/\alpha]}(0,y) \:d\alpha \\
&= \frac{1}{8 \pi} \int_1^\infty e^{\varepsilon \alpha t}\:d\alpha = -\frac{1}{8 \pi} \frac{1}{\varepsilon t} \: e^{\varepsilon t}
= \frac{1}{8 \pi} \frac{1}{\varepsilon \,|t|} \,\big( 1 + \O(\varepsilon) \big)
\end{align*}
(note that both~$-\xi^2$ and~$\alpha/\nu$ are positive, making it possible to leave out the
Heaviside function).
Similarly, if~$\xi$ is spacelike and~$t>0$, the leading contribution for small~$\varepsilon$ is given by
\begin{align*}
-2 & \int_{-\infty}^0 e^{\varepsilon \alpha t}\:S^{\bowtie, (n+1)}_{[\nu/\alpha]}(0,y) \:d\alpha 
= 2 \int_{-\infty}^0 e^{\varepsilon \alpha t}\:\Theta \Big(-\xi^2+\frac{\nu}{\alpha} \Big) \:H^{(1)}_{[\nu/\alpha]}(0,y) \:d\alpha \\
&= \frac{1}{8 \pi} \int_{-\infty}^0 e^{\varepsilon \alpha t}\:\Theta \Big(-\xi^2+\frac{\nu}{\alpha} \Big)\:d\alpha
= \frac{1}{8 \pi}\: \frac{1}{\varepsilon t} \: \exp \Big( \varepsilon t\: \frac{\nu}{\xi^2} \Big) 
= \frac{1}{8 \pi} \frac{1}{\varepsilon\, |t|} \,\big( 1 + \O(\varepsilon) \big) \:.
\end{align*}
This is in agreement with the formula of Lemma~\ref{lemmaexconv}, if one keeps in mind that,
for the leading contribution to the light-cone expansion under consideration,
we may set~$|t|=|\vec{\xi}|$.

We remark that, proceeding similarly to higher order on the light cone and to higher orders in~$\varepsilon$,
one could in principle re-derive Lemma~\ref{lemmaexconv} using light-cone expansion techniques.
However, due to the rather different structures of the formulas in Proposition~\ref{prpKfinal2}
and Lemma~\ref{lemmaexconv}, it does not seem possible to recover Lemma~\ref{lemmaexconv}
by applying a simple re-summation technique to~\eqref{exc1} and~\eqref{exc2}.
}} \QEDrem
\end{Example}

\section{Proof of the Mass Cone Expansions} \label{appproof}
In this appendix, we give the proof of Theorems~\ref{thmTVK}, \ref{thmTVK2}, \ref{thmKgK}, \ref{thmTVT},
and~\ref{thmconvolve}.
\Proof[Proof of Theorem~\ref{thmTVK}.] \label{proofTVK} Setting~$\nu=m^2$, by comparing~\eqref{Tm0def}
and~\eqref{Tmpdef} with the notions of Definition~\ref{defnu}, we see that
\beq \label{TSrel}
T^{[n]}_{\nu}(p) = -2 \pi\: S^{\wedge, (n)}_{[\nu]}(0,p) \quad \text{and} \quad
T^{[n]}_{\nu}(-p) = -2 \pi\: S^{\vee, (n)}_{[\nu]}(0,p) \qquad \text{for all~$n \in \N$}\:.
\eeq
Therefore, the convolution integral in~\eqref{TVK1} can be expanded
using Proposition~\ref{prpKfinal2} in the case~$r=0$, setting~$x=0$ and~$y=p$.
Using that~$p$ lies in the past of the origin, we obtain
\begin{align*}
2 &\pi i \int_{\hat{M}} S^{\wedge, (0)}_{[\nu]}(0,k)\:V(k)\: K^{(0)}(p-k) \\
&= -2\sum_{n=0}^\infty \frac{1}{n!} 
\int_1^\infty (\alpha - \alpha^2)^n \: (\Box^n V) \big|_{\alpha p}\:S^{\bowtie, (n+1)}_{[\nu/\alpha]}(0,p)  \:d\alpha \\
&\quad\, -\sum_{n=0}^\infty \frac{1}{n!} \int_0^1
(\alpha - \alpha^2)^n \: (\Box^n V) \big|_{\alpha p} \:
S^{\wedge, (n+1)}_{[\nu/\alpha]}(0,p)\:d\alpha \:.
\end{align*}
Using~\eqref{Sbanu} and~\eqref{Swanu}, we can reorganize the terms to get
\begin{align*}
2 &\pi i \int_{\hat{M}} S^{\wedge, (0)}_{[\nu]}(0,k)\:V(k)\: K^{(0)}(p-k) \\
&= -2\sum_{n=0}^\infty \frac{1}{n!} 
\int_1^\infty (\alpha - \alpha^2)^n \: (\Box^n V) \big|_{\alpha p}\:H^{(n+1)}_{[\nu/\alpha]}(0,p)  \:d\alpha \\
&\quad\, -\sum_{n=0}^\infty \frac{1}{n!} \int_0^\infty
(\alpha - \alpha^2)^n \: (\Box^n V) \big|_{\alpha p} \:
S^{\wedge, (n+1)}_{[\nu/\alpha]}(0,p)\:d\alpha \:.
\end{align*}
Again using~\eqref{TSrel}, the last sum gives precisely~\eqref{TVK1}.
In order to estimate the first sum, we note that~$V$ decays on the scale~$\varepsilon^{-1}$,
whereas~$\underline{\omega} \ll -\varepsilon^{-1}$.
Therefore the first sum is an error term of the order~$\O(1/(\varepsilon \underline{\omega}))$.

Let us verify that the mass cone expansion converges.
Again using that~$V$ decays on the scale~$\varepsilon^{-1}$, whereas~$\underline{\omega} \ll -\varepsilon^{-1}$,
it suffices to consider the integrand for
\[ \alpha = \frac{k^0}{\underline{\omega}} \lesssim \frac{1}{\varepsilon \,|\underline{\omega}|} \ll 1 \:. \]
Then, using~\eqref{Vbound}, we obtain
\begin{align}
\big| (\alpha - \alpha^2)^n \: (\Box^n V) \big|_{\alpha p} \big|
&\lesssim \big| \alpha^n \: (\Box^n V) \big|_{\alpha p} \big| 
\leq \bigg( c\:\Big( \alpha \varepsilon^2 +  \frac{\ell_{\min}}{|\underline{\omega}|} \Big) \bigg)^n \:e^{\varepsilon k^0} \\
&\leq \bigg( c'\:\Big( \frac{\varepsilon}{|\underline{\omega}|} +  \frac{\ell_{\min}}{|\underline{\omega}|} \Big) \bigg)^n \:e^{\varepsilon k^0} \lesssim \bigg( c'\:\frac{\ell_{\min}}{|\underline{\omega}|}\bigg)^n \label{scale1}
\end{align}
with a new constant~$c'>0$.
Moreover, according to~\eqref{Tmpdef}, 
\beq \label{scale2}
\big| S^{\wedge, (n+1)}_{m^2/\alpha}(0,p) \big| \lesssim \big| p^{2n} \big| \simeq \bigg( \frac{|\underline{\omega}|}{\ell_{\min}} \bigg)^n \:.
\eeq
Taking the product of~\eqref{scale1} and~\eqref{scale2}, only the factor~$(c')^n$ remains.
In view of the factorials in~\eqref{Tmpdef}, the mass cone expansion indeed converges.

In order to prove the expansions~\eqref{TVK3} and~\eqref{TVK4}, we proceed similarly and apply Proposition~\ref{prpKfinal2} for~$x=0$, but now setting~$y=-p$. We thus obtain
\begin{align*}
2 &\pi i \int_{\hat{M}} S^{\wedge, (0)}_{[\nu]}(0,-k)\:V(k)\: K_0(p-k) \\
&= 2 \sum_{n=0}^\infty \frac{1}{n!} 
\int_{-\infty}^0 (\alpha - \alpha^2)^n \: (\Box^n V) \big|_{\alpha p}\:S^{\bowtie, (n+1)}_{[\nu/\alpha]}(0,p)  \:d\alpha \\
&= \sum_{n=0}^\infty \frac{1}{n!} 
\int_{-\infty}^0 (\alpha - \alpha^2)^n \: (\Box^n V) \big|_{\alpha p}\:S^{\wedge, (n+1)}_{[\nu/\alpha]}(0,p)  \:d\alpha \\
&\quad\:-2 \sum_{n=0}^\infty \frac{1}{n!} 
\int_{-\infty}^0 (\alpha - \alpha^2)^n \: (\Box^n V) \big|_{\alpha p}\:H^{(n+1)}_{[\nu/\alpha]}(0,p)  \:d\alpha
\end{align*}
(in the last step we applied~\eqref{Sbanu} and~\eqref{Swanu}).
Again using~\eqref{TSrel}, the first sum gives~\eqref{TVK3}. Therefore, it remains to show
that the last sum is an error of the form~\eqref{TVK4}.
To this end, we return to the scaling behavior~\eqref{scale1} of the coefficient of the mass
cone expansion. Using that~$H$ is a polynomial, we obtain a power series with the scaling
\[ \lesssim \sum_{n=0}^\infty \Big| \alpha^n \: (\Box^n V) \big|_{\alpha p} \Big| 
\: \Big( p^2 - \frac{m^2}{\alpha} \Big)^n \:. \]
Differentiating in~$p$ and performing an index shift~$n+1 \rightarrow n$, 
we can apply the last inequality in~\eqref{scale1} to obtain a scaling factor
\[ \bigg( \frac{\ell_{\min}}{|\underline{\omega}|} \bigg)\: |\underline{\omega}| = \ell_{\min} \:. \]
Using the integration-by-parts argument~\eqref{integration-by-parts},
this scaling behavior of the derivatives in momentum space
proves the desired decay in position space on the scale~$\ell_{\min}$. This concludes the proof.
\QED

\Proof[Proof of Theorem~\ref{thmTVK2}.] \label{proofTVK2}
Follows exactly as Theorem~\ref{thmTVK} if we apply Proposition~\ref{prpKfinal2}
in the case~$r=0$ and differentiate both sides with respect to~$\nu$
using~\eqref{pneg} and~\eqref{Knudiff}.
\QED

\Proof[Proof of Theorem~\ref{thmKgK}.] \label{proofKgK}
We can apply the light-cone expansion of Corollary~\ref{corKK} in momentum space, setting~$x=0$ and~$y=p$,
\begin{align}
&\int \frac{d^4k}{(2 \pi)^4} \:K^{(-s)}(k) \: g(k) \: K_0\big( p -k \big) \notag \\
&= -\frac{1}{32 \pi^6} \sum_{n=0}^\infty \frac{1}{n!} 
\int_{-\infty}^\infty \alpha^{-s} \:
(\alpha - \alpha^2)^n \: (\Box^n g) \big|_{\alpha p} \:d\alpha
\:S^{\bowtie, (n-s+1)}(0,p) \label{KgK1pr} \\
&\quad\, -\frac{1}{32 \pi^6}  \sum_{n=0}^\infty \frac{1}{n!} 
\int_0^1 \alpha^{-2}\:
(\alpha - \alpha^2)^n \: (\Box^n g) \big|_{\alpha p} \:d\alpha
\:H^{(n-s+1)}(0,p) \:. \label{KgK2pr}
\end{align}
It remains to show that the summands in~\eqref{KgK2pr} can be absorbed into the
error term~\eqref{KgK4} and that replacing~$S^{\bowtie, (n-s+1)}$ by~$S^{\times, (n-s+1)}$
gives correction terms which are also of the form~\eqref{KgK4}.

In preparation, we determine the scaling behavior in~$n$. We first note that
the factors~$H^{(n-s+1)}$ and~$S^{\bowtie, (n-s+1)}-S^{\times, (n-s+1)}$
are polynomials in~$p^2$ of order~$n-s$ (see~\eqref{eq:bb} and~\eqref{eq:by}).
According to~\eqref{gbound}, each Laplacian gives a scaling factor~$\ell_{\min}/\omega_0$.
Next, the integrands in~\eqref{KgK1pr} and~\eqref{KgK2pr}
vanish unless~$\alpha$ lies in the range
\[ \alpha \simeq \frac{\omega_0}{\omega} \lesssim 1 \]
(in the last step we used the first error term in~\eqref{KgK4}).
Therefore, the line integrals scale in~$n$ like
\[ \bigg( c\: \frac{\ell_{\min}}{\omega} \bigg)^n \:. \]

In order to determine the decay of the error terms in position space, we need to
analyze the scaling behavior of their derivatives in momentum space. In fact,
we need to show that each derivative in momentum space gives a scaling factor~$\ell_{\min}$.
Differentiating in~$p$ gives two contributions: If the polynomial in~$p^2$ is differentiated, this
corresponds to incrementing~$n$ and multiplying by~$\omega$, giving the desired scaling factor
\[ \omega\: c\: \frac{\ell_{\min}}{\omega} =  c\: \ell_{\min} \:. \]
If, on the other hand, the potential~$\Box^n g$ is differentiated, we get a scaling factor
\[ \alpha \:\sqrt{c\: \frac{\ell_{\min}}{\omega}} \lesssim \sqrt{c\: \frac{\ell_{\min}\, \omega_0^2}{\omega^3}} 
\lesssim \ell_{\min}\:. \]
This concludes the proof of Theorem~\ref{thmKgK}.
\QED

\Proof[Proof of Theorem~\ref{thmTVT}.] \label{proofTVT}
Since we evaluate for~$p$ only with~$p^0 \ll -\varepsilon^{-1}$, keeping in mind the error term~$\O(1/(\varepsilon
\omega))$, we may replace the factor~$T^{[r]}_0$ by~$4 \pi^2 i K^{(r)}_0$ (see~\eqref{Kndef}).
Now we can proceed exactly as in the proof of Theorem~\ref{thmTVK} by applying Proposition~\ref{prpKfinal2}
for general~$r \geq -2$.
\QED

\Proof[Proof of Theorem~\ref{thmconvolve}.] \label{proofconvolve}
We write the convolution integral~\eqref{mcm1} in the form~\eqref{mcform}
with~$V$ according to~\eqref{mcV}.
Using the identification~\eqref{TSrel}, we can apply
Proposition~\ref{prpKfinal2} in the case~$r=-2$, setting~$x=0$ and~$y=p$. We thus obtain
\begin{align}
2 &\pi i \int_{\hat{M}} S^{\wedge, (0)}_{[\nu]}(0,k)\:V(k)\: K^{[-2]}_0(p-k) \label{convprelim} \\
&= -2\sum_{n=0}^\infty \frac{1}{n!} \notag \\
&\quad\qquad \times \left\{
\begin{array}{ll} 
\displaystyle \int_{-\infty}^0 \frac{1}{(1-\alpha)^2}\:
(\alpha - \alpha^2)^n \: (\Box^n V) \big|_{\alpha y + (1-\alpha) x}\:S^{\bowtie, (n-1)}_{[\nu/\alpha]}(0,p)  \:d\alpha
& \text{if~$p^0>0$} \\[1em]
\displaystyle \int_1^\infty \frac{1}{(1-\alpha)^2}\:
(\alpha - \alpha^2)^n \: (\Box^n V) \big|_{\alpha y + (1-\alpha) x}\:S^{\bowtie, (n-1)}_{[\nu/\alpha]}(0,p)  \:d\alpha
& \text{if~$p^0<0$}
\end{array} \right. \notag \\
&\quad\, -\sum_{n=0}^\infty \frac{1}{n!} \int_0^1
\frac{1}{(1-\alpha)^2}\: (\alpha - \alpha^2)^n \: (\Box^n V) \big|_{\alpha p} \:
S^{\wedge, (n-1)}_{[\nu/\alpha]}(0,p)\:d\alpha \:. \notag
\end{align}

We first consider the leading term for $n=0$ of the
resulting expansion (the terms for~$n>0$ will be estimated below). We denote it by~$B(p)$.
It is given by
\begin{align*}
B(p) &= \left\{
\begin{array}{ll} 
\displaystyle -2 \int_{-\infty}^0 \frac{1}{(1-\alpha)^2}\:\big( \alpha \slashed{p}+m \big) \: e^{\varepsilon \alpha \omega}\:
g\big((1-\alpha)\, p \big) \:S^{\bowtie, (-1)}_{[m^2/\alpha]}(0,p) \:d\alpha
& \text{if~$\omega\geq 0$} \\[1em]
\displaystyle -2 \int_1^\infty \frac{1}{(1-\alpha)^2}\:\big( \alpha \slashed{p}+m \big) \: e^{\varepsilon \alpha \omega}\:
g\big((1-\alpha)\, p \big) \:S^{\bowtie, (-1)}_{[m^2/\alpha]}(0,p)  \:d\alpha
& \text{if~$\omega<0$}
\end{array} \right. \\
&\quad\, -\int_0^1 (1-\alpha)^{-1}\:\big( \alpha \slashed{p}+m \big) \: e^{\varepsilon \alpha \omega}\:
g\big((1-\alpha)\, p \big) \:S^{\wedge, (-1)}_{[m^2/\alpha]}(0,p)\:d\alpha \:.
\end{align*}
If the parameter~$a$ vanishes, the Green's operator~$S^{\bowtie}_{a, [\nu]}$ coincides with
the symmetric Green's operator~$S^\times_{a, [\nu]}$
(see~\eqref{Sbanu} and~\eqref{Stanu}). Therefore, we may replace
the factor~$S^{\bowtie, (-1)}_\bullet$ by~$S^{\times, (-1)}_\bullet$.
Decomposing the latter Green's operator into
the advanced and retarded parts, we thus obtain
\begin{align*}
B(p) &= - \int_{-\infty}^0 \frac{1}{(1-\alpha)^2}\:\big( \alpha \slashed{p}+m \big) \: e^{\varepsilon \alpha \omega}\:
g\big((1-\alpha)\, p \big) \:S^{\vee,(-1)}_{[m^2/\alpha]}(0,p) \:d\alpha \\
&\quad\, -\int_0^\infty \frac{1}{(1-\alpha)^2}\:\big( \alpha \slashed{p}+m \big) \: e^{\varepsilon \alpha \omega}\:
g\big((1-\alpha)\, p \big) \:S^{\wedge,(-1)}_{[m^2/\alpha]}(0,p)  \:d\alpha \:.
\end{align*}
In view of the error term~$O (\omega_0^2/\omega^2)$, it suffices to consider the term~$|\omega|
\geq 2 \omega_0$. Then by~\eqref{etasupp}, the function~$g$ vanishes unless~$\alpha$
is close to one. Therefore, the first integral vanishes, and in the second integral we may
integrate over the whole real line. Using~\eqref{gneudef}, we obtain
\begin{align}
B(p) &= -\int_{-\infty}^\infty \frac{1}{(1-\alpha)^2}\:\big( \alpha \slashed{p}+m \big) \: e^{\varepsilon \alpha \omega}\:
g\big((1-\alpha)\, p \big) \:S^{\wedge, (-1)}_{[m^2/\alpha]}(0,p) \:d\alpha
\label{Afinal} \\
&= -\frac{2}{\pi}\: \omega^2\, \Phi(\hat{\vec{p}}\,) \int_{-\infty}^\infty \big( \alpha \slashed{p}+m \big) \: e^{\varepsilon \alpha \omega}\: \hat{\eta}_0\big((1-\alpha)\, \omega \big) \:S^{\wedge, (-1)}_{[m^2/\alpha]}(0,p)  \:d\alpha \:.
\end{align}
For the further analysis, it is preferable to introduce the integration variable
\[ 
\kappa := (1-\alpha)\, \omega \:,\quad \alpha = 1 - \frac{\kappa}{\omega} \]
(note that~$\kappa$ coincides with the variable~$k^0$ in~\eqref{convprelim}).
We thus obtain
\[ 
B(p) = \frac{2}{\pi}\: \omega\, \Phi(\hat{\vec{p}}\,) \int_{-\infty}^\infty 
\hat{\eta}_0(\kappa) \Big\{
\big( \alpha \slashed{p}+m \big) \: e^{\varepsilon \alpha \omega}\: \:S^{\wedge, (-1)}_{[m^2/\alpha]}(0,p)
\Big\} \Big|_{\alpha =  1 - \kappa/\omega } \:d\kappa \:. \]
We now expand the curly brackets in powers of~$\kappa/\omega$. The zero order term gives
\[ B(p) = \frac{2}{\pi}\: \omega\, \Phi(\hat{\vec{p}}\,)  \:
\big( \slashed{p}+m \big) \: e^{\varepsilon \omega}\: S^{\wedge, (-1)}_{[m^2]}(0,p) + \O \Big(
\frac{\kappa}{\omega} \Big) \:. \]
We thus obtain the contribution on the right of~\eqref{mc0}.
In order to organize the higher orders, we first apply the mean value theorem to conclude that there
is~$\kappa_0$ with~$|\kappa_0| < \omega_0$ such that
\begin{align*}
\int_{-\infty}^\infty &
\hat{\eta}_0(\kappa) \Big\{
\big( \alpha \slashed{p}+m \big) \: e^{\varepsilon \alpha \omega}\:S^{\wedge, (-1)}_{[m^2/\alpha]}(0,p)
\Big\} \Big|_{\alpha =  1 - \kappa/\omega } \:d\kappa \\
&= S^{\wedge, (-1)}_{[m^2/\alpha_0]}(0,p)
\int_{-\infty}^\infty 
\hat{\eta}_0(\kappa) \Big\{
\big( \alpha \slashed{p}+m \big) \: e^{\varepsilon \alpha \omega}\:
\Big\} \Big|_{\alpha =  1 - \kappa/\omega } \:d\kappa \:.
\end{align*}
Expanding the integral in powers of~$\kappa$ gives the error terms in~\eqref{mc0}.

It remains to consider the summands~$n>0$ of the light cone expansion in momentum space
and to relate them to the error terms in~\eqref{err1} and~\eqref{err2}.
We begin with the case~$n=1$.
In this case, the Laplacian of the function in the integrand in~\eqref{convprelim} comes up.
Moreover, increasing~$n$ by one gives an additional
factor~$\alpha-\alpha^2$ in the integrand of the line integral. Finally, the upper index
of the Green's operator is increased. Thus, compared to the line integral in~\eqref{Afinal}
we now get the integral
\beq \label{n1int}
\int_{-\infty}^\infty \frac{\alpha}{1-\alpha}\;
\Box_k \Big( \big( \slashed{p} - \slashed{k}+m \big) \: e^{\varepsilon (p^0-k^0)}\:g(k) \Big) \Big|_{k=(1-\alpha) p}
\:G^{(0)}_{[m^2/\alpha]}(0,p)\: d\alpha
\eeq
with~$G=S^\wedge$ or~$G=H$ and $I=(-\infty, 1]$ or~$[1, \infty)$.
Using~\eqref{gneudef}, it is most convenient to write the Laplacian as
\begin{align*}
&\Box_k \Big( \big( \slashed{p} - \slashed{k}+m \big) \: e^{\varepsilon (p^0-k^0)}\:g(k) \Big) \\
&=\Box_k \Big( \big( \slashed{p} - \slashed{k}+m \big) \: e^{\varepsilon (p^0-k^0)} \:
(k^0)^2 \: \hat{\eta}_0(k^0)\: \Phi\big( -\epsilon(k^0)\: \hat{\vec{k}}\,\big) \Big) \\
&=\frac{\partial^2}{\partial (k^0)^2} \Big( \big( \slashed{p} - \slashed{k}+m \big) \: e^{\varepsilon (p^0-k^0)} \:
(k^0)^2\: \hat{\eta}_0(k^0)\: \Phi\big( -\epsilon(k^0)\: \hat{\vec{k}}\,\big) \Big) \\ 
&\quad\: -\big( \slashed{p} - \slashed{k}+m \big) \: e^{\varepsilon (p^0-k^0)} \:
(k^0)^2 \,\hat{\eta}_0(k^0)\: \frac{1}{(k^0)^2}\: \Delta_{S^2} \Phi\big( -\epsilon(k^0)\:
\hat{\vec{k}}\,\big) \\ 
&\quad\: +2\,e^{\varepsilon (p^0-k^0)} \:
(k^0)^2\:\hat{\eta}_0(k^0)\: \vec{\gamma} \vec{\nabla} \Phi\big( -\epsilon(k^0)\: \hat{\vec{k}}\,\big) \:, 
\end{align*}
where~$\Delta_{S^2}$ is the spherical Laplacian in polar coordinates.
These contributions are bounded by
\[ \lesssim |\omega| \big(1+l^2) \: e^{\varepsilon \omega} \:. \]
Moreover, the factor~$\alpha/(1-\alpha)$ in the integrand in~\eqref{n1int} can be estimated by
\[ \frac{\alpha}{1-\alpha} \lesssim \frac{|\omega|}{\omega_0}\:. \]
Finally, integrating over~$\alpha$ gives a factor~$1/|\omega|$. We thus obtain the error term~\eqref{err1}.

In the remaining case~$n>1$, one must carefully distinguish between the
Green's operators~$S^\wedge_{[\nu]}$ and~$S^{\bowtie}_{a, [\nu]}$.
This explains why, instead of an integral over the real line~\eqref{Afinal}, one gets separate integrals
over the intervals~$(-\infty, 1]$ and~$[1, \infty)$. As a consequence, we can no longer use
symmetry properties of the function~$\hat{\eta}_0$ for cancellations of terms.
Instead, one needs to determine the scaling behavior of the integrand.
Systematically, the leading contribution for general~$n$ involves a scaling factor
\[ \sim \frac{(1 - \Delta^n_{S^2}) N(k)}{\omega_0^{2n}} \: \Big( \frac{\omega_0}{\omega} \Big)^n \]
(here we use that each factor~$(\alpha-\alpha^2)^n$ gives a scaling factor~$\omega_0/\omega$).
Moreover, the factors~$S^{\wedge, (n)}_{[\nu]}$ and~$S^{\bowtie, (n)}_{a, [\nu]}$
can be estimated by the polynomial~$|p^2|^{n-1}$.
This explains the error term~\eqref{err2}, concluding the proof of Theorem~\ref{thmconvolve}.
\QED

\Thanks{\em{Acknowledgments:}}
I am grateful to Claudio Dappiaggi, Niky Kamran and Moritz Reintjes for helpful discussions.


\begin{thebibliography}{10}

\bibitem{cfsweblink}
\emph{Link to web platform on causal fermion systems:
  \href{https://www.causal-fermion-system.com}{\textrm{www.causal-fermion-system.com}}}.

\bibitem{lagrange}
Y.~Bernard and F.~Finster, \emph{On the structure of minimizers of causal
  variational principles in the non-compact and equivariant settings},
  arXiv:1205.0403 [math-ph], Adv. Calc. Var. \textbf{7} (2014), no.~1, 27--57.

\bibitem{bjorken}
J.D. Bjorken and S.D. Drell, \emph{Relativistic {Q}uantum {M}echanics},
  McGraw-Hill Book Co., New York, 1964.

\bibitem{linhyp}
C.~Dappiaggi and F.~Finster, \emph{Linearized fields for causal variational
  principles: {E}xistence theory and causal structure}, arXiv:1811.10587
  [math-ph], Methods Appl. Anal. \textbf{27} (2020), no.~1, 1--56.

\bibitem{fockdynamics}
C.~Dappiaggi, F.~Finster, N.~Kamran, and M.~Reintjes, \emph{The {F}ock dynamics
  of causal fermion systems}, in preparation.

\bibitem{sea}
F.~Finster, \emph{Definition of the {D}irac sea in the presence of external
  fields}, arXiv:hep-th/9705006, Adv. Theor. Math. Phys. \textbf{2} (1998),
  no.~5, 963--985.

\bibitem{u22}
\bysame, \emph{Local {$\rm U(2,2)$} symmetry in relativistic quantum
  mechanics}, arXiv:hep-th/9703083, J. Math. Phys. \textbf{39} (1998), no.~12,
  6276--6290.

\bibitem{firstorder}
\bysame, \emph{Light-cone expansion of the {D}irac sea to first order in the
  external potential}, arXiv:hep-th/9707128, Michigan Math. J. \textbf{46}
  (1999), no.~2, 377--408.

\bibitem{light}
\bysame, \emph{Light-cone expansion of the {D}irac sea in the presence of
  chiral and scalar potentials}, arXiv:hep-th/9809019, J. Math. Phys.
  \textbf{41} (2000), no.~10, 6689--6746.

\bibitem{pfp}
\bysame, \emph{The {P}rinciple of the {F}ermionic {P}rojector}, hep-th/0001048,
  hep-th/0202059, hep-th/0210121, AMS/IP Studies in Advanced Mathematics,
  vol.~35, American Mathematical Society, Providence, RI, 2006.

\bibitem{reg}
\bysame, \emph{On the regularized fermionic projector of the vacuum},
  arXiv:math-ph/0612003, J. Math. Phys. \textbf{49} (2008), no.~3, 032304, 60.

\bibitem{continuum}
\bysame, \emph{Causal variational principles on measure spaces},
  arXiv:0811.2666 [math-ph], J. Reine Angew. Math. \textbf{646} (2010),
  141--194.

\bibitem{cfs}
\bysame, \emph{The {C}ontinuum {L}imit of {C}ausal {F}ermion {S}ystems},
  arXiv:1605.04742 [math-ph], Fundamental Theories of Physics, vol. 186,
  Springer, 2016.

\bibitem{positive}
\bysame, \emph{Positive functionals induced by minimizers of causal variational
  principles}, arXiv:1708.07817 [math-ph], Vietnam J. Math. \textbf{47} (2019),
  23--37.

\bibitem{perturb}
\bysame, \emph{Perturbation theory for critical points of causal variational
  principles}, arXiv:1703.05059 [math-ph], Adv. Theor. Math. Phys. \textbf{24}
  (2020), no.~3, 563--619.

\bibitem{elhom}
F.~Finster, M.~Frankl, and C.~Langer, \emph{The homogeneous causal action
  principle on a compact domain in momentum space}, arXiv:2205.04085 [math-ph],
  Adv. Calc. Var. (2023).

\bibitem{lqg}
F.~Finster and A.~Grotz, \emph{A {L}orentzian quantum geometry},
  arXiv:1107.2026 [math-ph], Adv. Theor. Math. Phys. \textbf{16} (2012), no.~4,
  1197--1290.

\bibitem{rrev}
F.~Finster, A.~Grotz, and D.~Schiefeneder, \emph{Causal fermion systems: {A}
  quantum space-time emerging from an action principle}, arXiv:1102.2585
  [math-ph], {Q}uantum {F}ield {T}heory and {G}ravity (F.~Finster, O.~M\"uller,
  M.~Nardmann, J.~Tolksdorf, and E.~Zeidler, eds.), Birkh\"auser Verlag, Basel,
  2012, pp.~157--182.

\bibitem{vacstab}
F.~Finster and S.~Hoch, \emph{An action principle for the masses of {D}irac
  particles}, arXiv:0712.0678 [math-ph], Adv. Theor. Math. Phys. \textbf{13}
  (2009), no.~6, 1653--1711.

\bibitem{review}
F.~Finster and M.~Jokel, \emph{Causal fermion systems: {A}n elementary
  introduction to physical ideas and mathematical concepts}, arXiv:1908.08451
  [math-ph], {P}rogress and {V}isions in {Q}uantum {T}heory in {V}iew of
  {G}ravity (F.~Finster, D.~Giulini, J.~Kleiner, and J.~Tolksdorf, eds.),
  Birkh\"auser Verlag, Basel, 2020, pp.~63--92.

\bibitem{fockbosonic}
F.~Finster and N.~Kamran, \emph{Complex structures on jet spaces and bosonic
  {F}ock space dynamics for causal variational principles}, arXiv:1808.03177
  [math-ph], Pure Appl. Math. Q. \textbf{17} (2021), no.~1, 55--140.

\bibitem{fockfermionic}
\bysame, \emph{Fermionic {F}ock spaces and quantum states for causal fermion
  systems}, arXiv:2101.10793 [math-ph], Ann. Henri Poincar\'{e} \textbf{23}
  (2022), no.~4, 1359--1398.

\bibitem{dirac}
F.~Finster, N.~Kamran, and M.~Oppio, \emph{The linear dynamics of wave
  functions in causal fermion systems}, arXiv:2101.08673 [math-ph], J.
  Differential Equations \textbf{293} (2021), 115--187.

\bibitem{gaugefix}
F.~Finster and S.~Kindermann, \emph{A gauge fixing procedure for causal fermion
  systems}, arXiv:1908.08445 [math-ph], J. Math. Phys. \textbf{61} (2020),
  no.~8, 082301.

\bibitem{intro}
F.~Finster, S.~Kindermann, and J.-H. Treude, \emph{An {I}ntroductory {C}ourse
  on {C}ausal {F}ermion {S}ystems}, in preparation,
  \href{https://causal-fermion-system.com/intro-public.pdf}{www.causal-fermion-system.com/intro-public.pdf}
  (2023).

\bibitem{dice2014}
F.~Finster and J.~Kleiner, \emph{Causal fermion systems as a candidate for a
  unified physical theory}, arXiv:1502.03587 [math-ph], J. Phys.: Conf. Ser.
  \textbf{626} (2015), 012020.

\bibitem{noether}
\bysame, \emph{Noether-like theorems for causal variational principles},
  arXiv:1506.09076 [math-ph], Calc. Var. Partial Differential Equations
  \textbf{55:35} (2016), no.~2, 41.

\bibitem{jet}
\bysame, \emph{A {H}amiltonian formulation of causal variational principles},
  arXiv:1612.07192 [math-ph], Calc. Var. Partial Differential Equations
  \textbf{56:73} (2017), no.~3, 33.

\bibitem{osi}
\bysame, \emph{A class of conserved surface layer integrals for causal
  variational principles}, arXiv:1801.08715 [math-ph], Calc. Var. Partial
  Differential Equations \textbf{58:38} (2019), no.~1, 34.

\bibitem{localize}
F.~Finster and M.~Kraus, \emph{Construction of global solutions to the
  linearized field equations for causal variational principles},
  arXiv:2210.16665 [math-ph], to appear in Methods Appl. Anal. (2023).

\bibitem{banach}
F.~Finster and M.~Lottner, \emph{Banach manifold structure and
  infinite-dimensional analysis for causal fermion systems}, arXiv:2101.11908
  [math-ph], Ann. Global Anal. Geom. \textbf{60} (2021), no.~2, 313--354.

\bibitem{norm}
F.~Finster and J.~Tolksdorf, \emph{Perturbative description of the fermionic
  projector: {N}ormalization, causality and {F}urry's theorem}, arXiv:1401.4353
  [math-ph], J. Math. Phys. \textbf{55} (2014), no.~5, 052301.

\bibitem{collapse}
F.~et~al Finster, \emph{Causal fermion systems as an effective collapse
  theory}, in preparation.

\bibitem{qftcorrect}
F.~Finster~et al, \emph{Corrections to quantum field theory in the theory of
  causal fermion systems}, in preparation.

\bibitem{friedlander2}
F.G. Friedlander, \emph{Introduction to the {T}heory of {D}istributions},
  second ed., Cambridge University Press, Cambridge, 1998, With additional
  material by M. Joshi.

\bibitem{landau2}
L.D. Landau and E.M. Lifshitz, \emph{The {C}lassical {T}heory of {F}ields},
  Revised second edition. Course of Theoretical Physics, Vol. 2. Translated
  from the Russian by M. Hamermesh, Pergamon Press, Oxford, 1962.

\bibitem{lax}
P.D. Lax, \emph{Functional {A}nalysis}, Pure and Applied Mathematics (New
  York), Wiley-Interscience [John Wiley \& Sons], New York, 2002.

\bibitem{lagrange-hoelder}
M.~Oppio, \emph{H\"older continuity of the integrated causal {L}agrangian in
  {M}inkowski space}, arXiv:2109.04728 [math-ph], to appear in Adv. Theor.
  Math. Phys. (2023).

\bibitem{peskin+schroeder}
M.E. Peskin and D.V. Schroeder, \emph{An {I}ntroduction to {Q}uantum {F}ield
  {T}heory}, Addison-Wesley Publishing Company Advanced Book Program, Reading,
  MA, 1995.

\bibitem{reed+simon}
M.~Reed and B.~Simon, \emph{Methods of {M}odern {M}athematical {P}hysics. {I},
  {F}unctional analysis}, second ed., Academic Press Inc., New York, 1980.

\bibitem{sakurai}
J.J. Sakurai and J.~Napolitano, \emph{Advanced {Q}uantum {M}echanics}, second
  ed., Addison-Wesley Publishing Company, 1994.

\bibitem{treude}
J.-H. Treude, \emph{{Decay in Outgoing Null Directions of Solutions of the
  Massive Dirac Equation in Certain Asymptotically Flat, Static Spacetimes}},
  Dissertation, Universit\"at Regensburg,
  \href{http://epub.uni-regensburg.de/32344/}{https://doi.org/10.5283/epub.32344/}
  (2015).

\bibitem{weinberg}
S.~Weinberg, \emph{The {Q}uantum {T}heory of {F}ields. {V}ol. {I}}, Cambridge
  University Press, Cambridge, 1996, Foundations, Corrected reprint of the 1995
  original.

\end{thebibliography}
\providecommand{\bysame}{\leavevmode\hbox to3em{\hrulefill}\thinspace}
\providecommand{\MR}{\relax\ifhmode\unskip\space\fi MR }
\providecommand{\MRhref}[2]{%
  \href{http://www.ams.org/mathscinet-getitem?mr=#1}{#2}
}
\providecommand{\href}[2]{#2}

\end{document}